\documentclass[11pt]{article} 
\usepackage[letterpaper,margin=0.8in]{geometry}

\usepackage[T1]{fontenc}
\usepackage{xcolor}
\usepackage{microtype}
\usepackage{xspace}
\usepackage{comment}

\def\colorful{1}
\ifnum\colorful=1

\else

\fi

\newif\ifred
\redtrue

\usepackage{titlesec}
\titleformat*{\paragraph}{\bfseries}

\usepackage{thmtools}
\usepackage{amsthm}
\usepackage{thm-restate}
\usepackage{amsmath,amssymb,amsfonts,mathtools}
\usepackage{empheq}
\usepackage{dsfont}
\usepackage{mathrsfs}
\usepackage{bbm}

\usepackage{graphicx}
\usepackage{enumitem}
\usepackage{caption}
\usepackage{framed}

\usepackage{tikz}
\usetikzlibrary{patterns}
\usetikzlibrary{arrows,shapes,automata,backgrounds,petri,positioning}
\usetikzlibrary{shadows}
\usetikzlibrary{calc}
\usetikzlibrary{spy}
\usetikzlibrary{angles,quotes}
\usetikzlibrary{matrix}

\usepackage{pgf,pgfplots}
\usepackage{pgfmath,pgffor}
\pgfplotsset{compat=1.17}

\usepackage{algorithm}
\usepackage[noend]{algpseudocode}

\definecolor[named]{ACMBlue}{cmyk}{1,0.1,0,0.1}
\definecolor[named]{ACMYellow}{cmyk}{0,0.16,1,0}
\definecolor[named]{ACMOrange}{cmyk}{0,0.42,1,0.01}
\definecolor[named]{ACMRed}{cmyk}{0,0.90,0.86,0}
\definecolor[named]{ACMLightBlue}{cmyk}{0.49,0.01,0,0}
\definecolor[named]{ACMGreen}{cmyk}{0.20,0,1,0.19}
\definecolor[named]{ACMPurple}{cmyk}{0.55,1,0,0.15}
\definecolor[named]{ACMDarkBlue}{cmyk}{1,0.58,0,0.21}

\usepackage{aliascnt}

\newtheorem{theorem}{Theorem}[section]

\newaliascnt{question}{theorem}

\aliascntresetthe{question}

\newaliascnt{cond}{theorem}

\aliascntresetthe{cond}

\newaliascnt{condition}{theorem}

\aliascntresetthe{condition}

\newaliascnt{lemma}{theorem}
\newtheorem{lemma}[lemma]{Lemma}
\aliascntresetthe{lemma}

\newaliascnt{corollary}{theorem}
\newtheorem{corollary}[corollary]{Corollary}
\aliascntresetthe{corollary}

\newaliascnt{corrollary}{theorem}

\aliascntresetthe{corrollary}

\newaliascnt{informal theorem}{theorem}
\newtheorem{informal theorem}[informal theorem]{Theorem (informal statement)}
\aliascntresetthe{informal theorem}

\newaliascnt{proposition}{theorem}
\newtheorem{proposition}[proposition]{Proposition}
\aliascntresetthe{proposition}

\newaliascnt{claim}{theorem}
\newtheorem{claim}[claim]{Claim}
\aliascntresetthe{claim}

\newaliascnt{fact}{theorem}
\newtheorem{fact}[fact]{Fact}
\aliascntresetthe{fact}

\newaliascnt{obs}{theorem}

\aliascntresetthe{obs}

\newaliascnt{observation}{theorem}

\aliascntresetthe{observation}

\newaliascnt{remark}{theorem}
\newtheorem{remark}[remark]{Remark}
\aliascntresetthe{remark}

\newaliascnt{infproposition}{theorem}

\aliascntresetthe{infproposition}

\newaliascnt{assumption}{theorem}

\aliascntresetthe{assumption}

\newaliascnt{definition}{theorem}
\newtheorem{definition}[definition]{Definition}
\aliascntresetthe{definition}

\usepackage[colorlinks,citecolor=blue,linkcolor=magenta,bookmarks=true]{hyperref}
\usepackage[nameinlink,capitalise]{cleveref}

\crefname{theorem}{Theorem}{Theorems}
\crefname{question}{Question}{Questions}
\crefname{conj}{Conjecture}{Conjectures}
\crefname{cond}{Condition}{Conditions}
\crefname{condition}{Condition}{Conditions}
\crefname{lemma}{Lemma}{Lemmas}
\crefname{corollary}{Corollary}{Corollaries}
\crefname{corrollary}{Corollary}{Corollaries}
\crefname{informal theorem}{Theorem}{Theorems}
\crefname{proposition}{Proposition}{Propositions}
\crefname{claim}{Claim}{Claims}
\crefname{fact}{Fact}{Facts}
\crefname{obs}{Observation}{Observations}
\crefname{observation}{Observation}{Observations}
\crefname{remark}{Remark}{Remarks}
\crefname{infproposition}{Proposition}{Propositions}
\crefname{assumption}{Assumption}{Assumptions}
\crefname{definition}{Definition}{Definitions}
\crefname{example}{Example}{Examples}

\crefname{algocf}{Algorithm}{Algorithms}
\Crefname{algocf}{Algorithm}{Algorithms}
\crefname{algorithm}{Algorithm}{Algorithms}
\Crefname{algorithm}{Algorithm}{Algorithms}

\crefname{ineq}{Inequality}{Inequalities}
\creflabelformat{ineq}{#2{\upshape(#1)}#3}

\crefname{sub}{Subsection}{Subsections}
\creflabelformat{sub}{#2{\upshape(#1)}#3}

\crefname{subsection}{Subsection}{Subsections}
\creflabelformat{subsection}{#2{\upshape(#1)}#3}

\creflabelformat{sub}{#2{\upshape #1}#3}
\creflabelformat{subsection}{#2{\upshape #1}#3}
\crefname{sdp}{SDP}{SDPs}
\creflabelformat{sdp}{#2{\upshape(#1)}#3}

\crefname{lp}{LP}{LPs}
\creflabelformat{lp}{#2{\upshape(#1)}#3}



\newcommand{\eqdef}{\coloneqq}

\newcommand{\lp}{\left}
\newcommand{\rp}{\right}
\newcommand\norm[1]{\left\| #1 \right\|}

\renewcommand\vec[1]{\mathbf{#1}}
\DeclareMathOperator*{\pr}{\mathbf{Pr}}
\DeclareMathOperator*{\E}{\mathbf{E}}
\newcommand{\proj}{\mathrm{proj}}

\def\d{\mathrm{d}}

\newcommand{\cN}{\mathcal{N}}

\DeclareMathOperator*{\argmax}{argmax}

\newcommand{\by}{\mathbf{y}}
\newcommand{\bv}{\mathbf{v}}
\newcommand{\bu}{\mathbf{u}}
\newcommand{\bz}{\mathbf{z}}
\newcommand{\bw}{\mathbf{w}}

\newcommand{\bc}{\mathbf{c}}

\newcommand{\bI}{\mathbf{I}}

\newcommand{\bA}{\mathbf{A}}

\newcommand{\err}{\mathrm{err}}

\newcommand{\R}{\mathbb{R}}
\newcommand{\s}{\mathbb{S}}
\newcommand{\Z}{\mathbb{Z}}
\newcommand{\N}{\mathbb{N}}
\newcommand{\eps}{\epsilon}

\newcommand{\poly}{\mathrm{poly}}

\newcommand{\polylog}{\mathrm{polylog}}

\newcommand{\bU}{\mathbf{U}}

\newcommand{\diam}{\mathrm{diam}}

\newcommand{\sign}{\mathrm{sign}}

\newcommand{\Ind}{\mathds{1}}

\newcommand{\ba}{\mathbf{a}}
\newcommand{\bb}{\mathbf{b}}
\newcommand{\be}{\mathbf{e}}

\newcommand{\bT}{\mathbf{T}}

\newcommand{\x}{\vec x}
\newcommand{\w}{\vec w}

\newcommand{\opt}{\mathrm{opt}}

\newcommand{\abs}[1]{\lp| #1 \rp|}
\newcommand{\A}{\mathcal{A}}

\newcommand{\clip}{\operatorname{clip}}

\renewcommand\Pr{\pr}

\date{}

\begin{document}

\title{Gaussian Partitioning for Reliable Learning of Multi-Index Models}

\author{
Ilias Diakonikolas\thanks{Supported by NSF Medium Award CCF-2107079, 
ONR Award number N00014-25-1-2268, and a Guggenheim Fellowship.}\\
University of Wisconsin-Madison\\
\texttt{ilias@cs.wisc.edu}
\and
Giannis Iakovidis\thanks{Supported in part by NSF Medium Award CCF-2107079 
and ONR Award number N00014-25-1-2268.}\\
University of Wisconsin-Madison\\
{\tt iakovidis@wisc.edu}\\
\and
Daniel M. Kane\thanks{Supported by NSF Medium Award CCF-2107547.}\\
University of California, San Diego\\
\texttt{dakane@cs.ucsd.edu}
\and
Mingchen Ma\thanks{Supported by ONR Award number N00014-25-1-2268.}\\
University of Wisconsin-Madison\\
\texttt{mingchen@cs.wisc.edu}
}

\maketitle

\vspace{-0.5cm}

\begin{abstract}
We develop \emph{Gaussian partitioning}, a general algorithmic primitive for distributions that are 
smooth with respect to a Gaussian. Given samples from such a distribution, our algorithm partitions 
the space into polyhedral regions so that almost all of the probability mass lies in thin strips:  
regions where the coordinates along a small number of mutually orthogonal directions lie in short 
intervals, while the remaining directions are unrestricted. 
Additionally, within each such region, the conditional distribution approximately matches low-degree 
Gaussian moments in the unrestricted directions. Our construction builds on the logconcave 
partitioning method of Diakonikolas, Kane, and Ma (FOCS 2026). 

We apply this primitive to develop new algorithms for learning Boolean
multi-index models. Our results are as follows: 
\begin{itemize}
\item \textbf{Agnostic and Testable Learning.}
Under input marginals smooth with respect to the standard Gaussian, we give learners with error 
$(1+c)\mathrm{opt}+\epsilon$ for broad classes of multi-index models, including functions of 
halfspaces, low-dimensional convex concepts, and low-dimensional polynomial threshold functions. For 
Boolean functions of $k$ halfspaces, the dimension dependence is $d^{\mathrm{poly}(k/c)}$, improving 
the previous exponential dependence on $k/c$ in the exponent. For functions of $k$ halfspaces, we 
also obtain a testable agnostic learner with the same error guarantee and dimension dependence 
$d^{\mathrm{poly}(k/c)}$.
    \item \textbf{Learning from Positive Examples with Imperfect Unlabeled Data.} In the PUI model, 
    the learner receives positive examples from an unknown input distribution that is smooth with 
    respect to the standard Gaussian reference. We give general algorithms and Statistical Query 
    (SQ) lower bounds that qualitatively characterize the complexity of PUI learning  for a broad 
    family of multi-index models through Gaussian moment matching. For fixed smoothness and positive-mass parameters, we obtain a PUI learner for intersections of $k$ halfspaces with dimension 
    dependence $d^{O(\log(k/\eps))}$ and matching SQ lower bounds, improving the previous 
    $d^{\mathrm{poly}(1/\epsilon)}$ upper bound even for a single halfspace.

\item \textbf{Agnostic Reliable Learning.}
We give a simple general reduction from agnostic reliable learning to PUI learning. Under input 
marginals smooth with respect to the standard Gaussian, this yields negative reliable learners for 
intersections of $k$ halfspaces with dimension dependence $d^{O(\log(k/\epsilon))}$ for fixed 
smoothness parameters.
\end{itemize}

Finally, for PQ learning and Testable learning with Distribution Shift (TDS), we give qualitatively 
optimal SQ lower bounds of $d^{\Omega(1/\epsilon^2)}$, even for intersections of two halfspaces with 
Gaussian training data and arbitrary test distributions. These bounds rule out, within the SQ 
framework, the logarithmic dependence on $1/\epsilon$ in the dimension exponent achieved by our PUI 
and one-sided reliable learners.
\end{abstract}

\thispagestyle{empty}

\newpage

\tableofcontents

\thispagestyle{empty}

\newpage

\setcounter{page}{1}

\section{Introduction}\label{introduction}

A broad range of supervised learning algorithms use properties of Gaussian inputs, such as
low-degree polynomial approximations and moment bounds. In several
learning settings, however, the observed examples need not follow a
Gaussian distribution. For example, boosting changes the weights of
examples during its execution, and a learner given only positive
examples observes the original distribution conditioned on an unknown
event. {Thus, even when the original input distribution is Gaussian,
the distribution observed by the learner need not be.}

Motivated by such applications in learning theory,
we develop a general algorithmic primitive for partitioning the
input space so that {Gaussian approximation bounds} become
applicable within the resulting regions.
Given a Gaussian reference distribution $P$ and samples from
a distribution $Q$, we assume that events of {sufficiently}
small probability under $P$ also have small probability under $Q$.
This \emph{smoothness} assumption, defined in
\Cref{def:relaxed-smoothness}, allows $Q$ to be otherwise very
different from $P$.
Our algorithm partitions $\R^d$ into polyhedral regions
{so that, on cells containing almost all of the $Q$-mass,
the conditional distribution approximately matches low-degree
Gaussian moments in the unrestricted directions.}
We call this primitive \emph{Gaussian partitioning}.

Our main result, \Cref{thm:intro-partitioning}, gives a partition
whose number of cells is independent of the ambient
dimension $d$. {Cells containing almost all of the $Q$-mass are
\emph{thin strips}: their projections onto the span of their
constraint normals have small diameter, while the orthogonal
coordinates remain unrestricted. The dimension of this span
is independent of $d$. Within each such cell $R$, the conditional
distribution $Q\mid R$ approximately matches low-degree
Gaussian moments in the unrestricted directions.}
The small diameter guarantee is crucial for our learning theory applications. 
For a target Boolean function of bounded Gaussian surface area, 
it allows us to fix the {constrained} coordinates
with small total classification error. We can therefore analyze the
target within a cell using Gaussian moment and approximation bounds
in the {unrestricted directions}.

{Our construction builds on the logconcave partitioning
method of~\cite{diakonikolas2026robust}. The earlier method
compares moments with conditional logconcave distributions;
our thin cells allow the learning step to use an untruncated
Gaussian reference in the unrestricted directions.}
This distinction is important because Gaussian approximation
bounds can require substantially fewer moments.

{Our Gaussian partitioning primitive gives new algorithms
for learning Boolean multi-index models in several settings:
agnostic learning, testable agnostic learning, learning from
positive examples with imperfect unlabeled data (PUI learning),
and one-sided reliable learning.}
In \Cref{sec:intro-partitioning}, we formally describe our 
partitioning result and 
in \Cref{sec:intro-applications} we present its learning theory applications.

\subsection{Gaussian Partitioning}\label{sec:intro-partitioning}

{Smoothness with respect to the Gaussian distribution is the
distributional assumption underlying our learning algorithms.
Intuitively, events that are sufficiently rare under the
reference distribution must also have small probability under
the input distribution.}
We use the following definition from~\cite{lee2026smooth},
specialized to distributions on $\R^d$.

\begin{definition}[$(q,\sigma)$-Smoothness~\cite{lee2026smooth}]
\label{def:relaxed-smoothness}
Let $P,Q$ be probability distributions on $\R^d$,
let $q\geq1$, and let $\sigma\in(0,1]$. We say that $Q$ is
\emph{$(q,\sigma)$-smooth with respect to $P$} if, for every measurable
event $E\subseteq\R^d$,
$
    Q(E)\leq (1/\sigma) \,  P(E)^{1/q} \;.
$
\end{definition}

For $q=1$, this condition is equivalent to obtaining $Q$ from $P$ by
rejection sampling with acceptance probability at least $\sigma$.
We also say that $Q$ has \emph{weight at least $\sigma$} with respect
to $P$. Allowing $q>1$ permits an unbounded density ratio while still
controlling the probability of events that are unlikely under $P$.
For example, $\cN(\boldsymbol\mu,\bI_d)$ is
$(2,e^{-\|\boldsymbol\mu\|_2^2/2})$-smooth with respect to
$\cN(\vec{0},\bI_d)$; see \Cref{lem:gaussian-mixture-smoothness}.
{More generally, bounded $\chi^2$ divergence or bounded
R\'enyi divergence of order greater than one implies smoothness;
see \cite[Appendix~D.4]{lee2026smooth}.}

\smallskip

{To formally state our main result, we need some additional terminology.}
For a subspace $V\subseteq\R^d$, let $\proj_V$ denote the orthogonal
projection onto $V$. For a polyhedral cell $R$, let $V_R$ be the span
of the normal vectors of its defining inequalities. 
{We measure the diameter of a cell in its constrained
directions by} 
$ \diam(R)\eqdef \sup_{\bu,\bv\in \proj_{V_R}(R)}\|\bu-\bv\|_2$. 
When this quantity is small, $R$ is a thin strip along $V_R^\perp$. 
{For a standard Gaussian reference $P$, the projection of
$P\mid R$ onto $V_R^\perp$ is still standard Gaussian:
conditioning on $R$ restricts only the $V_R$-coordinates.}

We say that $P\mid R$ and $Q\mid R$, the corresponding conditional distributions on $R$, 
match their first $k$ \emph{directional moments on $V_R^\perp$ with slack $\Delta$} if,
for every unit vector $\bv\in V_R^\perp$ and every $i\in[k]$,
$
    \big| \E_{\x\sim Q\mid R}[(\bv\cdot\x)^i]
       -\E_{\x\sim P\mid R}[(\bv\cdot\x)^i] \big|\leq\Delta \;.
$
This is the notion of approximate moment matching used throughout the paper;
see \Cref{def:moment-matching}.
{With this setup, we now state our main algorithmic result.}

\begin{theorem}[Gaussian Partitioning; Informal Version of \Cref{thm:relaxed-gaussian-partitioning}]
\label{thm:intro-partitioning}
Let $P=\cN(\vec{0},\bI_d)$, and let $Q$ be a distribution on $\R^d$
that is $(q,\sigma)$-smooth with respect to $P$.
For $k\in\Z_+$ and $\eta,\delta,\Delta\in(0,1/2)$, there is an
algorithm using at most $d^kB$ samples from $Q$ and $d^{O(k)}B$ time,
where
$
    B=\exp\!\left(
       \frac{(qk)^{O(k)}}{\Delta^{O(1)}}
       \log^{O(k)}\!\frac{2}{\sigma\eta}
       \log\frac{2}{\delta}
       \right) \;,
$
that, with probability at least $2/3$, returns a polyhedral partition
$\mathcal U$ of $\R^d$ and a subcollection $\mathcal G\subseteq\mathcal U$
such that $|\mathcal U|\leq B$ and
$Q(\bigcup_{R\in\mathcal G}R)\geq1-\eta$.
Moreover, every $R\in\mathcal G$ satisfies:
\begin{enumerate}[label=(\roman*), itemsep=-1pt]
    \item $Q(R)\geq B^{-1}$.
    \item $\displaystyle
       \diam(R)\leq\delta$.
    \item $P\mid R$ and $Q\mid R$ match their first $k$ directional
    moments on $V_R^\perp$ with slack $\Delta$.
\end{enumerate}
\end{theorem}

We call the cells in $\mathcal G$ \emph{certified}.
The parameter $\delta$ bounds the {constrained diameter}
of every certified cell, as stated in item~(ii).
For fixed $q,k,\Delta$, the complexity 
of our algorithm 
is polynomial in $d$ and quasi-polynomial in $1/(\sigma\eta\delta)$.
We emphasize that the underlying construction uses only samples from $Q$ 
and is independent of any target function or labels.

{We next explain why the partitioning theorem is useful
for learning.}
For concept classes of bounded Gaussian surface area, a
sufficiently small diameter bound allows us to fix the
$V_R$-coordinates in each certified cell with small total error.
The learning problem in each cell then involves only the
unrestricted coordinates, where the conditional distribution
approximately matches Gaussian moments.

\vspace{-0.2cm}

\paragraph{Comparison with~\cite{diakonikolas2026robust}.}
Recent work \cite{diakonikolas2026robust} developed a related partitioning subroutine
as part of a general algorithm for agnostically learning Multi-Index Models (MIMs) 
under logconcave marginals. Given a logconcave distribution
$P$ and a distribution $Q$ obtained by rejection sampling from $P$,
their algorithm partitions the space so that $P\mid R$ and $Q\mid R$
approximately match low-degree moments on cells containing almost all
of the $Q$-mass. The comparison is made after an affine normalization
of each cell. {This is often enough to construct a weak learner over these conditional distributions, and these can be combined by a boosting algorithm.}

The reference distribution in each cell is therefore $P\mid R$,
which is logconcave but not necessarily Gaussian, even when $P$
itself is Gaussian.
{Their analysis uses approximation bounds for general
logconcave distributions, which can require (exponentially) higher degree than
the corresponding Gaussian bounds. The thinness of our cells
allows us to fix the constrained coordinates with small total
error and apply Gaussian approximation and moment bounds in
the unrestricted directions.}

\subsection{Algorithmic Applications in Learning Theory}\label{sec:intro-applications}

Our learning applications concern Boolean multi-index models, namely
functions that depend only on the projection onto an unknown
low-dimensional subspace. Formally, we have the following definition.

\begin{definition}[Multi-Index Model]\label{def}\label{def:intro-mim}
A function $f:\R^d\to\{\pm1\}$ is called a $K$-MIM if there exists a
subspace $U\subseteq\R^d$ of dimension at most $K$ such that
$f(\x)=f(\proj_U\x)$ for every $\x\in\R^d$.
We call such a subspace $U$ a relevant subspace of $f$.
Equivalently, $f(\x)=g(\mathbf W\x)$ for a link function
$g:\R^K\to\{\pm1\}$ and a matrix $\mathbf W\in\R^{K\times d}$.
\end{definition}

The relevant subspace is unknown to the learner. The link function
may also be unknown, as for arbitrary Boolean functions of halfspaces;
in special cases, such as intersections of halfspaces, its form is
specified by the concept class. In the settings of interest, $K$ is much smaller than the
ambient dimension $d$. Examples include functions of a small number of
halfspaces, low-dimensional convex concepts, and low-dimensional
polynomial threshold functions.

A growing line of work studies algorithms for MIMs under
different distributional and noise assumptions. We discuss this work
in \Cref{sec:intro-related-work}. We now present the learning
problems addressed by our Gaussian partitioning primitive.

\vspace{-0.2cm}

\subsubsection{Semi-Agnostic Learning}\label{sec:intro-agnostic}

We start with the standard notion of distribution-specific agnostic learning 
with approximate error guarantees\footnote{We use ``semi-agnostic'' for the guarantee
$(1+c)\opt+\eps$; standard agnostic learning corresponds to
$c=0$.}.

\begin{definition}[Semi-Agnostic Learning under Smooth Marginals]
\label{def:intro-agnostic}
Let $\mathcal F$ be a class of Boolean functions on $\R^d$, and let $D$
be a distribution on $\R^d\times\{\pm1\}$ whose $\x$-marginal $D_\x$
is $(q,\sigma)$-smooth with respect to $\cN(\vec{0},\bI_d)$.
No assumptions are made on the labels.
For a hypothesis $h:\R^d\to\{\pm1\}$, define
$\err_D(h)\eqdef\Pr_{(\x,y)\sim D}[h(\x)\neq y]$ and 
$\opt_D(\mathcal F)\eqdef\inf_{f\in\mathcal F}\err_D(f)$.
Given i.i.d.\ samples from $D$ and parameters $c,\eps\in(0,1)$,
the goal of a $(1+c)$-approximate agnostic learner is to output a
hypothesis $h$ satisfying
$\err_D(h)\leq(1+c)\opt_D(\mathcal F)+\eps$ 
with probability at least $2/3$.
\end{definition}

Our general algorithmic result for this model applies to MIM classes that are closed under
affine restrictions, admit Gaussian sandwiching polynomials, and have
uniformly bounded Gaussian surface area. 

\begin{definition}[$(\ell,K)$-Well-Behaved MIMs]
\label{def:intro-wellbehaved}
Let $\ell:(0,1]\to\Z_+$ be non-increasing, and let $\mathcal F$ be a
class of Boolean $K$-MIMs on $\R^d$. We say that $\mathcal F$ is
$(\ell,K)$-well-behaved if the following hold.
\begin{enumerate}[label=(\roman*),itemsep=-2pt,topsep=0.5pt]
    \item \emph{Closure under affine restrictions.}
    For every $f\in\mathcal F$, every subspace $W\subseteq\R^d$,
    and every $\bb\in W^\perp$, the function
    $f_{W,\bb}(\x)\eqdef f(\bb+\proj_W\x)$ belongs to $\mathcal F$.
    \item \emph{Gaussian sandwiching.}
    For every $f\in\mathcal F$ and every $a\in(0,1)$, there are
    polynomials $p^l,p^u$ of degree at most $\ell(a)$ such that
    $p^l(\x)\leq f(\x)\leq p^u(\x)$ for every $\x\in\R^d$, and 
    $\E_{\x\sim\cN(\vec{0},\bI_d)}[p^u(\x)-p^l(\x)]\leq a.$
    \item \emph{Bounded Gaussian surface area.}
    The class satisfies
    $\Gamma(\mathcal F)\eqdef\sup_{f\in\mathcal F}\Gamma(f)<\infty$,
    where $\Gamma(f)$ denotes the Gaussian surface area of $f$; {see \Cref{def:gaussian-surface-area}}.
\end{enumerate}
\end{definition}

The three conditions in \Cref{def:intro-wellbehaved} serve different
purposes. Closure under affine restrictions keeps the target in
the class after the {constrained coordinates} are fixed.
Sandwiching polynomials transfer Gaussian approximation guarantees
to a distribution with approximately matching moments.
The surface-area bound controls the {total} classification
error introduced by fixing the {constrained coordinates}.

Our main agnostic learning result is the following: 
\begin{theorem}[Semi-Agnostic Learning of MIMs; Informal Version of
\Cref{thm:main-learning}]
\label{thm:intro-agnostic}
Let $\mathcal F$ be an $(\ell,K)$-well-behaved MIM class and 
$D$ be a distribution on $\R^d\times\{\pm1\}$ such that $D_\x$ is $(q,\sigma)$-smooth with
respect to $\cN(\vec{0},\bI_d)$.
For $c,\eps\in(0,1/2)$, set $\ell_c=\ell(c_0\cdot c)$ 
for a sufficiently small universal constant $c_0>0$. There is an algorithm that draws
$
    n=d^{O(\ell_c)}\exp\!\left(
      (qK\ell_c)^{O(\ell_c)}
      \log^{O(\ell_c)}\!\frac{2}{c\sigma\eps}
      \log(2+\Gamma(\mathcal F))
      \right)
$ 
samples from $D$, runs in $\poly(n)$ time, and, with probability at
least $2/3$, outputs a hypothesis $h:\R^d\to\{\pm1\}$ satisfying
$\err_D(h)\leq(1+c)\opt_D(\mathcal F)+\eps.$
\end{theorem}

The main point is that the degree determining the exponent of
$d$ depends on the approximation factor $c$, but not on the excess
error $\eps$. For fixed class parameters, $q$, and $c$, the algorithm
has polynomial dependence on $d$ and quasipolynomial dependence on
$1/(\sigma\eps)$. The error guarantee is the same as
in~\cite{diakonikolas2026robust}; the improvement comes from using
Gaussian sandwiching degrees in place of logconcave ones. 

Combining \Cref{thm:intro-agnostic} with known Gaussian sandwiching
bounds~\cite{kane2013learning, klivans2026sandwiching} gives
\Cref{tab:intro-agnostic}. 
The table records only the dependence on $d$; the theorem gives the remaining factors. 
In particular, for arbitrary Boolean functions of $k$ halfspaces, the
dimension dependence improves from $d^{\exp(O((k/c)^4))}$
in~\cite{diakonikolas2026robust} to
$d^{\poly(k/c)}$. 

{We also prove an SQ lower bound for agnostic learning with
error $(1+c)\opt+\eps$, even under standard Gaussian marginals;
see \Cref{sec:multiplicative-sq}.
When $\eps\leq c$, this bound shows that SQ algorithms require
either exponentially many queries or tolerance at most
$d^{-\Omega(\ell(c))}$, where $\ell(c)$ denotes the Gaussian
$L_1$ approximation degree at accuracy $O(c)$.
This lower bound is tight up to polynomial factors in the
dimension exponent for the classes considered in our applications.
In fact, for halfspaces, $L_1$ sandwiching and $L_1$ approximation
have the same degree $\Theta(1/c^2)$, yielding tightness up to
constant factors in the exponent.}

\begin{table}[t]
\centering
\renewcommand{\arraystretch}{1.2}
\begin{tabular}{@{}p{0.65\linewidth}p{0.29\linewidth}@{}}
\hline
\textbf{Concept class} & \textbf{Dimension dependence} \\
\hline
Intersections of $k$ halfspaces 
& $d^{\widetilde O(k^3/c^4)}$
\\
Arbitrary Boolean functions of $k$ halfspaces 
& $d^{\widetilde O(k^5/c^4)}$ \\
Convex concepts of intrinsic dimension at most $K$
& $d^{\widetilde O(K^5/c^4)}$ \\
Degree-at-most-$r$ PTFs of intrinsic dimension at most $K$
& $d^{\widetilde O(r^6K^5/c^4)}$ \\
\hline
\end{tabular}
\caption{Dimension dependence for $(1+c)$-approximate agnostic
learning under marginals that are $(q,\sigma)$-smooth with respect
to {the standard Gaussian}. {For fixed class parameters, $q$, and $c$, the remaining
factor is quasipolynomial in $1/(\sigma\eps)$.}}
\label{tab:intro-agnostic}
\vspace{-0.2cm}

\end{table}

\vspace{-0.2cm}

\paragraph{{Extension to Testable Agnostic Learning.}}
{For functions of $k$ halfspaces, our guarantees also
extend to testable agnostic learning. Given samples from an
arbitrary labeled distribution, the learner either rejects or
outputs a hypothesis with error at most $(1+c)\opt+\eps$,
with probability at least $2/3$. Whenever the input marginal
is $(q,\sigma)$-smooth with respect to the standard Gaussian,
the learner accepts and returns a hypothesis satisfying this
guarantee with probability at least $2/3$.
{The dimension dependence is
$d^{\widetilde O(k^5/c^4)}$ for arbitrary functions of
$k$ halfspaces and $d^{\widetilde O(k^3/c^4)}$ for intersections;
see \Cref{cor:testable-functions-halfspaces,cor:testable-intersections}.}}

\subsubsection{Learning from Positive Examples with Imperfect Unlabeled Data}
\label{sec:intro-pu}

As our second learning-theoretic application, we give new
algorithms for learning from positive examples with
imperfect unlabeled data (PUI learning). For intersections of
halfspaces and low-dimensional convex concepts, our algorithms have
logarithmic dependence on the inverse accuracy in the dimension
exponent. These results follow from {a qualitative
characterization of PUI learning complexity in terms of Gaussian
moment matching, via an algorithmic upper bound and an SQ lower bound.}

\begin{definition}[PUI Learning~\cite{lee2026smooth}]
\label{def:intro-pu}
Let $\mathcal F$ be a class of Boolean functions on $\R^d$, let $D$
be a reference distribution, and let $D^*$ be an unknown distribution
that is $(q,\sigma)$-smooth with respect to $D$.
For an unknown target $f\in\mathcal F$, the learner receives
the smoothness parameters $q,\sigma$, 
sample access to $D$ and $D_+^*\eqdef D^*\mid\{f=1\}$,
and a lower bound $p_+>0$ on $D^*(f=1)$.
Given $\eps\in(0,1)$, the goal is to output
$h:\R^d\to\{\pm1\}$ satisfying
$\Pr_{\x\sim D^*}[h(\x)\neq f(\x)]\leq\eps$
with probability at least $2/3$.
\end{definition}
We consider the standard Gaussian reference
$D=\cN(\vec{0},\bI_d)$.
For a class $\mathcal F$, {we define} the \emph{Gaussian PUI learning exponent}
$m^*(\eps,\Delta,\sigma,q)$ {to be} the least positive integer $m$
{such that, whenever the positive conditional distribution
$D_+^*$ matches the first $m$ Gaussian directional moments
to error $\Delta$, we have $D^*(f=-1)<\eps$.}
This implication must hold for every target $f\in\mathcal F$ and
every distribution $D^*$ that is $(q,\sigma)$-smooth with respect
to $D$ and satisfies $D^*(f=1)>0$.
Here, moment matching to additive error $\Delta$ means that
$|\E_{\x\sim D_+^*}[(\bv\cdot\x)^i] -\E_{z\sim\cN(0,1)}[z^i]|\leq\Delta$ 
for every unit vector $\bv\in\R^d$ and every $1\leq i\leq m$.
If no such degree exists, we set $m^*=\infty$;
see \Cref{def:puexponent}.
Thus, matching sufficiently many moments of the positive examples
certifies that predicting positive has small error.
Gaussian partitioning allows us to apply this criterion within
individual cells.

Our general complexity result for PUI learning is informally stated below 
(see \Cref{thm:pumim} and \Cref{thm:sq-pu-general} for detailed formal statements).

\begin{theorem}[PUI Learning; Informal]
\label{thm:intro-pu-complexity}
Let $\mathcal F$ be a class of Boolean functions on $\R^d$,
and consider PUI learning with reference
$D=\cN(\vec{0},\bI_d)$ and $(q,\sigma)$-smooth input distributions,
where $q\geq1$ and $\sigma\in(0,1]$.
\begin{enumerate}[label=(\roman*), leftmargin=*,itemsep=-1pt]
    \item {\bf Upper Bound.}
    Suppose $\mathcal F$ is closed under affine restrictions
    and $\Gamma(\mathcal F)<\infty$.
    For $\eps,\Delta\in(0,1/2)$ and a given lower bound $p_+>0$, let
    $m=m^*\!(c_1\eps,\Delta,\frac{c_2\eps\sigma p_+}{q},2q)<\infty,$
    where $c_1,c_2>0$ are sufficiently small universal constants.
    Then PUI learning to error $\eps$ is possible using
    $
        n=d^m\exp\!\left(
          \frac{(qm)^{O(m)}
          \log^{O(m)}\!\bigl(2/(\sigma\eps p_+)\bigr)
          \log(2+\Gamma(\mathcal F))}
          {\Delta^{O(1)}}
        \right)
    $
    samples and $\poly(n)$ time.

    \item {\bf SQ Lower Bound.}
    Suppose $\mathcal F$ contains all rotations of a $K$-MIM $f$
    admitting a $(q,\sigma)$-smooth input distribution with
    positive mass $p_+>0$ and negative mass at least $\rho>0$,
    whose positive conditional distribution matches Gaussian
    moments up to degree $r\geq2$.
    Then, PUI learning to error $p_+(\sigma\rho)^q/2$ requires either
    $2^{d^{\Omega(1)}}$ SQ queries or tolerance at most
    $d^{-\Omega(r)}$.
\end{enumerate}
\end{theorem}

Our algorithmic result does not require the concepts to be
MIMs. For the SQ lower bound, it suffices that the class contain
all rotations of a suitable low-dimensional target.
The characterization is qualitative: the upper and lower bounds
use different accuracy and smoothness parameters, and the lower
bound requires exact moment matching; see
\Cref{rem:sq-comparison}.

For fixed smoothness parameters and a fixed positive-mass
lower bound, the previous algorithms~\cite{lee2026smooth}
had dimension dependence $d^{\poly(1/\eps)}$, even for a
single halfspace. Our bounds reduce
the dependence on inverse accuracy in this exponent to logarithmic.

Specifically, for intersections of halfspaces we obtain qualitatively 
matching upper and SQ lower bounds.

\begin{theorem}[PUI Learning Intersections of Halfspaces;
Informal Versions of
\Cref{cor:pu-intersections,cor:sq-pu-intersections}]
\label{thm:intro-pu-intersections}
For intersections of $k$ halfspaces with a standard
Gaussian reference, the following hold.
\begin{enumerate}[label=(\roman*), leftmargin=*,itemsep=-1pt,topsep=0.5pt]
    \item For fixed smoothness parameters $q,\sigma$ and a
fixed lower bound $p_+>0$, there is a PUI learner
under input distributions that are $(q,\sigma)$-smooth
with respect to the standard Gaussian
achieving error $\eps$ with probability at least $2/3$,
with complexity $d^{O(\log(k/\eps))}$, suppressing factors independent of $d$.

    \item {For suitable constant smoothness parameters and a
constant positive-mass lower bound}, any SQ learner achieving error $\eps$ requires either
    $2^{d^{\Omega(1)}}$ queries or tolerance at most
    $d^{-\Omega(\log(k/\eps))}$.
\end{enumerate}
\end{theorem}

Our framework also gives PUI learners for low-dimensional
convex concepts; see \Cref{cor:pu-convex}.
\Cref{tab:intro-pu} summarizes our upper bounds.

\begin{table}[t]
\centering
\renewcommand{\arraystretch}{1.2}
\begin{tabular}{@{}p{0.48\linewidth}p{0.47\linewidth}@{}}
\hline
\textbf{Concept class} & \textbf{Dimension dependence} \\
\hline
Intersections of $k$ halfspaces
& $d^{O(\log(2k)+q\log(2q/(\sigma\eps p_+)))}$ \\
Convex concepts of intrinsic dimension $K$
& $d^{O(K+q\log(2q/(\sigma\eps p_+)))}$
\\ 
\hline
\end{tabular}
\vspace{-0.1cm}
\caption{Dimension dependence of our PUI learners with a
standard Gaussian reference. For each row, the sample
complexity is $n=d^m\exp(m^{O(m)})$, with $m$ a sufficiently large
constant multiple of the quantity inside $O(\cdot)$ in the displayed
exponent. The running time is $\poly(n)$.}
\label{tab:intro-pu}
\end{table}

\subsubsection{Agnostic Reliable Learning}\label{sec:intro-reliable}

Reliable learning models situations in which one type of mistake
is more costly than the other. For example, a spam filter should make
very few false positives while detecting as much spam as possible.
The agnostic reliable model of~\cite{rev-KKM12} formalizes this requirement by minimizing one
type of error subject to making almost none of the other. Unlike
ordinary agnostic learning, the comparator is required to make zero
errors of the protected type. We follow the convention
of~\cite{intro-DRZ24}.

\begin{definition}[Agnostic Reliable Learning]
\label{def:intro-reliable}
Let $\mathcal F$ be a class of Boolean functions on $\R^d$ containing
the constant function $-1$, and let $D$ be a distribution on
$\R^d\times\{\pm1\}$. Define
$
    \mathcal G_D\eqdef
    \{f\in\mathcal F:
      \Pr_{(\x,y)\sim D}[f(\x)=1,y=-1]=0\}.
$
A positive reliable learner, given i.i.d.\ samples from $D$ and
$\eps\in(0,1)$, outputs a hypothesis $h:\R^d\to\{\pm1\}$ such that,
with probability at least $2/3$,
$
    \Pr_{(\x,y)\sim D}[h(\x)=1,y=-1]\leq\eps
$ 
and
$
    \Pr_{(\x,y)\sim D}[h(\x)=-1,y=1]
    \leq
    \inf_{f\in\mathcal G_D}
    \Pr_{(\x,y)\sim D}[f(\x)=-1,y=1]+\eps.
$
Negative reliable learning is defined by interchanging the labels
$1$ and $-1$ and requiring that $\mathcal F$ contain the constant
function $1$.
\end{definition}

Our reliable-learning algorithms follow from a reduction to
PUI learning. The reduction applies to any class with a PUI learner;
it does not require a separate partitioning argument. 

\begin{theorem}[From PUI to Reliable Learning;
Informal Version of \Cref{lem:reliable-to-pu}]
\label{thm:intro-reliable-reduction}
Suppose that $\mathcal F$ contains the constant function $1$ and
has a PUI learner with sample complexity $N(\eps,\sigma,p_+,q)$,
running time $T(\eps,\sigma,p_+,q)$, and failure probability at most
$1/12$. Then $\mathcal F$ has a negative reliable learner under
$(q,\sigma)$-smooth Gaussian marginals using
$
 O\!\left(N(\eps/2,\sigma\eps/4,\eps/4,q)/\eps
          +1/\eps^2\right)
$
labeled samples and the analogous time bound with $N$ replaced by
$T$, with success probability at least $2/3$.
If $\mathcal F$ contains the constant function $-1$ and
$-\mathcal F$ admits the corresponding PUI learner, applying the
reduction to $-\mathcal F$ gives positive reliable learning of
$\mathcal F$.
\end{theorem} 

Combining this reduction with our general PUI learner gives
negative reliable learning for classes closed under affine
restrictions, containing $1$, and having finite Gaussian surface
area. The dimension exponent is controlled by
$m=m^*(c_0\eps,\Delta,c_0\sigma\eps^3/q,2q)<\infty$,
for $\eps,\Delta\in(0,1/2)$ and a sufficiently small universal
constant $c_0>0$; see \Cref{cor:reliable-mim}.
For intersections and unions of halfspaces, we obtain the following.

\begin{theorem}[Reliable Learning of Intersections and Unions;
Informal Version of \Cref{cor:reliable-halfspaces}]
\label{thm:intro-reliable}
Let $D$ be a distribution on $\R^d\times\{\pm1\}$ whose marginal
$D_\x$ is $(q,\sigma)$-smooth with respect to
$\cN(\vec{0},\bI_d)$. For $\eps\in(0,1/2)$, there is a negative
reliable learner for intersections of $k$ halfspaces (and a
positive reliable learner for unions of $k$ halfspaces) using
at most
$n=d^m\exp\!\left(m^{O(m)}\right)$ labeled samples and $\poly(n)$ time, 
where $m= O(\log(k)+q\log\frac{q}{\sigma\eps})$. 
\end{theorem}

A conceptual contribution of the above is that 
reliable learning can be qualitatively easier than fully
agnostic learning {for a range of concept classes}. 
For Gaussian halfspaces, this separation was established
by~\cite{intro-DRZ24}: reliable learning admits a logarithmic dimension
exponent in inverse accuracy, whereas fully agnostic learning to error $\opt+\eps$ 
has SQ complexity $d^{\Omega(1/\eps^2)}$~\cite{DKPZ21}.
Our result extends the algorithmic side of this separation to
intersections and unions and to marginals that are only smooth with
respect to a Gaussian. For a fixed number of halfspaces, the
logarithmic dependence on $1/\eps$ is already necessary in the SQ model, 
by the halfspace lower bound of~\cite{intro-DRZ24}.

Finally, we remark that our reduction also gives negative reliable learning of convex
concepts of intrinsic dimension $K$ and positive reliable learning
of their complements; see \Cref{lem:reliable-to-pu,cor:pu-convex}.

\paragraph{Computational Limits for PQ and TDS learning.}
\label{sec:intro-pq-tds}
Our preceding results raise the question of whether Gaussian partitioning
can also improve the complexity dependence in other models of reliable
learning with distribution shift. 
We consider the realizable setting, with a common target
function, a Gaussian training marginal, and an arbitrary test
marginal. 
In PQ learning~\cite{goldwasser20pq,kalai2021efficient},
the learner receives labeled training examples and unlabeled test
examples and may abstain. It must abstain rarely on the training
distribution and make few incorrect, non-abstaining predictions on
the test distribution. In testable learning with distribution shift
(TDS), the learner may reject the test distribution; whenever it
accepts, its hypothesis must have small test error, and it must accept
when the training and test distributions agree. The formal definitions
are in \Cref{def pq learning,def tds learning}. Recent work gives
algorithms and reductions for these models, including algorithms for
Gaussian halfspaces and their intersections based on polynomial
approximation and testing~\cite{rev-KSV24,rev-GSSV24,rev-CKKSV24,
klivans2026equivalence,klivans2026sandwiching}.

For {intersections of halfspaces without balance assumptions},
the known general guarantees have inverse-polynomial accuracy
dependence in the dimension exponent.
{We show that this dependence cannot be reduced to logarithmic
within the SQ framework.}
Even for intersections of two halfspaces and Gaussian training data,
any {learner new{with SQ access to both the labeled training distribution
and the unlabeled test distribution}} must make
$2^{d^{\Omega(1)}}$ queries or use tolerance at most
$d^{-\Omega(1/\eps^2)}$; see
\Cref{lm pq tds sq lower bound}. 
{This significantly improves the prior known SQ lower bound of $d^{\Tilde{\Omega}(\log(1/\eps))}$ from \cite{klivans2024learning}.}
For TDS, the accuracy exponent matches the known general bound
for {any fixed $k\geq2$}, up to logarithmic
factors~\cite{rev-KSV24}.
For PQ learning, it rules out the logarithmic dependence achieved
by our {PUI} and one-sided reliable learners.

\subsection{Detailed Overview of Techniques}\label{sec:intro-overview}
Our main technical contribution is a Gaussian partitioning
primitive that produces thin regions covering almost all of the
input mass. Within each such region, the conditional input
distribution approximately matches low-degree Gaussian moments
in the unrestricted directions.
We first explain the geometry of the partition
and how to construct it efficiently. We then describe the
learning applications and the moment-matching constructions
underlying our SQ lower bounds.

\vspace{-0.2cm}

\paragraph{Brief Overview of Approach.}
We start with the entire space and build the partition
recursively. On each current region, we check whether the
conditional input distribution approximately matches low-degree
Gaussian moments in every unrestricted direction. If we find a
moment discrepancy, we subdivide the region into narrow intervals
along the corresponding direction and continue on the resulting
regions. Regions where we certify moment matching are kept.
The main challenges are to find discrepancy directions efficiently
and to control the number of regions produced by the recursion.
We address the first through a low-dimensional search. For the
second, we use a potential function to measure progress under
subdivision, together with smoothness bounds that control the
recursion depth and the total mass discarded. 

The geometry of the regions is what makes the partition useful
for learning. Conditioning the Gaussian reference on a region
changes only its constrained coordinates, leaving the remaining
coordinates standard Gaussian. At the same time, the narrow
intervals ensure that the constrained coordinates vary little
within each retained region. 
{For a target function with bounded Gaussian surface area,
we can fix the constrained coordinates within each certified
region with small total error over the regions. The resulting
functions depend only on the unrestricted coordinates, where
the conditional input distributions approximately match
Gaussian moments.} 
This allows us to apply Gaussian approximation bounds within the
partition. 

For agnostic learning, moment matching lets us transfer
Gaussian polynomial upper and lower approximations to the
conditional input distributions. These approximations give a
weak learner, which can then be boosted. For PUI learning, we
use a different consequence of moment matching: for the classes
we consider, if the positive examples match sufficiently many
Gaussian moments, then the target has small negative probability.
Applying this certificate within the regions gives our PUI
learners, and a reduction then yields one-sided reliable learning.
Our SQ lower bounds use distributions that match many Gaussian
moments exactly. Embedding these distributions in unknown
low-dimensional subspaces produces instances that are difficult
for SQ algorithms to distinguish from a Gaussian. 

{We now turn to the partition itself. We first explain why
conditioning on a region creates a difficulty for applying
Gaussian approximation bounds, and how this motivates the
thinness condition.}

\subsubsection{From Logconcave Partitioning to Thin Gaussian Cells} 
The partitioning procedure of~\cite{diakonikolas2026robust} takes
samples from a distribution $Q$ obtained by rejection sampling from a
logconcave reference $P$. It recursively partitions the space until, on
cells containing almost all of the $Q$-mass, the conditional
distributions $Q\mid R$ and $P\mid R$ approximately match low-degree
directional moments after an affine normalization. For fixed moment
and discarded-mass parameters, the number of cells is polynomial in
the inverse acceptance probability. This reduces learning under $Q$
to weak learning under distributions whose moments are close to those
of a logconcave distribution. 

{A limitation of this approach is that its reference
distribution on a cell is $P\mid R$. General logconcave
distributions have weaker tail bounds than a Gaussian, and
the available polynomial approximation bounds can require
exponentially higher degree.}
For example, the approach in~\cite{diakonikolas2026robust}
agnostically learns arbitrary functions of $K$ halfspaces
within error $(1+c)\opt+\eps$ by matching moments up to degree
$\exp(O((K/c)^4))$.
{\cite{diakonikolas2026robust} uses this degree bound even when the
original reference is Gaussian: conditioning on a cell produces
a truncated Gaussian, and affine normalization does not in
general restore a Gaussian distribution. Thus, that earlier approach 
cannot leverage the sharper Gaussian approximation bounds.}

We instead construct cells whose coordinates along a set of
mutually orthogonal directions lie in short intervals, while
the remaining coordinates are unrestricted. Let $V_R$ denote
the span of these constrained directions. For every {certified cell} we
retain, the projection $\proj_{V_R}(R)$ has small diameter. 
The cell therefore resembles a thin
strip: its constrained coordinates lie in a small set, while its
orthogonal coordinates remain unrestricted. Under the Gaussian
reference $P=\cN(\vec{0},\bI_d)$, conditioning on $R$ changes only
the constrained coordinates; the marginal on $V_R^\perp$ remains
standard Gaussian. {Our moment guarantee states that $Q\mid R$ approximately
matches low-degree Gaussian moments in every direction in
$V_R^\perp$.}

The purpose of the diameter bound is to remove the target's
dependence on the constrained coordinates.
For a Boolean function $f$ of bounded Gaussian surface area,
we show that there are points $\bz_R\in\proj_{V_R}(R)$ such that
the affine restrictions $f_R(\x)=f(\bz_R+\x_{V_R^\perp})$
approximate $f$ on the certified cells with small total error
{under the Gaussian reference $P$}.
To prove this, within each certified cell we resample the
constrained coordinates from their conditional Gaussian
distribution and keep the orthogonal coordinates fixed.
Outside these cells we leave the point unchanged. This coupling
preserves the Gaussian marginal and moves each point by at most
the constrained diameter{; see \Cref{lem:pertubation} for
the Gaussian smoothing argument.}

{By smoothness, the same affine restrictions have small
total error under $Q$.} 
By choosing the grid resolution sufficiently small, we make
this error as small as required by the application. The guarantee is
averaged over the cells; it does not assert a small conditional error
in every cell. The points $\bz_R$ are used only in the analysis.
When the class is closed under affine restrictions, each $f_R$ remains
in the class and depends only on the coordinates where Gaussian
moment matching holds.

\subsubsection{Constructing the Partition}

{At each cell $R$, we search for a direction in $V_R^\perp$
along which one of the first $k$ moments of $Q\mid R$ differs
significantly from the corresponding Gaussian moment. We call
the procedure used for this search a \emph{moment discrepancy
finder}.}
A certificate that the first $k$ directional moments approximately
match their Gaussian values makes $R$ a certified cell; a
discrepancy supplies a new direction along which to subdivide $R$.
The procedure can also reject a cell when the checks needed for
an efficient search fail. Starting from the entire space, we apply
this procedure recursively. 
{We first explain how a moment discrepancy leads to a useful
subdivision and how to control the recursion depth. We then
describe the discrepancy finder and show how conditional
smoothness provides the guarantees needed by both steps.}

\vspace{-0.2cm}

\paragraph{Subdividing Cells and Controlling the Depth.}

\vspace{-0.1cm}

On a cell with the required conditional smoothness, a moment
discrepancy along {a direction} $\bv$ implies that
{there is a threshold $t$ for which
$\Pr_{\x\sim Q\mid R}[\bv\cdot\x\leq t]$ differs from
$\Pr_{z\sim\cN(0,1)}[z\leq t]$ by a nontrivial amount};
see \Cref{lem:relaxed-moment-good-split}.
We estimate these probabilities with fresh samples and use the split
only if its gap is large enough. {A single threshold would split $R$ into two cells.} To obtain thin cells, we also cut the central range
of the coordinate $\bv\cdot\x$ into intervals of width at most
$\delta_0$, including the chosen threshold among the endpoints.
{Here $\delta_0>0$ is the grid resolution, chosen sufficiently
small relative to the recursion-depth bound to ensure that each
retained cell has small diameter in its constrained directions.}
The two unbounded tail intervals are discarded, and we recurse on
the bounded intervals.

Each new direction lies in the orthogonal complement of all
previous constraint directions. Thus, after $r$ subdivisions along
a path, a remaining cell restricts $r$ orthonormal coordinates to
intervals of width at most $\delta_0$. Its constrained diameter is
at most $\delta_0\sqrt r$. In particular, a depth bound $L$ gives
both $\dim(V_R)\leq L$ and
$\diam(R)\leq\delta_0\sqrt L$. 
{The depth therefore controls both the number of constrained
directions and the diameter of the cell in those directions.}

We measure progress using the affinity potential
$
    \Psi_t=\sum_{R\in A_t}\sqrt{P(R)Q(R)},
$
where $A_t$ consists of the cells split in round $t$.
The validated probability gap makes the conditional laws of the
children under $P$ and $Q$ different. Quantitatively, their squared
Hellinger distance is bounded below, so a split reduces the total
affinity contribution of its parent by a factor $1-\kappa$ for a
positive $\kappa$ determined by the gap. Subdividing further into
short intervals cannot increase this contribution. Consequently,
$\Psi_{t+1}\leq(1-\kappa)\Psi_t$; see
\Cref{lem:relaxed-depth-analysis}. 

For $q=1$, the bounded density ratio converts small affinity
directly into small remaining $Q$-mass. For $q>1$, smoothness implies
that the density ratio is bounded outside a set of small $Q$-mass,
and the same argument applies on the remaining set; see
\Cref{lem:relaxed-smoothness-truncation}. 
{After sufficiently many rounds, the cells that remain
active have small total $Q$-mass. We stop the recursion at this
point and discard these cells. The required number of rounds
is polylogarithmic in the inverse smoothness and discarded-mass
parameters when the moment parameters and $q$ are fixed.}
We also
adapt the low-mass potential of~\cite{diakonikolas2026robust} to
subdivisions with many children. It bounds the total mass lost when
cells become too small for conditional sampling. Together with
Gaussian tail bounds and the conditional smoothness analysis below,
this accounts for the discarded cells and gives the stated
quasipolynomial complexity; see
\Cref{lem:relaxed-small-mass,lem:relaxed-tail-mass}. 

\vspace{-0.2cm}

\paragraph{Finding a Discrepancy from Hermite Moments.}
{We reduce the search for a moment discrepancy to a
low-dimensional subspace. Fix a cell $R$ and consider the
projection of $Q\mid R$ onto $V_R^\perp$. Let $\mathcal H_i$
be the normalized order-$i$ Hermite tensor, and let $\bA_i$
be its expectation under this projected distribution.
Every directional raw moment discrepancy through degree $k$
is a linear combination of contractions
$\bA_i\cdot\bv^{\otimes i}$.}

{We estimate these tensors from samples, flatten each
along one index, and let $W$ be the span of the left singular
vectors with sufficiently large singular values. Passing the
Frobenius-norm test bounds the dimension of $W$ by
$k^{O(k)}H^2/\Delta^2$, where $H$ is the norm cutoff and
$\Delta$ is the desired moment slack. The tensor components
involving directions orthogonal to $W$ are small. Consequently,
projecting a direction onto $W$ approximately preserves
the Hermite contractions that determine its moment discrepancies.}

{The projection need not have unit length. We therefore
adjoin one direction in $V_R^\perp$ orthogonal to $W$, when
such a direction exists, and use it to restore the lost norm.
This gives a unit vector whose directional moment discrepancies
are close to those of the original direction, without changing
their degrees. Searching a sufficiently fine net of the unit
sphere in this enlarged subspace then either finds a discrepancy
of size $\Omega(\Delta)$ or certifies moment matching in every
direction; see
\Cref{thm:standalone-capped-finder,claim:standalone-capped-reduction}.}

{Conditional smoothness supplies the required tensor norm
bounds. Contracting $\mathcal H_i$ with a tensor of unit
Frobenius norm gives a degree-$i$ polynomial of Gaussian
$L_2$ norm at most one. Gaussian hypercontractivity and
smoothness bound its expectation under the projected conditional
distribution. Taking the supremum over these tensors bounds
$\|\bA_i\|_F$; see \Cref{lem:relaxed-polynomial-control}.
The polynomial second-moment bound controls estimation error,
while the Frobenius-norm cutoff controls the dimension of
the search.}

The Hermite and singular-value approach already appears as
motivation in~\cite{diakonikolas2026robust}.
Our geometry allows us to use it throughout the recursion because
the reference on the remaining coordinates is always Gaussian.

\vspace{-0.2cm}

\paragraph{Controlling Conditional Smoothness.}
{Both the subdivision step and the discrepancy finder require
control of the conditional distribution.}
Without control of the
tails, a moment discrepancy could come from a tiny amount of mass
far from the origin, and no threshold need give a useful gap.
Conditional smoothness supplies the required tail bounds. It also
bounds the Hermite moment tensors used to find a discrepancy
efficiently. The difficulty is that conditioning on a cell may
weaken smoothness, so we need these guarantees on cells containing
most of the mass.

For $q=1$, the density ratio of $Q$ to $P$ is bounded. After
conditioning on a cell $R$, this bound {is multiplied by}
$P(R)/Q(R)$, but cells where this factor is too large have small
total $Q$-mass. For $q>1$, the original density ratio can be
unbounded, so this argument is insufficient. Instead, we prove an
averaging statement for conditional densities: for every 
{finite measurable partition}, cells containing at least $1-\eta$
of the $Q$-mass satisfy that $Q\mid R$ is 
$(2q, \sigma\eta/(2q))$-smooth with respect to $P\mid R$.
The proof bounds a suitable $L_p$ norm of the global density ratio
and then bounds the $Q$-weighted average of the corresponding
conditional norms; see \Cref{lem:conditional-density-averaging}.
We apply this statement to {the final partition}, so the exponent
does not double at every level of the recursion.

Note that we cannot directly test whether a conditional distribution has
these smoothness parameters. {Instead, we check conditions that
suffice for reliable moment estimation, efficient discrepancy finding,
and progress under subdivision.}
First, we estimate the mass of a cell and stop
processing it if obtaining enough conditional samples would be too
expensive. For every cell passing this test, global smoothness gives
a uniform second-moment bound for low-degree polynomials under
$Q\mid R$, {relative to their Gaussian second moments}.
This makes their empirical expectations reliable.
Next, we test whether the empirical Hermite moment tensors have
small Frobenius norms, which is the condition needed for an efficient
search for a discrepancy. Finally, before subdividing a cell we check
that the proposed split actually separates its conditional distribution
from the Gaussian reference by a sufficiently large probability gap.
{We discard cells that fail the norm or split test and leave
them unchanged in the final partition. On a cell with the required
conditional smoothness, the norm test passes and any detected moment
discrepancy yields a split with the required gap. Applying the
averaging statement to the final partition therefore bounds the
total $Q$-mass lost to these rejections.
The uniform second-moment bound ensures that every returned
moment certificate is sound, including on cells without the
stated conditional smoothness guarantee.}

\subsubsection{Learning Theory Applications via Reductions to Gaussian Partitioning}

\paragraph{Agnostic Learning.}
{Our weak learner first applies Gaussian partitioning to the
input marginal. On each cell with sufficiently large estimated
mass, it fits a low-degree polynomial to the observed labels
using $L_1$ regression and thresholds it to obtain a Boolean
classifier. It combines these classifiers into a single hypothesis,
predicting $1$ on the remaining cells. We use this weak learner
within smooth agnostic boosting, constructing a new partition
for each reweighted input distribution.}

The key implication of sandwiching is that correlation with
the target gives correlation with a low-degree polynomial.
{For the analysis, we replace the target on each certified
cell by its affine restriction $f_R$.}
Gaussian sandwiching polynomials $p_R^l,p_R^u$ satisfy the
pointwise inequalities $p_R^l\leq f_R\leq p_R^u$. Their expected
gap remains small under the actual conditional input distribution
because its low-degree moments match the Gaussian moments.
To make the required moment slack independent of the ambient
dimension, we average the polynomials over irrelevant Gaussian
coordinates. They then depend on at most $K$ directions, so their
coefficient bounds depend on the intrinsic dimension $K$ rather
than $d$; see \Cref{fact:moment-matching-polynomial-expectations}.

In particular, either sandwiching polynomial approximates
$f_R$ in $L_1$ under the conditional input distribution. Since
$|y|=1$, this gives
$
    \left|\E[yf_R(\x)\mid R]-\E[yp_R^u(\x)\mid R]\right|
    \leq\E[|f_R(\x)-p_R^u(\x)|\mid R].
$
Thus, a significant correlation between the labels and $f_R$
implies a correlation with a low-degree polynomial. 
{The conditional $L_1$ loss of $p_R^u$ against the labels
is at most twice the conditional classification error of $f_R$
plus the expected gap between $p_R^u$ and $p_R^l$ on $R$.}
We use $L_1$ polynomial regression and thresholding to find
cellwise classifiers, which can be viewed as a partial hypothesis
on the retained cells. Averaging over cells recovers the original
label correlation, up to the total freezing and approximation
errors; no individual cell is required to have a small noise rate.
Since the omitted cells have small mass, completing this hypothesis
gives the weak learner of \Cref{lem:weaklearner}.

As in~\cite{diakonikolas2026robust}, smooth agnostic boosting
then gives error $(1+c)\opt+\eps$. The sandwiching accuracy, and
hence its degree, depend on $c$ rather than $\eps$; boosting only
changes the smoothness parameter from $\sigma$ to $\sigma\eps$.
Our partition allows this reduction to use Gaussian sandwiching
degrees in place of those for general logconcave distributions;
see \Cref{fact:boosting,thm:main-learning}.

\vspace{-0.2cm}

\paragraph{{Testable Agnostic Learning.}}
{The preceding agnostic learning argument combines properties of the partition---bounds on cell masses and constrained diameters, together with moment matching on certified cells---with smoothness to control the error from fixing the constrained coordinates. The partition properties can be checked empirically. Testable learning, however, requires an error guarantee whenever the learner accepts, even if the input marginal is not smooth. Thus it suffices to replace the direction fixing argument with an argument that it works for general distributions that pass efficient tests. We do this by assuming a stronger sandwiching result that holds for well-known function classes.}

{For any Boolean function of $k$ halfspaces, we show that on each thin cell a single pair of low-degree polynomials in the unrestricted coordinates sandwiches the target for every choice of the constrained coordinates within the cell. Their expected Gaussian gap, averaged over the cells using Gaussian cell masses, is controlled by the constrained diameters and the polynomial approximation error; see \Cref{def:testable-wellbehaved}. Essentially, this holds because there exist
upper and lower bounding functions in the unrestricted
coordinates that are close and they belong in the class. Using Gaussian sandwiching bounds for these functions gives the required polynomials; 
see \Cref{lem:functions-halfspaces-cell-sandwiching}.}

{Combining this property with checks on cell diameters, empirical
cell masses relative to their Gaussian masses, and conditional
empirical moments gives a testable weak learner (\Cref{thm:testable-weak-learner}). Boosting then yields an efficient testable learner with error $(1+c)\opt+\eps$;
see \Cref{thm:testable-main}. Moreover, our result extends beyond the Gaussian setting: our testable learner applies to any input marginal that is $(q,\sigma)$-smooth with respect to the Gaussian. Our result gives efficient testable learning algorithms for functions of halfspaces
}

\vspace{-0.2cm}

\paragraph{PUI Learning via Moment Certificates.}
For PUI learning, we partition the observed positive
distribution $D_+^*=D^*\mid\{f=1\}$, using
$D=\cN(\vec{0},\bI_d)$ as the reference. 
{Our hypothesis is positive on certified cells that pass
the additional mass test described below, and negative elsewhere.}
{The partition guarantee ensures that almost all positive
examples lie in certified cells, so predicting negative on the
uncertified cells incurs only a small false-negative error.} 
To control false positives, we need a different consequence of moment matching: for the classes
of interest, if the positive conditional distribution matches
enough Gaussian moments, then the negative region has small
probability under the target distribution. This is the
\emph{moment certificate} quantified by the Gaussian PUI learning
exponent in \Cref{def:puexponent}. It replaces the sandwiching
assumption used for agnostic learning. 

For a halfspace, the certificate follows from a univariate
polynomial that is nonpositive on the positive halfspace but has
positive Gaussian expectation. Its degree and coefficients show
that a halfspace excluding Gaussian mass $\rho$ cannot support a
distribution that matches the first $O(\log(2/\rho))$ Gaussian
moments to a sufficiently small constant slack. For an intersection
of $k$ halfspaces excluding mass $\rho$, one defining halfspace
excludes mass at least $\rho/k$, giving degree
$O(\log(2k/\rho))$. Smoothness converts negative target mass into
a lower bound on negative Gaussian mass. This gives the
logarithmic degree bound used by the learner; see
\Cref{lem:halfspaces,cor:intersection-moments}. 

{Applying this certificate within a cell requires some care.
After fixing the constrained coordinates, the target becomes
$f_R$, but the moments supplied by the partition come from
examples conditioned on $f=1$, not on $f_R=1$. A small
unconditional disagreement between these functions could change
their positive conditional distributions substantially if the
positive event were too rare.
{We therefore estimate $D_+^*(R)$ on each certified cell
and discard the cell if this estimate is too small relative
to its Gaussian mass $D(R)$. The cells discarded by this test
have small total $D_+^*$-mass. On the remaining cells, apart
from a collection of small total $D^*$-mass, the mass test and
smoothness give a lower bound on $D^*(f=1\mid R)$. Conditional
smoothness also bounds the second moments of the directional
monomials used in the certificate. Together with the bound on
the disagreement between $f$ and $f_R$, these estimates control
how much the moments change when we condition on $f_R=1$
instead of $f=1$. We can then apply the certificate to $f_R$
to bound the false-positive error.}
The resulting algorithm and analysis appear in
\Cref{alg:gaussian-pu,thm:pumim}.}

\vspace{-0.2cm}

\paragraph{From PUI to Reliable Learning.}
For negative reliable learning, consider a comparator with
no false negatives and condition the labeled distribution on the
event that its label agrees with the comparator. The comparator
is realizable on this event, and all positive examples survive,
so their conditional distribution is unchanged. The conditioned
feature distribution remains smooth with respect to the Gaussian.
{We can therefore run the PUI learner directly on the
observed positive examples. The conditioning is used only
in the analysis.} {The resulting hypothesis has small false-negative error,
since all positive examples lie in the agreement event. Its
false-positive error is at most the comparator's false-positive
error plus the PUI learning error.}
A constant hypothesis handles the case of very few positive labels.
Complementing labels and concepts gives positive reliable learning
of unions; see \Cref{lem:reliable-to-pu,cor:reliable-halfspaces}.

\vspace{-0.2cm}

\subsubsection{SQ Lower Bounds via Moment Matching}

\vspace{-0.1cm}

\paragraph{PUI Learning.}

Our SQ lower bounds use the non-Gaussian component analysis
(NGCA) method introduced in~\cite{DKS17-sq}, in the form developed
in~\cite{diakonikolas2023sq}. The method embeds a non-Gaussian
distribution that matches many Gaussian moments in an unknown
low-dimensional subspace and uses an independent Gaussian on its
orthogonal complement. Distinguishing the resulting distribution
from a standard Gaussian requires either many SQ queries or very
small tolerance; see \Cref{fact:ngca-sq}. We show that a successful
PUI learner would perform this distinction, yielding
\Cref{thm:sq-pu-general}. 

For halfspaces and their intersections, the main construction
is a distribution supported on the positive region that matches
Gaussian moments exactly through logarithmic degree. In one
dimension, we delete a small Gaussian tail and repair its missing
moments by a density correction on a central interval.
{The correction is small enough that the resulting density
remains nonnegative and bounded by a small multiple of the
Gaussian density.}
Products of these distributions give the construction for
intersections; see
\Cref{lem:halfspace-moment-lower-bound,lem:intersection-moment-lower-bound,cor:sq-pu-intersections}.

\vspace{-0.2cm}

\paragraph{PQ and TDS Learning.}
The PQ/TDS lower bound uses the same NGCA method with a
different moment-matching distribution. The discrete construction
of~\cite{DKS17-sq} matches $\Theta(m)$ Gaussian moments while
placing mass $\Theta(1/\sqrt m)$ at the origin. We embed it in a
hidden direction and take the target to be a sufficiently narrow
slab around the corresponding hyperplane. {By taking the slab sufficiently narrow, we can simulate
the labeled training SQ oracle using the Gaussian distribution
with constant $-1$ labels, within the requested tolerances
and without knowing the hidden direction.} The test distribution still places substantial mass
inside the slab. Setting $m=\Theta(1/\eps^2)$ forces a successful
PQ or TDS learner to distinguish the planted test distribution
from a Gaussian. Applying the NGCA lower bound of
\Cref{fact:ngca-sq} gives the inverse-quadratic accuracy dependence
in \Cref{lm pq tds sq lower bound}.

\vspace{-0.2cm}

\subsection{Related Work}
\label{sec:intro-related-work}

\vspace{-0.2cm}

Multi-index models have a long history in statistics, including
projection pursuit regression and sufficient dimension
reduction~\cite{rev-FS81,rev-Li91,rev-XTLZ02}.
Here we focus on algorithmic results for learning Boolean MIMs. 
Several distribution-specific algorithms are known for Boolean
MIMs. \cite{rev-Vempala10} gave a PAC learner for
intersections of halfspaces under logconcave distributions when
the defining hyperplanes pass through the mean. Under Gaussian
inputs, polynomial regression and Gaussian surface-area bounds
give agnostic learners for broad classes of Boolean
functions~\cite{kalai2008agnostically,rev-KOS08}.
\cite{kane2013learning} extended this
approach to agnostically learn functions of a constant number
of halfspaces under logconcave distributions.

In the distribution-free realizable setting,
\cite{rev-APS26} recently gave a PAC learner
for arbitrary functions of $k$ halfspaces whose running
time is subexponential in the ambient dimension for constant $k$.
Under an additional margin assumption, \cite{rev-PV26} obtained improved 
bounds for learning intersections of halfspaces. These results require 
subexponential time even for $k=2$.

For general MIMs under Gaussian inputs, 
\cite{DKKTZ24} gave agnostic learners with error
$\opt+\eps$ using membership queries, with fixed-degree
polynomial dependence on the ambient dimension.
For learning from random examples,
\cite{DIKZ25} developed a general approach
for broad classes of finite-valued MIMs in the presence of
adversarial label noise. Their applications include approximate
agnostic learners for intersections of halfspaces and multiclass
linear classifiers.
\cite{DIKR25} (see also \cite{rev-DLB25} for a similar result) 
developed an analogous framework
for real-valued MIMs under square loss.
Both works use conditional low-degree moments to iteratively
approximate the relevant subspace, and establish qualitatively
matching SQ lower bounds on the dimension dependence.
Their conditional-moment criteria also generalize the
generative-exponent condition for single-index parameter
recovery studied in~\cite{rev-DPLB24}. 

Other work on real-valued SIMs and MIMs includes analyses of
gradient methods and moment-based subspace
recovery~\cite{rev-DLS22,rev-ABAM23, rev-LVM26},
as well as agnostic learners for
SIMs~\cite{rev-WZDD24,rev-WZDD25}. 

The closest prior work is~\cite{diakonikolas2026robust}, 
which developed a logconcave partitioning procedure and used it
to construct weak partial learners that can be boosted to
approximate agnostic and testable agnostic learners.

\paragraph{AI Disclosure.} 
The authors set the research direction, developed all the ideas,
and proved all the theorems in this paper.
AI was used to audit and edit the proofs and 
algorithms for minor errors, and to improve the presentation. 
The authors reviewed the feedback and incorporated minor corrections.

\paragraph{Organization.}
The paper is structured as follows.
In \Cref{sec:PRELIMINARIES}, we introduce the necessary
definitions and preliminaries.
In \Cref{sec:relaxed-gaussian-partitioning}, we develop our partitioning
primitive for smooth distributions and establish its guarantees.
In \Cref{sec:agnostic}, we use this primitive to obtain our
agnostic learning algorithm, establishing
\Cref{thm:main-learning}, and derive consequences for
concrete concept classes.
In \Cref{sec PU}, we present and prove our PUI learning result,
\Cref{thm:pumim}, as well as our statistical query lower bound,
\Cref{thm:sq-pu-general}.
{In \Cref{sec:struct-intersections}, we give PUI learners
for intersections of halfspaces and low-dimensional convex
concepts, together with an SQ lower bound for intersections.
In \Cref{sec:reliable}, we reduce one-sided agnostic reliable
learning to PUI learning. In \Cref{sec:tds-pu}, we show that
the logarithmic dependence on $1/\eps$ in the dimension exponent
is impossible for PQ and TDS learning in the SQ model, even for
intersections of two halfspaces. The appendix also contains an
SQ lower bound for agnostic learning with a multiplicative
approximation guarantee (\Cref{sec:multiplicative-sq}) and our
testable agnostic learning results (\Cref{sec:testable}).}

\section{Preliminaries and Notations}
\label{sec:PRELIMINARIES}

\paragraph{Basic notation.}
For $n\in\mathbb Z_+$, let $[n]:=\{1,\ldots,n\}$.
We use bold lowercase letters for vectors and bold uppercase letters
for matrices and tensors.
For $\mathbf x\in\mathbb R^d$, $x_i$ denotes its $i$-th coordinate,
and $\|\mathbf x\|_2=(\sum_{i=1}^d x_i^2)^{1/2}$ denotes its
Euclidean norm. We often omit the subscript and write $\|\mathbf x\|$.
We write $\mathbf x\cdot\mathbf y$ for the inner product and
$\mathbb S^{d-1}:=\{\mathbf x\in\mathbb R^d:\|\mathbf x\|_2=1\}$
for the unit sphere.
For a subspace $V\subseteq\mathbb R^d$, let $V^\perp$ denote its
orthogonal complement and write $\operatorname{proj}_V\mathbf x$,
or $\mathbf x_V$, for the orthogonal projection of $\mathbf x$ onto $V$.
We denote the linear span of a set of vectors $L$ by
$\operatorname{span}(L)$, and write
$V+W:=\{\mathbf v+\mathbf w:\mathbf v\in V,\mathbf w\in W\}$.
Let $\mathbf e_i$ be the $i$-th standard basis vector and
$\mathbf I_d$ the $d\times d$ identity matrix.
The Frobenius norm $\|\mathbf A\|_F$ of a matrix or tensor is the
square root of the sum of its squared entries.
For $A\subseteq\mathbb R^d$, define
$\operatorname{dist}(\mathbf x,A):=\inf_{\mathbf a\in A}
\|\mathbf x-\mathbf a\|_2$.
Let $R = \{\x \mid \bw^{(i)}\cdot\x \le b_i, i \in [m]\}$ be a polyhedral cell in $\R^d$ and let $V_R=\mathrm{span}(\bw^{(1)},\dots,\bw^{(m)})$ be the subspace spanned by the constraints that define $R$. We define $\diam(R) = \sup \{\norm{\bu-\bv} \mid \bu,\bv \in \proj_{V_R}(R)\}$ to be the diameter of $R$.

We use standard asymptotic notation, with $\widetilde O(\cdot)$
suppressing polylogarithmic factors.
We write $a\lesssim b$ when $a\leq Cb$ for an absolute constant $C>0$.
The notation $\operatorname{poly}(\cdot)$ and
$\operatorname{polylog}(\cdot)$ denotes polynomial and
polylogarithmic dependence, respectively.

\paragraph{Probability notation.}
We write $\E_{\mathbf x\sim D}[g(\mathbf x)]$ for expectation
under $D$ and $\Pr_{\mathbf x\sim D}[\mathcal E]$ for the probability
of an event $\mathcal E$.
We omit the distribution when it is clear from context and write
$D(A):=\Pr_{\mathbf x\sim D}[\mathbf x\in A]$.
For a joint distribution $D$ over $(\mathbf x,y)$, let
$D_{\mathbf x}$ denote its feature marginal.
We write $\mathcal N(\boldsymbol\mu,\mathbf I_d)$ for the Gaussian
distribution with mean $\boldsymbol\mu$ and identity covariance;
$\mathcal N(\mathbf0,\mathbf I_d)$ is the standard Gaussian.
We write $Q\ll P$ if $Q$ is absolutely continuous with respect to $P$, meaning that $P(E)=0$ implies $Q(E)=0$ for every measurable set $E$. This guarantees the existence of the density ratio $r=\mathrm dQ/\mathrm dP$, uniquely defined $P$-almost everywhere.
For discrete probability distributions $P=(p_i)_{i=1}^m$ and $Q=(q_i)_{i=1}^m$, their squared Hellinger distance is $H^2(P,Q):=1-\sum_{i=1}^m\sqrt{p_iq_i}$.
For a distribution $P$, a measurable function $f:\R^d\to\R$,
and $1\leq p<\infty$, we define the $L_p(P)$ norm of $f$ by
$\|f\|_{L_p(P)}:=(\E_{\x\sim P}[|f(\x)|^p])^{1/p}$.

\begin{definition}[Degree-$\ell$ $L_1$ sandwiching polynomials]
\label{def:l1-sandwiching-polynomials}
Let $\mathcal H$ be a class of Boolean functions on $\R^K$, let
$\mathcal D$ be a class of distributions on $\R^K$, and let
$\ell:(0,1]\to\Z_+$ be integer-valued. We say that $\mathcal H$ admits
degree-$\ell$ $L_1$ sandwiching polynomials over $\mathcal D$ if, for every
$c\in(0,1)$, every $g\in\mathcal H$, and every $\mu\in\mathcal D$, there
exist polynomials $p^l,p^u$ of degree at most $\ell(c)$ such that
$p^l(\bz)\leq g(\bz)\leq p^u(\bz)$ for every $\bz\in\R^K$, and  
$\E_{\bz\sim\mu}\!\left[|p^u(\bz)-p^l(\bz)|\right]\leq c.$
\end{definition}

\begin{definition}[Directional Moment matching]
\label{def:moment-matching}
Let $P$ and $Q$ be distributions over $\R^d$, and let
$W\subseteq\R^d$ be a subspace. We say that $P$ and $Q$ match their
first $k$ directional moments on $W$ with slack $\Delta$ if, for every unit vector
$\bv\in W$ and every $i\in[k]$,
$
\left|
\E_{\x\sim P}[(\bv\cdot \x)^i]
-
\E_{\x\sim Q}[(\bv\cdot \x)^i]
\right|
\leq \Delta.
$
When $W=\R^d$, we simply say that $P$ and $Q$ match their first $k$ directional
moments with slack $\Delta$.
\end{definition}

We refer the reader to \Cref{sec:omitted-preliminaries} for the omitted preliminaries.

\section{Gaussian Space Partitioning in Quasi-Polynomial Time}
\label{sec:relaxed-gaussian-partitioning}

In this section, we give the main algorithmic result in this paper, \Cref{thm:relaxed-gaussian-partitioning}, which formalizes our Gaussian Partitioning primitive. First, we give an overview of our approach. 
In \Cref{sec:partitioning-alg-black-box-disc}, we present the algorithm and prove the main helper lemmas supporting the analysis. In \Cref{sec:partitioninalg-proof}, we provide the complete analysis of our algorithm, and in \Cref{sec:relaxed-discrepancy-finder}, we prove the guarantees of the Moment Discrepancy Finder subroutine  (\Cref{thm:standalone-capped-finder}), which is needed for the analysis.

As mentioned in the introduction, our algorithm, given samples from a distribution $Q$ that is smooth with respect to $\cN(\vec 0,\vec I_d)$, efficiently partitions the space and marks a subcollection of cells, which we will refer to as \emph{certified}. Each marked cell $R$ satisfies a mass lower bound, a diameter bound (within the subspace $V_R$ spanned by its constraint normals), and a directional moment matching guarantee in $V_R^{\perp}$, as stated below. These certified cells together contain almost all of the probability mass of $Q$.
We state the formal result below.

\begin{theorem}[Gaussian Partitioning Algorithm]
\label{thm:relaxed-gaussian-partitioning}
Let $k,d\in\N_+$, $q\geq1$, $\sigma\in(0,1]$.
Let $P=\cN(\vec{0},\bI_d)$ and suppose
that $Q$ is a distribution on $\R^d$ that is $(q,\sigma)$-smooth with respect to $P$.
There is an algorithm such that given 
$\eta,\delta,\Delta\in(0,1/2)$, it draws 
at most $d^k B$
samples from $Q$,
runs in time $d^{O(k)}B$ for  $B=\exp\left(
       \frac{O((qk)^{O(k)})}{\Delta^{O(1)}}
       \log^{O(k)}(\frac{2}{\sigma\eta})\,
       \log(\frac{2}{\delta})\right)$, and, with probability at least $2/3$,
returns a polyhedral partition $\mathcal U$ of $\R^d$ with
$|\mathcal U|\leq B$ and a  collection
$\mathcal G\subseteq\mathcal U$ satisfying
$Q\left(\bigcup_{R\in\mathcal G}R\right)\geq1-\eta$.
Every $R\in\mathcal G$ satisfies:
\begin{enumerate}[label=(\roman*)]
 \item $Q(R)\geq1/B$.
 \item
 $\diam(R)\leq\delta$,
 where the diameter is measured in the constraint subspace $V_R$.
 \item $P\mid R$ and $Q\mid R$
 match their first $k$ directional moments on $V_R^\perp$ with slack $\Delta$.
\end{enumerate}
\end{theorem}
To comment on the complexity bound, it is useful to consider how the parameters of \Cref{thm:relaxed-gaussian-partitioning} are chosen in our subsequent learning applications.
For example, in our agnostic learning application (\Cref{thm:main-learning}), the moment order $k$, moment slack $\Delta$, and discarded mass $\eta$ depend only on the complexity of the model class and the relative approximation parameter $c$. For a class of bounded complexity, such as Boolean functions of a constant number of halfspaces, and constant $c$, these parameters can therefore be held constant. The desired additive accuracy $\eps$ enters through boosting, which uses distributions with smoothness parameter $\sigma\eps$. The diameter parameter $\delta$ can be taken polynomially small in $\sigma\eps$, contributing only a factor of $\log(2/\delta)$ to the exponent. These choices imply that, for fixed $q$, the complexity is polynomial in $d$ and quasi-polynomial in $1/(\sigma\eps)$. 

\begin{remark}[Comparison with Logconcave Partitioning \cite{diakonikolas2026robust}]
{\em A quasipolynomial-complexity logconcave partitioning result also appears in
\cite{diakonikolas2026robust}, but our theorem provides several
improvements in the Gaussian setting. The main improvement is
the diameter bound in outcome~(ii) of
\Cref{thm:relaxed-gaussian-partitioning}.
As discussed in the introduction, for targets with bounded
Gaussian surface area, small cell diameter allows us to fix
the coordinates in each cell's constraint subspace with only
a small total approximation error.
The resulting function on each cell depends only on $V_R^\perp$,
where the directional moment-matching guarantee applies.
This allows us to transfer certain Gaussian learning guarantees
to the conditional distributions on the certified cells,
which is vital for our improved complexity results.
Second, our partitioning theorem applies to general
$(q,\sigma)$-smooth distributions, whereas
\cite{diakonikolas2026robust} considers only the case $q=1$.
Finally, our complexity bound has an improved dependence on
the moment order $k$.}
\end{remark}

Before diving into the details of \Cref{thm:relaxed-gaussian-partitioning}, 
we first give a rough overview of the intuition behind the algorithm. 
As in \cite{diakonikolas2026robust}, our starting point is a \emph{moment discrepancy finder}.
This is a routine that, given sample access to a distribution $Q$, either certifies that $Q$ approximately matches the first $k$ directional moments of a standard Gaussian or returns a direction $\bv$ along which some moment of order $i\in[k]$ does not match.

We show in~\Cref{sec:relaxed-discrepancy-finder} that if $Q$ is $(q,\sigma)$-smooth with respect to a Gaussian distribution, or, more generally, if the expected Hermite tensors under $Q$ have small Frobenius norms, then such a routine can be implemented efficiently. In fact, the latter condition gives us a way to test whether the routine can perform the required moment discrepancy task reliably within a fixed computational budget. Formally, we state the following result for the task of moment discrepancy finding, whose proof and analysis are deferred to \Cref{sec:relaxed-discrepancy-finder}.

\begin{proposition}[Moment Discrepancy Finder]
\label{thm:standalone-capped-finder}
Let $k,d\in\Z_+$, $\Delta\in(0,1/2)$, and $V,H\geq1$.
Let $P=\cN(\vec{0},\bI_d)$ and let $Q$ be a distribution over $\R^d$
satisfying
$\E_{\x\sim Q}[p(\x)^2]
    \leq V\|p\|_{L_2(P)}^2
    $  for every polynomial $p$ of degree at most $k$.

\Cref{alg:standalone-moment-discrepancy-finder} draws 
$d^k O(k^{O(k)})\frac{V}{\Delta^2}
$
samples from $Q$, runs in 
$d^{O(k)}
      \exp\left(O(k^{O(k)})\frac{H^2}{\Delta^2}
                         \log\frac{H}{\Delta}\right)V^{O(1)}
$ time 
and returns $\textsc{Reject}$, $\textsc{None}$, or a unit direction.
With probability at least $2/3$, all of the following hold:
\begin{enumerate}[leftmargin=*]
    \item If the algorithm returns a direction $\bw$, then for some
    $\ell\in[k]$,
    \[
        \left|\E_{\x\sim Q}[(\bw\cdot\x)^\ell]
             -\E_{\x\sim P}[(\bw\cdot\x)^\ell]\right|
        \geq3\Delta/8.
    \]
    \item If the algorithm returns $\textsc{None}$, then for every unit
    vector $\bv$ and every $\ell\in[k]$,
    \[
        \left|\E_{\x\sim Q}[(\bv\cdot\x)^\ell]
             -\E_{\x\sim P}[(\bv\cdot\x)^\ell]\right|<\Delta.
    \]
    \item If every Hermite moment tensor
    $\bA_i=\E_{\x\sim Q}[\mathcal H_i(\x)]$, $i\in[k]$, satisfies
    $\|\bA_i\|_F \le  H$, the algorithm does not return
    $\textsc{Reject}$.
\end{enumerate}
\end{proposition}

We now give the intuition for constructing the desired partition using a moment discrepancy finder as a black box. We focus on the case $q=1$.
Suppose that we find a direction $\bv$ from our moment discrepancy finding algorithm. 
Since $Q$ has a moment discrepancy along $\bv$ and is smooth with respect to the standard Gaussian $P$, the argument in \cite{diakonikolas2026robust} yields a threshold $t\in\R$ such that the Bernoulli distributions of $\Ind\{\bv\cdot\x\ge t\}$ under $\x\sim Q$ and $\x\sim P$ have large Kolmogorov distance. This threshold splits the space into two halfspaces.
To obtain a partition into cells of small diameter, in addition to the split used in \cite{diakonikolas2026robust}, we further refine each half into short intervals along $\bv$.
The observation is that in expectation, the weight of $Q\mid R$ with respect to $P\mid R$ 
is not smaller than the weight of $Q$ with respect to $P$.
This implies that, if we recursively run this procedure over each $R$, we would expect the weight of $Q\mid R$ with respect to $P\mid R$ to become either extreme (very small or very large)  for most cells. Thus, the procedure should stop after a small number of rounds, with the desired moment matching holding on cells covering all but a small amount of $Q$-mass.

However, such a naive analysis does not immediately imply \Cref{thm:relaxed-gaussian-partitioning}. 
There are still several important challenges. First, we must ensure that the total number of cells does not grow exponentially in $1/\sigma$.
Second, although each step introduces a new direction $\bv$ and a fine grid along $\bv$, the diameter of a cell $R$ also depends on the dimension of the subspace defining $R$,
hence we must show that this dimension does not grow too large.
Furthermore, the intuition above requires $Q\mid R$ to remain sufficiently smooth with respect to $P\mid R$ as we recursively apply the procedure to each cell $R$.
For $q>1$, we must therefore show that this smoothness property holds on cells containing most of the $Q$-mass.

\subsection{Partitioning Algorithm} \label{sec:partitioning-alg-black-box-disc}
In this subsection, we present the Gaussian partitioning algorithm,
using a moment discrepancy finder as a black box.
We first show how moment discrepancies yield useful one-dimensional
splits that allow us to keep the partition depth small.
We then describe the algorithm and establish the lemmas needed
to bound the total $Q$-mass of uncertified cells.

Following the discussion above, we introduce our goals when splitting cells along each direction. Namely, we aim to obtain a large separation between the probabilities assigned by $P$ and $Q$ to the two sides of the split (which will help us control the number of cells), and to subdivide all but the tail regions into intervals of width at most $\delta_0$ ($\delta_0$ will be set much smaller than $\delta$ so that the overall diameter is at most $\delta$).

\begin{definition}[Gaussian Good Split and Refinement]
\label{def:relaxed-split-refinement}
Let $\kappa\in(0,1)$ and $T,\delta_0>0$ be parameters.
Let $P=\cN(0,1)$ and let $Q$ be a distribution on $\R$.
A threshold $t$ is a $\kappa$-Gaussian 
good split if the Bernoulli distributions with parameters
$p_t:=\Pr_{z\sim P}[z\geq t]$ and
$q_t:=\Pr_{z\sim Q}[z\geq t]$ satisfy
$H^2(p_t,q_t)=1-\sqrt{p_tq_t}
-\sqrt{(1-p_t)(1-q_t)}\geq\kappa$, where $H^2$ is the squared Hellinger distance.
For $t\in[-T,T]$, subdivide each nonempty interval $[-T,t)$ and
$[t,T)$ into consecutive half-open intervals of length $\delta_0$,
with possibly shorter end intervals, and include the two intervals
$(-\infty,-T)$ and $[T,\infty)$. The resulting partition of $\R$ refines the binary split at $t$ and contains at most $\lceil2T/\delta_0\rceil+3$ intervals.
\end{definition}

We start by showing that, if $Q$ is $(q,\sigma)$-smooth with respect to $P$, then a moment discrepancy of size $\Delta$ between $Q$ and $P$ implies the existence of a $\kappa$-Gaussian good split, with $\kappa$ inverse-polylogarithmic in $1/\sigma$ for fixed $q,k,\Delta$.
Furthermore, such a good split can be efficiently found and refined.
The intuition here is that, for a general distribution $Q$, a moment discrepancy does not necessarily imply the existence of a quantitatively good split, because a small amount of mass far in the tails can create a large moment discrepancy. However, the smoothness of $Q$ controls the contribution of the tails to its low-degree moments. Thus, once a discrepancy exists, there must be a threshold $t$ in a bounded interval such that $Q(z>t)$ differs quantitatively from $P(z>t)$, which also gives a quantitative lower bound on the squared Hellinger distance. By checking empirical tail probabilities at candidate thresholds, we can efficiently find such a good split and refine it.

\begin{lemma}[Moment Discrepancy implies a Gaussian Good Split]
\label{lem:relaxed-moment-good-split}
Let $P=\cN(0,1)$, and let $Q$ be a distribution over $\R$ {that is $(q,\sigma)$-smooth
with respect to $P$, where $q\geq1$ and $\sigma\in(0,1]$.}
Let $k\in\N_+$ and $\Delta\in(0,1)$, and suppose that
{$\abs{\E_{z\sim P}[z^i]-\E_{z\sim Q}[z^i]}\geq\Delta$
for some $i\in[k]$.}
Define
{$\Lambda:=q\bigl(k\log(2qk)+\log(16/(\sigma\Delta))\bigr)$}.
{There are universal constants $C_0,C>1$ such that, for every
$b\geq C_0\sqrt{\Lambda}$, there exists a threshold $t\in[-b,b]$
satisfying
\[
 \abs{\Pr_{z\sim P}[z\geq t]-\Pr_{z\sim Q}[z\geq t]}
 \geq\frac{\Delta}{4b^k}.
\]
The threshold can also be chosen to be a $\kappa$-Gaussian good
split, where $\kappa:=\Delta^2/(C\Lambda)^k$.}
Moreover, for every $\delta\in(0,1)$, there is an algorithm that draws
$
    O\left(\frac{(C\Lambda)^k}{\Delta^2}
    {\log\frac{2}{\delta}}\right)
    ={\polylog_{q,k,\Delta}(2/\sigma)\log(2/\delta)}
$
samples from $Q$, runs in sample-polynomial time, and
{outputs a threshold satisfying these conclusions}
with probability at least $1-\delta$.
\end{lemma}

\begin{proof}[Proof of \Cref{lem:relaxed-moment-good-split}]
{Fix any $i\in[k]$ witnessing the moment discrepancy.
Smoothness implies $Q\ll P$, since $P(E)=0$ implies $Q(E)=0$.
Thus, $Q$ is absolutely continuous with respect to Lebesgue measure.}
We first show that truncating to an interval of radius
{$O(\sqrt{\Lambda})$} changes the moment discrepancy by at most
$\Delta/8$.
Let $L=C_0\sqrt{\Lambda}\geq1$, where $C_0>1$ is a sufficiently
large universal constant. 
Cauchy--Schwarz, the bound $\E_{z\sim P}[|z|^{2i}]\leq(2i)^i$,
and the standard Gaussian tail bound give
\begin{align*}
    \E_{z\sim P}\left[|z|^i\Ind\{|z|>L\}\right]
    &\leq
    \left(\E_{z\sim P}[|z|^{2i}]\right)^{1/2}
    \Pr_{z\sim P}[|z|>L]^{1/2}\\
    &\leq \sqrt{2}(2i)^{i/2}e^{-L^2/4}
    \leq {\Delta/16}.
\end{align*}
For $t>0$, smoothness and the Gaussian tail bound give
\[
    \Pr_{z\sim Q}[|z|>t]
    \leq\frac{\Pr_{z\sim P}[|z|>t]^{1/q}}{\sigma}
    \leq\frac{2^{1/q}}{\sigma}e^{-t^2/(2q)}
    \leq\frac2{\sigma}e^{-t^2/(2q)}.
\]
Writing the moment as an integral of tail probabilities, we obtain
\begin{align*}
    \E_{z\sim Q}[|z|^{2i}]
    =\int_0^\infty 2it^{2i-1}\Pr_Q[|z|>t]\,\d t
    \leq\frac2{\sigma}\int_0^\infty
        2it^{2i-1}e^{-t^2/(2q)}\,\d t
    =\frac2{\sigma}(2q)^i i!
    \leq\frac2{\sigma}(2qi)^i.
\end{align*}
In particular, all the moments used below are finite.
Applying Cauchy--Schwarz under $Q$ gives
\begin{align*}
    \E_{z\sim Q}\left[|z|^i\Ind\{|z|>L\}\right]
    \leq
    \left(\E_Q[|z|^{2i}]\right)^{1/2}
    \Pr_Q[|z|>L]^{1/2}
    \leq\frac2{\sigma}(2qi)^{i/2}e^{-L^2/(4q)}
    \leq\frac{\Delta}{16}.
\end{align*}
The last inequalities in both tail estimates hold for a sufficiently
large universal $C_0$, uniformly over $i\in[k]$, since
\[
    \frac{L^2}{q}
    =C_0^2\left(k\log(2qk)+\log\frac{16}{\sigma\Delta}\right)
    \geq C_0^2\left(i\log(2qi)+\log\frac{16}{\sigma\Delta}\right).
\]
Let $\clip_L(z):=\max\{-L,\min\{z,L\}\}$ and define
$g(z):=\clip_L(z)^i$.
Since $|z^i-g(z)|\leq |z|^i\Ind\{|z|>L\}$, the assumed moment
discrepancy and the preceding two tail estimates yield
\begin{equation}\label{eq:relaxed-truncated-moment-gap}
    \abs{\E_{z\sim P}[g(z)]-\E_{z\sim Q}[g(z)]}
    \geq \Delta-{\frac{\Delta}{16}}-\frac{\Delta}{16}
    =\frac{7\Delta}{8} \;.
\end{equation}
Define the restricted Kolmogorov discrepancy
\[
    d_{\mathrm K,L}(P,Q)
    :=\max_{t\in[-L,L]}
    \abs{\Pr_{z\sim P}[z\geq t]-\Pr_{z\sim Q}[z\geq t]}.
\]
The maximum exists because both tail probabilities are continuous.

\begin{claim}\label{claim:relaxed-moment-to-kolmogorov-gap}
We have $d_{\mathrm K,L}(P,Q)\geq 7\Delta/(16L^k)$.
\end{claim}
\begin{proof}[Proof of \Cref{claim:relaxed-moment-to-kolmogorov-gap}]
For every $x\in\R$, the clipped monomial satisfies the tail-integral
identity
\[
    g(x)
    =\int_0^L it^{i-1}
    \left(
        \Ind\{x\geq t\}+(-1)^i\Ind\{x\leq-t\}
    \right)\d t.
\]
Taking expectations and applying Fubini's theorem gives
\begin{align*}
    \E_P[g]-\E_Q[g]
    ={}&\int_0^L it^{i-1}
        \left(\Pr_P[z\geq t]-\Pr_Q[z\geq t]\right)\d t+(-1)^i\int_0^L it^{i-1}
        \left(\Pr_P[z\leq-t]-\Pr_Q[z\leq-t]\right)\d t.
\end{align*}
Since $P$ and $Q$ are absolutely continuous, a lower-tail discrepancy
at $-t$ equals the corresponding upper-tail discrepancy in absolute
value. All thresholds in these integrals lie in $[-L,L]$, so
\begin{align*}
    \abs{\E_P[g]-\E_Q[g]}
    &\leq 2d_{\mathrm K,L}(P,Q)\int_0^L it^{i-1}\,\d t\\
    &=2L^i d_{\mathrm K,L}(P,Q)
    \leq2L^k d_{\mathrm K,L}(P,Q),
\end{align*}
where the last inequality uses $i\leq k$ and $L\geq1$.
Combining this with \eqref{eq:relaxed-truncated-moment-gap}
gives the claimed lower bound.
This completes the proof of \Cref{claim:relaxed-moment-to-kolmogorov-gap}.
\end{proof}

Set $\rho:=\Delta/(4L^k)$. By
\Cref{claim:relaxed-moment-to-kolmogorov-gap}, there is a threshold
$t_\star\in[-L,L]$ such that $|p_\star-q_\star|\geq7\Delta/(16L^k)=7\rho/4$,
where $p_\star:=\Pr_{z\sim P}[z\geq t_\star]$ and
$q_\star:=\Pr_{z\sim Q}[z\geq t_\star]$.
For Bernoulli distributions, $H^2(\operatorname{Bern}(p),\operatorname{Bern}(q))\geq(p-q)^2/2$.
Since $L^{2k}=C_0^{2k}\Lambda^k$, choosing the universal constant
$C$ sufficiently large gives
\[
    H^2(\operatorname{Bern}(p_\star),\operatorname{Bern}(q_\star))
    \geq\frac{\rho^2}{2}
    \geq\frac{\Delta^2}{(C\Lambda)^k}
    =\kappa \;.
\]
For every $b\geq L$, the same threshold belongs to $[-b,b]$ and has gap at least $\rho\geq\Delta/(4b^k)$, proving that such a threshold exists whenever some degree $i\in[k]$ witnesses the moment discrepancy.

It remains to find such a threshold efficiently. We draw $n$ samples from $Q$ and sort them as $z_{(1)}\leq\cdots\leq z_{(n)}$. For each sample point $z_{(a)}\in[-L,L]$, the empirical probabilities of $z\geq z_{(a)}$ and $z>z_{(a)}$ are $(n-a+1)/n$ and $(n-a)/n$, respectively. We compare both with the Gaussian upper-tail probability $1-\Phi(z_{(a)})$. We also check the endpoints $-L$ and $L$ using their empirical upper-tail probabilities, and choose the threshold $\widehat t$ giving the largest absolute difference. Since $Q$ is absolutely continuous, it assigns zero probability to each point, so $Q(z\geq t)=Q(z>t)$.

Between consecutive sample points, the empirical upper-tail
probability is constant and the Gaussian upper-tail probability
is monotone.
Thus, checking both endpoints and both sides of each sample point
in the interval suffices to maximize the empirical discrepancy
over $[-L,L]$.

By the DKW inequality (see \Cref{fact:dkw}), for
\[
    n=\left\lceil8\rho^{-2}\log\frac2\delta\right\rceil
    =O\left(\rho^{-2}\log\frac2\delta\right) \;,
\]
with probability at least $1-\delta$, all one-sided empirical
tail probabilities are within $\rho/4$ of their population values.
On this event, the empirical discrepancy at $t_\star$ is at least
$7\rho/4-\rho/4=3\rho/2$.
Consequently, the maximizing threshold satisfies
\[
 |p_{\widehat t}-q_{\widehat t}|
 \geq3\rho/2-\rho/4=5\rho/4
 \geq\rho\geq\frac{\Delta}{4b^k}.
\]
It follows that
\[
    H^2\left(
        \operatorname{Bern}(p_{\widehat t}),
        \operatorname{Bern}(q_{\widehat t})
    \right)
    \geq\frac{\rho^2}{2}
    \geq\frac{\Delta^2}{(C\Lambda)^k}
    =\kappa.
\]
Moreover, $|\widehat t|\leq L\leq b$.
Since $\rho=\Delta/(4L^k)$ and
$L^{2k}=C_0^{2k}\Lambda^k$, the claimed sample bound follows.
Sorting the samples gives sample-polynomial running time.
All estimates use the same $k$-dependent radius and apply to any
witnessing degree $i\in[k]$; the algorithm does not need to know $i$.
This completes the proof of \Cref{lem:relaxed-moment-good-split}.
\end{proof}

\begin{algorithm}
\caption{\textsc{GaussianSpacePartition}}
\label{alg:relaxed-gaussian-partitioning}
\begin{algorithmic}[1]
 \State \textbf{Input:} Samples from $Q$ and parameters
         $d,k,q,\sigma,\eta,\delta,\Delta$,  
         a moment discrepancy finder
 $\A_m^\beta$ (\Cref{cor:relaxed-finder-amplification}).
 \State \textbf{Output:} A partition $\mathcal U$ and certified cells
         $\mathcal G$.
\medskip
 \State Set parameters  according to (\hyperref[par:relaxed-partition-parameters]{Parameter Choices of \Cref{alg:relaxed-gaussian-partitioning}}), 
 $\mathcal U_1=\{\R^d\}$;
         $\mathcal G=\mathcal B=\mathcal T=\mathcal E
           =\emptyset$.
 \For{$t=1,\ldots,L$}
  \State $\mathcal U_{t+1}=\emptyset$.
  \State Draw $\lceil8\xi^{-2}\log(2N_*/\beta)\rceil$ fresh samples
         from $Q$; let $\widehat q_R$ be their empirical mass
         in each $R\in\mathcal U_t$.
  \For{$R\in\mathcal U_t$}
   \State \textbf{if} $R$ is not terminal, classify it into
          $\mathcal B$ if $\widehat q_R\leq\xi/2$.
   \If{$R\in\mathcal G\cup\mathcal B\cup
                    \mathcal T\cup\mathcal E$}
      Add $R$ to $\mathcal U_{t+1}$.
   \Else
    \State Run $\A_m^\beta$ on the $V_R^\perp$ projection of
           $Q\mid R$, with parameters $k,\Delta,V,H$; 
           if $V_R^\perp=\{\vec{0}\}$, take the output to be
           $\textsc{None}$.
    \If{the output is $\textsc{None}$} 
       Add $R$ to $\mathcal G$ and $\mathcal U_{t+1}$.
    \ElsIf{the output is $\textsc{Reject}$}
       Add $R$ to $\mathcal E$ and $\mathcal U_{t+1}$.
    \Else
     \State Let $\bv\in V_R^\perp$ be the returned unit direction.
     \State Obtain $n_{\rm CDF}$ fresh samples from $Q\mid R$.
    \State Find a threshold $s\in[-T,T]$ maximizing the empirical
       Gaussian upper-tail gap in direction $\bv$, as in
       \Cref{lem:relaxed-moment-good-split}. \If{this maximum is below $2g$}
        Add $R$ to $\mathcal E$ and $\mathcal U_{t+1}$.
     \Else
      \State Refine $R$ along a maximizing threshold using
             \Cref{def:relaxed-split-refinement}.
      \State Add every child to $\mathcal U_{t+1}$;
             Add the two tail children to $\mathcal T$.
     \EndIf
    \EndIf
   \EndIf
  \EndFor
 \EndFor
 \State \Return $(\mathcal U_{L+1},\mathcal G)$.
\end{algorithmic}
\end{algorithm}

Given \Cref{lem:relaxed-moment-good-split}, we provide an overview of our main partitioning algorithm, \Cref{alg:relaxed-gaussian-partitioning}. The algorithm maintains a partition $\mathcal U_t$ of $\R^d$. At each iteration $t$, we select a subcollection $A_t\subseteq\mathcal U_t$ of cells that have not yet been certified to satisfy the requirements of \Cref{thm:relaxed-gaussian-partitioning}. For each $R\in A_t$, we call the moment discrepancy finder to find a direction $\bv\perp V_R$ and then subdivide $R$ into  small intervals along $\bv$.

Equivalently, the algorithm builds a tree rooted at $\R^d$, with each node corresponding to a cell created during its execution. The active cells in $A_t$ are those split at iteration $t$, and their children lie one level deeper in the tree. The final partition consists of the cells corresponding to the leaves of this tree.   Below, we give an overview of the analysis, which also explains some additional steps needed beyond this general template.

\paragraph{Bounding the Recursion Depth.}
The above structure implies that the key quantity to control during the analysis
is the depth $L$ of the tree at termination. Since each split creates
many new cells, the total number of cells can grow exponentially
with $L$. The depth also controls the cell diameters: if a cell is
defined by intervals of width at most $\mu$ along at most $L$ mutually
orthogonal directions, then its diameter in its constraint subspace
is at most $\mu\sqrt{L}$.

The way we keep track of the depth is to keep track of a potential function 
$\Psi_t:=\sum_{R \in A_t}\sqrt{p_R q_R}$, where $p_R = \Pr_{P}[R]$ and $q_R = \Pr_Q[R]$.
As we will see later, under the smoothness condition, for most of the cells, 
the ratio $q_R/p_R$ can be controlled. Thus, $\sum_{R \in A_t} q_R$ can be controlled 
as a linear function of $\Psi_t$ up to a small additive error.
This implies that if we can show $\Psi_t$ drops fast with respect to $t$, 
then we can stop when $\Psi_t$ is small enough to make the tree have a reasonable depth.

To see why a fast decreasing rate is possible, 
suppose we split a single cell $R$ into new cells $R_1,\dots,R_M$.  
Then we have $\sum_{j=1}^M\sqrt{p_{R_j}q_{R_j}}
        =\sqrt{p_Rq_R}\sum_{j=1}^M\sqrt{P_R(R_j)Q_R(R_j)},$ 
where $P_R(Q_R)$ is the distribution of $P(Q)$ conditioned over $R$.
Notice that $1-\sum_{j=1}^M\sqrt{P_R(R_j)Q_R(R_j)}$ is exactly the squared Hellinger distance 
between the distributions supported over $R_i, i \in [M]$ with probability $P_R,Q_R$. 
This implies that as long as the intervals we construct along 
a direction $\bv$ can create a Hellinger distance at least $\kappa$, 
$\Psi_t$ drops at a rate of $\exp(-\kappa t)$.
This is exactly the outcome of \Cref{lem:relaxed-moment-good-split}.

\paragraph{Selecting Cells to Split.}
Thus, the criteria for selecting the cells in $A_t$ to be split are clear. First, $Q\mid R$ must have a moment discrepancy relative to $P\mid R$ in some direction in $V_R^\perp$. Second, $q_R$ must be sufficiently large to allow efficient sampling from $Q\mid R$. Third, $R$ must not have any tail constraints, which would make its diameter unbounded. Finally, $Q\mid R$ must be sufficiently smooth with respect to $P\mid R$, so that we can reliably run the moment discrepancy finder and find a good split that provides the progress needed to control the recursion depth.

\paragraph{Conditional Smoothness.}
Notice that the first three conditions can be checked using empirical estimates. However, verifying that $Q\mid R$ is sufficiently smooth with respect to $P\mid R$ can be much harder, even when $Q$ itself is smooth with respect to $P$.
To get an intuition, we consider the special case of $q=1$, where the smoothness is equivalent to the weight between $Q\mid R$ and $P \mid R$. In this case, since $Q$ has $\sigma$ weight with respect to $P$, the weight between $Q \mid R$ and $P \mid R$ is exactly $\sigma q_R/p_R$. This implies that we only need to check the ratio between $q_R$ and $p_R$. And clearly, the total $Q$ mass of the cells that are not smooth cannot be too large. 
However, for the case of $q>1$, the situation is different.
Since we only know that $Q(S) \le P(S)^{1/q}/\sigma$, for a set $S \subseteq R$, even if $q_R/p_R$ is large, $(Q\mid_R(S))/(P\mid_R(S)^{1/q})$ still has a linear dependence on $p_R^{1/q-1}$. So, if $p_R$ is small, potentially $Q \mid R$ can be highly unsmooth with respect to $P \mid R$ under the $q$-smooth definition. This yields two undesired consequences. 
First, without a smoothness guarantee, we may not be able to run the moment discrepancy finder reliably within the given computational budget. In particular, failure to return a direction does not necessarily imply that $Q\mid R$ and $P\mid R$ approximately match moments.
So potentially, the moment discrepancy finder may miss the cells that are not good enough. Second, even if we have found a direction $\bv$, along which the moment discrepancy exists, as discussed in \Cref{lem:relaxed-moment-good-split}, a quantitative good split could not exist due to the non-smoothness, which prevents us from controlling the progress and the depth of \Cref{alg:relaxed-gaussian-partitioning}.
To avoid this, we notice that the success of \Cref{lem:relaxed-moment-good-split} can be checked by looking at the empirical probability. If a moment discrepancy exists but we do not find a good split, then we can simply discard this cell, because of the unsmoothness. Furthermore, \Cref{thm:standalone-capped-finder} indeed succeeds under a broader testable condition, where the expectation of Hermite tensor is not very large under $Q$. Since a smooth distribution must satisfy this condition, when we find the algorithm may not reliably succeed, we can simply discard this cell.
The only remaining issue is to show that we will not discard too much $Q$-mass. Fortunately, in \Cref{lem:conditional-density-averaging}, we show that if $Q$ is $(q,\sigma)$-smooth with respect to $P$, then {outside a collection of cells of small total $Q$-mass}, $Q \mid R$ is smooth with respect to $P \mid R$, with a slightly worse parameter, which means we will only discard a tiny $Q$ mass in this way.

To conclude the analysis, it remains to show that when \Cref{alg:relaxed-gaussian-partitioning} stops, the fraction of cells that have low $Q$-mass or contain a tail inequality is small. In \Cref{lem:relaxed-tail-mass} and \Cref{lem:relaxed-small-mass}, we provide separate analyses.

Before the formal analysis, we first state 
the parameter choices of \Cref{alg:relaxed-gaussian-partitioning}.
\paragraph{Parameter Choices of \Cref{alg:relaxed-gaussian-partitioning}.}\phantomsection\label{par:relaxed-partition-parameters}
Choose universal constants $C_0,C>1$ sufficiently large, with
$C_0$ at least the constant required by
\Cref{lem:relaxed-moment-good-split}.
Define the error tolerances
$\lambda=\varepsilon=\tau_0=\eta/100$,
the conditional smoothness parameter
$\sigma_*=\sigma\eta/(400q)$,
the polynomial norm threshold
$H=e\left(4q\log(4eq/\sigma_*)\right)^{k/2}$,
the auxiliary tail parameter
$\Lambda_*=2q\left(k\log(4qk)+\log(64/(\sigma_*\Delta))\right)$,
the truncation bound
$b=C_0\sqrt{\Lambda_*}$,
the probability-gap parameter
$g=\Delta/(100b^k)$,
and the potential contraction parameter
$\kappa=g^2/8$.

Define the potential bound
$M=1/(\sigma^q\tau_0^{q-1})$,
the recursion depth bound
$L=1+\left\lceil\kappa^{-1}\log(100\sqrt M/\eta)\right\rceil$,
the grid width
$\delta_0=\delta/\sqrt L$,
the tail cutoff
$T=\max\left\{b,\sqrt{2\log(200L/(\sigma\eta/100)^q)}\right\}$,
the grid-size bound
$m=\lceil2T/\delta_0\rceil+3$,
and the small-mass potential exponent
$s_0=\kappa/(2(\kappa+\log m))$.

Finally, define the $Q$-mass cutoff
$\xi=\left((\sigma\varepsilon)^q
\lambda^{1/2+s_0}\right)^{1/(2s_0)}$,
the cell-count bound
$N_*=(m+1)^{L+1}$,
the per-call failure probability
$\beta=1/(1000(N_*+L+1))$,
the polynomial variance bound
$V=8q(4q-1)^k/(\sigma\xi)$,
and the CDF-estimation sample size
$n_{\rm CDF}=\left\lceil8g^{-2}\log(2/\beta)\right\rceil$.

\paragraph{Mass Analysis of Active Cells.}

We first use a potential analysis in 
\Cref{lem:relaxed-depth-analysis} to show that the $Q$-mass 
of the cells that are active during \Cref{alg:relaxed-gaussian-partitioning} 
drops exponentially fast. We remark that instead of directly working 
with a smooth distribution $Q$, we relax the assumption 
and work with a distribution $Q$ such that most of the $Q$-mass 
is supported over points whose density ratio $dQ/dP$ is bounded. 
As we show later in \Cref{lem:relaxed-smoothness-truncation}, 
any smooth distribution $Q$ satisfies this assumption.

\begin{lemma}[Mass of Active Cells Decreases Exponentially]
\label{lem:relaxed-depth-analysis}
Let $P=\cN(\vec{0},\bI_d)$, let $Q$ be $(q,\sigma)$-smooth with
respect to $P$, where $q\geq1$ and $\sigma\in(0,1]$, and write
$r=\mathrm dQ/\mathrm dP$. Assume $M\geq 1$ and $\tau_0\in[0,1]$ such that $Q(r>M)\leq\tau_0$.
Let $g>0$ be the split-validation parameter used by
\Cref{alg:relaxed-gaussian-partitioning}.
Let $L\in\N_+$ and $\kappa\in(0,1)$ satisfy
$\kappa\leq g^2/2$.
Consider \Cref{alg:relaxed-gaussian-partitioning} through
iteration $L$.
Assume the event that all empirical CDFs used to validate splits in these iterations differ from
their population CDFs by at most $g/4$ at every threshold.

For $t\in[L]$, denote by $A_t$ the cells that are split in round $t$,
and write $p_R=P(R)$ and $q_R=Q(R)$. Then
\[
    {\sum_{R\in A_t}q_R
    \leq\sqrt M\,e^{-\kappa(t-1)}+\tau_0.}
\]
In particular, for
$\eta\in(0,1)$, setting
$
    L=1+\lceil\kappa^{-1}
                   \log({100\sqrt M}/{\eta})\rceil
$
makes their total mass at most $\eta/100+\tau_0$.
\end{lemma}

\begin{proof}[Proof of \Cref{lem:relaxed-depth-analysis}]
For $t\in[L]$, define $\Psi_t:=\sum_{R\in A_t}\sqrt{p_Rq_R}$.
We first show that $\Psi_{t+1}\leq(1-\kappa)\Psi_t$ for $t<L$.

By assumption all empirical probabilities used to validate splits are within $g/4$ of their true values. Consider a split of a cell $R$ in a unit direction $\bv\in V_R^\perp$ at threshold $s$. Since the constraints defining $R$ depend only on $\proj_{V_R}\x$, the distribution of $\bv\cdot\x$ under $P\mid R$ remains $\cN(0,1)$.  The algorithm makes this split only if the empirical estimate of $\Pr_{\x\sim Q\mid R}[\bv\cdot\x\geq s]$ differs from $\Pr_{z\sim\cN(0,1)}[z\geq s]$ by at least $2g$. The corresponding true probabilities therefore differ by at least $2g-g/4\geq g$. The squared Hellinger distance between the resulting Bernoulli distributions is consequently at least $     \frac{g^2}{2}\geq\kappa$.  Thus, every split made by the algorithm is $\kappa$-Gaussian good.

Fix $R\in A_t$, and let $R_1,\ldots,R_m$ be all of its children,
including the two terminal tail  children. 
Let $P_R,Q_R$ be the distributions on these children, defined by $
    P_R(R_j)=P(R_j\mid R),
    Q_R(R_j)=Q(R_j\mid R), j\in[m].$
Both distributions have total mass one. Since the full child
partition refines the $\kappa$-good binary split,
\[
    H^2(P_R,Q_R)
    =1-\sum_{j=1}^m\sqrt{P_R(R_j)Q_R(R_j)}
    \geq\kappa.
\]

Indeed, the $\kappa$-good split partitions the space $\R^d$ restricted on $R$ into two parts, creating two Bernoulli distributions, with a squared Hellinger distance at least $\kappa$. Since $P_R,Q_R$ come from a refinement of the good split, their Hellinger distance cannot be smaller.
Consequently,
\[
    \sum_{j=1}^m\sqrt{p_{R_j}q_{R_j}}
    =\sqrt{p_Rq_R}
      \sum_{j=1}^m\sqrt{P_R(R_j)Q_R(R_j)}
    \leq(1-\kappa)\sqrt{p_Rq_R}.
\]
The sum over children that are split in the next round is no larger.
{All stopped or certified cells remain unchanged and are never
retested. Thus, every cell in $A_{t+1}$ is a child of a cell in $A_t$.}
Summing over $R\in A_t$ gives $\Psi_{t+1}\leq(1-\kappa)\Psi_t$.
Since $\Psi_1\leq1$, we obtain
\[
    \Psi_t\leq(1-\kappa)^{t-1}\leq e^{-\kappa(t-1)}.
\]
{Now we will use the above potential to bound the $Q$-mass of $A_t$. Let $b_R$ be the $Q$-mass of points of $R$ with a small density ratio, i.e.,  
\[
    b_R:=\int_R r\Ind\{r\leq M\}\,\mathrm dP \;.
\]
Since $b_R\leq Mp_R$ and $b_R\leq q_R$, we have
$b_R^2\leq Mp_Rq_R$. Therefore,
\begin{align*}
    \sum_{R\in A_t}q_R
    &=\sum_{R\in A_t}b_R
      +Q\left(\left(\bigcup_{R\in A_t}R\right)
                                      \cap\{r>M\}\right)\\
    &\leq\sqrt M\sum_{R\in A_t}\sqrt{p_Rq_R}+\tau_0 \leq\sqrt M\,e^{-\kappa(t-1)}+\tau_0.
\end{align*}
Every unprocessed leaf remaining after round $L$ is a child of a
cell in $A_L$, because every previously processed cell was either
split or stopped permanently. The children partition their parents,
so the total mass of these leaves is at most
\[
    \sum_{R\in A_L}q_R
    \leq\sqrt M\,e^{-\kappa(L-1)}+\tau_0.
\]
For the displayed choice of $L$,
$\sqrt M\,e^{-\kappa(L-1)}\leq\eta/100$.
This proves the lemma.}
\end{proof}
The following lemma shows that the density-ratio assumption in
\Cref{lem:relaxed-depth-analysis} follows from $(q,\sigma)$-smoothness.
For every $\eps\in(0,1)$, choosing
$M=1/(\sigma^q\eps^{q-1})$ ensures that the density ratio is
at most $M$ outside a set of $Q$-mass at most $\eps$.

\begin{lemma}[Approximately Bounded Density Ratio]
\label{lem:relaxed-smoothness-truncation}
Let $P,Q$ be distributions on the same measurable space, with $Q$
$(q,\sigma)$-smooth with respect to $P$, where $q\geq1$ and
$\sigma\in(0,1]$. Write $f=\mathrm dQ/\mathrm dP$.
For every $\varepsilon\in(0,1)$, the set
\[
    G=\left\{\x:f(\x)\leq
        \frac{1}{\sigma^q\varepsilon^{q-1}}\right\}
\]
satisfies $Q(G)\geq1-\varepsilon$. Moreover, $Q\mid G$ is
$\bigl(1,(1-\varepsilon)\sigma^q\varepsilon^{q-1}\bigr)$-smooth
with respect to $P$.
\end{lemma}
We defer the proof of \Cref{lem:relaxed-smoothness-truncation}
to \Cref{app:partitioning-details}.

\paragraph{Mass Analysis of Discarded Cells.}
We next analyze the $Q$-mass of cells that will be discarded
during the execution of \Cref{alg:relaxed-gaussian-partitioning}.
In particular, these are cells whose defining constraints include
an unbounded interval and cells with small $Q$-mass.

First, we bound the total $Q$-mass of cells whose defining
constraints include a tail segment (the collection $\mathcal{T}$
in \Cref{alg:relaxed-gaussian-partitioning}).
\begin{lemma}[Negligible Mass of Gaussian Tail Cells]
\label{lem:relaxed-tail-mass}
Let $P=\cN(\vec{0},\bI_d)$, let $Q$ be $(q,\sigma)$-smooth with
respect to $P$, where $q\geq1$ and $\sigma\in(0,1]$,
and let $L\in\N_+$ and $\eta\in(0,1)$.
Choose $T>0$ satisfying
$
 T^2\geq2\log( {200L}/{(\sigma\eta/100)^q}).
$
Consider \Cref{alg:relaxed-gaussian-partitioning} through
iteration $L$. Then, for every $t\leq L+1$,
\[
 \sum_{R\in\mathcal T_t}q_R
 \leq\frac{(2Le^{-T^2/2})^{1/q}}{\sigma}
 \leq\frac{\eta}{100}.
\]
\end{lemma}

\begin{proof}[Proof of \Cref{lem:relaxed-tail-mass}]
Tail cells are carried forward unchanged, so
$\mathcal T_t\subseteq\mathcal T_{t+1}$.
It therefore suffices to bound the total mass of
$\mathcal T_{L+1}$.

Let $A_s$ be the cells
split in round $s$. Consider a set that will be split in round $s$, $ R\in A_s$  and
let $\bv_{R}$ be its split unit direction. 
Denote the tail children by 
$
    E_{R}:=R\cap
       \bigl(\{\x:\bv_{R}\cdot \x<-T\}
                    \cup\{\x:\bv_{R}\cdot \x\geq T\}\bigr).
$

Since $\bv_R\in V_R^{\perp}$ by Gaussian independence  
\[
    P(E_{R})
    =2P(R)\Phi(-T)
    \leq2P(R)e^{-T^2/2}.
\] Let 
$E=\bigcup_{R\in\mathcal T_{L+1}}R$. Using the above bound we have 
\begin{align*}
    P(E)
    &\leq\sum_{s=1}^L\sum_{R\in A_s}P(E_{R})\\
    &\leq2e^{-T^2/2}
             \sum_{s=1}^L\sum_{R\in A_s}P(R)
    \leq2Le^{-T^2/2}\;,
\end{align*}
where we used that the cells in $A_s$ are disjoint. 
Applying smoothness to the  event $E$,
\[
    \sum_{R\in\mathcal T_{L+1}}q_R
    =Q(E)
    \leq\frac{P(E)^{1/q}}{\sigma}
    \leq\frac{(2Le^{-T^2/2})^{1/q}}{\sigma}.
\]
The choice of $T$ gives
\[
    2Le^{-T^2/2}
    \leq\frac1{100}\left(\frac{\sigma\eta}{100}\right)^q
    \leq\left(\frac{\sigma\eta}{100}\right)^q.
\]
Consequently, $Q(E)\leq\eta/100$, which concludes the proof.
\end{proof}

Second, we bound the total $Q$-mass of cells with small
individual $Q$-mass (the collection $\mathcal{B}$
in \Cref{alg:relaxed-gaussian-partitioning}).
\begin{lemma}[Negligible Total Mass of Low-$Q$-Mass Cells]
\label{lem:relaxed-small-mass}
Let $Q$ be $(q,\sigma)$-smooth with respect to $P$.
For $t\leq L+1$, consider the partition $\mathcal U_t$ of \Cref{alg:relaxed-gaussian-partitioning},
and let $\mathcal B_t=\{R\in\mathcal U_t:Q(R)\leq\xi\}$.
Let $m\geq2$ bound the number of children in every refinement, and set
$
 s={\kappa}/{(2(\kappa+\log m))}.
$
Assuming that every executed split is $\kappa$-Gaussian good,
for every $\lambda\in(0,1)$,
\[
 \sum_{R\in\mathcal B_t}Q(R)
 \leq\lambda+\frac1\sigma
       \left(\frac{\xi^{2s}}{\lambda^{1/2+s}}\right)^{1/q}.
\]
In particular, for $\varepsilon\in(0,1)$, the choice
$
 \xi=\left((\sigma\varepsilon)^q
                   \lambda^{1/2+s}\right)^{1/(2s)}
$
makes this total $Q$-mass at most $\lambda+\varepsilon$.
\end{lemma}

\begin{proof}[Proof of \Cref{lem:relaxed-small-mass}]
Write $p_R=P(R)$ and $q_R=Q(R)$.
For each iteration $t$, define the potential
\[
 F_t\eqdef\sum_{R\in\mathcal U_t}(p_Rq_R)^{1/2-s}.
\]
Since $\mathcal U_1=\{\R^d\}$, we have $F_1=1$.
We show that a refinement cannot increase this potential.
Consider a cell $R$ split into $R_1,\ldots,R_j$, where $j\leq m$,
and put $p_i=P(R_i\mid R)$ and $q_i=Q(R_i\mid R)$.
All children, including both terminal tails, are retained in the
partition, so $(p_i)_{i=1}^j$ and $(q_i)_{i=1}^j$ are probability
vectors. The refinement contains the validated binary split and
hence by \Cref{fact:hellinger-refinement} satisfies
\[
 \sum_{i=1}^j\sqrt{p_iq_i}\leq1-\kappa.
\]
Since $0<1-2s<1$, concavity gives
\begin{align*}
 \sum_{i=1}^j(p_iq_i)^{1/2-s}
 \leq j^{2s}
             \left(\sum_{i=1}^j\sqrt{p_iq_i}\right)^{1-2s}\leq m^{2s}(1-\kappa)^{1-2s}\leq1.
\end{align*}
The last inequality follows since by the definition of $s$, 
$
 2s\log m+(1-2s)\log(1-\kappa)
 \leq2s\log m-(1-2s)\kappa=0.
$
Consequently,
\[
 \sum_{i=1}^j(p_{R_i}q_{R_i})^{1/2-s}
 \leq(p_Rq_R)^{1/2-s}.
\]

Stopped and certified cells are carried forward unchanged.
Applying the preceding inequality to each split in an iteration
proves $F_{t+1}\leq F_t$, hence $F_t\leq1$ for every $t$.

{Let $\mathcal B_t^-$ consist of the cells in $\mathcal B_t$ with
$q_R\leq\lambda p_R$, and put
$\mathcal B_t^+=\mathcal B_t\setminus\mathcal B_t^-$.
Disjointness implies
\[
 \sum_{R\in\mathcal B_t^-}q_R
 \leq\lambda\sum_{R\in\mathcal B_t^-}p_R\leq\lambda.
\]
For each $R\in\mathcal B_t^+$ we have
$p_R<q_R/\lambda\leq\xi/\lambda$ and $(p_Rq_R)^{1/2-s}\geq\lambda^{1/2-s}p_R^{1-2s}$.
Therefore,
\begin{align*}
 P\left(\bigcup_{R\in\mathcal B_t^+}R\right)
 =\sum_{R\in\mathcal B_t^+}p_R^{1-2s}p_R^{2s}\leq(\xi/\lambda)^{2s}
                 \sum_{R\in\mathcal B_t^+}p_R^{1-2s}\leq(\xi/\lambda)^{2s}\lambda^{-(1/2-s)}F_t
 \leq\frac{\xi^{2s}}{\lambda^{1/2+s}}.
\end{align*}
Apply global smoothness once to this union of cells to obtain
\[
 \sum_{R\in\mathcal B_t^+}q_R
 \leq\frac1\sigma
      \left(\frac{\xi^{2s}}{\lambda^{1/2+s}}\right)^{1/q}.
\]
Adding the bounds for the two collections proves the claim.
Substitution of the choice of $\xi$ completes the proof.}
\end{proof}

\paragraph{Conditional Smoothness and Polynomial Moment Bounds.}
Finally, conditioning on a cell $R$ can worsen the smoothness
parameters, so the moment discrepancy finder may return
$\textsc{Reject}$.
We prove two lemmas to show that cells rejected in this way have
small total $Q$-mass.
First, \Cref{lem:conditional-density-averaging} shows that if $Q$
is smooth with respect to $P$, then, for every partition, most of
the $Q$-mass lies in cells $R$ such that $Q\mid R$ is smooth with
respect to $P\mid R$, with slightly worse smoothness parameters.
Second, \Cref{lem:relaxed-polynomial-control} bounds the second
moments of low-degree polynomials and the norms of the Hermite
moment tensors under a smooth distribution, providing the bounds
required by the moment discrepancy finder.

\begin{lemma}[Conditional smoothness on most cells]
\label{lem:conditional-density-averaging}
Let $P,Q$ be probability distributions on $\R^d$, let
$q\geq1$ and $\sigma,\eta\in(0,1)$, and suppose that $Q$ is
$(q,\sigma)$-smooth with respect to $P$.
For every finite  measurable partition $\mathcal P$
of $\R^d$, there is a collection $\mathcal G\subseteq\mathcal P$
of cells of positive $Q$-mass such that
$    \sum_{R\in\mathcal G}Q(R)\geq1-\eta,
$
and, for every $R\in\mathcal G$, the conditional distribution
$Q\mid R$ is $(2q,\sigma\eta/(2q))$-smooth with respect to
$P\mid R$.
\end{lemma}

\begin{proof}[Proof of \Cref{lem:conditional-density-averaging}]
Smoothness implies that $Q\ll P$, hence the density ratio $r=\mathrm dQ/\mathrm dP$ is defined. 
Let $a=2q/(2q-1)$. We first show that $r$ has bounded $L_a$ norm under $P$.
\begin{claim}\label{label:norm}
     It holds that $\|r\|_{L_a(P)}\leq2q/\sigma$.
\end{claim}
\begin{proof}[Proof of \Cref{label:norm}]
For $q=1$, this follows from $r\leq1/\sigma$ almost surely.
For $q>1$, smoothness gives
\[
    tP(r>t)\leq Q(r>t)
    \leq\frac{P(r>t)^{1/q}}{\sigma}\;.
\]
Hence $P(r>t)\leq\min\{1,(\sigma t)^{-b}\}$, for $b=q/(q-1)$.
Integrating this tail bound yields
\begin{align*}
    \E_P[r^a]
    = a\int_0^\infty t^{a-1}P(r>t)\,\mathrm d t\leq a\int_0^{1/\sigma}t^{a-1}\,\mathrm d t
       +a\sigma^{-b}\int_{1/\sigma}^\infty t^{a-b-1}\,\mathrm d t=\sigma^{-a}\left(1+\frac{a}{b-a}\right)
      =(2q-1)\sigma^{-a}.
\end{align*}
This proves that $\|r\|_{L_a(P)}\leq2q/\sigma$ and completes the proof of \Cref{label:norm}.
\end{proof}
For each cell with $q_R:=Q(R)>0$, we have that $p_R:=P(R)>0$. Define the conditional density ratio
$s_R(x):=\mathrm d(Q\mid R)/\mathrm d(P\mid R)(x)
=(p_R/q_R)r(x)$ for $x\in R$.

By H\"older's inequality and disjointness of the cells,
\begin{align*}
    \sum_{R:q_R>0}q_R\|s_R\|_{L_a(P\mid R)}
    &=\sum_{R:q_R>0}
      p_R^{1-1/a}\left(\int_R r^a\,\mathrm dP\right)^{1/a}\\
    &\leq
      \left(\sum_{R:q_R>0}p_R\right)^{1-1/a}
      \left(\sum_{R:q_R>0}\int_R r^a\,\mathrm dP\right)^{1/a}\\
    &\leq\|r\|_{L_a(P)}
    \leq\frac{2q}{\sigma}.
\end{align*}
Consequently, the cells satisfying
$\|s_R\|_{L_a(P\mid R)}>2q/(\sigma\eta)$ have total $Q$-mass
at most $\eta$. 

Let $\mathcal G$ consist of the remaining cells
of positive $Q$-mass.
For every $R\in\mathcal G$ and measurable event $F$, another
application of H\"older's inequality gives
\begin{align*}
    (Q\mid R)(F)
    &=\int_F s_R\,\mathrm d(P\mid R)\leq
      \left(\int_F s_R^a\,\mathrm d(P\mid R)\right)^{1/a}
      \left(\int_F 1\,\mathrm d(P\mid R)\right)^{1-1/a}\\
    &\leq \|s_R\|_{L_a(P\mid R)}
          (P\mid R)(F)^{1-1/a}\leq \frac{2q}{\sigma\eta}
          (P\mid R)(F)^{1/(2q)}.
\end{align*}
This completes the proof of \Cref{lem:conditional-density-averaging}.
\end{proof}

We next show that under the smoothness assumption, we can bound the norm of a fixed degree polynomial $p$ under $Q$ in terms of its norm under the standard Gaussian distribution.

\begin{lemma}[Polynomial Norm under Smoothness]
\label{lem:relaxed-polynomial-control}
Let $P=\cN(\vec{0},\bI_d)$, let $Q$ be $(q,\sigma)$-smooth with
respect to $P$, where $q\geq1$ and $\sigma\in(0,1]$, and let
$k\in\N_+$. For every polynomial $p$ of degree at most $k$,
\[
    \E_Q[p^2]\leq\frac{2q}{\sigma}(4q-1)^k\|p\|_{L_2(P)}^2.
\]
Moreover, for the normalized Hermite tensors $\mathcal H_i$,
$i\in[k]$, we have 
\[
    \|\E_Q[\mathcal H_i]\|_F
    \leq e\left(2q\log\frac{2eq}{\sigma}\right)^{i/2}
    \leq e\left(2q\log\frac{2eq}{\sigma}\right)^{k/2}.
\]
\end{lemma}
\begin{proof}[Proof of \Cref{lem:relaxed-polynomial-control}]
Smoothness implies $Q\ll P$, therefore we can define the density ratio
$r=\mathrm dQ/\mathrm dP$. 
Let $a=2q/(2q-1)$. By \Cref{label:norm}, $\|r\|_{L_a(P)}\leq2q/\sigma$. 
Therefore, H\"older's inequality and
Gaussian hypercontractivity give
\begin{align*}
    \E_Q[p^2]
    &=\E_P[rp^2]\\
    &\leq \|r\|_{L_a(P)}\|p^2\|_{L_{2q}(P)}
      =\|r\|_{L_a(P)}\|p\|_{L_{4q}(P)}^2\\
    &\leq \frac{2q}{\sigma}(4q-1)^k\|p\|_{L_2(P)}^2,
\end{align*}
where we used that  $1/a+1/(2q)=1$, 
which proves the first assertion.

For the tensor bound, fix $i\in[k]$. For every order-$i$ tensor
$\bT$ with $\|\bT\|_F=1$, define
$p_{\bT}(\x)=\bT\cdot\mathcal H_i(\x)$.
This polynomial has degree at most $i$ and
$\|p_{\bT}\|_{L_2(P)}\leq1$.
By duality of the Frobenius norm,
\[
    \|\E_Q[\mathcal H_i]\|_F
    =\sup_{\|\bT\|_F=1}
       |\bT\cdot\E_Q[\mathcal H_i]|
    =\sup_{\|\bT\|_F=1}|\E_Q[p_{\bT}]|.
\]
Thus, it suffices to bound expectations under $Q$ of polynomials
of degree at most $i$ whose Gaussian $L_2$ norm is at most $1$.

Since $\|r\|_{L_1(P)}=1$ and
$\|r\|_{L_a(P)}\leq2q/\sigma$, interpolation gives, for every
$t\geq2q$,
\[
    \|r\|_{L_{t/(t-1)}(P)}
    \leq
    \|r\|_{L_1(P)}^{1-2q/t}
    \|r\|_{L_a(P)}^{2q/t}
    \leq\left(\frac{2q}{\sigma}\right)^{2q/t}.
\]
For completeness, if $c=t/(t-1)$ and
$\theta=(c-1)/(a-1)$, then
$c=(1-\theta)+a\theta$. H\"older's inequality applied to
$r^c=r^{1-\theta}(r^a)^\theta$ gives this interpolation inequality
after taking $c$th roots, since $a\theta/c=2q/t$.

Choose $t=2q\log(2eq/\sigma)\geq2q\geq2$.
Then $(2q/\sigma)^{2q/t}
=\exp(\log(2q/\sigma)/\log(2eq/\sigma))\leq e$.
For every polynomial $p$ of degree at most $i$,
H\"older's inequality and Gaussian hypercontractivity give
\begin{align*}
    |\E_Q[p]|
    &\leq\E_Q[|p|]
      =\E_P[r|p|]\\
    &\leq\|r\|_{L_{t/(t-1)}(P)}\|p\|_{L_t(P)}\\
    &\leq e(t-1)^{i/2}\|p\|_{L_2(P)}
      \leq et^{i/2}\|p\|_{L_2(P)}.
\end{align*}
Applying this bound to $p_{\bT}$ and taking the supremum over
$\|\bT\|_F=1$ yields
\[
    \|\E_Q[\mathcal H_i]\|_F
    \leq et^{i/2}
    =e\left(2q\log\frac{2eq}{\sigma}\right)^{i/2}.
\]
This concludes the proof of \Cref{lem:relaxed-polynomial-control}.
\end{proof}

\subsection{Proof of the Partitioning Theorem}
\label{sec:partitioninalg-proof}
\begin{proof}[Proof of \Cref{thm:relaxed-gaussian-partitioning}]
Run \Cref{alg:relaxed-gaussian-partitioning}, and write
$p_R\eqdef P(R)$ and $q_R\eqdef Q(R)$.
Let $\mathcal U_{L+1}$ denote the final partition
and $\mathcal G$ the collection of certified cells.
Let $\mathcal B,\mathcal T$, and $\mathcal E$
denote the collections of cells stopped by the mass test,
tail rule, and either finder rejection or failed
split validation, respectively. Finally, let
$\mathcal D=\mathcal U_{L+1}\setminus
(\mathcal G\cup\mathcal B\cup\mathcal T\cup\mathcal E)$
denote the leaves still unprocessed after round $L$.

We prove the result by first showing that every certified cell
$R \in \mathcal G$ 
satisfies the moment-matching and small-diameter
properties and has nontrivial $Q$-mass, with the parameters specified
in the statement. It then remains to show that the remaining 
cells in $\mathcal U_{L+1}\setminus \mathcal{G}$ have total $Q$-mass at most $\eta$.

We work on the simultaneous success event $\Omega$ on which
all empirical cell masses have error at most $\xi/4$, every
amplified finder satisfies the conclusions of
\Cref{cor:relaxed-finder-amplification}, every fresh empirical
CDF has uniform error at most $g/4$, and every conditional
sampling budget succeeds. We verify later in the proof
that $\Pr[\Omega]\geq99/100$. %

\begin{claim}[Guarantees on Certified Cells]
\label{claim:relaxed-certified-leaves}
Every $R\in\mathcal G$ satisfies $q_R>\xi/4$
and $\diam(R)\leq\delta_0\sqrt L$.
Moreover, $P\mid R$ and $Q\mid R$ match their first $k$ moments
on $V_R^\perp$ with slack $\Delta$.
\end{claim}

\begin{proof}[Proof of \Cref{claim:relaxed-certified-leaves}]
A cell reaches the finder only after passing the empirical
mass test. On $\Omega$, this implies $q_R\geq\widehat q_R-\xi/4>\xi/4$.
We establish the finder's second-moment assumption uniformly
for every such cell. For a polynomial $p$ on $V_R^\perp$, its
extension $p(\proj_{V_R^\perp}\x)$ has the same Gaussian $L_2$
norm. 
By \Cref{lem:relaxed-polynomial-control}, for every such
polynomial of degree at most $k$,
\begin{align*}
 \E_{Q\mid R}[p^2]
 &\leq \frac{\E_Q[p^2]}{q_R}
 \leq V\|p\|_{L_2(\cN(\vec{0},\bI_{V_R^\perp}))}^2,
\end{align*}
where $V=\frac{8q}{\sigma\xi}(4q-1)^k$, using $q_R>\xi/4$.
This single bound applies to every finder input at every depth.
Membership in $R$ depends only on $V_R$, so Gaussian
independence implies that the projection of $P\mid R$ onto
$V_R^\perp$ is standard Gaussian. A cell is certified only when
the amplified finder returns $\textsc{None}$; hence
\Cref{cor:relaxed-finder-amplification} gives the asserted
moment matching for the full distribution $Q\mid R$.
If $V_R^\perp=\{\vec{0}\}$, the assertion is vacuous, as in
\Cref{def:moment-matching}.

Every new constraint direction is a unit vector orthogonal to
the preceding constraint subspace. The directions along a
root-to-leaf path are therefore orthonormal. By
\Cref{def:relaxed-split-refinement}, each constrained coordinate
of a nontail leaf lies in an interval of width at most $\delta_0$.
Certified leaves are nontail and have at most $L$ constrained
coordinates, giving $\diam(R)\leq\delta_0\sqrt L$.
This concludes the proof of \Cref{claim:relaxed-certified-leaves}.
\end{proof}

It remains to bound the mass of the uncertified final leaves.
We first control the unprocessed, low-mass, and tail leaves.

\begin{claim}[Mass of $\mathcal D$, $\mathcal B$, $\mathcal T$]
\label{claim:relaxed-unprocessed-leaves}
The final collections satisfy
$
 \sum_{R\in\mathcal D\cup\mathcal B\cup\mathcal T}q_R
 \leq\frac{5\eta}{100}.
$
\end{claim}

\begin{proof}[Proof of \Cref{claim:relaxed-unprocessed-leaves}]
Every executed split has empirical CDF gap at least $2g$.
On $\Omega$, its true gap is at least $2g-g/4\geq g$.
The reference marginal is standard Gaussian, and the squared
Hellinger distance between Bernoulli distributions is at least
half the squared difference of their parameters. Thus every
executed split is $\kappa$-Gaussian good, since
$g^2/2\geq\kappa$; see \Cref{def:relaxed-split-refinement}.

Write $r=\mathrm dQ/\mathrm dP$. By
\Cref{lem:relaxed-smoothness-truncation},
$M=1/(\sigma^q\tau_0^{q-1})$ satisfies $Q(r>M)\leq\tau_0$.
Every leaf in $\mathcal D$ is a child of a cell split in round $L$.
Hence \Cref{lem:relaxed-depth-analysis} and the choice of $L$ give
\[
 \sum_{R\in\mathcal D}q_R
 \leq\sqrt M\,e^{-\kappa(L-1)}+\tau_0
 \leq\eta/100+\tau_0=2\eta/100.
\]

On $\Omega$, every cell in $\mathcal B$ has
$q_R\leq3\xi/4<\xi$. Since all executed splits are
$\kappa$-Gaussian good, \Cref{lem:relaxed-small-mass} applies
to $\mathcal U_{L+1}$ with $s=s_0$. The choices of $\xi$,
$\lambda$, and $\varepsilon$ therefore give
\[
 \sum_{R\in\mathcal B}q_R
 \leq\lambda+\varepsilon=2\eta/100.
\]
Applying  \Cref{lem:relaxed-tail-mass} with our chosen value of $T$ gives
$\sum_{R\in\mathcal T}q_R
\leq \sigma^{-1}(2Le^{-T^2/2})^{1/q}
\leq\eta/100$.
Combining the above bounds concludes the proof of \Cref{claim:relaxed-unprocessed-leaves}.
\end{proof}
\begin{claim}[Mass Stopped by Rejection or Failed Split Validation]
\label{claim:relaxed-rejected-leaves}
The remaining terminal collection satisfies
$\sum_{R\in\mathcal E}q_R\leq\eta/100$.
\end{claim}

\begin{proof}[Proof of \Cref{claim:relaxed-rejected-leaves}]
Apply \Cref{lem:conditional-density-averaging} to the final
partition $\mathcal U_{L+1}$, with parameters
$(q,\sigma/2,\eta/100)$. This is valid because $Q$ is also
$(q,\sigma/2)$-smooth and $\sigma/2<1$. It gives a collection
$\mathcal S\subseteq\mathcal U_{L+1}$ such that
\[
 \sum_{R\in\mathcal S}q_R\geq1-\eta/100,
 \qquad
 Q\mid R\text{ is }(2q,\sigma_*)\text{-smooth with respect to }
 P\mid R\quad(R\in\mathcal S),
\]
where $\sigma_*\eqdef \sigma\eta/(400q)$.
We show that $\mathcal E\cap\mathcal S=\emptyset$.

Suppose $R\in\mathcal E\cap\mathcal S$. The cell was tested
at an earlier round and then carried forward unchanged, so
its conditional distribution and constraint subspace at that
test are the same as in the final partition. By \Cref{fact:smoothness-projection}  smoothness is
preserved by projection. 
Since the projection of $P\mid R$
onto $V_R^\perp$ is standard Gaussian, the corresponding
projection of $Q\mid R$ is $(2q,\sigma_*)$-smooth.
\Cref{lem:relaxed-polynomial-control} therefore gives, for
each $i\leq k$,
\[
 \left\|\E_{Q\mid R}
       [\mathcal H_i(\proj_{V_R^\perp}\x)]\right\|_F
 \leq e\left(4q\log\frac{4eq}{\sigma_*}\right)^{i/2}
 \leq H.
\]
By the third conclusion of
\Cref{cor:relaxed-finder-amplification}, the finder cannot
reject this cell.

The cell must therefore have been stopped by failed split
validation after the finder returned a direction. By
\Cref{cor:relaxed-finder-amplification}, this direction has
raw moment discrepancy at least $\Delta/4$ in some degree
in $[k]$. Applying \Cref{lem:relaxed-moment-good-split} with parameters
$(2q,\sigma_*)$ and discrepancy $\Delta/4$ gives a threshold
$u_*\in[-b,b]$ whose Gaussian upper-tail gap is at least
$\Delta/(16b^k)\geq6g$.
Because $T\geq b$ and the empirical CDF error is at most
$g/4$, the empirical maximum searched by
\Cref{alg:relaxed-gaussian-partitioning} is at least
$6g-g/4>2g$. Thus split validation also succeeds, a
contradiction. We conclude that
$\mathcal E\subseteq\mathcal U_{L+1}\setminus\mathcal S$,
which proves \Cref{claim:relaxed-rejected-leaves}.
\end{proof}

All uncertified final cells belong to
$\mathcal B\cup\mathcal T\cup\mathcal E
\cup\mathcal D$. The preceding bounds give
\[
 Q\left(\bigcup_{R\in\mathcal U_{L+1}\setminus\mathcal G}R\right)
 \leq\frac{(2+2+1+1)\eta}{100}
 =\frac{6\eta}{100}<\eta.
\]
Together with \Cref{claim:relaxed-certified-leaves}, this
establishes the structural and moment-matching guarantees.
By \Cref{def:relaxed-split-refinement}, every split creates
at most $m$ children, so at most
$N_*\eqdef (1+m)^{L+1}$ cells are created.
We now verify the probability of $\Omega$ and the quantitative
mass, diameter, and complexity bounds.

\begin{claim}[Probability of Success Event]
\label{claim:relaxed-success-event}
With the parameters in (\hyperref[par:relaxed-partition-parameters]{Parameter Choices of \Cref{alg:relaxed-gaussian-partitioning}}), the event $\Omega$
satisfies
$
 \Pr[\Omega]\geq1-(L+4N_*)\beta\geq99/100.
$
\end{claim}

\begin{proof}[Proof of \Cref{claim:relaxed-success-event}]
There are at most $N_*$ calls to the moment discrepancy finding algorithm,
$N_*$ CDF validation stages to check whether the found split is good enough, and $2N_*$ conditional
sampling stages from the cells $R$ created during the implementation.

Condition successively on the preceding execution, up to
the first failure. Fresh samples and Hoeffding's inequality
give failure probability at most $\beta/N_*$ per mass
estimate, hence at most $L\beta$ in total. While these
estimates are accurate, the uniform $V$-bound established
in \Cref{claim:relaxed-certified-leaves} applies to every
finder input. Thus \Cref{cor:relaxed-finder-amplification}
bounds finder failures by $N_*\beta$ in total.
The Dvoretzky--Kiefer--Wolfowitz inequality similarly
bounds CDF estimation failures by $N_*\beta$.

Couple each conditional sampling stage with an unlimited
rejection-sampling stream, whose accepted observations
are independent samples from $Q\mid R$. Since every tested
cell has $q_R>\xi/4$, a binomial lower-tail bound and the
prescribed sampling budget give probability at most
$\beta$ of obtaining too few observations. These failures
total at most $2N_*\beta$. Summing the failure bounds
proves \Cref{claim:relaxed-success-event}.
\end{proof}

\medskip\noindent\textbf{Complexity.}
Set
\[
 \mathsf E=
 \frac{O((qk)^{O(k)})}{\Delta^3}
 \log^{k+2}\frac2{\sigma\eta}\log\frac2{\delta_0}.
\]
Expanding (\hyperref[par:relaxed-partition-parameters]{Parameter Choices of \Cref{alg:relaxed-gaussian-partitioning}}) gives
\[
 L\leq
 \frac{O((qk)^{O(k)})}{\Delta^3}
 \log^{k+1}\frac2{\sigma\eta},
 \qquad
 \log\frac{N_*V}{\xi g\beta}
 +k^{O(k)}\frac{H^2}{\Delta^2}\log\frac H\Delta
 \leq\mathsf E.
\]

By \Cref{def:relaxed-split-refinement}, at most $N_*$
cells are created, each requiring at most one finder call
and one CDF validation stage.
By \Cref{cor:relaxed-finder-amplification}, each finder
call uses at most $d^k\exp(O(\mathsf E))$ conditional samples
and $d^{O(k)}\exp(O(\mathsf E))$ time, for every $H\geq1$.
Each CDF stage uses $O(g^{-2}\log(2/\beta))$ conditional
samples. Rejection sampling contributes a factor
$O(1/\xi)$ and logarithmic overhead, while mass estimation
uses $O(L\xi^{-2}\log(2N_*/\beta))$ additional draws.
Routing samples, computing Gaussian cell probabilities,
and scanning empirical CDFs add polynomial overhead.
Thus, the total sample complexity is at most
$d^k\exp(O(\mathsf E))$ and the running time is at most
$d^{O(k)}\exp(O(\mathsf E))$.

Finally, by the choice $\delta_0=\delta/\sqrt L$,
$\diam(R)\leq\delta_0\sqrt L=\delta$ for every $R\in\mathcal G$,
and the preceding bounds give the stated sample and running-time
complexities, $N_*\leq B$, and $\xi/4\geq1/B$, upon choosing the
implicit constants in $B$ sufficiently large.
Together with \Cref{claim:relaxed-success-event}, this
establishes all the stated guarantees.
This completes the proof of \Cref{thm:relaxed-gaussian-partitioning}.
\end{proof}
\subsection{Efficient Moment Discrepancy Finding Algorithm}
\label{sec:relaxed-discrepancy-finder}

We now describe the moment {discrepancy} 
finding algorithm for \Cref{thm:standalone-capped-finder}. 
In \Cref{cor:relaxed-finder-amplification}, we {show how to} 
boost \Cref{thm:standalone-capped-finder} 
into a high probability regime.

\begin{algorithm}[htbp]
    \caption{\textsc{GaussianDiscrepancyFinder}}
    \label{alg:standalone-moment-discrepancy-finder}
    \begin{algorithmic}[1]
\State \textbf{Input:} Sample access from $Q$ {over $\R^d$},
    {parameters $k\in\Z_+$, $\Delta\in(0,1/2)$, and $V,H\geq1$}
\State \textbf{Output:} {$\textsc{Reject}$, $\textsc{None}$,
or a direction $\bw\in\s^{d-1}$, with the guarantees of
\Cref{thm:standalone-capped-finder}.}

\medskip

\State {Set $F_k=(2k)^{2k}$, $\tau=\Delta/(100k2^kF_k)$, and
    $n=\lceil3V(d+k)^k/\tau^2\rceil$.}
\State {Draw $\x^{(1)},\ldots,\x^{(n)}$ independently from $Q$.}
\For{$i\in[k]$}
    \State Compute $\widehat{\bA}_i=(1/n)\sum_{s=1}^n\mathcal H_i(\x^{(s)})$.
    \If{{$\|\widehat{\bA}_i\|_F>2H$}}
        \State \Return {$\textsc{Reject}$}
    \EndIf
    \State Flatten $\widehat{\bA}_i$ as a matrix and let $U_i$ be the span of its left singular vectors with singular value at least $\tau$.
\EndFor
\State {Let $W=\operatorname{span}(U_1,\ldots,U_k)$.}
\State {If $W\neq\R^d$, choose any $\be\in W^\perp\cap\s^{d-1}$ and set $W'=W+\operatorname{span}\{\be\}$; otherwise set $W'=W$.}
\State Let $\mathcal C$ be a $\rho$-net of {$W'\cap\s^{d-1}$} for
    {$\rho=\Delta/(100kF_kH)$}.
\State Let
\[
{(\ell_*,\widehat{\bw})\in
\argmax_{\ell\in[k],\,\bw\in\mathcal C}
\left|
\frac1n\sum_{s=1}^n(\bw\cdot\x^{(s)})^\ell
-\E_{Z\sim\cN(0,1)}[Z^\ell]
\right|.}
\]
\If{
\[
{\left|
\frac1n\sum_{s=1}^n(\widehat{\bw}\cdot\x^{(s)})^{\ell_*}
-\E_{Z\sim\cN(0,1)}[Z^{\ell_*}]
\right|\geq\Delta/2}
\]
}
    \State \Return $\widehat{\bw}$
\Else
    \State \Return $\textsc{None}$
\EndIf
    \end{algorithmic}
\end{algorithm}

{When a nonempty sample is supplied explicitly to
\Cref{alg:standalone-moment-discrepancy-finder}, set $n$ to its size
and use the supplied points in all empirical averages, omitting the
sample-size selection and sampling steps. Repeated observations
are counted with their multiplicities.}

The intuition behind the algorithm is as follows. Though for an arbitrary distribution $Q$, finding a moment discrepancy between $Q$ and $P$ might require us to do a brute force search over $\R^d$, if $Q$ satisfies the condition of \Cref{thm:standalone-capped-finder}, we would like to find some low-dimensional subspace $W$ such that the moment discrepancy is preserved over $W$. Once this is done, we only need to run the brute force search over $W$, which significantly reduces the complexity of the algorithm.

To see why such a low-dimensional subspace exists, we consider the term $\E_{\x\sim Q}[p] - \E_{\x\sim P}[p]$, for an arbitrary polynomial $p$ with degree at most $k$. Using Hermite polynomial decomposition, we can show there exists some degree $k$ polynomial $G$, such that $\E_{\x\sim Q}[p] - \E_{\x\sim P}[p] = \E_{\x \sim P}[G(\x)p(\x)]$. 
Recall that we can always expand $G(\x) = \sum_{i=1}^k \bA_i\cdot\mathcal H_i(\x)$ using Hermite tensors. If we choose $p(\x) = (\w\cdot\x)^k$, then $\E_{\x \sim P}[G(\x)p(\x)]$ is a linear combination of $\w^{\otimes i} \cdot \bA_i$.
So if we flatten $\bA_i$ as a bilinear form over $\R^{d \times d^{i-1}}$ and choose $W$ to be the span of large singular vectors of all $\bA_i$, {then adjoining one unit direction from $W^\perp$ gives a subspace $W'$ that approximately
preserves the moment discrepancy.}
However, it remains to show that $W$ has a low dimension.

The assumption on $Q$ implies that $G$ has a small $L_2$ norm, which further implies that each of the $\bA_i$ must have small Frobenius norm. So there is no $\bA_i$ that can have many large singular vectors and $W$ must have low dimension.

\begin{proof}[Proof of \Cref{thm:standalone-capped-finder}]
{We first construct a polynomial representing the low-degree
moments of $Q$. We use the normalized probabilists' Hermite
polynomials $h_j(t)=\mathrm{He}_j(t)/\sqrt{j!}$, with $h_0=1$.
For a multi-index $\beta$, put
$H_\beta(\x)=\prod_{a=1}^d h_{\beta_a}(\x_a)$.
The polynomials $H_\beta$, $|\beta|\leq k$, form an orthonormal basis
for the degree-at-most-$k$ polynomials in $L_2(P)$.
Define
\[
    g_k(\x)=\sum_{|\beta|\leq k}
                   \E_{\by\sim Q}[H_\beta(\by)]H_\beta(\x).
\]
All its coefficients are finite: the $V$-assumption and
Cauchy--Schwarz give
$|\E_Q H_\beta|\leq(\E_Q H_\beta^2)^{1/2}\leq\sqrt V$.
For every polynomial $p$ of degree at most $k$, expansion in this
orthonormal basis gives}
\begin{align*}
    {\E_{\x\sim Q}[p(\x)]}
    &{=\sum_{|\beta|\leq k}
           \langle p,H_\beta\rangle_{L_2(P)}\,\E_Q[H_\beta]}\\
    &{=\E_{\x\sim P}[p(\x)g_k(\x)].}
\end{align*}
{In particular, $\E_P[g_k]=\E_Q[1]=1$. Moreover, by duality in
the finite-dimensional polynomial space and the $V$-assumption,}
\begin{align*}
    {\|g_k\|_{L_2(P)}}
    &{=\sup_{\deg p\leq k,\ \|p\|_{L_2(P)}=1}
                      |\E_P[p g_k]|}\\
    &{=\sup_{\deg p\leq k,\ \|p\|_{L_2(P)}=1}
                      |\E_Q[p]|}\\
    &{\leq
      \sup_{\deg p\leq k,\ \|p\|_{L_2(P)}=1}
                      (\E_Q[p^2])^{1/2}
      \leq\sqrt V.}
\end{align*}
{Thus, $G=g_k-1$ has degree at most $k$, has Gaussian mean zero,
and satisfies $\|G\|_{L_2(P)}^2\leq V-1$. }
For every polynomial $p(\x)$ of degree no larger than $k$, it holds
\begin{align*}
    \E_{\x\sim Q}[p(\x)]-\E_{\x\sim P}[p(\x)]
    =\E_{\x\sim P}[G(\x)p(\x)].
\end{align*}
Let $\mathcal H_i$ be the $i$th order Hermite tensor.
We expand $G(\x)$ using the Hermite decomposition
$G(\x)=\sum_{i=1}^k\bA_i\cdot\mathcal H_i(\x)$, 
where for $i\in[k]$, $\bA_i$ is an order-$i$ symmetric tensor over
$\R^d$.
{
Parseval's identity therefore gives
\[
    \|\bA_i\|_F^2\leq\sum_{j=1}^k\|\bA_j\|_F^2
    =\|G\|_{L_2(P)}^2\leq V-1.
\]}
Moreover, for every $\bw\in\s^{d-1}$, we write
{simultaneously for $\ell\in[k]$}
\[
    {(\bw\cdot\x)^\ell
       =\sum_{i=0}^\ell a_{\ell,i}
                  \bw^{\otimes i}\cdot\mathcal H_i(\x),}
\]
{where $\sum_{i=0}^{\ell}|a_{\ell,i}|\leq F_k\eqdef(2k)^{2k}=O(k^{O(k)})$ for every $\ell\in[k]$.}

{Explicitly, when $\ell-i$ is nonnegative and even,
\[
 a_{\ell,i}
 =\frac{\ell!}
        {2^{(\ell-i)/2}((\ell-i)/2)!\sqrt{i!}},
\]
and $a_{\ell,i}=0$ otherwise. This follows by expanding a
monomial in normalized probabilists' Hermite polynomials.
Every displayed coefficient is at most $k!$, so
$\sum_{i=0}^{\ell}|a_{\ell,i}|\leq(k+1)k!\leq(2k)^{2k}$.
Thus the bound also includes $k=1$.}

Define the degree $\ell$ moment {discrepancy} at direction $\bw$ as
\begin{align*}
    {F_\ell(\bw)}
    &{:=\E_Q[(\bw\cdot\x)^\ell]
                   -\E_P[(\bw\cdot\x)^\ell]}\\
    &{=\E_P[G(\x)(\bw\cdot\x)^\ell]
      =\sum_{i=1}^\ell a_{\ell,i}\bA_i\cdot\bw^{\otimes i},
      \qquad \ell\in[k].}
\end{align*}
Let $\tau\in(0,1)$ be a parameter that will be determined later.
Also for $i\in[k]$ let
\[
    \widehat{\bA}_i:=\frac1n\sum_{s=1}^n\mathcal H_i(\x^{(s)})
\]
be the empirical Hermite tensor, where the $\x^{(s)}$ are i.i.d.
samples from $Q$.

\begin{claim}[Empirical tensor estimation]
\label{claim:standalone-capped-estimation}
If $n\geq3V(d+k)^k/\tau^2$, with probability at least $2/3$
we have simultaneously for every $i\in[k]$ that
$
    \norm{\widehat{\bA}_i-\bA_i}_F\leq\tau.
$
\end{claim}

\begin{proof}
We estimate the tensors $\bA_i$ directly from samples.
By our normalization of the Hermite tensors,
$\bA_i=\E_{\x\sim Q}[\mathcal H_i(\x)]$, and hence
$\E[\widehat{\bA}_i]=\bA_i$. Moreover, by the assumption that $\E_{\x\sim Q}[p(\x)^2]\leq V\E_{\x\sim P}[p(\x)^2]$ for every polynomial $p$ of degree at most $k$, we have
\begin{align*}
    \E\left[\norm{\widehat{\bA}_i-\bA_i}_F^2\right]
    \leq\frac1n\E_{\x\sim Q}
                      \left[\norm{\mathcal H_i(\x)}_F^2\right]
    \leq\frac Vn\E_{\x\sim P}
                      \left[\norm{\mathcal H_i(\x)}_F^2\right] \;.
\end{align*}
{For the second inequality, apply the assumed $V$-bound to
each entry of $\mathcal H_i$, which is a polynomial of degree $i$,
and sum. The tensor normalization gives the exact count
\[
    \E_P\|\mathcal H_i(\x)\|_F^2
    =\sum_{|\beta|=i}
       \frac{i!}{\beta!}\frac{\beta!}{i!}\E_P[H_\beta^2]
    =\binom{d+i-1}{i}.
\]
Consequently,}
\begin{align*}
    \E\left[\sum_{i=1}^k
          \norm{\widehat{\bA}_i-\bA_i}_F^2\right]
    &{\leq\frac Vn\sum_{i=1}^k\binom{d+i-1}{i}}\\
    &{=\frac Vn\left(\binom{d+k}{k}-1\right)
      \leq\frac{V(d+k)^k}{n}.}
\end{align*}
It follows from Markov's inequality that, for
\[
    {n\geq\frac{3V(d+k)^k}{\tau^2},}
\]
with probability at least $2/3$ we have
$\norm{\widehat{\bA}_i-\bA_i}_F\leq\tau$ simultaneously for every $i\in[k]$.
{Indeed, the probability that the sum of squared errors exceeds
$\tau^2$ is at most $1/3$, and controlling this sum controls every
individual tensor.}
This completes the proof of \Cref{claim:standalone-capped-estimation}.
\end{proof}

For the remainder of the proof we condition on the event in
\Cref{claim:standalone-capped-estimation}.
{The algorithm first checks whether some
$\|\widehat{\bA}_i\|_F>2H$. If so, it returns $\textsc{Reject}$.
If every $\|\bA_i\|_F\leq H$, then
\[
    \|\widehat{\bA}_i\|_F\leq\|\bA_i\|_F+\tau
                        \leq H+\tau\leq2H,
\]
since $H\geq1$ and $\tau<1$. Thus rejection is impossible in that
case on the event under consideration. 
In all subsequent steps assume that the algorithm does not reject, so
$\|\widehat{\bA}_i\|_F\leq2H$ for every $i\in[k]$.}

For $i\in[k]$, flatten $\widehat{\bA}_i$ as a matrix in
$\R^{d\times d^{i-1}}$, and let $U_i$ be the subspace spanned by its
left singular vectors with corresponding singular value at least
$\tau$.
{Let $W=\operatorname{span}(U_1,\ldots,U_k)$.
If $W\neq\R^d$, fix any $\be\in W^\perp\cap\s^{d-1}$ and set
$W'=W+\operatorname{span}\{\be\}$; otherwise set $W'=W$.}

\begin{claim}[{Low-dimensional Reduction}]
\label{claim:standalone-capped-reduction}
{On the estimation event, assuming that the algorithm does not reject,
\[
    \dim(W')\leq1+\frac{4kH^2}{\tau^2}.
\]
Moreover, for every $\bu\in\s^{d-1}$ there exists a unit vector
$\bar\bu\in W'$ such that, simultaneously for every $\ell\in[k]$,
\[
    |F_\ell(\bar\bu)-F_\ell(\bu)|
    \leq4F_k2^k\tau.
\]
In particular,
$|F_\ell(\bar\bu)|\geq|F_\ell(\bu)|-4F_k2^k\tau$.}
\end{claim}

\begin{proof}
For $i\in[k]$, we flatten the empirical tensor
$\widehat{\bA}_i$ as a matrix in $\R^{d\times d^{i-1}}$
and let $U_i$ be the subspace spanned by its left singular vectors
with corresponding singular value at least $\tau$.
{Since the algorithm does not reject, $\|\widehat{\bA}_i\|_F\leq2H$, so}

\begin{align*}
    \dim(U_i)
    \leq\frac{\|\widehat{\bA}_i\|_F^2}{\tau^2}
    {\leq\frac{4H^2}{\tau^2}.}
\end{align*}
{Let $W=\operatorname{span}(U_1,\ldots,U_k)$. Then
\[
    \dim(W)\leq\sum_{i=1}^k\dim(U_i)
             \leq\frac{4kH^2}{\tau^2}.
\]
Since $\dim(W')\leq\dim(W)+1$, this proves the dimension bound.}
For a unit vector $\bz$, write
\[
    {F_\ell(\bz)
       =\E_P[G(\x)(\bz\cdot\x)^\ell]
       =\sum_{i=1}^\ell a_{\ell,i}\bA_i\cdot\bz^{\otimes i},
       \qquad \ell\in[k],}
\]
{and, for $\by\in W$, define the auxiliary expression}
\[
    {\widetilde F_\ell(\by)
          :=\sum_{i=1}^\ell a_{\ell,i}\bA_i\cdot\by^{\otimes i}.}
\]
{Write $\bu=\bv+\bw$, with $\bv=\proj_W\bu$ and
$\bw=\proj_{W^\perp}\bu$.}
Then we have
\begin{align*}
    \bu^{\otimes j}
    =(\bv+\bw)^{\otimes j}
    =\bv^{\otimes j}
      +\sum_{\substack{S\subseteq[j]\\S\neq\emptyset}}
                  \bigotimes_{i=1}^j\bz_i(S),
\end{align*}
where for $S\subseteq[j]$ and $i\in[j]$, $\bz_i(S)=\bw$ if $i\in S$
and $\bz_i(S)=\bv$ otherwise. This implies that
\begin{align*}
    \bA_j\cdot\bu^{\otimes j}-\bA_j\cdot\bv^{\otimes j}
    =\bA_j\cdot
       \sum_{\substack{S\subseteq[j]\\S\neq\emptyset}}
                    \bigotimes_{i=1}^j\bz_i(S).
\end{align*}
By symmetry, we may without loss of generality assume that
$\bz_1(S)=\bw$.
{Since $\bw\in W^\perp$ and $U_j\subseteq W$, it is
orthogonal to $U_j$.}
By the definition of $U_j$, the restriction of the flattening of
$\widehat{\bA}_j$ to $U_j^\perp$ has operator norm at most $\tau$.
{Equivalently, if $\widehat{\vec{M}}_j$ denotes that flattening, then
$\|\proj_{U_j^\perp}\widehat{\vec{M}}_j\|_{\mathrm{op}}\leq\tau$.}
Therefore, for a fixed $S$,
\begin{align*}
    \abs{\bA_j\cdot\bigotimes_{i=1}^j\bz_i(S)}
    &\leq
    \abs{\widehat{\bA}_j\cdot\bigotimes_{i=1}^j\bz_i(S)}
    +\abs{(\bA_j-\widehat{\bA}_j)
                    \cdot\bigotimes_{i=1}^j\bz_i(S)}\\
    &\leq\tau+\norm{\bA_j-\widehat{\bA}_j}_F
    \leq2\tau,
\end{align*}
where we used that $\norm{\bv}_2,\norm{\bw}_2\leq1$.
This implies that
\begin{align*}
    {\abs{F_\ell(\bu)-\widetilde F_\ell(\bv)}}
    &{\leq2\sum_{i=1}^\ell
                       |a_{\ell,i}|(2^i-1)\tau}\\
    &{\leq2F_k2^k\tau,\qquad\ell\in[k].}
\end{align*}
{If $W=\R^d$, take $\bar\bu=\bu$. Otherwise, using the fixed
unit vector $\be\in W^\perp$, set} $\bar\bu=\bv+\sqrt{1-\norm{\bv}_2^2}\,\be$.
{In this latter case, put
$\bw'=\sqrt{1-\|\bv\|_2^2}\,\be$. Then
$\bw'\in W^\perp$, $\|\bw'\|_2\leq1$, and
$\bar\bu=\bv+\bw'\in W'\cap\s^{d-1}$.
Expanding $(\bv+\bw')^{\otimes j}$ gives the same explicit sum over
nonempty subsets as above, with each factor equal to $\bv$ or $\bw'$.
Each such term has contraction at most $2\tau$, since
$\bw'\perp U_j$ and the same operator-norm and estimation bounds
apply. Therefore,}
\[
    {|F_\ell(\bar\bu)-\widetilde F_\ell(\bv)|
           \leq2F_k2^k\tau,\qquad\ell\in[k].}
\]
The triangle inequality now gives
\[
    {|F_\ell(\bar\bu)-F_\ell(\bu)|
           \leq4F_k2^k\tau,\qquad\ell\in[k],}
\]
which establishes the required bound.
This completes the proof of \Cref{claim:standalone-capped-reduction}.
\end{proof}

{Choose
\[
    \tau=\frac{\Delta}{100k2^kF_k},
    \qquad
    n=\left\lceil\frac{3V(d+k)^k}{\tau^2}\right\rceil.
\]
Then $\tau\in(0,1)$,
$4F_k2^k\tau=\Delta/(25k)\leq\Delta/25$, and
\Cref{claim:standalone-capped-reduction} shows that a discrepancy
of size $\Delta$ in any degree has a unit representative in $W'$
with discrepancy at least $24\Delta/25$ in that degree.
Since $F_k=O(k^{O(k)})$ and $(d+k)^k\leq d^k(k+1)^k$,
the sample size is $d^k O(k^{O(k)})V/\Delta^2$.}

Moreover, on the event in
\Cref{claim:standalone-capped-estimation}, for every unit vector $\bz$,
\begin{align*}
    {\left|
       \sum_{i=1}^\ell a_{\ell,i}
                          \widehat{\bA}_i\cdot\bz^{\otimes i}
       -\sum_{i=1}^\ell a_{\ell,i}
                          \bA_i\cdot\bz^{\otimes i}
    \right|}
    {\leq\sum_{i=1}^\ell|a_{\ell,i}|\tau}
    {\leq F_k\tau\leq\Delta/100,\qquad\ell\in[k].}
\end{align*}
{Define
\[
    \widehat F_\ell(\bz)
       :=\sum_{i=1}^\ell a_{\ell,i}
                         \widehat{\bA}_i\cdot\bz^{\otimes i}.
\]
The polynomial expansion also shows that
\[
    \widehat F_\ell(\bz)
       =\frac1n\sum_{s=1}^n(\bz\cdot\x^{(s)})^\ell
          -\E_{Z\sim\cN(0,1)}Z^\ell.
\]
Thus the same samples estimate all directional discrepancies
uniformly, including directions selected from those samples.}
The search over {$W'$} can therefore be performed using the
empirical tensors $\widehat{\bA}_i$.

\begin{claim}[Search over the Reduced Subspace]
\label{claim:standalone-capped-search}
{On the estimation event, assuming that the algorithm does not reject,
a search over
$W'\cap\s^{d-1}$ can be implemented in time
\[
    d^{O(k)}
    \exp\left(O(k^{O(k)})\frac{H^2}{\Delta^2}
                         \log\frac{H}{\Delta}\right).
\]
It returns a direction with discrepancy at least $3\Delta/8$
in some degree, or certifies that all directional discrepancies in
degrees $1,\ldots,k$ are smaller than $\Delta$.}
\end{claim}

\begin{proof}
{Let $r=\dim(W')$. The dimension estimate and the chosen value
of $\tau$ give
\[
    r\leq1+\frac{4kH^2}{\tau^2}
      \leq O(k^{O(k)})\frac{H^2}{\Delta^2}.
\]
On the branch under consideration,
$\|\widehat{\bA}_i\|_F\leq2H$ for all $i\in[k]$.}
Therefore, for unit vectors $\bu,\bv\in{W'}$ we have
\begin{align*}
    {\abs{\sum_{i=1}^\ell a_{\ell,i}
                         \widehat{\bA}_i\cdot\bu^{\otimes i}
          -\sum_{i=1}^\ell a_{\ell,i}
                         \widehat{\bA}_i\cdot\bv^{\otimes i}}} 
    &{\leq
       \sum_{i=1}^\ell|a_{\ell,i}|
          \|\widehat{\bA}_i\|_F
          \|\bu^{\otimes i}-\bv^{\otimes i}\|_F}\\
    &{\leq2kHF_k\|\bu-\bv\|_2.}
\end{align*}
{For completeness, the last step uses the telescoping identity
\[
    \bu^{\otimes i}-\bv^{\otimes i}
       =\sum_{a=1}^i
          \bu^{\otimes(a-1)}\otimes(\bu-\bv)
                       \otimes\bv^{\otimes(i-a)},
\]
which implies
$\|\bu^{\otimes i}-\bv^{\otimes i}\|_F\leq i\|\bu-\bv\|_2$.
The bound holds simultaneously for all $\ell\in[k]$.}
Let $\mathcal C$ be a $\rho$-net of
{$W'\cap\s^{d-1}$}, where {$\rho=\Delta/(100kF_kH)$.}
Such a net can be chosen with $|\mathcal C|\leq(3/\rho)^r$.
{An explicit construction also fits the claimed time bound:
identify $W'$ with $\R^r$, round coordinates of the unit sphere to
a grid of spacing $\rho/(4\sqrt r)$, and normalize every nonzero
grid point. Rounding changes a unit vector by at most $\rho/8$,
and normalization adds at most the same error. Enumerating the
grid in $[-1-\rho,1+\rho]^r$ therefore produces a $\rho$-net with
at most $(C\sqrt r/\rho)^r$ points.
We may use this constructible net in the algorithm.
For its size,}
\[
    {\log|\mathcal C|
       \leq r\log\frac{C\sqrt r}{\rho}.}
\]
The choice of $\rho$ gives
\[
    {\log\frac{C\sqrt r}{\rho}
      =O\left(k\log(2k)+\log\frac H\Delta\right)
      \leq O(k\log(2k))\log\frac{H}{\Delta},}
\]
{where we used $F_k=(2k)^{2k}$, $H/\Delta>2$, and
$r\leq O(k^{O(k)})H^2/\Delta^2$.}
Consequently,
\[
    {\log|\mathcal C|
       \leq O(k^{O(k)})\frac{H^2}{\Delta^2}
                          \log\frac{H}{\Delta},}
\]
and hence
\[
    {|\mathcal C|
       \leq\exp\left(O(k^{O(k)})\frac{H^2}{\Delta^2}
                          \log\frac{H}{\Delta}\right).}
\]

By \Cref{claim:standalone-capped-reduction},
{if $|F_\ell(\bu)|\geq\Delta$ for some unit $\bu$ and some
$\ell\in[k]$, there is a unit vector $\bar\bu\in W'$ satisfying} {$|F_\ell(\bar\bu)|\geq24\Delta/25$.}
{The uniform empirical approximation bound then gives}
\[
    {|\widehat F_\ell(\bar\bu)|
       \geq24\Delta/25-\Delta/100=19\Delta/20.}
\]
Let $\bz\in\mathcal C$ satisfy
$\norm{\bz-\bar\bu}_2\leq\rho$. By the choice of $\rho$
and the Lipschitz bound,
\[
    {|\widehat F_\ell(\bz)|
       \geq19\Delta/20-2kHF_k\rho
       =93\Delta/100>\Delta/2.}
\]
{Enumerate $\mathcal C$ and all $\ell\in[k]$.
If $\max_{\ell\in[k],\,\bz\in\mathcal C}
|\widehat F_\ell(\bz)|\geq\Delta/2$, return a maximizing direction
$\bw$. Otherwise return $\textsc{None}$.}
The same empirical approximation bound then gives,
{whenever a direction $\bw$ is returned,}
\[
    {|F_\ell(\bw)|
       \geq|\widehat F_\ell(\bw)|-\Delta/100
       \geq49\Delta/100\geq3\Delta/8}
\]
{for its selected degree $\ell$.
Using the defining property of $F_\ell$, this is precisely
\[
    \left|\E_Q[(\bw\cdot\x)^\ell]
           -\E_P[(\bw\cdot\x)^\ell]\right|\geq3\Delta/8.
\]
Conversely, if a discrepancy of size at least $\Delta$ existed in
any direction and degree, the preceding net argument would force
the empirical maximum above $\Delta/2$. Thus returning
$\textsc{None}$ certifies that every such discrepancy is smaller
than $\Delta$.}

Evaluating the empirical expression at one point of the net takes
$d^{O(k)}$ time, {with factors depending only on $k$ absorbed
into the displayed exponential bound}. Hence the total search time is
\[
    {d^{O(k)}|\mathcal C|
       \leq d^{O(k)}
          \exp\left(O(k^{O(k)})\frac{H^2}{\Delta^2}
                              \log\frac{H}{\Delta}\right),}
\]
which establishes the required running-time bound.
This completes the proof of \Cref{claim:standalone-capped-search}.
\end{proof}

{To finish, the algorithm draws the stated $n$ samples, computes
the empirical Hermite tensors, rejects if some empirical tensor norm
exceeds $2H$, and otherwise constructs the singular subspaces and performs the net search
just specified. Computing the tensors, their norms, the flattenings,
and the singular subspaces takes
$\poly(n,(d+k)^k)$ arithmetic operations.
Together with the last claim, this gives the stated running time.
There are no additional sampling stages in this finder.

The single event in \Cref{claim:standalone-capped-estimation} has
probability at least $2/3$. On this event the preceding arguments
simultaneously prove the returned-direction guarantee, the
$\textsc{None}$ certificate, and the absence of rejection whenever
all $\|\bA_i\|_F\leq H$. } 
 This completes the proof of \Cref{thm:standalone-capped-finder}.
\end{proof}

\section{Efficient Agnostic Learning via Gaussian Partitioning} \label{sec:agnostic}

As our first application of \Cref{thm:relaxed-gaussian-partitioning},
we give an agnostic learning algorithm for a broad class of
multi-index models under the standard Gaussian and, more generally,
under marginal distributions that are $(q,\sigma)$-smooth with
respect to it. We first present our result and the intuition behind our approach. 
We begin by defining the class of multi-index models we consider.

\begin{definition}[$(\ell,K)$-well-behaved MIMs]\label{def:wellbehaved}
Let $\ell:(0,1]\to\mathbb Z_+$ be non-increasing, $K\in \Z_+$. We say that a class
$\mathcal F\subseteq\{f:\mathbb R^d\to\{\pm1\}\}$ of $K$-MIMs is
$(\ell,K)$-well-behaved if the following hold.

\begin{enumerate}
    \item \emph{(Closure under affine restrictions)}
    For every $f\in\mathcal F$, every subspace $W\subseteq\mathbb R^d$,
    and every $\bb\in W^\perp$, the function
    $
        f_{W,\bb}(\x):=f\bigl(\bb+\proj_W \x\bigr)
    $
    belongs to $\mathcal F$.

    \item \emph{(Gaussian sandwiching)}
  $\mathcal F$ admits degree-$\ell$ $L_1$ sandwiching
polynomials under the standard Gaussian.
    \item \emph{(Bounded Gaussian surface area)}
    The class satisfies
    $\Gamma(\mathcal F)\eqdef\sup_{f\in\mathcal F}\Gamma(f)<\infty$,
    where $\Gamma(f)$ denotes the Gaussian surface area of $f$.

\end{enumerate}
\end{definition}
For this class of functions, we prove the existence of an agnostic
QPTAS under any marginal distribution that is smooth
with respect to the standard Gaussian. Specifically, our main result is the following:
\begin{theorem}[Agnostic QPTAS under Smoothness Relative to a Gaussian]
\label{thm:main-learning}
{Fix $q\geq1$, $c,\eps\in(0,1/2)$, and $\sigma\in(0,1]$.} Let
$\mathcal F\subseteq\{\R^d\to\{\pm1\}\}$ be an
$(\ell,K)$-well-behaved MIM class.
Let $D$ be a distribution over
$\R^d\times\{\pm1\}$ whose $\x$-marginal $D_\x$ {is
$(q,\sigma)$-smooth with respect to $\cN(\vec{0},\bI_d)$}.
Let {$\ell_c=\ell(c'c)$} for a sufficiently small universal constant $c'$.
There exists an algorithm that draws
\[
n
=
d^{O(\ell_c) }\exp\left( {(qK\ell_c)^{O(\ell_c)}\log^{O(\ell_c)}(2/(c\sigma\eps))\log(2+\Gamma(\mathcal F))}   \right)
\]
samples from $D$, runs in $\poly(n)$ time, and, with probability at least
$2/3$, outputs a hypothesis $h:\R^d\to\{\pm1\}$ such that
\[
\err_D(h)
\leq
(1+c)\opt_D(\mathcal F)+\eps.
\]
\end{theorem}

We now briefly outline the intuition behind our result before we move towards the proof.

As in \cite{diakonikolas2026robust}, a core ingredient of our agnostic learning algorithm is that,
given labeled samples whose $\x$-marginal $Q$ is smooth with
respect to a Gaussian distribution, we can agnostically learn
a weak hypothesis over $Q$ in $d^{O(1)}$ time as long as the
label noise is not too large.
By \Cref{fact:boosting}, such a weak learner suffices to obtain
our agnostic approximation guarantee through boosting.

Let $D$ be a distribution over $\R^d\times\{\pm1\}$ whose
$\x$-marginal $D_\x=Q$ is $(q,\sigma)$-smooth with respect to
$\cN(\vec{0},\bI_d)$. The intuition here is that no matter how the label distribution changes, as long as we have sample access to $Q$, we use \Cref{thm:relaxed-gaussian-partitioning} to create a partition $\mathcal{U}$ of $\R^d$ such that, ignoring a tiny amount of $Q$ mass, all the cells $R \in \mathcal{U}$ satisfy the three properties:
\begin{enumerate}
    \item The $Q$ mass of each cell $R$ is not too small and thus can be efficiently sampled from.
    \item The radius of each cell $R$ is sufficiently small.
    \item The low-degree moments of $Q \mid R$ approximately match those of a Gaussian for every direction in $V_R^\perp$.
\end{enumerate}
In fact, any partition of $\R^d$ that satisfies the above three properties is sufficient for us to design an efficient weak learning algorithm. We can apply our partitioning algorithm throughout boosting, since
boosting only uses distributions obtained by mild conditioning,
which preserves smoothness with slightly worse parameters.

Now given such a partition,  since the diameter of a cell $R$ is very small, if we additionally assume that our target function has some noise sensitivity property (such as bounded Gaussian surface area), 
then we would be able to show that our function behaves approximately 
like a function over $V^{\perp}_R$ restricted to each $R$.
Indeed, fixing a representative point $\bz_R\in V_R$, we may approximate the target concept on $R$ by the function $f_R(\x_{V_{R}^\perp}):=f(\bz_R+\x_{V_{R}^\perp})$ (see \Cref{lem:pertubation}). 

Recall that our partition guarantees that $Q\mid R$ approximately
matches the low-degree directional moments of the standard
Gaussian on $V_R^\perp$, with slack $\Delta$.
Since $f_R$ depends only on this subspace and our class is closed
under affine restrictions, we can use Gaussian sandwiching
polynomials for $f_R$ with the degree guaranteed for $\mathcal F$.
By choosing the diameter sufficiently small, we also ensure that
$f_R$ is close to $f$ on average over the partition.
Using these sandwiching polynomials and moment matching, standard
polynomial regression \cite{kalai2008agnostically} yields error
bounds on each cell.
Averaging these bounds over the partition controls the overall
error in terms of the original noise rate and the total
approximation error.
Choosing the Gaussian approximation degree large enough to make
the approximation error a sufficiently small constant then yields
a weak learner, provided that the noise rate is not too large.

Note that our Gaussian Partition guarantees small cell diameter
and moment matching in the orthogonal, unconstrained directions,
allowing us to retain the Gaussian approximation degree.
This degree can be exponentially smaller than the logconcave
approximation degree used in \cite{diakonikolas2026robust},
giving an exponentially smaller exponent on $d$.

We structure the remaining section as follows:
In \Cref{sec:weak-learning}, we construct a weak learner given a Gaussian Partition with suitable parameters.
In \Cref{sec:agnostic-learning}, we combine this
learner with boosting to prove our main agnostic learning result \Cref{thm:main-learning}.
Finally, in \Cref{sec agnostic mixture}, we extend the result
to mixtures of shifted standard Gaussians and to smooth
distributions with an unknown Gaussian reference mean.

\subsection{Efficient Weak Learning Algorithm for Smooth Distributions}\label{sec:weak-learning}
In this section, we show that, given a Gaussian partition satisfying
the guarantees of \Cref{thm:relaxed-gaussian-partitioning} and a
well-behaved function class (\Cref{def:wellbehaved}), we can construct
a weak learner (\Cref{lem:weaklearner}).

Before stating this result, we recall the following boosting
procedure. It yields a relative error guarantee using a weak
learner that achieves error nontrivially below $1/2$ whenever
the noise rate is below a suitable constant.
\begin{fact}[{Corollary of Theorem 3.5 and Remark 3.6 in \cite{Feldman:10ab}}; see \cite{diakonikolas2026robust}]
\label{fact:boosting}
{Fix $a,\beta\in(0,1/2)$.}
Let $\mathcal H$ be a hypothesis class over $\R^d$ and let $\mathcal A$ be an
algorithm with the following guarantee: for every distribution $D$ over
$\R^d\times\{\pm1\}$ such that
$\opt_D(\mathcal H)\leq 1/2-a$, the algorithm draws $m$ examples
from $D$, runs in time $T$, and outputs a hypothesis $\widehat h$ such that
$\err_D(\widehat h)\leq 1/2-\beta$, with probability at least $99\%$. Then
there is an algorithm $\mathcal B$ with the following  guarantee:
for every distribution $D$ over $\R^d\times\{\pm1\}$ and for every $\eps>0$,
$\mathcal B$ makes {$\poly(\beta^{-1},\eps^{-1})$} many calls to $\mathcal A$, it
runs in time $\poly(T,\beta^{-1},\eps^{-1})$, and outputs a hypothesis
$\widehat h$ such that
$\err_D(\widehat h)\leq\opt_D(\mathcal H)/(1-2a)+\eps$, with probability
$99\%$. In particular, each call to $\mathcal A$ is defined over a
distribution $D'$ satisfying
$D'(\x,y)=D(\x,y)w(\x,y)$ for some weight function
$w:\R^d\times\{\pm1\}\to[0,1/\eps]${ with $\E_D[w]=1$}.
{For a fixed input distribution $D$ and accuracy $\eps$, the same conclusion holds if the weak-learning guarantee is available only for all normalized reweightings $D'$ of this form, since the proof uses the weak learner only on these distributions. }
\end{fact}
We are now ready to state our weak learning result for well-behaved
MIMs under the same weak-learning assumption.
\begin{theorem}[Weak Learner]\label{lem:weaklearner}
{Fix $q\geq1$, $\sigma\in(0,1]$, $\eta,\delta,\Delta,\tau\in(0,1/2)$, and $k,d\in\Z_+$.}
{Let $D$ be a distribution on $\mathbb{R}^d\times\{\pm1\}$ whose
$\x$-marginal $Q=D_\x$ is $(q,\sigma)$-smooth with respect to
$\cN(\vec{0},\bI_d)$.}
Let $\mathcal{F}\subseteq \{\R^d\to \{\pm 1\}\}$ be a class that is $(\ell,K)$-well-behaved {and assume that $\opt_D(\mathcal{F})\leq 1/(2(1+c)),c\in(0,1/2)$}. 
Let $\mathcal{P}$ be a partition that satisfies the guarantees of \Cref{thm:relaxed-gaussian-partitioning} for $Q$ with {parameters $(q,\sigma,\eta,\delta,k,\Delta)$, where $\eta=\tau=c/C_1$, $k=\ell(c/C_1)$, $\Delta=(2Kk)^{-C_1k}$, and $\delta=(c\sigma/C_1)^{3q}/(1+\Gamma(\mathcal{F}))^2$ for a sufficiently large constant $C_1>1$. }

Then there is   a weak learner  that, given the partition $\mathcal P$, draws
\[
    n=d^{O(k)}
    \exp\left(
        (qKk)^{O(k)}
        \log^{O(k)}(2/(c\sigma))
        \log(2+\Gamma(\mathcal F))
    \right),
\]
samples from $D$ and runs in $
\poly(n)
$ time 
and with probability at least $99\%$, it outputs a hypothesis $h:\R^d\to \{\pm 1\}$ such that
{$\Pr_{(\x,y)\sim D}[h(\x)\neq y]\leq
1/2-c/12$}.
\end{theorem}

The weak learner from \Cref{lem:weaklearner} is given explicitly in \Cref{alg weak learner}.
\begin{algorithm}[htbp]
    \caption{\textsc{WeakLearner}}
    \label{alg weak learner}
    \begin{algorithmic}[1]

        \State \textbf{Input:}
        Sample access to $D$, a partition $\mathcal P$ satisfying the guarantees of
        \Cref{thm:relaxed-gaussian-partitioning}, the bound $B$, degree $\ell$, and regression error $\tau$

        \State \textbf{Output:}
        A hypothesis $h:\R^d\to\{\pm1\}$

        \medskip

        \State Set $m=d^{O(\ell)}\poly(1/\tau,\log(12B))$.

        \State Draw $N_0=O(B^2\log B)$ fresh samples from $D$.
        \For{$R\in\mathcal P$}
            \State Let $\widehat q_R$ be the empirical fraction of these samples whose $\x$-coordinate lies in $R$.
        \EndFor
        \State Let $\mathcal H=\{R\in\mathcal P:\widehat q_R\geq3/(4B)\}$.

        \State Draw an additional fresh batch of $N_1=O(B(m+\log B))$ samples from $D$ and assign each sample to its cell in $\mathcal P$.

        \For{$R\in\mathcal H$}
            \If{$R$ contains at least $m$ samples from the second batch}
                \State Run the $L_1$ polynomial regression routine of \Cref{fact:polynomial-regression} on $m$ of these samples with degree $\ell$, regression error $\eps=\tau$, and failure probability $1/(12B)$, obtaining $q_R$ and $t_R\in[-1,1]$.
                \State Set $h_R(\x)=\sign(q_R(\x)-t_R)$.
            \Else
                \State Set $h_R(\x)=1$.
            \EndIf
        \EndFor

        \State Define $h(\x)=h_R(\x)$ when $\x\in R\in\mathcal H$, and define $h(\x)=1$ on the remaining cells.
        \State \Return $h$.
    \end{algorithmic}
\end{algorithm}
Before we move on to the proof of \Cref{lem:weaklearner}, we first prove the corresponding perturbation bound that allows us to approximate a function on a small-diameter cell by fixing its constraint coordinates.
\begin{lemma}[Perturbation on a Subcollection]\label{lem:pertubation}
Let $f:\R^d\to\{\pm 1\}$ be a concept with Gaussian surface area $\Gamma(f)$, let $\mathcal{P}$ be a partition of $\R^d$, and let $\mathcal{G}\subseteq\mathcal{P}$ be such that $\diam(R)\leq\delta$ for every $R\in\mathcal{G}$. Then, for every $\rho\in(0,1)$, there exist points $\bz_R\in\proj_{V_R}(R)$, one for each $R\in\mathcal{G}$, such that
\[
\sum_{R\in\mathcal{G}}
\Pr_{\x\sim\cN(\vec{0},\bI)}[\x\in R]
\Pr_{\x\sim\cN(\vec{0},\bI)}\left[
    f(\x)\neq f\left(\bz_R+\x_{V_R^\perp}\right)
    \,\middle|\,\x\in R
\right]
\lesssim
\sqrt{\rho}\,\Gamma(f)+\frac{\delta}{\rho}.
\]
\end{lemma}

\begin{proof}
For each cell $R\in\mathcal{G}$, let $\bu_R$ be an independent draw from the conditional distribution of $\x_{V_R}$ given $\x\in R$, where $\x\sim\cN(\vec{0},\bI)$. Given $\x\sim\cN(\vec{0},\bI)$, define
\[
\x' :=
\begin{cases}
\bu_R+\x_{V_R^\perp}, & \text{if }\x\in R\text{ for some }R\in\mathcal{G},\\
\x, & \text{if }\x\in R\text{ for some }R\in\mathcal{P}\setminus\mathcal{G}.
\end{cases}
\]
If $R\in\mathcal{G}$, then $R$ is determined only by its coordinates in $V_R$, so conditional on $\x\in R$ the random variables $\x_{V_R}$ and $\x_{V_R^\perp}$ are independent, and $\x_{V_R^\perp}$ remains a standard Gaussian on $V_R^\perp$. Consequently, conditional on $\x\in R$, the random variable $\x'$ has the same distribution as $\x$. The same conclusion holds trivially when $R\in\mathcal{P}\setminus\mathcal{G}$, since then $\x'=\x$. Since $\mathcal{P}$ is a partition, it follows that $\x'\sim\cN(\vec{0},\bI)$ marginally.

Let $g=T_\rho f$, where $T_\rho$ denotes the Gaussian noise operator. By the Gaussian noise-sensitivity bound for functions of Gaussian surface area $\Gamma(f)$, we have 
$\E_{\x\sim\cN(\vec{0},\bI)}\left[|f(\x)-g(\x)|\right]
\lesssim \sqrt{\rho}\,\Gamma(f).
$
Moreover, $g$ is $O(1/\rho)$-Lipschitz. If $\x\in R$ for some $R\in\mathcal{G}$, then $\x,\x'\in R$ and their $V_R^\perp$ components are identical, so
\[
\|\x-\x'\|_2
=
\|\x_{V_R}-\bu_R\|_2
\leq \diam(R)
\leq \delta.
\]
If $\x$ lies in a cell of $\mathcal{P}\setminus\mathcal{G}$, then $\x'=\x$. Thus, $\|\x-\x'\|_2\leq\delta$ in all cases. Therefore,
\begin{align*}
\Pr[f(\x)\neq f(\x')]
&=\frac12\E\left[|f(\x)-f(\x')|\right]\\
&\leq \frac12\E\left[|f(\x)-g(\x)|\right]
   +\frac12\E\left[|g(\x)-g(\x')|\right]
   +\frac12\E\left[|g(\x')-f(\x')|\right]\\
&\lesssim \sqrt{\rho}\,\Gamma(f)+\frac{\delta}{\rho},
\end{align*}
where in the last inequality we used that $\x'\sim\cN(\vec{0},\bI)$.

For each $R\in\mathcal{G}$ and $\bz\in\proj_{V_R}(R)$, define
\[
e_R(\bz)
:=
\Pr_{\x\sim\cN(\vec{0},\bI)}\left[
    f(\x)\neq f\left(\bz+\x_{V_R^\perp}\right)
    \,\middle|\,\x\in R
\right].
\]
By construction,
\[
\E_{\bu_R}\left[e_R(\bu_R)\right]
=
\Pr\left[f(\x)\neq f(\x')\,\middle|\,\x\in R\right].
\]
Hence,  there exists $\bz_R\in\proj_{V_R}(R)$ such that
\[
e_R(\bz_R)
\leq
\Pr\left[f(\x)\neq f(\x')\,\middle|\,\x\in R\right].
\]
Since $\x'=\x$ whenever $\x$ lies in a cell of $\mathcal{P}\setminus\mathcal{G}$, these cells contribute no disagreement. Therefore, multiplying by $\Pr_{\x\sim\cN(\vec{0},\bI)}[\x\in R]$ and summing over $R\in\mathcal{G}$ gives
\begin{align*}
&\sum_{R\in\mathcal{G}}
\Pr_{\x\sim\cN(\vec{0},\bI)}[\x\in R]e_R(\bz_R)\\
&\qquad\leq
\sum_{R\in\mathcal{G}}
\Pr_{\x\sim\cN(\vec{0},\bI)}[\x\in R]
\Pr\left[f(\x)\neq f(\x')\,\middle|\,\x\in R\right]\\
&\qquad=
\Pr[f(\x)\neq f(\x')]\\
    &\qquad\lesssim
\sqrt{\rho}\,\Gamma(f)+\frac{\delta}{\rho},
\end{align*}
which proves the claim.
\end{proof}

We are now ready to prove that Gaussian sandwiching approximations along with a Gaussian Partition can 
yield a weak learner when the noise rate is bounded away from $1/2$.
The intuition is as follows.
Let $g$ be the piecewise function defined on each certified cell $R$
by $g(\x):=f(\bz_R+\x_{V_R^\perp})$, where $\bz_R$ is supplied
by the previous lemma.
By that lemma, $g$ is close to $f$ on average over the partition.
Moreover, on each certified cell $R$, the restriction of $g$
depends only on $\x_{V_R^\perp}$ and, by the assumptions on
$\mathcal F$, admits Gaussian sandwiching polynomials.
Together with moment matching, these polynomials give a lower
bound on the correlation with the labels on each certified cell.

A particular cell, however, may have a substantially larger
conditional noise rate than the original distribution, and $f$
need not be close to $g$ on that cell.
We therefore average the correlation bounds over the partition.
This bounds the noise contribution by the global noise rate of $f$
and the approximation contribution by the total error between
$f$ and $g$, yielding a nontrivial overall correlation with the labels.

\begin{proof}[Proof of \Cref{lem:weaklearner}]
{Let $\mathcal{P}$ be a partition that satisfies the guarantees of \Cref{thm:relaxed-gaussian-partitioning} for $Q$ with parameters $(q,\sigma,\eta,\delta,k,\Delta)$.
We bound the error and complexity of \Cref{alg weak learner} given a partition with the stated parameters and then set those parameters accordingly to get the stated weak learning guarantee.
}

First, we establish the existence of a function that is a low-degree polynomial on each cell of the partition and whose approximation error is close to the stated bound. We then apply $L_1$ polynomial regression and thresholding on sufficiently high-mass cells to efficiently obtain a Boolean hypothesis.

Fix $\zeta>0$ and choose $f\in\mathcal F$ such that
$\err_D(f)\leq\opt_D(\mathcal F)+\zeta$.

Let $B$ {be an upper bound on the quantity} supplied by the partition theorem, 
\[B=
    \exp\left({(qk)^{O(k)}
        \log^{O(k)}(2/(\sigma\eta))} \log(1/\delta)/\Delta^{O(1)}\right) \;.\]
Let $\mathcal G\subseteq\mathcal P$ be a collection of good cells whose existence is guaranteed by \Cref{thm:relaxed-gaussian-partitioning}. By item~(ii) of the partitioning theorem,
$\diam(R)\leq\delta$ for every $R\in\mathcal G$.

Let $S=\bigcup_{R\in\mathcal{G}}R$. For each fixed cell $R\in \mathcal{G}$, define the function $f_{R}(\x)=f(\bz_R+\x_{V_R^\perp})$ where $\bz_{R}\in \proj_{V_R}(R)$ is as defined in \Cref{lem:pertubation}, and let $g:S\to \R$ be given by $g(\x)\eqdef \sum_{R\in \mathcal{G}} f_R(\x)\Ind\{\x\in R\}$. 
By {$(q,\sigma)$-smoothness and} \Cref{lem:pertubation} we have that
\begin{align*}
    \Pr_{\x\sim D_\x}[f(\x)\neq g(\x),\x\in S]
    &\leq {(1/\sigma)\Pr_{\x\sim \cN(\vec{0},\bI)}[f(\x)\neq g(\x),\x\in S]^{1/q}}\\
    &\lesssim {\frac{(\delta(1+\Gamma(f))^2)^{1/(3q)}}{\sigma}},
\end{align*}
where we apply \Cref{lem:pertubation} with diameter bound $\delta$ and choose {$\rho=(\delta/(1+\Gamma(f)))^{2/3}\in(0,1)$}.
As a result,
$\Pr_{D}[y\neq g(\x),\x\in S]\leq \err_D(f)+{O\left((\delta(1+\Gamma(f))^2)^{1/(3q)}/\sigma\right)}$.

Fix a cell $R\in\mathcal G$. Since $f$ is a $K$-MIM, so is
$f_R(\x):=f(\bz_R+\x_{V_R^\perp})$, and hence there exists a subspace
$U_R\subseteq V_R^\perp$, with $\dim(U_R)\le K$, on which $f_R$ depends.
Moreover, since $\mathcal F$ is $(\ell,K)$-well-behaved, $f_R\in\mathcal F$,
and therefore there exist degree-$\ell=\ell(\tau)$ polynomials
$p_R^\pm:\R^d\to\R$ such that
$p_R^-(\x)\le f_R(\x)\le p_R^+(\x)$ for all $\x\in\R^d$ and
$\E_{\x\sim\cN(\vec{0},\bI_d)}[p_R^+(\x)-p_R^-(\x)]\le\tau$.
Averaging each $p_R^\pm$ over the standard Gaussian on $U_R^\perp$
preserves both the sandwiching inequalities and their Gaussian error,
since $f_R$ depends only on $U_R$. Thus, after relabeling, we may assume
that $p_R^+$ and $p_R^-$, and hence $s_R:=p_R^+-p_R^-$, depend only on $U_R$.

Let $Q_R$ be the distribution of $\x_{U_R}$ under $D_\x\mid R$. Since $R$ is a good cell and $k\geq\ell$, $Q_R$ and $\cN(\vec{0},\bI_{U_R})$ match their first $\ell$ moments with slack $\Delta$. Moreover, $s_R\geq0$ and $\|s_R\|_{L_1(\cN(\vec{0},\bI_{U_R}))}\leq\tau$. Therefore, \Cref{fact:moment-matching-polynomial-expectations} gives
\[
\left|\E_{\x\sim Q_R}[s_R(\x)]-\E_{\x\sim\cN(\vec{0},\bI_{U_R})}[s_R(\x)]\right|
\leq (K\ell)^{O(\ell)}\Delta\tau.
\]
As a result,
\begin{align*}
\E_D[|y-p_R^+(\x_{V_{R}^{\perp}})|\mid R]
&\leq 2\Pr_D[y\neq g(\x)\mid R]
    +\E_D[p_R^+(\x_{V_{R}^{\perp}})-p_R^-(\x_{V_{R}^{\perp}})\mid R]\\
&\leq 2\Pr_D[y\neq g(\x)\mid R]
    +\E_{\cN(\vec{0},\bI)}[p_R^+(\x_{V_{R}^{\perp}})-p_R^-(\x_{V_{R}^{\perp}})]
    +(K\ell)^{O(\ell)}\Delta\tau\\
&\leq 2\Pr_D[y\neq g(\x)\mid R]+\tau+(K\ell)^{O(\ell)}\Delta\tau,
\end{align*}
where we used moment matching and the sandwiching error.

For $R\in \mathcal G$ define $p_R(\x)\eqdef p_R^+(\x_{V_R^\perp})$ and for $R\notin \mathcal G$
define $p_R(\x)=0$.
Let $p(\x)=\sum_{R\in \mathcal{P}}p_R(\x)\Ind\{\x\in R\}$.
By averaging the preceding cellwise bound and using
$\Pr_D[\x\notin\bigcup_{R\in\mathcal G}R]\leq\eta$, we obtain
\begin{align*}
    \E_D[|y-p(\x)|]
    &=\sum_{R\in\mathcal P}\Pr_D[R]\,
      \E_D[|y-p_R(\x)|\mid R]\\
    &\leq 2\err_D(f)+{O\left(\frac{(\delta(1+\Gamma(f))^2)^{1/(3q)}}{\sigma}\right)}
        +\eta+\tau+(K\ell)^{O(\ell)}\Delta\tau.
\end{align*}

We next analyze \Cref{alg weak learner}, which replaces the existential polynomials $p_R$ by efficiently computable
classifiers without using $\mathcal G$ in the algorithm. Draw
$N_0=O(B^2\log B)$ fresh samples from $D$ and, for each $R\in\mathcal P$, let
$\widehat q_R$ be the empirical fraction of samples whose $\x$-coordinate lies in $R$.
Since $|\mathcal P|\leq B$, Hoeffding's inequality and a union bound imply that, with probability at least $11/12$,
$|\widehat q_R-\Pr_D[R]|\leq1/(4B)$ simultaneously for every $R\in\mathcal P$.
Let
$\mathcal H:=\{R\in\mathcal P:\widehat q_R\geq3/(4B)\}$.
On this event, every $R\in\mathcal G$ belongs to $\mathcal H$, since $\Pr_D[R]\geq1/B$, and every $R\in\mathcal H$ satisfies $\Pr_D[R]\geq1/(2B)$.

For every $R\in\mathcal H$, apply \Cref{fact:polynomial-regression} to $D\mid R$, with degree $\ell$, regression error $\eps=\tau$, and failure probability $1/(12B)$. Each invocation needs
$m=d^{O(\ell)}\poly(1/\tau,\log(12B))$ samples from $D\mid R$ and the same amount of time. Draw an additional fresh batch of
$N_1=O(B(m+\log B))$ samples from $D$ and assign each sample to its cell. Conditional on the first sampling event, every $R\in\mathcal H$ has mass at least $1/(2B)$, so a Chernoff bound and a union bound imply that, with probability at least $11/12$, every cell in $\mathcal H$ contains enough conditional samples. Since $|\mathcal H|\leq B$, the regression calls take total time
$Bd^{O(\ell)}\poly(1/\tau,\log(12B))$.
Another union bound over their failure probabilities shows that, with probability at least $11/12$, all of these regression calls succeed. In particular, since $\mathcal G\subseteq\mathcal H$, simultaneously for all $R\in\mathcal G$ we obtain polynomials $q_R$ and thresholds $t_R\in[-1,1]$ such that
\[
\Pr_D[\sign(q_R(\x)-t_R)\neq y\mid R]
\leq \frac12\E_D[|y-p_R(\x)|\mid R]+\tau.
\]
The intersection of the preceding three events has probability at least $3/4$, and hence at least $2/3$.
Define $h(\x)=\sign(q_R(\x)-t_R)$ when $\x\in R\in\mathcal H$, and define it
arbitrarily on the remaining cells. In the analysis we only use the guarantees on $R\in\mathcal G$. Since
$\Pr_D[\x\notin S]\leq\eta$, averaging the preceding guarantees gives
\begin{align*}
\Pr_D[h(\x)\neq y]
&\leq \Pr_D[\x\notin S]
 +\frac12\sum_{R\in\mathcal G}\Pr_D[R]
   \E_D[|y-p_R(\x)|\mid R]+\tau\\
&\leq \frac12\E_D[|y-p(\x)|]
  +\frac12\Pr_D[\x\notin S]+\tau\\
&\leq \err_D(f)
 +{O\left(\frac{(\delta(1+\Gamma(f))^2)^{1/(3q)}}{\sigma}\right)}
 +\eta+2\tau+(K\ell)^{O(\ell)}\Delta\tau\\
&\leq \opt_D(\mathcal F)+\zeta
 +{O\left(\frac{(\delta(1+\Gamma(\mathcal F))^2)^{1/(3q)}}{\sigma}\right)}
 +\eta+2\tau+(K\ell)^{O(\ell)}\Delta\tau.
\end{align*}
Since $\zeta>0$ is arbitrary, this gives the stated guarantee. 
Since $|\mathcal P|\leq B$, the first sampling stage uses $O(B^2\log B)$ samples and the second uses
$O(Bd^{O(\ell)}\poly(1/\tau,\log B))$ samples. These quantities are {bounded by}
$d^{O(\ell)}\poly(1/\tau) \exp\left({(qk)^{O(k)}
        \log^{O(k)}(2/(\sigma\eta))} \log(1/\delta)/\Delta^{O(1)}\right)$
for sufficiently large constants.

{Let $c\in(0,1/2)$, and suppose that
$\opt_D(\mathcal F)\leq1/(2(1+c))$.
Choose a sufficiently large universal constant $C_1$ and set
$\eta=\tau=c/C_1$, $\ell=\ell(\tau)$, $k=\ell$, and
$
\Delta:=(2Kk)^{-C_1k},
\delta:=
(c\sigma/C_1)^{3q}/(1+\Gamma(\mathcal F))^2.
$
Then
\[
    \err_D(h)
    \leq \frac1{2(1+c)}+O\left(\frac{c}{C_1}\right)
    \leq \frac12-\frac c6.
\]
Repeating the procedure a constant number of times and selecting a
hypothesis using fresh validation samples with error estimates accurate
to $c/24$ yields, with probability
at least $99\%$, a hypothesis with error at most
$1/2-c/12$.
The sample complexity, including amplification, is
\[
    n=d^{O(k)}
    \exp\left(
        (qKk)^{O(k)}
        \log^{O(k)}(2/(c\sigma))
        \log(2+\Gamma(\mathcal F))
    \right),
\]
and the running time is $\poly(n)$.
}
\end{proof}

\subsection{Efficient Agnostic Learning under Gaussian-Smooth Distributions}
\label{sec:agnostic-learning}
In this section, we use the weak learner designed in the previous section in order to obtain our main agnostic learning algorithm (\Cref{thm:main-learning}).
We later apply our results to several MIM classes, and  
\Cref{tab:agnostic-applications} summarizes our concrete applications.

\begin{table}[h]
\centering
\small
\renewcommand{\arraystretch}{1.4}
\begin{tabular}{@{}p{0.62\linewidth}c@{}}
\hline
\textbf{Concept class}
& \textbf{Complexity} \\
\hline
$(\ell,K)$-well-behaved MIMs
& $d^{O(\ell(c'c))}$ \\

Intersections or unions of $k$ halfspaces
& $d^{\widetilde O(k^3/c^4)}$ \\

Arbitrary Boolean functions of $k$ halfspaces
& $d^{\widetilde O(k^5/c^4)}$ \\

Convex concepts of intrinsic dimension $K$
& $d^{\widetilde O(K^5/c^4)}$ \\

Degree-$r$ PTFs of intrinsic dimension $K$
& $d^{\widetilde O(r^6K^5/c^4)}$ \\
\hline
\end{tabular}
\caption{Agnostic learning results.
All results hold for marginal distributions that are
$(q,\sigma)$-smooth with respect to the standard Gaussian.
The displayed complexities have an additional factor
quasi-polynomial in $1/(\sigma\eps)$ for fixed class parameters,
$q$, and $c$.}
\label{tab:agnostic-applications}
\end{table}
Combining the boosting procedure in \Cref{fact:boosting}
with our weak learner from \Cref{lem:weaklearner},
we can now prove \Cref{thm:main-learning}.

\begin{proof}[Proof of \Cref{thm:main-learning}]
Set $\alpha=\sigma\eps$.
Every distribution $D'$ used by \Cref{fact:boosting} satisfies
$D'(\x,y)=D(\x,y)w(\x,y)$ for some
$w:\R^d\times\{\pm1\}\to[0,1/\eps]$ with $\E_D[w]=1$.
Writing $G=\cN(\vec{0},\bI_d)$, for every measurable
$E\subseteq\R^d$ we have
\[
    D'_\x(E)\leq D_\x(E)/\eps
    \leq G(E)^{1/q}/(\sigma\eps)
    =G(E)^{1/q}/\alpha.
\]
Thus $D'_\x$ is $(q,\alpha)$-smooth with respect to $G$.

Apply \Cref{thm:relaxed-gaussian-partitioning} to $D'_\x$
with the parameters specified in \Cref{lem:weaklearner},
using smoothness parameter $\alpha$.
Whenever $\opt_{D'}(\mathcal F)\leq1/(2(1+c))$, the resulting
partition and \Cref{lem:weaklearner} give a hypothesis
with error at most $1/2-c/12$ with  probability at least $99\%$.
This supplies the weak learner required
by \Cref{fact:boosting}. Consequently,
\[
    \err_D(h)
    \leq
    \frac{\opt_D(\mathcal F)}
         {1-2(1/2-1/(2(1+c)))}+\eps
    =(1+c)\opt_D(\mathcal F)+\eps.
\]

To verify the complexity, take $c'\leq1/C_1$.
Since $\ell$ is non-increasing,
$\ell(c/C_1)\leq\ell(c'c)=\ell_c$.
The partition construction and the amplified weak learner
therefore use at most
\[
    n_{\rm w}
    =d^{O(\ell_c)}
    \exp\left(
        (qK\ell_c)^{O(\ell_c)}
        \log^{O(\ell_c)}(2/(c\sigma\eps))
        \log(2+\Gamma(\mathcal F))
    \right)
\]
samples per call and polynomial time.

Samples from $D'$ are simulated by accepting samples from $D$
with probability $\eps w(\x,y)$, whose expectation is $\eps$.
Let $R=\poly(1/c,1/\eps)$ bound the number of boosting calls.
Drawing 
$O((n_{\rm w}+\log(100R))/\eps)$ samples for each weak learner call supplies the weak learner with $n_w$ samples with 
failure probability at most $1/100$, by the Hoeffding and union bounds.
Combining this with the $99\%$ success probability of boosting
gives overall success probability at least $98\%$.
Noticing that the number of weak learner calls as well as the time complexity of the weak learner are polynomial justifies our sample and time complexity bounds, completing the proof. 
\end{proof}

{
We first apply \Cref{thm:main-learning} to intersections of halfspaces.
\begin{corollary}[Agnostic Learning of Intersections of Halfspaces]
\label{cor:agnostic-intersections}
Let $q\geq1$, $k,d\in\N_+$, $c,\eps\in(0,1/2)$, and
$\sigma\in(0,1]$.
Let $\mathcal F_k$ be the class of intersections of at most
$k$ affine halfspaces over $\R^d$.
Let $D$ be a distribution over $\R^d\times\{\pm1\}$ whose
$\x$-marginal is $(q,\sigma)$-smooth with respect to
$\cN(\vec{0},\bI_d)$.
There is an algorithm that draws
\[
    n=d^{\widetilde O(k^3/c^4)}
    \exp\left(
        (qk/c)^{(k/c)^{O(1)}}
        \log^{(k/c)^{O(1)}}(2/(\sigma\eps))
    \right)
\]
samples from $D$, runs in time $\poly(n)$, and, with
probability at least $2/3$, outputs a hypothesis
$h:\R^d\to\{\pm1\}$ satisfying
$\err_D(h)\leq(1+c)\opt_D(\mathcal F_k)+\eps$.
\end{corollary}

\begin{proof}[Proof of \Cref{cor:agnostic-intersections}]
The class $\mathcal F_k$ consists of $k$-MIMs and is closed
under affine restrictions.
By
\Cref{fact:gaussian-sandwiching-smooth-boundary,fact:gaussian-boundary-smoothness},
it admits Gaussian $L_1$ sandwiching polynomials of degree
$\ell(\tau)=\widetilde O(k^3/\tau^4)$ and satisfies
$\Gamma(\mathcal F_k)=O(\sqrt{\log(2k)})$.
Thus $\mathcal F_k$ is $(\ell,k)$-well-behaved.
Applying \Cref{thm:main-learning} with $K=k$ and
$\ell_c=\ell(\Omega(c))=\widetilde O(k^3/c^4)$ gives the result.
\end{proof}
}

We next apply \Cref{thm:main-learning} to arbitrary Boolean
functions of halfspaces.
\begin{corollary}[Agnostic Learning of Functions of Halfspaces]
\label{cor:agnostic-functions-halfspaces}
{Let $q\geq1$, $k,d\in\mathbb N_+$, $c,\eps\in(0,1/2)$, and
$\sigma\in(0,1]$.}
Let $\mathcal H_k$ denote the class of arbitrary Boolean
functions of at most $k$ affine halfspaces over $\R^d$.
Let $D$ be a distribution over $\R^d\times\{\pm1\}$ whose
$\x$-marginal {$D_\x$ is $(q,\sigma)$-smooth with respect to
$\cN(\vec{0},\bI_d)$}.

There exists an algorithm that draws
\[
    n=
    d^{\widetilde O(k^5/c^4)}
    \exp\left(
    {(qk/c)}^{(k/c)^{{O(1)}}}
    \log^{(k/c)^{{O(1)}}}{(2/(\sigma\eps))}
    \right)
\]
samples from $D$, runs in time polynomial in $n$, and,
with probability at least $2/3$, outputs a hypothesis
$h:\R^d\to\{\pm1\}$ satisfying
\[
    \err_D(h)
    \leq (1+c)\opt_D(\mathcal H_k)+\eps.
\]
\end{corollary}

\begin{proof}
The class $\mathcal H_k$ is closed under affine restrictions,
since restricting an affine halfspace gives an affine
halfspace or a constant function.
Moreover, every $f\in\mathcal H_k$ depends only on the span
of the normals of its defining halfspaces, and hence is
a $k$-MIM.

The decision boundary of $f$ is contained in the union of
at most $k$ hyperplanes.
Gaussian anti-concentration therefore gives an $O(k)$
boundary-smoothness bound, and
$\Gamma(\mathcal H_k)=O(k)$.
By \Cref{fact:gaussian-sandwiching-smooth-boundary},
$\mathcal H_k$ admits Gaussian $L_1$ sandwiching polynomials
of degree
$
    \ell(\tau)=\widetilde O(k^5/\tau^4).
$
Thus $\mathcal H_k$ is $(\ell,k)$-well-behaved.
Applying \Cref{thm:main-learning} with {smoothness parameters $(q,\sigma)$ and} $K=k$ and substituting
$\ell(\Omega(c))=\widetilde O(k^5/c^4)$ and
$\Gamma(\mathcal H_k)=O(k)$ yields the stated sample
complexity, running time, and error guarantee.
\end{proof}
Applying our general theorem for Low-Dimensional convex sets gives the following corollary.  
\begin{corollary}[Agnostic Learning of Low-Dimensional Convex Sets]
\label{cor:agnostic-convex}
Let $q\geq1$, $K,d\in\mathbb N_+$, $c,\eps\in(0,1/2)$, and
$\sigma\in(0,1]$.
Let $\mathcal C_K$ denote the class of Boolean $K$-MIMs over
$\R^d$ whose positive regions are convex.
Let $D$ be a distribution over $\R^d\times\{\pm1\}$ whose
$\x$-marginal $D_\x$ is $(q,\sigma)$-smooth with respect to
$\cN(\vec{0},\bI_d)$.

There exists an algorithm that draws
\[
    n=
    d^{\widetilde O(K^5/c^4)}
    \exp\left(
        (qK/c)^{(K/c)^{O(1)}}
        \log^{(K/c)^{O(1)}}(2/(\sigma\eps))
    \right)
\]
samples from $D$, runs in time polynomial in $n$, and,
with probability at least $2/3$, outputs a hypothesis
$h:\R^d\to\{\pm1\}$ satisfying
\[
    \err_D(h)
    \leq (1+c)\opt_D(\mathcal C_K)+\eps.
\]
\end{corollary}

\begin{proof}
The class $\mathcal C_K$ is closed under affine restrictions,
since affine preimages preserve convexity and do not increase
the intrinsic dimension.

By \Cref{fact:gaussian-boundary-smoothness},
$\sigma(\mathcal C_K)=O(K\log(2K))$ and
$\Gamma(\mathcal C_K)\leq\sigma(\mathcal C_K)$.
By \Cref{fact:gaussian-sandwiching-smooth-boundary},
$\mathcal C_K$ admits Gaussian $L_1$ sandwiching polynomials
of degree $\ell(\tau)=\widetilde O(K^5/\tau^4)$.
Thus $\mathcal C_K$ is $(\ell,K)$-well-behaved.
Applying \Cref{thm:main-learning} with smoothness parameters
$(q,\sigma)$ and substituting
$\ell(\Omega(c))=\widetilde O(K^5/c^4)$ and the above
surface area bound yields the stated sample complexity,
running time, and error guarantee.
\end{proof}
Applying our general theorem also gives a result for Low-Dimensional PTFs.
\begin{corollary}[Agnostic Learning of Low-Dimensional PTFs]
\label{cor:agnostic-low-dimensional-ptfs}
Let $q\geq1$, $r,K,d\in\mathbb N_+$, $c,\eps\in(0,1/2)$, and
$\sigma\in(0,1]$.
Let $\mathcal P_{K,r}$ denote the class of polynomial threshold
functions over $\R^d$ of degree at most $r$ that are $K$-MIMs.
Let $D$ be a distribution over $\R^d\times\{\pm1\}$ whose
$\x$-marginal $D_\x$ is $(q,\sigma)$-smooth with respect to
$\cN(\vec{0},\bI_d)$.

There exists an algorithm that draws
\[
    n=
    d^{\widetilde O(r^6K^5/c^4)}
    \exp\left(
        (qrK/c)^{(rK/c)^{O(1)}}
        \log^{(rK/c)^{O(1)}}(2/(\sigma\eps))
    \right)
\]
samples from $D$, runs in time polynomial in $n$, and,
with probability at least $2/3$, outputs a hypothesis
$h:\R^d\to\{\pm1\}$ satisfying
\[
    \err_D(h)
    \leq (1+c)\opt_D(\mathcal P_{K,r})+\eps.
\]
\end{corollary}

\begin{proof}
The class $\mathcal P_{K,r}$ is closed under affine restrictions,
since composing a polynomial with an affine map does not increase
its degree, and the intrinsic dimension remains at most $K$.

By \Cref{fact:gaussian-boundary-smoothness},
$\sigma(\mathcal P_{K,r})=O(r^3K)$ and
$\Gamma(\mathcal P_{K,r})\leq\sigma(\mathcal P_{K,r})$.
By \Cref{fact:gaussian-sandwiching-smooth-boundary},
$\mathcal P_{K,r}$ admits Gaussian $L_1$ sandwiching polynomials
of degree $\ell(\tau)=\widetilde O(r^6K^5/\tau^4)$.
Thus $\mathcal P_{K,r}$ is $(\ell,K)$-well-behaved.
Applying \Cref{thm:main-learning} with smoothness parameters
$(q,\sigma)$ and substituting
$\ell(\Omega(c))=\widetilde O(r^6K^5/c^4)$ and the above
surface area bound yields the stated sample complexity,
running time, and error guarantee.
\end{proof}

\begin{remark}[Applications to Natural Distribution Families]
{\em Our agnostic learning algorithm works for all $(q,\sigma)$-smooth distributions with respect to the standard Gaussian. 
We are able to show that this includes some natural distribution families. In fact, 
mixtures of Gaussians with identity covariance and component means
of norm at most $M$ are $(2,e^{-M^2/2})$-smooth with respect to
$\cN(\vec{0},\bI_d)$
(see \Cref{lem:gaussian-mixture-smoothness}), so
\Cref{thm:main-learning} applies directly.
Moreover, for classes closed under translations, the same learning
guarantee holds for marginals that are $(q,\sigma)$-smooth with
respect to $\cN(\boldsymbol{\mu},\bI_d)$ for an unknown
$\boldsymbol{\mu}$, without sample access to the reference Gaussian.
See \Cref{sec agnostic mixture} for details.}
\end{remark}

\section{Learning with Positive and Imperfect Unlabeled Examples}\label{sec PU}

In this section, we give another application of \Cref{thm:relaxed-gaussian-partitioning} to a different reliable learning problem, \cite{lee2026smooth}, learning from positive examples and imperfect unlabeled data.

\begin{definition}[PUI learning with Imperfect Data]
    Let $\mathcal{F} \subseteq \{f: \R^d \to \{\pm 1\}\}$ be a hypothesis class. Let $D^*$ be a distribution over $\R^d \times  \{\pm 1\}$ such that there exists some $f^* \in \mathcal{F}$ with $\Pr_{(\x,y)\sim D^*}[f^*(\x) \neq y] = 0$ and $\Pr_{(\x,y)\sim D^*}[y = 1] \ge p_+$, where $p_+ \in (0,1)$. Suppose that there exists a reference distribution $D$ over $\R^d$ such that $D^*_\x$ is $(q,\sigma)$-smooth with respect to $D$. Given sample access to $D$ and $D^*_+$, where $D^*_+:=D^*_\x \mid y =1$, and $\eps \in (0,1)$, the goal is to output a hypothesis $h:\R^d \to \{\pm 1\}$ such that $\err_{D^*}(h):= \Pr_{(\x,y)\sim D^*}[h(\x) \neq y]\le \eps$.
\end{definition}
The algorithm proposed by \cite{lee2026smooth} uses an $L_1$-approximating polynomial developed in the agnostic learning literature as a subroutine. Roughly speaking, their algorithm requires an $L_1$-approximate polynomial to functions in $\mathcal{F}$ to an accuracy of $(\eps\sigma p_+)^{O(q)}$. Even if the reference distribution $D = \cN(\vec 0,\vec I)$ and $\mathcal{F}$ is the class of halfspaces, this results in a time complexity of $d^{(\eps\sigma p_+)^{-O(q)}}$. We would like to mention that if the function class $\mathcal{F}$ is adversarially constructed, for example, as the one used in \cite{diakonikolas2024truncation}, then according to \cite{lee2026smooth}, a time complexity of $d^{\Omega(1/\eps^2)}$ is unavoidable. 
However, as we show in this section, for many natural concept classes, one can potentially achieve a much better complexity.
In fact, we will see that the Gaussian partitioning algorithm, \Cref{thm:relaxed-gaussian-partitioning}, directly gives a much more efficient algorithm for the problem when the reference distribution $D = \cN(\vec 0,\vec I)$. Our starting point is to introduce the following definition called the Gaussian PUI learning Exponent, which qualitatively determines the computational complexity of PUI learning with a Gaussian reference.

\begin{definition}[Gaussian PUI Learning Exponent] \label{def:puexponent}
Let $\mathcal{F}\subseteq\{f:\R^d\to\{\pm1\}\}$ be a class of functions closed under affine transformations and let $D\eqdef \cN(\vec{0},\bI_d)$. We define the Gaussian PUI learning exponent {$m^*(\eps,\Delta,\sigma,q)$} to be the minimum integer $k\in\Z_+$ such that, for every $f\in\mathcal{F}$ and every distribution {$D^*$ that is $(q,\sigma)$-smooth with respect to $D$} satisfying $D^*(f(\x)=-1)\geq\eps$ and $D^*(f(\x)=1)>0$ , the positive conditional distribution $D_+^*\eqdef D^*\mid f(\x)=1$ cannot match the first $k$ moments of $\cN(\vec{0},\bI_d)$ with slack $\Delta$ in the sense of \Cref{def:moment-matching}. {If no such integer exists, set $m^*(\eps,\Delta,\sigma,q)=\infty$. } \end{definition}

The exponent $m^*$ allows us to both design efficient PUI learning algorithms and analyze the hardness of the learning problem. Specifically, the main results in this section are the following pair of theorems that quantify the computational complexity of PUI learning (for multi-index models).

Our main algorithmic result in this section is the following:

\begin{theorem}[PUI Learning with Gaussian References]\label{thm:pumim} Let $\mathcal{F}$ be a class of functions that is closed under affine restrictions {and has bounded Gaussian surface area $\Gamma(\mathcal{F})<\infty$} (satisfies items (1) and (3) of \Cref{def:wellbehaved}), let the reference distribution be the standard Gaussian $D=\cN(\vec{0},\bI_d)$, and let $D^*$ be a target distribution {that is $(q,\sigma)$-smooth with respect to $D$,  $q\geq1$ and $\sigma\in(0,1]$.}
Let $f\in\mathcal F$ be an unknown function, let $p_+>0$ satisfy $D^*(f=1)\geq p_+$, and set $D_+^*=D^*\mid f(\x)=1$.
Let $\eps,\Delta\in(0,1/2)$ and set {$k=m^{*}(c\eps,\Delta,c'\eps\sigma p_+/q,2q)$ for sufficiently small universal constants $c,c'>0$, and assume $k<\infty$.} There exists a PUI learning algorithm that, given {$\eps,\sigma,p_+,\Delta,q$} and {$d^{k}\exp((qk)^{O(k)}\log^{O(k)}(2/(\sigma\eps p_+))\log(2+\Gamma(\mathcal F))/\Delta^{O(1)})$} samples from $D_+^*$,  runs in time polynomial in its sample complexity and with probability at least $2/3$ outputs a hypothesis $h$ such that $\Pr_{\x\sim D^*}[h(\x)\neq f(\x)]\leq\eps$. 
\end{theorem}

Perhaps surprisingly, we show a qualitatively matching SQ lower bound:

\begin{theorem}[SQ Lower Bounds for PUI Learning of Gaussian MIMs]
\label{thm:sq-pu-general}
Let $d,K,m^*\in\Z_+$ with $m^*\geq2$ and
$\max\{K,m^*\}\leq d^c$, for a sufficiently small absolute
constant $c>0$ and sufficiently large $d$.
Let $\sigma,\eps\in(0,1]$ and $q\geq1$.
Let $\mathcal F\subseteq\{\R^d\to\{\pm1\}\}$ be a class
of Boolean functions. Suppose there exists $f\in\mathcal F$
such that:
\begin{enumerate}[leftmargin=*]
    \item $f$ depends only on the projection onto a subspace
    $V\subseteq\R^d$ of dimension $K$, i.e.,
    $f(\x)=g(\vec{W}\x)$ for $\vec{W}\in\R^{K\times d}$ satisfying
    $\vec{W} \vec{W}^\top=\bI_K$.

    \item $\mathcal F$ contains every rotation of $f$, i.e.,
    $\x\mapsto g(\bU\x)$ belongs to $\mathcal F$ for every
    $\bU\in\R^{K\times d}$ satisfying $\bU \bU^\top=\bI_K$.

    \item There exists a distribution
    $D^*$ that is $(q,\sigma)$-smooth with respect to $\cN(\vec{0},\bI_d)$ such that
    $p_+:=\Pr_{\x\sim D^*}[f(\x)=1]\in(0,1)$,
    $\Pr_{\x\sim D^*}[f(\x)=-1]\geq\eps$, and
    $(\sigma\eps)^q,p_+\geq d^{-cm^*}$.
    Moreover, $D_+^*:=D^*\mid\{f(\x)=1\}$ and $\cN(\vec{0},\bI_d)$
    match moments through degree $m^*$.
\end{enumerate}
Then any SQ algorithm that learns $\mathcal F$ over
distributions that are $(q,\sigma)$-smooth with respect to $\cN(\vec{0},\bI_d)$,
using SQ access only to the positive conditional distribution $D_+^*$,
to error $p_+(\sigma\eps)^q/2$ with probability at least $2/3$,
requires either $2^{d^{\Omega(1)}}$ queries or a query
of tolerance at most $d^{-\Omega(m^*)}$.
\end{theorem}

\begin{remark}[Qualitative Tightness]\label{rem:sq-comparison}
{\em The upper and lower bounds establish a qualitative correspondence between moment matching and SQ complexity, rather than a quantitative characterization by the same value of the PUI learning exponent. The proof of \Cref{thm:pumim} achieves error $\eps$ with dimension exponent
$
    m^*\!\left({\eps}/{8},\Delta,
               {\eps\sigma p_+}/{(12800q)},2q\right).
$
In contrast, under the assumptions of \Cref{thm:sq-pu-general}, an instance whose positive conditional distribution matches Gaussian moments exactly through degree $m$, and whose negative mass is at least $\rho$, yields SQ complexity $d^{\Omega(m)}$ for learning error $p_+(\sigma\rho)^q/2$. Such a witness implies $m^*(\rho,\Delta,\sigma,q)>m$ for every $\Delta>0$; the inequality for a fixed positive slack alone does not suffice to invoke the lower bound.

To compare the bounds at learning error $\eps$, one therefore sets $\rho=(2\eps/p_+)^{1/q}/\sigma$, provided the resulting parameters satisfy the lower-bound theorem's assumptions. The parameter differences arise from conditioning in the upper-bound proof, which changes the smoothness parameters, and from the accuracy required by the reduction from PUI learning to distinguishing in the lower-bound proof.

If $q$ is fixed and $\sigma^q p_+\geq\eps^{1-a}$ for some fixed $a\in(0,1)$, then $\rho\leq2^{1/q}\eps^{a/q}$, so the accuracy rescaling changes $\eps$ only by a fixed power. For logarithmic moment thresholds, these parameter changes can preserve the dimension exponent up to constant factors depending on $q$. This occurs for $q=1$ in the regime of \Cref{cor:sq-pu-intersections}.}
\end{remark}

\paragraph{Roadmap of \Cref{sec PU}}
In \Cref{sec PU learning general}, we give the proof of \Cref{thm:pumim}, giving the general PUI learning algorithm with a Gaussian reference. In \Cref{sec:lowerboundspu}, we prove \Cref{thm:sq-pu-general}, showing that the $d^{m^*}$ dependence is qualitatively optimal for multi-index models.
In \Cref{sec:struct-intersections}, we focus on natural concept classes such as halfspaces, intersections of halfspaces, and general $ K$-dimensional convex sets. In particular, for the class of intersections of $k$ halfspaces $\mathcal{F}_k$, we show that $m^* = \Theta(\log(k/(\sigma\eps)))$, implying these natural classes have much more efficient PUI learning algorithms.

\subsection{Faster Algorithm for PUI Learning with Gaussian Reference}\label{sec PU learning general}

\begin{algorithm}[htbp]
    \caption{\textsc{GaussianPUILearner}}
    \label{alg:gaussian-pu}
    \begin{algorithmic}[1]
        \State \textbf{Input:} Sample access to $D_+^*$ and 
        parameters {$\eps,\Delta,\sigma,q$}, and a lower bound $0<p_+\leq D^*(f=1)$. 
        \State \textbf{Output:} A hypothesis $h$ satisfying the guarantees of \Cref{thm:pumim}.
        \medskip
        \State Set $\Gamma=1+\Gamma(\mathcal F)$,
        $\alpha=\sigma p_+/4$, and {$\eta=\eps/8$}.
        \State Set {$\sigma_*=\alpha\eta/(400q)$} and
        {$k=m^*(\eta,\Delta,\sigma_*,2q)$}.
        \State Set $M_k=\sqrt{(2k-1)!!}$ and
        $\Delta_0=\Delta/(2^{k+2}M_k)$.
                \State Set
        {$\delta=\Gamma^{-2}
        \left(\frac{\alpha\eta\Delta}{Cqk}\right)^{Cq(k+q)}$},
        {for a sufficiently large universal constant $C>1$.}
        
        \State Run the Gaussian partitioning procedure of
        \Cref{thm:relaxed-gaussian-partitioning} on $D_+^*$ with parameters
        {$(q,\alpha,\eta,\delta,k,\Delta_0)$}, obtaining a partition
        $\mathcal P$ and its certified subcollection $\mathcal G_0$.
        \State {Set $\lambda=\eta/100$, and let $B$ be the cell-count bound in
        \Cref{thm:relaxed-gaussian-partitioning}.}
        \State {Draw $\lceil8B^2\log(200B)\rceil$ fresh samples from $D_+^*$, 
        let $\widehat q_R$ be their empirical mass in each $R\in\mathcal G_0$.}\label{line:empirical-ratio}
        \State {Set $\mathcal G=\{R\in\mathcal G_0:\widehat q_R>2\lambda D(R)\}$.} \label{line:ratiotest}

        \State \Return the hypothesis
        \[
            h(\x)=
            \begin{cases}
                1, & \x\in\displaystyle\bigcup_{R\in\mathcal G}R,\\
                -1, & \text{otherwise}.
            \end{cases}
        \]
    \end{algorithmic}
\end{algorithm}

In this section, we give the proof of \Cref{thm:pumim}.
Our learning algorithm is stated as \Cref{alg:gaussian-pu}. The definition of $m^*$ suggests a very simple PUI learning algorithm. For intuition, we first consider a one-dimensional PUI learning problem.
We will use the contrapositive part of $m^*$. Suppose that $D^*_+$ matches its first $m^*$ moments with respect to a standard Gaussian distribution within a slack $\Delta$. Then this implies that the fraction of negative examples under $D^*$ must be at most $\eps$. Thus, we can safely output a hypothesis that labels every example to be positive.
However, the moment matching condition in general fails for a general distribution $D^*_+$ over $\R^d$. 
Fortunately, since $D^*$ is smooth with respect to a Gaussian, as long as $p_+$ is not too small, $D^*_+$ is also smooth with respect to the Gaussian distribution. We therefore apply our Gaussian partitioning algorithm with $Q = D^*_+$ and $k = m^*$.

Recall \Cref{thm:relaxed-gaussian-partitioning}, our Gaussian partitioning algorithm partitions $\R^d$ into cells such that most of $D^*_+$ mass falls into some $R$ with small radius and satisfies the moment-matching condition over $V_R^\perp$. 
Furthermore, $D^*_+ \mid R$ is still smooth.
The small radius condition together with a mild noise sensitivity condition (Gaussian surface area) implies that we can assume, restricted at $R$, the target function $f$ only depends on directions in $V_R^\perp$.

Following our discussion above,
the moment matching condition together with the smoothness condition implies that if we label all examples in $R$ as positive, we introduce at most $\eps$ conditional error. Thus, if we label all examples over these moment matching cells $R$ as positive and the other examples as negative, then the fraction of false positives is at most $O(\eps)$. On the other hand, since the fraction of false negatives under $D^*_+$ is at most $O(\eps)$, it is also at most $O(\eps)$ under $D^*$. Thus, we get a PUI learning algorithm with complexity $d^{m^*}$.

\begin{proof}[Proof of \Cref{thm:pumim}]
Set $\Gamma\eqdef 1+\Gamma(\mathcal F)$, so that $\Gamma(f)\leq\Gamma$. 
Let $\mathcal P$ and $\mathcal G_0$ be the partition and certified
subcollection returned by \Cref{thm:relaxed-gaussian-partitioning} for
$D_+^*$ with parameters $(q,\alpha,\eta,\delta,k,\Delta_0)$ as in
\Cref{alg:gaussian-pu}, and let $\mathcal G\subseteq\mathcal G_0$
be the cells retained by its ratio test (Line \ref{line:ratiotest}).
We take $\alpha=\sigma p_+/4\leq\sigma D^*(f=1)$, so that
{$D_+^*$ is $(q,\alpha)$-smooth with respect to $D$} and $\alpha<1/2$, as required by
\Cref{thm:relaxed-gaussian-partitioning}. We use {$\eta=\eps/8$} and
{$k=m^*(\eta,\Delta,\sigma_*,2q)$} with
{$\sigma_*=\alpha\eta/(400q)$}.
{For the analysis, set $M_k:=\sqrt{(2k-1)!!}$, $M:=1/(\sigma^q\eta^{q-1})$,
$\pi_0:=p_+\eta/(100M)$,
$A_{q,k}:=8\sqrt{4q}(8q-1)^{k/2}M_k$, and
$r_0:=(\sigma_*\pi_0\Delta/A_{q,k})^{4q}$.}
Let $B$ be the bound in \Cref{thm:relaxed-gaussian-partitioning}
for these parameters. The theorem guarantees that
$\diam(R)\leq\delta$ for every $R\in\mathcal G_0$ and
$|\mathcal P|\leq B$.
{Every $R\in\mathcal G_0$ also satisfies $D_+^*(R)\geq1/B$.}

{Let $\lambda=\eta/100$. Hoeffding's inequality and a union bound give
$|\widehat q_R-Q(R)|\leq1/(4B)\leq Q(R)/4$ simultaneously
for all $R\in\mathcal G_0$, with probability at least $99/100$, for the estimates of Line \ref{line:empirical-ratio}.
On this event, every retained cell satisfies
$Q(R)\geq(4/5)\widehat q_R>\lambda D(R)$.
Every discarded cell satisfies
$Q(R)\leq(4/3)\widehat q_R\leq(8/3)\lambda D(R)\leq3\lambda D(R)$.
Therefore, summing the aforementioned inequality we have that  
on its estimation event, every retained cell satisfies
$D_+^*(R)/D(R)\geq\eta/100$, and
$D_+^*(\bigcup_{R\in\mathcal G_0\setminus\mathcal G}R)\leq3\lambda$.}

{The Gaussian masses $D(R)$ are computable from the orthogonal interval
constraints defining each cell.
In what follows, work on the joint success event of the partitioning routine
and the estimation event.}

Let $h$ be defined by $h(\x)=1$ if $\x\in R$ for some $R\in\mathcal G$, and $h(\x)=-1$ otherwise.
\begin{claim}[Small Mass of False Negatives]\label{claim:pu-false-negatives}
We have $\Pr_{\x\sim D^*}[h(\x)=-1,f(\x)=1]\leq\eta+3\lambda.$
\end{claim}
\begin{proof}
Note that by \Cref{thm:relaxed-gaussian-partitioning} we had that the total mass of $D^*_+$ outside $\mathcal{G}_0$ is at most $\eta$. 
Therefore, since only $3\lambda$ mass of $D_+^*$  belongs in $\mathcal{G}_0\setminus \mathcal{G}$ we have that 
$D_+^*(S)\geq1-\eta-3\lambda$, for  $S\eqdef\bigcup_{R\in\mathcal G}R$.

Since $h(\x)=-1$ exactly when $\x\notin S$ and $D_+^*=D^*\mid f(\x)=1$, it follows that
$\Pr_{\x\sim D^*}[h(\x)=-1,f(\x)=1]=D^*(f(\x)=1)D_+^*(S^c)
{\leq(\eta+3\lambda)D^*(f(\x)=1)\leq\eta+3\lambda.}$
\end{proof}

\begin{lemma}[Small Mass of False Positives]\label{claim:pu-false-positives}
{We have} {$\Pr_{\x\sim D^*}[h(\x)=1,f(\x)=-1]\leq3\eta
    +O\left(\frac1\sigma\left(\frac{(\delta\Gamma^2)^{1/3}}{r_0}\right)^{1/q}\right).$}
\end{lemma}

\begin{proof}[Proof of \Cref{claim:pu-false-positives}]Applying \Cref{lem:pertubation} with $\rho=(\delta/\Gamma)^{2/3}$ gives points $\bz_R\in\proj_{V_R}(R)$ such that, for $f_R(\x)\eqdef f(\bz_R+\proj_{V_R^\perp}\x)$, we have $\sum_{R\in\mathcal G}\Pr_{\x\sim D}[f(\x)\neq f_R(\x)\mid R]D(R)=O((\delta\Gamma^2)^{1/3})$.
Moreover, by closure under affine restrictions, $f_R\in\mathcal F$.

{Write $A:=\sum_{R\in\mathcal G}D(R)\Pr_D[f\ne f_R\mid R]=O((\delta\Gamma^2)^{1/3})$. Applying $(q,\sigma)$-smoothness once to the union $\bigcup_{R\in\mathcal G}(R\cap\{f\ne f_R\})$, we have}
{\begin{align*}
    \Pr_{\x\sim D^*}[h(\x)=1,f(\x)=-1]
    &\leq \sum_{R\in\mathcal G}
    \Pr_{\x\sim D^*}[f_R(\x)=-1\mid R]D^*(R)
    +\frac{A^{1/q}}{\sigma}.
\end{align*}}

Note that, moreover, by \Cref{thm:relaxed-gaussian-partitioning}, $D_+^*\mid R$ matches moments with $D\mid R$ on $V_R^\perp$.
Also, $D|R$ on $V_R^{\perp}$ is a standard Gaussian. 

We aim to apply the definition of $m^*$ for the functions $f_R$ for cells $R\in \mathcal G$. 
In order to do that we  show that 
for almost all cells in $\mathcal G$, measured
by their $D^*$-mass, the distribution $Q_R$ obtained by projecting
$D^*\mid R$ onto $V_R^\perp$ satisfies that $Q_R\mid f_R=1$
approximately matches the first $k$ moments of the standard Gaussian
on $V_R^\perp$. 
We obtain this by using the fact that $f$ is very close to $f_R$ on average.

\begin{claim}[$Q_R|f_R(\x)=1$ matches moments with $\cN(\vec{0},\bI_{V_R^{\perp}})$]\label{cl:fRmatchesmoments}
Suppose the partition is constructed with moment slack {$\Delta_0\leq\Delta/4$}.
There exists a subcollection $\mathcal H\subseteq\mathcal G$ such that
\[
    \sum_{R\in\mathcal G\setminus\mathcal H}D^*(R)
    {\leq 2\eta+\frac1\sigma\left(\frac A{r_0}\right)^{1/q}.}
\]
For every $R\in\mathcal H$, let $Q_R$ be the distribution of
$\proj_{V_R^\perp}\x$ for $\x\sim D^*\mid R$.
Then $Q_R\mid f_R=1$ matches the first $k$ moments of the standard
Gaussian on $V_R^\perp$ with slack $\Delta$. {Moreover, $Q_R$ is $(2q,\sigma_*)$-smooth with respect to this Gaussian, and $Q_R(f_R=1)\geq\pi_0/2>0$.}
\end{claim}

\begin{proof}[Proof of \Cref{cl:fRmatchesmoments}]
{Let $\mathcal H$ be the collection of certified cells satisfying the following three conditions:}
{\[
\mathcal H:=\{R\in\mathcal G:\ D^*\mid R\text{ is }(2q,\sigma_*)\text{-smooth w.r.t. }D\mid R,\ \Pr_{D^*}[f=1\mid R]\geq\pi_0,\ \Pr_{D^*}[f\ne f_R\mid R]\leq r_0^{1/q}\}.
\]}
First we show that only a small mass of certified cells does not belong in $\mathcal{H}$ and afterwards we show that for all $R\in \mathcal{H},$ $Q_R|f_R=1$ matches moments with the corresponding standard Gaussian.

\smallskip

\noindent\textbf{Mass of excluded certified cells.}
For the first condition, apply \Cref{lem:conditional-density-averaging} to $D^*$ and $D$ on the entire partition, with parameters $(q,\sigma/2,\eta)$. Since $\sigma\eta/(4q)\geq\sigma_*$, the certified cells failing this condition have total $D^*$-mass at most $\eta$.

For the second condition, recall $M=1/(\sigma^q\eta^{q-1})$. The laws of the cell index under $D^*$ and $D$ satisfy the same $(q,\sigma)$-smoothness inequality. Applying \Cref{lem:relaxed-smoothness-truncation} to these laws shows that cells with $D^*(R)/D(R)>M$ have total $D^*$-mass at most $\eta$. Every {retained} cell with $D^*(R)/D(R)\leq M$ satisfies $\Pr_{D^*}[f=1\mid R]=D^*(f=1)D_+^*(R)/D^*(R)\geq p_+\eta/(100M)=\pi_0$. Therefore, the {retained} cells failing the second condition have total $D^*$-mass at most $\eta$.

For the third condition, the union of the cellwise disagreement events has $D$-mass $A$, so global $(q,\sigma)$-smoothness gives
\[
    \sum_{R\in\mathcal G}D^*(R)
        \Pr_{D^*}[f\ne f_R\mid R]
    \leq \frac{A^{1/q}}{\sigma}.
\]
Markov's inequality therefore bounds the total $D^*$-mass of certified cells failing this condition by
\[
    r_0^{-1/q}\sum_{R\in\mathcal G}D^*(R)
        \Pr_{D^*}[f\ne f_R\mid R]
    \leq \frac1\sigma\left(\frac A{r_0}\right)^{1/q}.
\]
Combining the three conditions by a union bound yields
\[
    \sum_{R\in\mathcal G\setminus\mathcal H}D^*(R)
    \leq 2\eta+\frac1\sigma\left(\frac A{r_0}\right)^{1/q}.
\]

\smallskip

\noindent\textbf{Moment matching on retained cells.}
Now fix $R\in\mathcal H$. 
It suffices to show that $(\proj_{V_R^\perp}(D^*\mid R))\mid f_R=1$
has approximately the same first $k$ moments as
$\proj_{V_R^\perp}(D^*\mid(R,f=1))$.
Since $f_R$ depends only on the $V_R^\perp$ coordinates, we have
\(
    (\proj_{V_R^\perp}(D^*\mid R))\mid f_R=1
    =
    \proj_{V_R^\perp}(D^*\mid(R,f_R=1)).
\)

Let $P=D^*\mid R$, $p=P(f=1)$, {$p'=P(f_R=1)$}, and fix a unit
vector $\bv\in V_R^\perp$ and $j\in[k]$.
Expanding the conditioning indicators gives
\[
\begin{aligned}
    \E_{P\mid f_R=1}[(\bv\cdot \x)^j]
    -\E_{P\mid f=1}[(\bv\cdot \x)^j]
    &=
    \frac{\E_P[(\bv\cdot \x)^j\Ind\{f_R(\x)=1\}]}{{p'}}
    -\frac{\E_P[(\bv\cdot \x)^j\Ind\{f(\x)=1\}]}{p}\\
    &=
    \frac{\E_P[(\bv\cdot \x)^j
        (\Ind\{f_R(\x)=1\}-\Ind\{f(\x)=1\})]}{{p'}}
    +\frac{{p-p'}}{{p'}}\E_{P\mid f=1}[(\bv\cdot \x)^j].
\end{aligned}
\]
We bound each term separately. 
{Since $R\in\mathcal H$, $p\geq\pi_0$ and $P(f\ne f_R)\leq r_0^{1/q}$. Therefore,}
{\[
    |p-p'|\leq \Pr_{D^*}[f(\x)\neq f_R(\x)\mid R]
    \leq r_0^{1/q},
    \qquad
    p'\geq\pi_0-r_0^{1/q}
    \geq\frac{\pi_0}{2},
\]}
{where the last inequality follows from
$r_0^{1/q}=(\sigma_*\pi_0\Delta/A_{q,k})^4\leq\pi_0/2$.}

{By \Cref{fact:smoothness-projection}, the projection of $P$ onto $V_R^\perp$ is $(2q,\sigma_*)$-smooth with respect to the standard Gaussian. Hence \Cref{lem:relaxed-polynomial-control} and Cauchy--Schwarz give}
{\begin{align*}
    \left|\E_P\left[
        (\bv\cdot \x)^j
        (\Ind\{f_R(\x)=1\}-\Ind\{f(\x)=1\})
    \right]\right|
    &\leq \sqrt{\E_P[(\bv\cdot \x)^{2j}]}\,
            \sqrt{P(f\ne f_R)}\\
    &\leq \frac{\sqrt{4q}(8q-1)^{k/2}M_k}{\sqrt{\sigma_*}}\,
            r_0^{1/(2q)}
     =\frac{A_{q,k}}{8\sqrt{\sigma_*}}\,r_0^{1/(2q)},
\end{align*}}
since $\bv\cdot \x$ under $D\mid R$ is standard Gaussian.
The certified moment-matching guarantee with slack $\Delta/4$
also implies
\[
    \left|\E_{P\mid f=1}[(\bv\cdot \x)^j]\right|
    \leq M_k+\Delta/4\leq 2M_k.
\]
Substituting these bounds into the preceding identity yields
{\begin{align*}
    \left|
        \E_{P\mid f_R=1}[(\bv\cdot \x)^j]
        -\E_{P\mid f=1}[(\bv\cdot \x)^j]
    \right|
    &\leq \frac{2}{\pi_0}
        \left(\frac{A_{q,k}}{8\sqrt{\sigma_*}}r_0^{1/(2q)}
            +2M_kr_0^{1/q}\right)\\
    &\leq \frac{A_{q,k}}{2\sigma_*\pi_0}r_0^{1/(4q)}
    =\frac{\Delta}{2},
\end{align*}}
where we used
{$r_0\leq1$, $\sigma_*\leq1$, and $A_{q,k}\geq16M_k$.}
Combining this with the partition's moment slack $\Delta/4$
shows that $Q_R\mid f_R=1$ matches the first $k$ Gaussian
moments with slack $\Delta$. {The conditional smoothness of $Q_R$ follows from \Cref{fact:smoothness-projection}, and the bound on $p'$ proves its positive mass assertion.}
This completes the proof of \Cref{cl:fRmatchesmoments}.
\end{proof}
{Fix $R\in\mathcal H$.} {By \Cref{cl:fRmatchesmoments}, $Q_R$ is $(2q,\sigma_*)$-smooth with respect to $\cN(\vec{0},\bI_{V_R^\perp})$.}
To apply the definition of $m^*$ on $\R^d$, let $T_R$ be the
distribution of $\bu+\bz$, where $\bu\sim Q_R$ and
$\bz\sim\cN(\vec{0},\bI_{V_R})$ are independent. {Independent Gaussian extension preserves the same smoothness parameters: if $E_\bz:=\{\bu\in V_R^\perp:\bu+\bz\in E\}$, then smoothness and Jensen's inequality give}
{\[
    T_R(E)\leq\frac1{\sigma_*}\E_\bz[\cN(\vec{0},\bI_{V_R^\perp})(E_\bz)^{1/(2q)}]
    \leq\frac{D(E)^{1/(2q)}}{\sigma_*}.
\]}

Expanding directional monomials shows that the corresponding lift
of {$\proj_{V_R^\perp}(D_+^*\mid R)$} matches the first $k$ moments of $D$
with slack at most $2^kM_k\Delta_0=\Delta/4$.
The indicator comparison in the proof of
\Cref{cl:fRmatchesmoments} applies to these lifted moments as well:
under the reference law $\x\sim D\mid R$, independently of $\bz$,
the vector {$\proj_{V_R^\perp} \x+\bz$} is standard Gaussian, and
{$f_R(\proj_{V_R^\perp} \x+\bz)=f_R(\x)$}.
{For the moment-comparison bound, use Cauchy--Schwarz under $(\x,\bz)\sim(D^*\mid R)\otimes\cN(\vec{0},\bI_{V_R})$, bound the disagreement probability by $r_0^{1/q}$, and bound the polynomial second moment using the $(2q,\sigma_*)$-smooth law $T_R$. Thus the same bounds show that $T_R\mid f_R=1$ matches the first $k$ moments of $D$ with slack $\Delta$.}

Since $f_R\in\mathcal F$, {$T_R$ is $(2q,\sigma_*)$-smooth with respect to $D$ and $T_R(f_R=1)>0$}, and
{$k\geq m^*(\eta,\Delta,\sigma_*,2q)$}, the definition of $m^*$ implies
$T_R(f_R=-1)<{\eta}$. As $f_R$ depends only on {$V_R^\perp$}, we obtain
\[
    \Pr_{\x\sim D^*}[f_R(\x)=-1\mid R]
    =Q_R(f_R=-1)=T_R(f_R=-1)<{\eta}.
\]
Consequently, using the initial false-positive bound and
\Cref{cl:fRmatchesmoments},
{\begin{align*}
    \Pr_{\x\sim D^*}[h(\x)=1,f(\x)=-1]
    &\leq
    \sum_{R\in\mathcal H}
        D^*(R)\Pr_{\x\sim D^*}[f_R(\x)=-1\mid R]
    +\sum_{R\in\mathcal G\setminus\mathcal H}D^*(R)
    +\frac{A^{1/q}}{\sigma}\\
    &\leq 3\eta+\frac1\sigma\left(\frac A{r_0}\right)^{1/q}
          +\frac{A^{1/q}}\sigma\\
    &\leq3\eta+
    O\left(\frac1\sigma\left(\frac{(\delta\Gamma^2)^{1/3}}{r_0}\right)^{1/q}\right) \;.
\end{align*}}
This completes the proof of \Cref{claim:pu-false-positives}.
\end{proof}
Set
\(
{    \eta:=\eps/8,
    \sigma_*:=\alpha\eta/(400q),
    k:=m^*(\eta,\Delta,\sigma_*,2q).}
\)
Take the internal moment slack to be $\Delta_0:=\Delta/(2^{k+2}M_k)$.
The partition guarantees $\diam(R)\leq\delta$ on every certified
cell and has at most $B$ cells. {The choice in \Cref{alg:gaussian-pu} satisfies}
{\[
    \delta=\Gamma^{-2}
    \left(\frac{\alpha\eta\Delta}{Cqk}\right)^{Cq(k+q)}
    \leq \frac{c_0r_0^3}{\Gamma^2}
    \left(\frac{\sigma\eps}{8}\right)^{3q},
\]}
{where $c_0>0$ is a sufficiently small fixed universal constant and $C>1$ is sufficiently large. Indeed, $M_k^2\leq(2k)^k$, $\alpha\leq\sigma$, and
$\sigma_*\pi_0=(\alpha\eta)^2(\sigma\eta)^{q-1}/(10^4q)$ give}
{\begin{align*}
    \log\frac{1}{c_0r_0^3(\sigma\eta)^{3q}}
    &=\log\frac{1}{c_0}
      +12q\log\frac{A_{q,k}}{\sigma_*\pi_0\Delta}
      +3q\log\frac1{\sigma\eta}\\
    &=O\left(q(k+q)\log\frac{2qk}{\alpha\eta\Delta}\right).
\end{align*}}
{For sufficiently large $C$, this is at most
$Cq(k+q)\log(Cqk/(\alpha\eta\Delta))$, proving the displayed inequality.}
{The Gaussian freezing error $A$ from the proof above satisfies $A=O((\delta\Gamma^2)^{1/3})$. By the preceding inequality, a sufficiently small universal $c_0$ ensures $A\leq r_0(\sigma\eps/8)^q$. As $r_0\leq1$, this gives}
{\[
    \frac1\sigma\left(\frac A{r_0}\right)^{1/q}
    +\frac{A^{1/q}}\sigma\leq\eps/4.
\]}
Combining \Cref{claim:pu-false-negatives,claim:pu-false-positives},
on the {joint success event}, gives
\[
    \Pr_{\x\sim D^*}[h(\x)\neq f(\x)]
    {\leq 4\eta+3\lambda+\eps/4<\eps,}
\]
{since $\eta=\eps/8$ and $\lambda=\eta/100$.
By \Cref{claim:relaxed-success-event}, the partitioning routine succeeds
with probability at least $99/100$; conditional on its successful output, the estimation event fails with probability at most $1/100$.
Thus the joint event has probability at least $98/100>2/3$.}

\paragraph{Complexity.}
By \Cref{thm:relaxed-gaussian-partitioning}, applied to $D_+^*$ with
moment slack $\Delta_0$ and diameter bound $\delta$, the {partitioning stage}
uses at most $d^kB$ samples and runs in time $d^{O(k)}B$, with $B$ as above.
{The ratio test uses an additional $O(B^2\log(2B))$ positive samples
and time polynomial in $B$ and $d$. These costs are absorbed by increasing
the universal constants in the exponential below.}
Substituting the parameter choices and simplifying the
dimension-independent factors, a sufficient sample bound is
\[
    N=
   d^{k}
{\exp\left(
\frac{(qk)^{O(k)}}{\Delta^{O(1)}}
\log^{O(k)}\frac2{\sigma\eps p_+}\,
\log(2+\Gamma(\mathcal F))\right),}
\]
with running time polynomial in $N$.
\end{proof}

Similar to agnostic learning that we discussed in \Cref{sec agnostic mixture}, for PUI learning we can also remove the assumption of having sample access to distribution $D$. In particular, assuming $D^*$ is smooth with respect to some unknown mixture of Gaussians or some unknown Gaussian, we can even perform efficient PUI learning with only sample access to $D^*_+$. 

\begin{corollary}[PUI learning under Gaussian mixtures and unknown means]
\label{cor:pu-mixtures-unknown-mean}
Let $\mathcal F$ satisfy the hypotheses of \Cref{thm:pumim},
let $f\in\mathcal F$, and suppose that
$D^*(f=1)\geq p_+>0$.
If
$D^*\leq Q/\sigma$, where
$Q=\sum_{i=1}^s\lambda_i\cN(\boldsymbol{\mu}_i,\bI_d)$
and $\|\boldsymbol{\mu}_i\|_2\leq M$ for every $i$, then
the guarantees of \Cref{thm:pumim} hold with $q=2$ and
$\sigma$ replaced by $\bar\sigma=\sigma e^{-M^2/2}$.

Suppose additionally that $\mathcal F$ is closed under translations.
If $D^*\leq\cN(\boldsymbol{\mu},\bI_d)/\sigma$ for an unknown
$\boldsymbol{\mu}$, the same guarantees hold with $q=2$ and
$\bar\sigma=e^{-1}\sigma^3p_+^2$.
The algorithm requires sample access only to $D_+^*$;
neither knowledge of $\boldsymbol{\mu}$ nor samples from
$\cN(\boldsymbol{\mu},\bI_d)$ are needed.
\end{corollary}

\begin{proof}
In the mixture case, \Cref{lem:gaussian-mixture-smoothness} gives
\[
    D^*(E)
    \leq\frac{Q(E)}{\sigma}
    \leq\frac{e^{M^2/2}}{\sigma}
        \cN(\vec{0},\bI_d)(E)^{1/2}
\]
for every measurable event $E$. Thus \Cref{thm:pumim} applies
with parameters $(2,\bar\sigma)$.

For the unknown-mean case, conditioning on the positives gives
$D_+^*\leq\cN(\boldsymbol{\mu},\bI_d)/(\sigma p_+)$.
The Gaussian moment-generating formula and Jensen's inequality
imply
\[
    \left\|\E_{\x\sim D_+^*}[\x]-\boldsymbol{\mu}\right\|_2
    \leq\sqrt{2\log\frac1{\sigma p_+}}.
\]
Gaussian tail bounds also give
$\E_{\x\sim D_+^*}[\|\x-\boldsymbol{\mu}\|_2^2]
=O(d\log(2/(\sigma p_+)))$.
Consequently, using $O(d\log(2/(\sigma p_+)))$ positive samples,
we can estimate their mean by $\vec{m}$ with Euclidean error
at most $1$, with high constant probability. On this event,
\[
    \|\boldsymbol{\mu}-\vec{m}\|_2^2
    \leq4\log\frac1{\sigma p_+}+2.
\]
The centered target distribution is dominated by
$\cN(\boldsymbol{\mu}-\vec{m},\bI_d)/\sigma$.
By \Cref{lem:gaussian-mixture-smoothness}, it is therefore
$(2,\bar\sigma)$-smooth with respect to the standard Gaussian,
since
\[
    \sigma e^{-\|\boldsymbol{\mu}-\vec{m}\|_2^2/2}
    \geq e^{-1}\sigma^3p_+^2=\bar\sigma.
\]

Apply \Cref{thm:pumim} to fresh centered positive samples.
Translation closure ensures that the centered target concept
belongs to $\mathcal F$. If the learner returns $h$, output
$H(\x)=h(\x-\vec{m})$, which preserves the target error.
Choosing the constant failure probabilities in mean estimation
and partitioning sufficiently small gives the stated success
guarantee. The additional mean-estimation cost is absorbed into
the sample and running-time bounds.
\end{proof}

\subsection{SQ Lower Bound for PUI Learning of MIMs with Gaussian Reference}\label{sec:lowerboundspu}

Here we show that a dependence similar to $d^{m^*}$  in
\Cref{thm:pumim} is essentially unavoidable for SQ learning
algorithms when $\mathcal{F}$ is a class of $K$-multi-index models.
The intuition comes directly from the definition of $m^*$.
In particular, there exists a $K$-dimensional instance $D^*$
whose positive conditional distribution $D^*_+$ matches
the first $m^*-1$ moments of a standard Gaussian, even though
$D^*$ assigns at least $\eps$ probability mass to negative
examples.

We use this moment-matching instance to construct a hard
instance of non-Gaussian component analysis (NGCA).
Specifically, we randomly choose a $K$-dimensional subspace
$\bU$ of $\R^d$ and construct a distribution $Q$ such that
its marginal over $\bU$ is $D^*_+$, while its marginal over
$\bU^\perp$ is an independent standard Gaussian.
Since $D^*_+$ matches the first $m^*-1$ Gaussian moments,
distinguishing $Q$ from $\cN(\vec 0,\vec I)$ is an instance of
the SQ-hard NGCA problem studied in
\cite{DKS17-sq,diakonikolas2023sq}. We also refer the reader to \Cref{sec:omitted-preliminaries} for background on SQ learning.

To obtain the lower bound, suppose that there exists an
SQ algorithm that learns the corresponding PUI learning
problem to sufficiently small error.
Given an unknown input distribution that is either $Q$
or $\cN(\vec 0,\vec I)$, we run this learner using the input
distribution as the positive-example distribution and
$\cN(\vec 0,\vec I)$ as the unlabeled distribution.
Let $\widehat h$ be the resulting hypothesis.
We then compare the probability that $\widehat h$ predicts
a negative label under the input distribution with the
corresponding probability under $\cN(\vec 0,\vec I)$.

If the input distribution is $\cN(\vec 0,\vec I)$, these two
probabilities are identical.
On the other hand, if the input distribution is $Q$,
the learning guarantee forces $\widehat h$ to predict
a significant fraction of negative examples under
$\cN(\vec 0,\vec I)$, since the underlying instance $D^*$
has at least $\eps$ negative mass.
At the same time, because $Q$ is the positive conditional
distribution, the learner must predict negative labels
on only a small fraction of $Q$.
Consequently, the two probabilities differ significantly,
allowing us to distinguish $Q$ from $\cN(\vec 0,\vec I)$.
This contradicts the SQ lower bound for NGCA and
establishes the claimed dependence on $d$ and $m^*$.

\begin{proof}[Proof of \Cref{thm:sq-pu-general}]
Write $G_j:=\cN(\vec{0},\bI_j)$.
Let $\mu$ be the distribution of $\vec{W}\x$ for $\x\sim D^*$,
and let $\mu_+:=\mu\mid\{g(\bz)=1\}$.
Since the conditioning event depends only on $\vec{W}\x$,
$\Pr_{\bz\sim \mu}[g(\bz)=-1]=\Pr_{\x\sim D^*}[f(\x)=-1]\geq\eps$,
$\Pr_{\bz\sim \mu}[g(\bz)=1]=p_+$, and $\mu_+$ is the distribution of $\vec{W}\x$
for $\x\sim D_+^*$.

Every polynomial in $\bz\in\R^K$ of degree at most $m^*$
becomes a polynomial in $\x$ of the same degree after
substituting $\bz=\vec{W}\x$. Therefore, $\mu_+$ and $G_K$ match
moments through degree $m^*$.
By \Cref{fact:smoothness-projection}, $\mu$ is $(q,\sigma)$-smooth with respect to $G_K$. Conditioning gives $(q,\sigma p_+)$-smoothness of $\mu_+$.

We will use the following fact about the hardness of NGCA.

\begin{fact}[NGCA SQ Lower Bound
{\cite[Theorem~1.5]{diakonikolas2023sq}}]
\label{fact:ngca-sq}
Fix $\lambda\in(0,1)$ and $a\in(0,(1-\lambda)/4)$.
Let $d,K,r\in\Z_+$, with $r\geq2$ even and
$\max\{K,r\}\leq d^\lambda/\log d$.
Let $A$ be a distribution over $\R^K$ matching the moments
of $G_K$ through degree $r$.

For $\bU\in\R^{K\times d}$ satisfying $\bU\bU^\top=\bI_K$,
let $P_{\bU,A}$ denote the distribution of
$
\bU^\top\bz+(\bI_d-\bU^\top\bU)\vec{g},
$
where $\bz\sim A$ and $\vec{g}\sim G_d$ are independent.
For sufficiently large $d$, any SQ algorithm distinguishing
$G_d$ from $P_{\bU,A}$ for a uniformly random $\bU$, with
probability at least $2/3$, requires either
$2^{d^{\Omega(a)}}$ queries or a query of tolerance at most
$C_{K,r}d^{-((1-\lambda)/4-a)r}$, where
$C_{K,r}=O\bigl(\Gamma((K+r)/2)/\Gamma(K/2)\bigr)$.
\end{fact}

For each $\bU\in\R^{K\times d}$ satisfying $\bU \bU^\top=\bI_K$,
define $f_\bU(\x):=g(\bU\x)$, $D_\bU^*:=P_{\bU,\mu}$, and
$D_{+,\bU}^*:=P_{\bU,\mu_+}$.
By the rotation assumption, $f_\bU\in\mathcal F$.
Since $\bU\x=\bz$ in the definition of $P_{\bU,A}$,
\[
\begin{aligned}
    \Pr_{\x\sim D_\bU^*}[f_\bU(\x)=-1]
    &=\Pr_{\bz\sim\mu}[g(\bz)=-1]\geq\eps,\\
    \Pr_{\x\sim D_\bU^*}[f_\bU(\x)=1]&=p_+,\\
    D_{+,\bU}^*&=D_\bU^*\mid\{f_\bU(\x)=1\}.
\end{aligned}
\]
Note that $D_\bU^*$ is $(q,\sigma)$-smooth with respect to $G_d=\cN(\vec{0},\bI_d)$. Indeed, for any measurable event $E\subseteq\R^d$, smoothness and Jensen's inequality give
\begin{align*}
    \Pr_{\x\sim D_\bU^*}[\x\in E]
    &=\E_{\vec{g}\sim G_d}\!\left[\Pr_{\bz\sim \mu}\bigl[\bU^\top\bz+(\bI_d-\bU^\top \bU)\vec{g}\in E\bigr]\right]\\
    &\leq\frac1\sigma\E_{\vec{g}\sim G_d}\!\left[\Pr_{\bz\sim G_K}\bigl[\bU^\top\bz+(\bI_d-\bU^\top \bU)\vec{g}\in E\bigr]^{1/q}\right]\\
    &\leq\frac1\sigma
        \left(\E_{\vec{g}\sim G_d}\!\left[\Pr_{\bz\sim G_K}\bigl[\bU^\top\bz+(\bI_d-\bU^\top \bU)\vec{g}\in E\bigr]\right]\right)^{1/q}\\
    &=\frac{\Pr_{\x\sim P_{\bU,G_K}}[\x\in E]^{1/q}}{\sigma}
     =\frac{\Pr_{\x\sim G_d}[\x\in E]^{1/q}}{\sigma}.
\end{align*}
Here $P_{\bU,G_K}=G_d$ because, when $\bz\sim G_K$, the independent Gaussian components have covariance matrices summing to $\bI_d$. Conditioning gives $\Pr_{\x\sim D_{+,\bU}^*}[\x\in E]\leq \Pr_{\x\sim G_d}[\x\in E]^{1/q}/(\sigma p_+)$, so $D_{+,\bU}^*$ is $(q,\sigma p_+)$-smooth.

Set $r:=2\lfloor m^*/2\rfloor$.
Then $r\geq2$ is even and $m^*/2\leq r\leq m^*$.
Thus $\mu_+$ matches Gaussian moments through degree $r$.
Apply \Cref{fact:ngca-sq} with $A=\mu_+$,
$\lambda=1/4$, and $a=1/16$.
For sufficiently small $c$ and sufficiently large $d$,
the assumption $\max\{K,m^*\}\leq d^c$ ensures that
$\max\{K,r\}\leq d^{1/4}/\log d$.
Furthermore,
\[
    \frac{\Gamma((K+r)/2)}{\Gamma(K/2)}
    =
    \prod_{j=0}^{r/2-1}\left(\frac K2+j\right)
    \leq
    \left(\frac{K+r}{2}\right)^{r/2}
    \leq d^{cr/2}.
\]
Consequently, distinguishing $G_d$ from $D_{+,\bU}^*$
for a uniformly random $\bU$ requires either
$2^{d^{\Omega(1)}}$ queries or a query of tolerance at most
$\tau_d:=C_{K,r}d^{-r/8}$, where
$\tau_d=O\bigl(d^{-(1/8-c/2)r}\bigr)
=d^{-\Omega(m^*)}$.

\begin{claim}[Reduction from PUI Learning to Distinguishing]
\label{claim:pu-to-distinguishing}
Suppose an SQ learner, given access only to $D_{+,\bU}^*$,
uses at most $Q$ queries and, for every $\bU$, outputs
$h:\R^d\to\{\pm1\}$ satisfying
$\Pr_{\x\sim D_\bU^*}[h(\x)\neq f_\bU(\x)]\leq p_+(\sigma\eps)^q/2$
with probability at least $2/3$.
Then it yields an SQ algorithm distinguishing $G_d$
from $D_{+,\bU}^*$ with probability at least $2/3$.
Using the known Gaussian expectation, the algorithm makes
the learner's queries and one additional query of tolerance
$(\sigma\eps)^q/8$.
\end{claim}

\begin{proof}[Proof of \Cref{claim:pu-to-distinguishing}]
Let $P$ denote the unknown distribution, which is either
$G_d$ or $D_{+,\bU}^*$.
Run the learner using the SQ oracle for $P$, and let $h$
be its output. Define $a:=\Pr_{\x\sim G_d}[h(\x)=-1]$.
Make one additional SQ query to obtain $v$ satisfying
$|v-\Pr_{\x\sim P}[h(\x)=-1]|\leq(\sigma\eps)^q/8$.
Declare the planted case if $a-v>(\sigma\eps)^q/4$,
and otherwise declare the Gaussian case.

If $P=G_d$, then $\Pr_{\x\sim P}[h(\x)=-1]=a$, regardless of the learner's
output. Hence $|a-v|\leq(\sigma\eps)^q/8$, and the test
declares the Gaussian case.

Now suppose $P=D_{+,\bU}^*$ and the learner succeeds.
Write $e_+:=\Pr_{\x\sim D_\bU^*}[f_\bU(\x)=1,h(\x)=-1]$ and
$e_-:=\Pr_{\x\sim D_\bU^*}[f_\bU(\x)=-1,h(\x)=1]$.
Then $e_++e_-\leq p_+(\sigma\eps)^q/2$ and
$\Pr_{\x\sim D_{+,\bU}^*}[h(\x)=-1]=e_+/p_+$.
By smoothness and the negative-mass bound,
\[
    \Pr_{\x\sim G_d}[h(\x)=-1]
    \geq \bigl(\sigma \Pr_{\x\sim D_\bU^*}[h(\x)=-1]\bigr)^q
    \geq \bigl(\sigma(\eps-e_-)\bigr)^q.
\]
Since $p_+\leq1-\eps$, we have
$p_+q\sigma^q\eps^{q-1}\leq q(1-\eps)\eps^{q-1}\leq1-\eps^q\leq1$.
Convexity therefore gives
\begin{align*}
    \Pr_{\x\sim G_d}[h(\x)=-1]-\Pr_{\x\sim D_{+,\bU}^*}[h(\x)=-1]
    &\geq(\sigma\eps)^q-q\sigma^q\eps^{q-1}e_--\frac{e_+}{p_+}\\
    &\geq(\sigma\eps)^q-\frac{e_-+e_+}{p_+}\\
    &\geq\frac{(\sigma\eps)^q}{2}.
\end{align*}
Consequently, $a-v\geq3(\sigma\eps)^q/8$, so the test
declares the planted case whenever the learner succeeds.
\end{proof}

It remains to ensure that the additional query cannot
itself supply the precision required by
\Cref{fact:ngca-sq}.
The parameter assumptions give
$(\sigma\eps)^q/8\geq d^{-cm^*}/8$.
On the other hand, $r\geq m^*/2$ implies
$\tau_d=O\bigl(d^{-(1/16-c/4)m^*}\bigr)$.
Choosing $c$ sufficiently small therefore ensures
$(\sigma\eps)^q/8>2\tau_d$ for sufficiently large $d$.

Suppose the learner uses fewer than exponentially many
queries and every query has tolerance greater than $2\tau_d$.
By \Cref{claim:pu-to-distinguishing}, this yields an NGCA
distinguisher using the learner's queries and one additional
query, all with tolerance greater than $\tau_d$, contradicting
\Cref{fact:ngca-sq}.
Thus the learner requires either $2^{d^{\Omega(1)}}$ queries
or a query of tolerance at most
$2\tau_d=d^{-\Omega(m^*)}$.
\end{proof}
Combining \Cref{thm:sq-pu-general} with a moment-matching construction
for a class of MIMs yields a PUI learning lower bound for that class.

\subsection{PUI Learning Intersections of Halfspaces and Beyond}\label{sec:struct-intersections}

As concrete applications of \Cref{thm:pumim} and \Cref{thm:sq-pu-general}, we derive PUI learning algorithms for the class of halfspaces, intersections of halfspaces, and general $K$-dimensional convex sets.
We also derive matching SQ lower bounds for the class of halfspaces and intersections of halfspaces. Let $\mathcal F_k$ denote the class of intersections of at most
$k$ halfspaces, including the constant functions.
The main results in this section are the following theorems.

We next apply our PUI framework to intersections of halfspaces,
obtaining a learner whose exponent in $d$ depends logarithmically
on $k$ and $1/\eps$.
\begin{theorem}[PUI Learning Intersections of Halfspaces]
\label{cor:pu-intersections}
Let $k,d\in\mathbb N_+$, $\eps\in(0,1/2)$, {$q\geq1$,} and
$\sigma\in(0,1]$.
Let $D=\cN(0,I_d)$ and let $D^*$ be a distribution
{that is $(q,\sigma)$-smooth with respect to $D$}.
Let $f\in\mathcal F_k$ be an unknown target intersection of $k$ halfspaces, and set
$
    p_+:=D^*(f=1)>0,
    D_+^*:=D^*\mid\{f=1\}.
$
Set {$m:=\lceil C(\log(2k)+q\log(2q/(\sigma\eps p_+)))\rceil$} for a
sufficiently large universal constant $C>0$.
There is an algorithm which, given {$\eps,\sigma,p_+,q$} and sample
access to $D_+^*$, draws at most
\[
    N=d^m\exp\!\left((2m)^{O(m)}\right)
\]
samples.
The algorithm runs in time polynomial in $N$ and, with
probability at least $2/3$, outputs a hypothesis
$h:\R^d\to\{\pm1\}$ satisfying
$
    \Pr_{x\sim D^*}[h(\x)\neq f(\x)]\leq\eps.
$
\end{theorem}

For convex concepts of intrinsic dimension at most $K$,
our PUI general algorithm yields a learner whose exponent in $d$
is linear in $K$ and logarithmic in $1/\eps$.

\begin{theorem}[PUI Learning Low-Dimensional Convex Sets]
\label{cor:pu-convex}
Let $K,d\in\mathbb N_+$, $\eps\in(0,1/2)$, {$q\geq1$,} and
$\sigma\in(0,1]$.
Let $D=\cN(0,I_d)$ and let $D^*$ be a distribution
{that is $(q,\sigma)$-smooth with respect to $D$}.
Let $f\in\mathcal C_K$ be an unknown target $K$-dimensional convex set, and set
$
    p_+:=D^*(f=1)>0,
    D_+^*:=D^*\mid\{f=1\}.
$
Set {$m:=\lceil C(K+q\log(2q/(\sigma\eps p_+)))\rceil$} for a
sufficiently large universal constant $C>0$.
There is an algorithm which, given {$\eps,\sigma,p_+,q$} and sample
access to $D_+^*$, draws at most
\[
    N=d^m\exp\!\left((2m)^{O(m)}\right)
\]
samples.
The algorithm runs in time polynomial in $N$ and, with
probability at least $2/3$, outputs a hypothesis
$h:\R^d\to\{\pm1\}$ satisfying
$
    \Pr_{x\sim D^*}[h(\x)\neq f(\x)]\leq\eps.
$
\end{theorem}

We next give an SQ lower bound of $d^{\Omega(\log(k/\eps))}$,
matching the dependence on $d$ of our sample complexity
upper bound when the smoothness parameters are constant
and $p_+$ is bounded below by a constant.
\begin{theorem}[SQ Lower Bound for PUI Learning Intersections of Halfspaces]
\label{cor:sq-pu-intersections}
There exists a sufficiently small universal constant $c>0$ such that
the following holds for sufficiently large $d$.
Let $k,d\in\Z_+$, $\eps\in(0,1/2)$ and $m^*=\log(k/\eps)$. Assume that
$
    \max\{k,m^*\}\leq d^c,
     \eps\leq c/k^3.
$
Then any SQ algorithm that learns $\mathcal F_k$ over distributions that are $(1,1/2)$-smooth with respect to $\cN(\vec{0},\bI_d)$, using SQ access only to the positive
conditional distribution, to error $\eps(1-\eps)/4$ with
probability at least $2/3$, requires either $2^{d^{\Omega(1)}}$ queries or a query
of tolerance at most $d^{-\Omega(m^*)}$.
\end{theorem}

The main technical ingredient is to show that $m^*(\eps,\Delta,\sigma,q) = O(\log(k/(\sigma \eps)^q))$, while on the other hand even for $q = 1$ and $\sigma = 1/2$, $m^* = \Omega(\log(k/\eps))$. Together, these show that $d^{\Theta(\log(k/\eps))}$ time complexity is both sufficient and necessary for PUI learning intersections of halfspaces with a Gaussian reference.

\paragraph{Upper Bounds for PUI Learning Exponents of Intersections of Halfspaces}

We start with the proof of \Cref{cor:pu-intersections}. The idea of upper bounding $m^*$ for intersections of $k$ halfspaces is to first derive an upper bound for a single halfspace. Specifically, given a one-dimensional halfspace $\Ind\{x < t\}$, if $t = \Theta(\sqrt{\log(1/\eps)})$, then under the Gaussian distribution, the fraction of negative examples is roughly $\eps$. We show that any distribution supported on $(-\infty,t]$ cannot match its first $O(t^2)$ moments within a slack of $1/4$ with those of a standard Gaussian. 
Since an intersection of $k$ halfspaces is a conjunction of $k$ halfspaces, we know that if the fraction of negative examples under Gaussian is at least $\eps$, there must be at least one halfspace that contains at least $\eps/k$ negative examples under Gaussian. Combining this fact with the structural results on a single halfspace, we know that any distribution sampled from the positive region of the intersection cannot match $\log(k/\eps)$ moments up to a constant slack.
Notice that once this is proven, a smoothness assumption immediately gives an upper bound of $m^* = O(\log(1/(\sigma \eps)^q))$ for a single halfspace.

\begin{lemma}[Structural Result for Halfspaces]\label{lem:halfspaces}
Let \(G=\cN(0,1)\), let $t\geq0$, and let \(P\) be supported on \((-\infty,t]\)
with finite moments through degree \(2\lceil 1+t^2\rceil+1\).
Then $P$ and $G$ do not match their first \(2\lceil 1+t^2\rceil+1\)
moments to slack less than $1/4$.

Let $\eps<1/2$. If $\Pr_{x\sim G}[x\ge t]\ge\epsilon$, then
$t= O(\sqrt{\log(1/\epsilon)})$.
Hence,  $P$ and $G$ cannot match all moments
through degree $\lfloor C\log(1/\epsilon)\rfloor$ to slack less than
$1/4$ for a sufficiently large constant $C>0$.
\end{lemma}

\begin{proof}[Proof of \Cref{lem:halfspaces}]
Let $r=\lceil 1+t^2\rceil, a=\sqrt{2r-1}$, \(n=2r-1=a^2\) and define the polynomial
$q(x)=(x-t)x^{2r-2}(x+a)^2$.

Since \(q\le 0\) on the support of \(P\), we have \(\E_{x\sim P}[q(x)]\le0\).
Denote $\mu_r\eqdef \E_{x\sim G}[x^{2r-2}]
\ge1$.
Expanding $q$ and using that the odd Gaussian moments vanish, we get
\[
\begin{aligned}
\E_{x\sim G}[q(x)]
&=(2a-t)\E_{x\sim G}[x^{2r}]
  -ta^2\E_{x\sim G}[x^{2r-2}]\\
&=\bigl[(2a-t)(2r-1)-ta^2\bigr]\mu_r\\
&=\bigl[n(2a-t)-ta^2\bigr]\mu_r\\
&=2a^2(a-t)\mu_r.
\end{aligned}
\]
Here the second line uses
$\E_{x\sim G}[x^{2r}]=(2r-1)\E_{x\sim G}[x^{2r-2}]
=(2r-1)\mu_r$, and the last line uses $n=a^2$.
Write \(B=2a^2(a-t)\). Since \(n\ge 2t^2+1\), we have
\(|t|/a\le 1/\sqrt2\), so \(B>0\).

Note that expanding $q$ gives that the sum of the absolute values of its non-constant monomial coefficients is
\[
C=1+|2a-t|+|a^2-2at|
  +\Ind\{r\ge2\}|t|a^2.
\]
Therefore $P$ and $G$ cannot match their first $2r+1$ moments to slack
less than $(B/C)\mu_r$.

It remains to bound \(B/C\). If \(t\le0\), set \(v=-t/a\). Since \(a\ge1\),
\[
C\le 1+(2+v)a+(1+2v)a^2+va^3
 \le 4(1+v)a^3=2B.
\]
If \(t>0\), then \(r\ge2\), so \(a\ge\sqrt3\). Setting
\(v=t/a\le1/\sqrt2\), we obtain
\[
C\le1+2a+a^2+\frac{a^3}{\sqrt2}
  <\frac94a^3,
\qquad
B=2(1-v)a^3\ge(2-\sqrt2)a^3>\frac9{16}a^3.
\]
Thus \(B/C>1/4\) in this case as well. Since $\mu_r\ge1$,
we have $(B/C)\mu_r\ge1/4$, proving the claim.
\end{proof}
For an intersection of \(k\) halfspaces, the union bound implies that one defining halfspace excludes at least a \(1/k\) share of that mass. Applying the single-halfspace result to the projection onto its normal vector then gives a moment discrepancy in that direction.
Now we state our result for intersections of halfspaces.
\begin{corollary}[Structural Result for Intersections of Halfspaces]\label{cor:intersection-moments}
There exists a sufficiently large constant $C>0$ such that the following holds.
Let \(G=\cN(\vec{0},\bI_d)\), and let $f$ be an intersection of  $k$ halfspaces such that \(\Pr_{\x\sim G}[f(\x)=-1]\geq\epsilon\), where \(0<\epsilon<1/2\).
Let \(P\) be a distribution  supported on $\{\x:f(\x)=1\}$ that has finite moments through degree $m=C\lceil\log(k/\eps)\rceil$.
Then, \(P\) and \(G\) do not match moments through degree
\(m\) to slack less than \(1/4\) in the
directional sense of \Cref{def:moment-matching}.
\end{corollary}

\begin{proof}[Proof of \Cref{cor:intersection-moments}]
Denote by 
\[
H=\bigcap_{i=1}^k
\{\x:\bv_i\cdot \x\leq t_i\},
\qquad \|\bv_i\|_2=1
\]
the positive region of $f$.
Since
\[
H^c=\bigcup_{i=1}^k
\{\x: \bv_i\cdot \x>t_i\},
\]
the union bound implies that some \(i\in[k]\) satisfies
\[
\Pr_{\x\sim G}[ \bv_i\cdot \x >t_i]\geq\frac{\epsilon}{k}.
\]
Assume first that $t_i\geq0$.
Under \(G\), the random variable \( \bv_i\cdot \x\) has
distribution \(\cN(0,1)\). Under \(P\), it is supported on
\((-\infty,t_i]\).

Therefore, by \Cref{lem:halfspaces}, the distribution of
$\bv_i\cdot \x$ under $P$ cannot match the Gaussian moments through
degree $m=C\lceil\log(k/\eps)\rceil$ with slack less than $1/4$,
for a sufficiently large universal constant $C>0$.

If $t_i\leq0$, then the distribution of $\bv_i\cdot \x$ under $P$
is also supported on $(-\infty,0]$. Applying \Cref{lem:halfspaces}
with $t=0$ rules out matching the first three Gaussian moments
with slack less than $1/4$. Since $m\geq3$ for a sufficiently large
$C$, this also rules out moment matching through degree $m$.
In either case, $P$ and $G$ cannot match moments through degree $m$
with slack less than $1/4$ in the sense of
\Cref{def:moment-matching}, completing the proof.
\end{proof}

We apply \Cref{thm:pumim} to intersections of halfspaces.
Let $\mathcal F_k$ denote the class of intersections of at most
$k$ halfspaces, including the constant functions.
By \Cref{cor:intersection-moments}, for every fixed
$\Delta\in(0,1/4)$,
{$
    m^*(\eps,\Delta,\sigma,q)
    = O\left(\log(2k/(\sigma\eps)^q)\right).
$}
Indeed, {$(q,\sigma)$-smoothness} implies that if
$D^*(f=-1)\geq\eps$, then  {$D(f=-1)\geq(\sigma\eps)^q$}.
Combining this bound with \Cref{thm:pumim} yields the following
corollary.

\begin{proof}[Proof of \Cref{cor:pu-intersections}]
The class $\mathcal F_k$ is closed under affine restrictions
and satisfies $\Gamma(\mathcal F_k)=O(k)$.
Taking $\Delta=1/8$, \Cref{cor:intersection-moments} gives
{\[
    m^*\!\left(\frac{\eps}{8},\frac18,
               \frac{\eps\sigma p_+}{12800q},2q\right)
    =O\!\left(\log(2k)+q\log\frac{2q}{\sigma\eps^2p_+}\right)
    \leq m.
\]}
Run \Cref{alg:gaussian-pu} with moment order $m$; its proof only
requires the order to be at least the displayed PUI learning exponent.
Since $\Gamma(\mathcal F_k)=O(k)$ and
{$q,\log(2/(\sigma\eps p_+)),\log(2k)=O(m)$},
the exponential factor in \Cref{thm:pumim} is at most
\[
    {\exp\!\left(
        (qm)^{O(m)}\log^{O(m)}\!\left(\frac2{\sigma\eps p_+}\right)
        \log(2k)
    \right)}
    \leq\exp\!\left((2m)^{O(m)}\right).
\]
Thus the sample complexity is $d^m\exp((2m)^{O(m)})$ and the
running time is polynomial in this quantity. The error guarantee
follows from \Cref{thm:pumim}. Here $k$ counts halfspaces and $m$
is the moment order used by the PUI algorithm.
\end{proof}

\paragraph{Upper Bounds for PUI Learning Exponents of Low Dimensional Convex Sets}

Similar analysis also holds for general $K$-dimensional convex sets. The idea is as follows. Suppose that $C^* \subseteq \R^K$ is a $K$-dimensional convex set such that $\Pr_{\x \sim \cN(0,\bI_K)}[\x \not \in C^*]>\eps$. Then we claim that there is some halfspace $h = \{\x \mid \bu\cdot\x \le t\}$ for some moderate $t>0$ that contains $C^*$. 
In other words, $C^*$ must exclude some point that is not extremely far from the origin. 
By \Cref{lem:halfspaces}, we know that any distribution sampled from $h$ cannot match $\Omega(1+t^2)$ many moments with respect to a Gaussian, and therefore the distribution sampled from $C^*$ cannot. In fact, we can show that the threshold $t$ can be chosen as $O(\sqrt{K}+\sqrt{\log(1/\eps)})$, we know that the PUI learning exponent of a $K$-dimensional convex set is at most $O(K+ q\log(1/(\sigma\eps)))$.

\begin{lemma}[Structural Results for $K$-Dimensional Convex Sets]\label{lm structure convex set}
    There exists a sufficiently large constant $C>0$ such that the following holds.
Let \(G=\cN(\vec{0},\bI_d)\), and let $C^*$ be a $K$-dimensional convex set such that \(\Pr_{\x\sim G}[\x \not\in C^*]\geq\epsilon\), where \(0<\epsilon<1/2\).
Let \(P\) be a distribution supported on $C^*$ that has finite moments through degree $m=C\lceil(K + \log(1/\eps))\rceil$.
Then, \(P\) and \(G\) do not match moments through degree
\(m\) to slack less than \(1/4\) in the
directional sense of \Cref{def:moment-matching}.
\end{lemma}

\begin{proof}[Proof of \Cref{lm structure convex set}]
    Without loss of generality, we assume that $C^* \subseteq \R^K$. Let $R^2 = 2K\log(2) + 4\log(2/\eps)$. Using the bound on the norm of a $K$ dimensional Gaussian vector, we know that $\Pr_{\x \sim \cN(0,\bI_K)}[\norm{\x}>R] < \eps/2$.
    Since $\Pr_{\x\sim G}[\x \not\in C^*]\geq\epsilon$, we know that there must be some point $\x^*$ such that $\norm{\x^*} \le R$, while $\x^* \not \in C^*$. Thus, applying the separating hyperplane theorem to $C^*$ and $\x^*$, we know that there exists a halfspace $h = \{\x \mid \bu\cdot \x \le t\}$, where $\bu \in \s^{K-1}$ and $0\le t< R$ such that $C^* \subseteq h$.

    Since $P$ has support $C^*$, we know that $P$ is also supported on $h$. By \Cref{lem:halfspaces}, we know that $P$ cannot match $O(1+t^2) = O(K + \log(2/\eps))$ moments with a standard Gaussian within a slack of $1/4$. 
\end{proof}

\begin{proof}[Proof of \Cref{cor:pu-convex}]
    The class $\mathcal C_K$ is closed under affine restrictions
and satisfies $\Gamma(\mathcal C_K)=O(K^{1/4})$.
Taking $\Delta=1/8$, \Cref{lm structure convex set} gives
{\[
    m^*\!\left(\frac{\eps}{8},\frac18,
               \frac{\eps\sigma p_+}{12800q},2q\right)
    =O\!\left(K+q\log\frac{2q}{\sigma\eps^2p_+}\right)
    \leq m.
\]}
Run \Cref{alg:gaussian-pu} with moment order $m$; its proof only
requires the order to be at least the displayed PUI learning exponent.
Since $\Gamma(\mathcal C_K)=O(K^{1/4})$
the exponential factor in \Cref{thm:pumim} is at most
\[
    {\exp\!\left(
        (qm)^{O(m)}\log^{O(m)}\!\left(\frac2{\sigma\eps p_+}\right)
        \log(2K)
    \right)}
    \leq\exp\!\left((2m)^{O(m)}\right).
\]
Thus the sample complexity is $d^m\exp((2m)^{O(m)})$ and the
running time is polynomial in this quantity. The error guarantee
follows from \Cref{thm:pumim}.
\end{proof}

\paragraph{Lower Bounds for PUI Learning Exponents of Intersections of Halfspaces}

We next investigate the tightness of \Cref{cor:intersection-moments}. We seek an intersection \(f\) of \(k\) halfspaces and a distribution \(D^\star\leq2G\) (i.e., \(\sigma=1/2\)), such that \(\Pr_{\x\sim D^\star}[f(\x)=-1]\geq\eps\) while the positive conditional distribution \(D^\star_+=D^\star\mid f(\x)=1\) matches \(G\) through degree \(\Omega(\log(k/\eps))\). The construction below shows that the logarithmic moment bound is tight up to constants in the regime \(\eps\leq c/k^3\), for a sufficiently small universal constant \(c>0\).

We first settle the case of a single halfspace.
In particular we prove that there exists a moment matching distribution supported on a halfspace that is very close to the Gaussian (the domination parameter approaches $1$) of nontrivial negative mass. 
The construction deletes the Gaussian right tail and repairs its
missing moments by a bounded correction on a central interval.

\begin{lemma}[Moment Matching Lower bound for Halfspaces]
\label{lem:halfspace-moment-lower-bound}
Let $G=\cN(0,1)$, let $t\geq6$, and set
$\eps:=\Pr_{x\sim G}[x>t]$ and $m:=\lfloor t^2/32\rfloor$.
There exists a probability distribution $Q$ supported on $(-\infty,t]$
such that $Q\leq(1+2e^{-5t^2/32})G$ and $Q$ matches the Gaussian moments through degree $m$.
Moreover, for the halfspace $f_t(x):=2\Ind\{x\leq t\}-1$,
there exists a probability distribution
$D^\star\leq (1+2e^{-5t^2/32})G$ such that
$
    \Pr_{x\sim D^\star}[f_t(x)=-1]=\eps,
   D^\star_+:=D^\star\mid\{f_t(x)=1\}=Q.
$
\end{lemma}

\begin{proof}[Proof of \Cref{lem:halfspace-moment-lower-bound}]

Define the vector of Hermite polynomials $\bv(x):=(h_0(x),\ldots,h_m(x))^\top$. 
Removing the Gaussian tail $(t,\infty)$ loses the Hermite moments
\[
    \bb:=\E_{x\sim G}[\Ind\{x>t\}\bv(x)].
\]
Since $h_0=1$, the first coordinate of $\bb$ is the lost
probability mass $\eps$.

We will restore these moments by adjusting the density on the
interval $I:=[-R,R]$, where $R:=\sqrt{8(m+1)}$.
Since $t\geq6$, the choice of $m$ gives $m\geq1$ and
$R^2\leq t^2/4+8<t^2$, so $I\subseteq(-\infty,t]$.

In particular, we seek a correction polynomial of the form
$r(x)=\bv(x)^\top\mathbf c$ satisfying
\[
    \E_{x\sim G}[\Ind\{x\in I\}r(x)\bv(x)]=\bb.
\]
Substituting $r$ shows that it suffices to solve the linear system $\bA \bc=\bb$, where \[
    \bA:=\E_{x\sim G}[\Ind\{x\in I\}\bv(x)\bv(x)^\top] \;.
\]

The following claim shows that such a system is solvable.
\begin{claim}[Invertibility of $\bA$]
\label{claim:halfspace-gram-invertibility}
We have $\bA\succeq\frac12\bI_{m+1}$. In particular, $\bA$ is invertible and
$\|\bA^{-1}\|_{\mathrm{op}}\leq2$.
\end{claim}

\begin{proof}[Proof of \Cref{claim:halfspace-gram-invertibility}]
Fix $\ba\in\R^{m+1}$ and write $p(x):=\ba^\top \bv(x)$.
By orthonormality, $\|p\|_{L_2(G)}=\|\ba\|_2$. 
We will use the following recurrence about Hermite polynomials $ xh_j(x)=\sqrt{j+1}\,h_{j+1}(x)+\sqrt j\,h_{j-1}(x)$.
Applying the recurrence, the triangle inequality, and
orthonormality within each of the two sums gives
\begin{align*}
    \|xp\|_{L_2(G)}
    \leq
    \left(\sum_{j=0}^m(j+1)\ba_j^2\right)^{1/2}
    +\left(\sum_{j=1}^m j\ba_j^2\right)^{1/2}
    \leq2\sqrt{m+1}\,\|p\|_{L_2(G)}.
\end{align*}
On $\{|x|>R\}$, we have $1\leq x^2/R^2$. Hence
\[
    \E_{x\sim G}[\Ind\{|x|>R\}p(x)^2]
    \leq\frac{\E_{x\sim G}[x^2p(x)^2]}{R^2}
    \leq\frac{4(m+1)}{R^2}\|\ba\|_2^2
    =\frac12\|\ba\|_2^2.
\]
Subtracting this bound from $\E_{x\sim G}[p(x)^2]=\|\ba\|_2^2$ yields
$\ba^\top \bA\ba\geq\frac12\|\ba\|_2^2$, proving \Cref{claim:halfspace-gram-invertibility}.
\end{proof}
By \Cref{claim:halfspace-gram-invertibility}, we may take
$\mathbf c=\bA^{-1}\bb$. With this choice of $r$, define
$w(x):=\Ind\{x\leq t\}+\Ind\{x\in I\}r(x)$. 
We show that reweighting $G$ by $w$ preserves its moments through degree $m$.
\begin{claim}[Moment Matching]
\label{claim:halfspace-moment-matching}
For every polynomial $p$ of degree at most $m$,
\[
    \E_{x\sim G}[w(x)p(x)]=\E_{x\sim G}[p(x)].
\]
In particular, $\E_{x\sim G}[w(x)]=1$.
\end{claim}

\begin{proof}[Proof of \Cref{claim:halfspace-moment-matching}]
The definitions of $\bA$ and $r$ give
\[
    \E_{x\sim G}[\Ind\{x\in I\}r(x)\bv(x)]
    =\E_{x\sim G}[\Ind\{x\in I\}\bv(x)\bv(x)^\top]\bA^{-1}\bb
    =\bb.
\]
Consequently,
\[
    \E_{x\sim G}[w(x)\bv(x)]
    =\E_{x\sim G}[\bv(x)]-\bb+\bb
    =\E_{x\sim G}[\bv(x)].
\]
Every polynomial of degree at most $m$ is a linear combination of
$h_0,\ldots,h_m$, so the asserted equality follows by linearity.
Taking $p=1$ gives normalization. This concludes the proof of \Cref{claim:halfspace-moment-matching}.
\end{proof}
It remains to bound the correction term $r$ to show that $w$ is nonnegative and hence that $\mathrm dQ=w\,\mathrm dG$ defines a probability distribution.
\begin{claim}[Bounded Correction]
\label{claim:halfspace-bounded-correction}
The polynomial $r$ satisfies $\sup_{x\in I}|r(x)|\leq 2e^{-5t^2/32}$.
\end{claim}

\begin{proof}[Proof of \Cref{claim:halfspace-bounded-correction}]
We start with the bound
\[
    \sup_{x\in I}|r(x)|
    \leq\sup_{x\in I}\|\bv(x)\|_2\,
         \|\bA^{-1}\|_{\mathrm{op}}\,\|\bb\|_2.
\]
By \Cref{claim:halfspace-gram-invertibility}, the middle factor is at most
$2$. It remains to bound $\|\bb\|_2$ and $\sup_{x\in I}\|\bv(x)\|_2$.

For every $\ba\in\R^{m+1}$ with $\|\ba\|_2=1$, Cauchy--Schwarz gives
\[
    |\ba^\top \bb|
    \leq\sqrt{\Pr_{x\sim G}[x>t]}\,
          \sqrt{\E_{x\sim G}[(\ba^\top \bv(x))^2]}
    =\sqrt\eps.
\]
Thus $\|\bb\|_2\leq\sqrt\eps$.
We next bound $\bv$ on $I$. Set $B:=R+\sqrt m$.
We show by induction that $|h_j(x)|\leq B^j/\sqrt{j!}$
for $x\in I$ and $0\leq j\leq m$.
The cases $j=0,1$ follow from $h_0=1$ and $h_1(x)=x$.
For the induction step, the  Hermite recurrence gives
\begin{align*}
    |h_{j+1}(x)|
    \leq
      \frac{|x|\,|h_j(x)|+\sqrt j\,|h_{j-1}(x)|}
           {\sqrt{j+1}}
    \leq
      \frac{RB^j+jB^{j-1}}{\sqrt{(j+1)!}}
    \leq
      \frac{B^{j+1}}{\sqrt{(j+1)!}},
\end{align*}
where we used $RB+j\leq RB+m\leq B^2$.
Since $B\geq\sqrt m$, the sequence $B^j/\sqrt{j!}$ is nondecreasing
for $0\leq j\leq m$. Also, $R\leq4\sqrt m$, so $B\leq5\sqrt m$.
Using $m!\geq(m/e)^m$, we obtain
\[
    |h_j(x)|
    \leq\frac{B^m}{\sqrt{m!}}
    \leq(5\sqrt e)^m
    \qquad(x\in I,\ 0\leq j\leq m).
\]
As $m+1\leq2^m$ for $m\geq1$, this implies
\[
    \sup_{x\in I}\|\bv(x)\|_2
    \leq\sqrt{m+1}(5\sqrt e)^m
    \leq(5\sqrt{2e})^m
    \leq e^{3m}.
\]
Combining the three bounds gives
$\sup_{x\in I}|r(x)|\leq2e^{3m}\sqrt\eps$.
The Gaussian tail bound gives $\eps\leq e^{-t^2/2}$. Therefore,
using $m\leq t^2/32$, we obtain 
$
    2e^{3m}\sqrt\eps
    \leq2e^{3t^2/32-t^2/4}
    =2e^{-5t^2/32}\;.
$
This concludes the proof of \Cref{claim:halfspace-bounded-correction}.
\end{proof}
Note that for $t\geq 6$, $\sup_{x\in I}|r(x)|<1$. Hence, since $I\subseteq(-\infty,t]$, \Cref{claim:halfspace-bounded-correction}
implies $0\leq w\leq 1+2e^{-5t^2/32}$ and $w=0$ on $(t,\infty)$.
By \Cref{claim:halfspace-moment-matching}, the measure $Q$ defined by
$\mathrm dQ=w\,\mathrm dG$ is therefore a probability distribution
matching the Gaussian moments through degree $m$. 

To construct the full target distribution, set
$
    D^\star:=(1-\eps)Q+\eps\,G(\,\cdot\mid x>t).
$
Since $0<\eps<1$, this is a probability distribution. Its density ratio with respect to $G$ is
\begin{align*}
    \frac{\mathrm dD^\star}{\mathrm dG}(x)
    &=(1-\eps)w(x)
      +\eps\frac{\Ind\{x>t\}}{\eps}\\
    &=(1-\eps)w(x)+\Ind\{x>t\}\\
    &\leq \max\left\{
        (1-\eps)(1+2e^{-5t^2/32}),\,1
    \right\}\\
    &\leq 1+2e^{-5t^2/32},
\end{align*}
because the two summands have disjoint supports. The two mixture components
are supported on $\{f_t(x)=1\}$ and $\{f_t(x)=-1\}$, respectively. Thus
$\Pr_{x\sim D^\star}[f_t(x)=-1]=\eps$, $\Pr_{x\sim D^\star}[f_t(x)=1]=1-\eps$, and
$D^\star_+=Q$, completing the proof.
\end{proof}

Now we generalize this lower bound to intersections of halfspaces.
\begin{lemma}[Moment Matching Lower Bound for Intersections of Halfspaces]
\label{lem:intersection-moment-lower-bound}
There exists a universal constant $c>0$ such that the following holds.
Let $k,d\in\Z_+$ with $k\leq d$ and $\eps\in(0,1/2)$.
Assume that $\eps\leq c/k^3$.
There exists an intersection of $k$ halfspaces
$f:\R^d\to\{\pm1\}$ and a probability distribution
$D^*\leq2\cN(\vec{0},\bI_d)$ such that $\Pr_{\x\sim D^*}[f(\x)=-1]=\eps$ and
$D^*_+\eqdef D^*\mid\{f(\x)=1\}$ matches the Gaussian moments through
total degree $\Omega(\log(k/\eps))$.
\end{lemma}

\begin{proof}
Set $\tau:=1-(1-\eps)^{1/k}$ and let $t$ satisfy
$\Pr_{z\sim\cN(0,1)}[z>t]=\tau$.
Since $\eps/k\leq\tau\leq2\eps/k\leq2c$, taking $c$ sufficiently
small ensures that $t\geq6$.
Set $\delta:=2e^{-5t^2/32}$ and $m:=\lfloor t^2/32\rfloor$.
Applying \Cref{lem:halfspace-moment-lower-bound} with tail mass $\tau$,
we obtain a probability distribution $D^\star$ satisfying
$D^\star\leq(1+\delta)\cN(0,1)$ and $\Pr_{z\sim D^\star}[z>t]=\tau$,
such that $D^\star\mid\{z\leq t\}$ matches the Gaussian moments
through degree $m$.

Define
\[
    D_k^\star:=(D^\star)^{\otimes k}\times\cN(\vec{0},\bI_{d-k}),
    \qquad
    H:=\bigcap_{i=1}^k\{\x\in\R^d:\x_i\leq t\},
    \qquad
    f(\x):=2\Ind\{\x\in H\}-1.
\]
We will show that $D_k^\star$ satisfies the lemma's conditions.
Write $D_{k,+}^\star:=D_k^\star\mid\{f(\x)=1\}$.
First, we prove the moment matching condition.

\begin{claim}\label{cl:intersectionsmomentmathcing}
$D_{k,+}^\star$ matches the moments of $\cN(\vec{0},\bI_d)$ through
total degree $m$.
\end{claim}

\begin{proof}[Proof of \Cref{cl:intersectionsmomentmathcing}]
Since $H$ is a coordinatewise intersection, independence gives
$D_{k,+}^\star=(D^\star\mid\{z\leq t\})^{\otimes k}
\times\cN(\vec{0},\bI_{d-k})$.
Fix a multi-index $\alpha\in\Z_{\geq0}^d$ with $|\alpha|\leq m$.
Each $\alpha_i\leq m$, so one-dimensional moment matching and
independence give
\[
    \E_{\x\sim D_{k,+}^\star}[\x^\alpha]
    =
    \left(
        \prod_{i=1}^k
        \E_{z\sim D^\star}[z^{\alpha_i}\mid z\leq t]
    \right)
    \prod_{i=k+1}^d\E_{z\sim\cN(0,1)}[z^{\alpha_i}]
    =
    \prod_{i=1}^d\E_{z\sim\cN(0,1)}[z^{\alpha_i}]
    =
    \E_{\x\sim\cN(\vec{0},\bI_d)}[\x^\alpha].
\]
The claim follows by linearity of expectation.
\end{proof}

Second, we bound the density ratio of $D_k^\star$ with respect
to the Gaussian.

\begin{claim}\label{cl:gaussiandomination}
It holds that
$D_k^\star\leq(1+\delta)^k\cN(\vec{0},\bI_d)\leq2\cN(\vec{0},\bI_d)$.
\end{claim}

\begin{proof}[Proof of \Cref{cl:gaussiandomination}]
Since $D^\star\leq(1+\delta)\cN(0,1)$, taking products gives
$D_k^\star=(D^\star)^{\otimes k}\times\cN(\vec{0},\bI_{d-k})
\leq(1+\delta)^k\cN(\vec{0},\bI_d)$.

For $t\geq6$, the Gaussian tail lower bound gives
$\tau\geq \frac{t}{\sqrt{2\pi}(t^2+1)}e^{-t^2/2}
\geq e^{-5t^2/8}$.
Consequently, $\delta=2e^{-5t^2/32}\leq2\tau^{1/4}$.
Using $\tau\leq2\eps/k$ and $\eps\leq c/k^3$, we obtain
$k\delta\leq2k\tau^{1/4}
\leq2(2k^3\eps)^{1/4}\leq2(2c)^{1/4}$.
Taking $c$ sufficiently small therefore gives
$(1+\delta)^k\leq e^{k\delta}\leq2$, proving the claim.
\end{proof}

Finally, we compute the negative mass of $f$ under $D_k^\star$.

\begin{claim}\label{cl:intersectionnegatives}
It holds that $\Pr_{\x\sim D_k^\star}[f(\x)=-1]=1-(1-\tau)^k$.
\end{claim}

\begin{proof}[Proof of \Cref{cl:intersectionnegatives}]
Each of the first $k$ coordinates independently satisfies
$\Pr_{\x\sim D_k^\star}[\x_i\leq t]=1-\tau$.
Thus $\Pr_{\x\sim D_k^\star}[f(\x)=1]=(1-\tau)^k$, and taking
the complement proves the claim.
\end{proof}

By the choice $\tau=1-(1-\eps)^{1/k}$,
\Cref{cl:intersectionnegatives} gives
$\Pr_{\x\sim D_k^\star}[f(\x)=-1]=\eps$.
By \Cref{cl:gaussiandomination},
$D_k^\star\leq2\cN(\vec{0},\bI_d)$.
Moreover, Gaussian tail estimates give
$t^2=\Theta(\log(1/\tau))$.
Since $t\geq6$ and $\eps/k\leq\tau\leq2\eps/k$, we have
$m=\lfloor t^2/32\rfloor=\Omega(\log(k/\eps))$.
Together with \Cref{cl:intersectionsmomentmathcing}, this shows
that $D_k^\star$ is the required distribution.
\end{proof}

\begin{proof}[Proof of \Cref{cor:sq-pu-intersections}]
By \Cref{lem:intersection-moment-lower-bound}, there exist an intersection
$f$ of $k$ halfspaces and a distribution $D^*\leq2\cN(\vec{0},\bI_d)$
such that $\Pr_{\x\sim D^*}[f(\x)=-1]=\eps$ and $D_+^*$ matches Gaussian moments
through degree $m=\Theta(m^*)$. The construction depends only on the
first $k$ coordinates, and $\mathcal F_k$ contains every rotation of $f$.
Moreover, $p_+=1-\eps\geq1/2$. For sufficiently small $c$ and
sufficiently large $d$, the stated parameter assumptions satisfy the
dimension and mass conditions of \Cref{thm:sq-pu-general}, with
$q=1$, $\sigma=1/2$, $K=k$, and moment degree $m$. Applying that theorem gives the claimed
lower bound.
\end{proof}

\section{Agnostic Reliable Learning} \label{sec:reliable}
As a third application, we show that the PUI learning algorithm through Gaussian partitioning also directly gives efficient learning algorithms for one-sided agnostic reliable learning defined as follows.

\begin{definition}[Agnostic Reliable learning]
Let $\mathcal{F}\subseteq \{\R^d\to \{\pm 1\}\}$ be a class of Boolean functions containing the function $f(\x)=-1$ and let $D$ be a distribution on $\R^d\times \{\pm 1\}$. We say that an algorithm positively reliably learns $\mathcal{F}$ under $D$ if it returns a hypothesis $h$ such that $\Pr_{(\x,y)\sim D}[h(\x)=1,y=-1]\leq \eps$ and
$$
\Pr_{(\x,y)\sim D}[h(\x)=-1,y=1]\leq
\inf_{f\in \mathcal{G}_D}
\Pr_{(\x,y)\sim D}[f(\x)=-1,y=1]+\eps,
$$
where $\mathcal{G}_D=\{f\in\mathcal{F}:
\Pr_{(\x,y)\sim D}[f(\x)=1,y=-1]=0\}.
$

Negative reliable learning is defined analogously, with the roles of the labels $1$ and $-1$ interchanged and with $\mathcal{F}$ containing the constant function $f(\x)=1$.%
\end{definition}

Our main result is that if a concept class $\mathcal{F}$ has a bounded PUI learning exponent $m^*$, then it also implies a reliable learning algorithm in time $d^{O(m^*)}$ under Gaussian marginal distribution.

\begin{theorem}[Agnostic Reliable Learning under Gaussian Distribution]
\label{cor:reliable-mim}
Let $\mathcal F$ be a class of functions that is closed under affine
restrictions {and has bounded Gaussian surface area $\Gamma(\mathcal{F})< \infty$} (satisfies items (1) and (3) of \Cref{def:wellbehaved}) and contains
the constant function $1$.
Let $D$ be a distribution over $\R^d\times\{\pm1\}$ whose
$\x$-marginal {is $(q,\sigma)$-smooth with respect to $\cN(\vec{0},\bI_d)$}, where
{$q\geq1$ and $\sigma\in(0,1]$}.
Let $\eps\in(0,1/2)$ and $\Delta\in(0,1/2)$, and set
$
    {m:=m^*(c\eps,\Delta,c\sigma\eps^3/q,2q)}
$
for a sufficiently small universal constant $c>0$. {Assume $m<\infty$.}
There is an algorithm which, given {$\eps,\sigma,\Delta,q$} and sample
access to $D$, uses at most
\[
    N=d^m\exp\!\left(
        {(qm)^{O(m)}}\log^{O(m)}(1/(\sigma\eps))
        \log(2+\Gamma(\mathcal F))/\Delta^{O(1)}
    \right)
\]
labeled samples, runs in time polynomial in $N$, and negatively
reliably learns $\mathcal F$ under $D$ with probability at least $2/3$.
\end{theorem}

The proof of \Cref{cor:reliable-mim} relies on the following reduction from agnostic reliable learning to PUI learning.

\begin{lemma}[Reduction from Reliable Learning to Smoothed PUI Learning]
\label{lem:reliable-to-pu}
Let $G=\cN(\vec{0},\bI_d)$, let $D$ be a distribution over
$\R^d\times\{\pm1\}$ whose $\x$-marginal {is $(q,\sigma)$-smooth with respect to $G$},
where {$q\geq1$ and $\sigma\in(0,1]$}, and let $\eps\in(0,1/2)$.
Assume that $\mathcal F$ contains the constant function $1$ and admits
a PUI learner $\mathcal A$ with failure probability at most $1/12$,
sample complexity {$N(\eps,\sigma,p_+,q)$}, and running time
{$T(\eps,\sigma,p_+,q)$}, where $p_+>0$ is a given lower bound on the
probability that the target function equals $1$.

Then there exists an algorithm that negatively reliably learns
$\mathcal F$ under $D$ with probability at least $2/3$, using
$O({N(\eps/2,\sigma\eps/4,\eps/4,q)}/\eps+1/\eps^2)$
labeled samples and
 $O({T(\eps/2,\sigma\eps/4,\eps/4,q)}/\eps+1/\eps^2)$
time.

Consequently, if $\mathcal F$ contains the constant function $-1$
and $-\mathcal F$ admits such a PUI learner, then $\mathcal F$ is
positively reliably learnable under $D$, with the same complexity
bounds for the learner of $-\mathcal F$.
\end{lemma}

The intuition of \Cref{lem:reliable-to-pu} is simple. To perform agnostic reliable learning, it is sufficient to learn a hypothesis with error $\eps$ over the distribution $D^*$, which is the distribution conditioned on the examples with $f^*(\x) = y$. However, we do not have sample access to $D^*$ directly. Fortunately, the assumption of reliable learning guarantees that $D \mid \{y=1\} = D^* \mid \{f^*(\x) = 1\}$. Since $D^*$ is smooth, by running our PUI learning algorithm over $D \mid \{y=1\}$, we immediately get a desired hypothesis.

\begin{proof}[Proof of \Cref{lem:reliable-to-pu}]
We prove the negative reliable case.
Let {$\rho:=\Pr_{(\x,y)\sim D}[y=1]$}.
Using $O(\eps^{-2})$ labeled samples, obtain an estimate {$\widehat\rho$}
such that {$|\widehat\rho-\rho|\leq\eps/4$} with probability at least $11/12$.
Condition on this event.
If {$\widehat\rho\leq\eps/2$}, return the constant hypothesis $h=-1$.
Its false-negative error is {$\rho\leq3\eps/4$}, and its false-positive
error is zero, so it satisfies the required guarantees.

Otherwise, {$\rho>\eps/4$}.
Let $\mathrm{OPT}$ denote the negative reliable optimum, and choose
$f^*\in\mathcal F$ such that
\[
    \Pr_{(\x,y)\sim D}[f^*(\x)=-1,y=1]=0,
    \qquad
    \Pr_{(\x,y)\sim D}[f^*(\x)=1,y=-1]\leq\mathrm{OPT}+\eps/2.
\]
Define $A:=\{y=f^*(\x)\}$, write $w:=\Pr_{(\x,y)\sim D}[(\x,y)\in A]$, and let $D^*$ be
the distribution of $\x$ under $D$ conditioned on $A$.
Every positive example belongs to $A$, so {$w\geq\rho>\eps/4$}.
Consequently, {for every measurable $E\subseteq\R^d$,}
{\[
    \Pr_{\x\sim D^*}[\x\in E]\leq \frac{\Pr_{\x\sim D_\x}[\x\in E]}{w}
    \leq \frac{\Pr_{\x\sim G}[\x\in E]^{1/q}}{\sigma w}
    \leq \frac{\Pr_{\x\sim G}[\x\in E]^{1/q}}{\sigma\eps/4}.
\]}
{Thus $D^*$ is $(q,\sigma\eps/4)$-smooth with respect to $G$.}
Moreover,
$
    D^*\mid\{f^*(\x)=1\}
    =D|y=1$ and $
    {\Pr_{\x\sim D^*}[f^*(\x)=1]=\rho/w\geq\rho>\eps/4\;.}
$
Thus we can supply the PUI learner with the required conditional
samples by rejection sampling from $D$, without knowing $f^*$,
$D^*$, or $\mathrm{OPT}$.

Write
$
    n:=\left\lceil {N(\eps/2,\sigma\eps/4,\eps/4,q)}\right\rceil.
$
Draw fresh labeled samples from $D$ and retain their $\x$-coordinates
exactly when $y=1$, until $n$ samples have been accepted or
$M:=\lceil C(n+1)/\eps\rceil$ labeled samples have been drawn.
If fewer than $n$ samples are accepted, return $h=-1$.
On the event considered above, each draw is accepted independently
with probability {$\rho>\eps/4$}.
The number $X$ of accepted samples among $M$ draws has mean
{$\mu=M\rho>C(n+1)/4$}.
For a sufficiently large absolute constant $C$, a Chernoff bound gives
\[
    \Pr[X<n]
    \leq \Pr[X\leq\mu/2]
    \leq \exp(-\mu/8)
    \leq \exp(-C(n+1)/32)
    \leq 1/12.
\]
The accepted points are independent samples from
$D|y=1$, including conditional on obtaining
$n$ of them before the cutoff, since the cutoff depends only on
the labels.
This procedure uses $O((n+1)/\eps)$ labeled samples and time.

Run $\mathcal A$ at accuracy $\eps/2$, with {smoothness parameters
$(q,\sigma\eps/4)$} and positive-mass lower bound $\eps/4$, using
the accepted samples.
Enforce its stated sample and running-time limits, returning
$h=-1$ if either is exceeded.
On the valid branch, these limits do not affect the PUI guarantee.
The learner yields a hypothesis $h$ satisfying
$
    \Pr_{\x\sim D^*}[h(\x)\neq f^*(\x)]\leq\eps/2
$
with probability at least $11/12$.
On this success event,
\[
    \Pr_{(\x,y)\sim D}[h(\x)=-1,y=1]
    =w\Pr_{\x\sim D^*}[h(\x)=-1,f^*(\x)=1]
    \leq w\eps/2\leq\eps/2.
\]
Furthermore,
\[
    \Pr_{(\x,y)\sim D}[h(\x)\neq y]
    \leq \Pr_{(\x,y)\sim D}[(\x,y)\in A^c]
        +w\Pr_{\x\sim D^*}[h(\x)\neq f^*(\x)]
    \leq\mathrm{OPT}+\eps.
\]
In particular, the false-positive error is at most
$\mathrm{OPT}+\eps$.
The mass estimate, rejection sampling, and PUI learner each have
failure probability at most $1/12$, so the overall success
probability is at least $3/4$, and hence at least $2/3$.

The preliminary mass estimate uses $O(\eps^{-2})$ samples and time.
Together with the rejection-sampling cutoff, this gives the stated
sample bound.
The total running time is
\[
    O\!\left(
        {T(\eps/2,\sigma\eps/4,\eps/4,q)}
        +\frac{{N(\eps/2,\sigma\eps/4,\eps/4,q)}+1}{\eps}
        +\frac{1}{\eps^2}
    \right).
\]
Since $T\geq N$ and $\eps<1$, this is bounded by the running time
in the statement.

For the positive reliable case, apply the negative reliable
reduction to the relabeled distribution $(\x,-y)$ and the class
$-\mathcal F$, which contains the constant function $1$.
The samples supplied to its PUI learner are drawn from $D$
conditioned on $y=-1$.
If that learner returns $\widehat g$, return $h=-\widehat g$.
The two reliable guarantees follow by interchanging the labels.
\end{proof}

Combining \Cref{thm:pumim} with \Cref{lem:reliable-to-pu} yields the proof of \Cref{cor:reliable-mim}.

\begin{proof}[Proof of \Cref{cor:reliable-mim}]
First amplify the success probability of the PUI learner in
\Cref{thm:pumim} to at least $11/12$ by taking the pointwise majority
of a sufficiently large constant number of independent runs at
accuracy $\eps/24$.
By the Hoeffding bound, with probability at least $11/12$, at least
$7/12$ of the runs succeed.
Whenever the majority is wrong, at least $1/12$ of all runs are
successful runs that are wrong at that point; hence the majority's
error is at most $12(\eps/24)=\eps/2$.

\Cref{lem:reliable-to-pu} invokes this learner with smoothness
parameter $\sigma'=\sigma\eps/4$ and positive-mass lower bound
$p'_+=\eps/4$. Each independent PUI run uses accuracy
$\eps'=\eps/24$. Thus the PUI algorithm uses
$\alpha=\sigma'p'_+/4=\sigma\eps^2/64$ and {conditional smoothness parameter}
{$\sigma_*=\alpha\eps'/(3200q)=\Theta(\sigma\eps^3/q)$}.
For sufficiently small $c$, the moment order required by each run
is at most $m$.

The parameter substitution gives
\[
    \log\frac1{\sigma'\eps'p'_+}
    =O\!\left(\log\frac1{\sigma\eps}\right).
\]
Applying the explicit PUI bounds from the proof of \Cref{thm:pumim},
and using $\log(1/\Delta)\leq1/\Delta$, yields a sufficient sample bound
\[
    d^m\exp\!\left(
        {(qm)^{O(m)}}\log^{O(m)}(1/(\sigma\eps))
        \log(2+\Gamma(\mathcal F))/\Delta^{O(1)}
    \right).
\]
Here the exponent $O(1)$ of $1/\Delta$ is universal and independent
of $m$. Logarithms of $1/(\sigma\eps)$ from the resolution are
absorbed into $\log^{O(m)}(1/(\sigma\eps))$, while the surface-area
factor remains $\log(2+\Gamma(\mathcal F))$.
The constant number of amplification runs, the rejection-sampling
factor $1/\eps$, and the $O(\eps^{-2})$ estimation cost are also
absorbed into this exponential factor. The total running time is
polynomial in the displayed sample bound.
\end{proof}

As an application of \Cref{cor:reliable-mim} and \Cref{cor:pu-intersections}, we immediately obtain reliable learning algorithms for learning intersections and unions of halfspaces.

\begin{corollary}[Reliable Learning of Intersections and Unions of Halfspaces]
\label{cor:reliable-halfspaces}
Let $k,d\in\mathbb N_+$, $\eps\in(0,1/2)$, {$q\geq1$,} and $\sigma\in(0,1]$.
Let $D$ be a distribution over $\R^d\times\{\pm1\}$ whose
$\x$-marginal {is $(q,\sigma)$-smooth with respect to $\cN(\vec{0},\bI_d)$}.
Set {$m:=\lceil C(\log(2k)+q\log(2q/(\sigma\eps)))\rceil$} for a sufficiently
large universal constant $C>0$.
There is an algorithm which, given {$\eps,\sigma,q$} and sample access
to $D$, uses at most
\[
    N=d^m\exp\!\left((2m)^{O(m)}\right)
\]
labeled samples, runs in time polynomial in $N$, and negatively
reliably learns the class $\mathcal F_k$ of intersections of at most
$k$ halfspaces with probability at least $2/3$.
With the same sample and running-time bounds, unions of at most
$k$ halfspaces are positively reliably learnable under $D$.

\end{corollary}

\begin{proof}[Proof of \Cref{cor:reliable-halfspaces}]
The class $\mathcal F_k$ contains the constant functions, is closed
under affine restrictions, and satisfies $\Gamma(\mathcal F_k)=O(k)$.
Taking $\Delta=1/8$, \Cref{cor:intersection-moments} gives
{\[
\begin{aligned}
    m^*(c\eps,1/8,c\sigma\eps^3/q,2q)
    &=O\left(\log(2k)+q\log\frac{2q}{\sigma\eps^4}\right)\\
    &=O\left(\log(2k)+q\log\frac{2q}{\sigma\eps}\right)
    \leq m.
\end{aligned}
\]}
Apply \Cref{cor:reliable-mim} with $\Delta=1/8$ and moment
order $m$. Since $\Gamma(\mathcal F_k)=O(k)$ and
{$q,\log(1/(\sigma\eps)),\log(2+k)=O(m)$}, its sample bound simplifies to
\[
    d^m\exp\!\left(
        {(qm)^{O(m)}}\log^{O(m)}(1/(\sigma\eps))\log(2+k)
    \right)
    \leq d^m\exp\!\left((2m)^{O(m)}\right).
\]
This proves the stated complexity and the negative reliable guarantee.
The complement of a union of at most $k$ halfspaces is an intersection
of at most $k$ halfspaces, so the positive reliable guarantee follows
from the last part of \Cref{lem:reliable-to-pu}.
\end{proof}

\section{SQ Lower Bound for Realizable TDS and PQ Learning under the Gaussian Distribution} \label{sec:tds-pu}

\Cref{cor:reliable-mim} yields efficient algorithms for agnostic
reliable learning.
For intersections of $k$ halfspaces under the standard Gaussian,
our algorithm has dimension dependence $d^{O(\log(k/\eps))}$.
This raises a natural question: can we obtain the same logarithmic
dependence on $1/\eps$ in the exponent of $d$ for closely related reliable learning 
models, such as PQ learning \cite{goldwasser20pq} and TDS learning
\cite{rev-KSV24}?
We first recall their definitions.

\begin{definition}[PQ learning]\label{def pq learning}
Let $\mathcal{F}\subseteq\{\R^d\to\{\pm1\}\}$,
and let $P,Q$ be distributions on $\R^d$.
We say that an algorithm $(\eps,\delta)$-PQ learns
$\mathcal{F}$ under $P$ if, for every target
$f\in\mathcal{F}$ and every test distribution $Q$,
given labeled samples $(\x,f(\x))$ with $\x\sim P$
and unlabeled samples $\x\sim Q$, it returns a
selective hypothesis
$h:\R^d\to\{\pm1,\bot\}$ such that, with
probability at least $1-\delta$, $\Pr_{\x\sim P}[h(\x)=\bot]\le\eps$
and $\Pr_{\x\sim Q}
[h(\x)\ne f(\x),\,h(\x)\ne\bot]\le\eps.$
\end{definition}

\begin{definition}[TDS learning]\label{def tds learning}
Let $\mathcal{F}\subseteq\{\R^d\to\{\pm1\}\}$
and let $P$ be a distribution on $\R^d$.
We say that an algorithm $(\eps,\delta)$-TDS
learns $\mathcal{F}$ under $P$ if, for every
target $f\in\mathcal{F}$ and every test
distribution $Q$, given labeled samples
$(\x,f(\x))$ with $\x\sim P$ and unlabeled
samples $\x\sim Q$, it either rejects or
accepts and returns a hypothesis
$h:\R^d\to\{\pm1\}$ satisfying:
\begin{enumerate}
    \item (Completeness.) If $Q=P$, the algorithm accepts with
probability at least $1-\delta$.
\item (Soundness.) With probability at least $1-\delta$, whenever
the algorithm accepts, its hypothesis satisfies $\Pr_{\x\sim Q}[h(\x)\ne f(\x)]\le\eps.$
\end{enumerate}

\end{definition}

Recent works \cite{kalai2021efficient,klivans2026equivalence}
have shown that, in the distribution-free setting, these models
are equivalent to agnostic reliable learning.
Under a Gaussian training marginal, however, known algorithms
based on polynomial regression \cite{rev-KSV24}
have dimension dependence $d^{\poly(k/\eps)}$.

We show that an improvement analogous to our reliable learning
result is impossible in the SQ model, even for realizable learning
of intersections of two halfspaces.
Specifically, any PQ or TDS learner with SQ access to both the labeled training distribution
and the unlabeled test distribution must either make $2^{d^{\Omega(1)}}$ queries or use
tolerance at most $d^{-\Omega(1/\eps^2)}$
(see \Cref{lm pq tds sq lower bound}).
This establishes a separation between the SQ complexities of
agnostic reliable learning and realizable PQ/TDS learning
in the Gaussian setting, despite their distribution-free equivalence.
We remark that \cite{klivans2024learning} also obtained an SQ lower bound for the 
class of intersections of two halfspaces in the TDS learning setting. However,
they only established a $d^{\Tilde{\Omega}(\log(1/\eps))}$ lower bound, which is not sufficient to establish the separation for our purpose.

\begin{theorem}[SQ Lower Bound for PQ and TDS learning]
\label{lm pq tds sq lower bound}
Let $\mathcal{F}$ be the class of intersections of
two halfspaces over $\R^d$. There exist
universal constants $c,c'>0$ such that, for every
sufficiently small $\eps>0$ and sufficiently
large $d$ depending on $\eps$, the following holds.
{Any algorithm $(\eps,1/20)$-PQ learns or
$(\eps,1/20)$-TDS learns $\mathcal F$ under
$\cN(\vec 0,\vec I)$, using SQ access to the labeled
training distribution and the unlabeled test distribution, must either make at least $2^{d^{c'}}$ SQ queries
in total or make a query to either oracle of tolerance
at most $d^{-c/\eps^2}$.}
\end{theorem}

Our construction follows directly from the discrete Gaussian distribution used in \cite{DKS17-sq}, which constructs a one-dimensional discrete distribution $D_\bv$ that matches $\Omega(1/\eps^2)$ moments with a one-dimensional standard Gaussian, but puts $\Omega(\eps)$ mass at 0.
Using standard NGCA techniques from \cite{diakonikolas2023sq}, any SQ-algorithm would take $d^{\Omega(1/\poly(\eps))}$ time to distinguish a standard Gaussian distribution from $Q_\bv$. But on the other hand, if we set $f^* = 2\Ind\{\x \mid \abs{\bv \cdot \x} \le \delta\}-1$, for some $\delta$ that is arbitrarily close to $0$, then the training distribution would just be $\cN(\vec{0},\bI) \times \{-1\}$, while a successful PQ/TDS learning algorithm would need to distinguish $Q_\bv$ from $\cN(\vec{0},\bI)$ so that the trained hypothesis has $\eps$ error over the test distribution.

\begin{proof}[Proof of \Cref{lm pq tds sq lower bound}]
We reduce the problem of non-Gaussian component
analysis (NGCA) to PQ or TDS learning.
Fix $r=\lfloor c_0/\eps^2\rfloor$ for a
sufficiently small universal constant $c_0>0$.
By \cite{DKS17-sq}, there exists a
discrete distribution $B$ that matches the
first $4r+1$ moments of $\cN(0,1)$
and assigns mass
\[
w_0=\frac{4^r}{(2r+1)\binom{2r}{r}}
    =\Theta(r^{-1/2})\ge 20\eps
\]
to zero.
For an unknown unit vector $\bv$, let
$Q_\bv$ denote the distribution whose
marginal along $\bv$ is $B$ and
whose orthogonal marginal is an independent
standard Gaussian.
By \Cref{fact:ngca-sq}, distinguishing
$\cN(\vec 0,\vec I)$ from $Q_\bv$, for a uniformly
random hidden direction $\bv$, requires either
$2^{d^{\Omega(1)}}$ SQ queries or tolerance
at most $d^{-\Omega(r)}$.

Suppose that there exists a PQ or TDS SQ learner
using at most $T$ SQ queries, each of tolerance
at least $\tau>0$.
Choose $t>0$ sufficiently small that
\[
\rho:=
\Pr_{\x\sim\cN(\vec 0,\vec I)}
[|\bv\cdot\x|\le t]
\le \min\{\eps,\tau/8\}.
\]
Consider the target
\[
f_\bv(\x)=
\begin{cases}
1,&|\bv\cdot\x|\le t,\\
-1,&\text{otherwise}.
\end{cases}
\]
This is an intersection of two general
halfspaces. Moreover,
\[
\Pr_{\x\sim Q_\bv}[f_\bv(\x)=1]
\ge w_0\ge 20\eps.
\]
We construct an SQ distinguisher for NGCA
using the assumed learner.

We show that the performance of the SQ learner over the training distribution
can be the same as one trained over a Gaussian distribution with examples all negative.
Specifically, for any SQ query
$\phi:\R^d\times\{\pm1\}\to[-1,1]$,
we return an estimate of
\[
a_\phi=
\E_{\x\sim\cN(\vec 0,\vec I)}[\phi(\x,-1)].
\]
Since $f_\bv(\x)=-1$ outside the slab
$|\bv\cdot\x|\le t$, we have
\[
\begin{aligned}
\left|
a_\phi-
\E_{\x\sim\cN(\vec 0,\vec I)}
[\phi(\x,f_\bv(\x))]
\right|\le
2\Pr_{\x\sim\cN(\vec 0,\vec I)}
[f_\bv(\x)=1]
\le 2\rho\le\tau/4.
\end{aligned}
\]

This implies that the performance of the learning algorithm can be seen as identical to a distribution whose label is always negative and the training stage provides no information to the learner.

We provide the unknown NGCA distribution
$Q$ as the learner's unlabeled test SQ oracle.

First suppose that the learner is a TDS
learner. If it rejects, declare that the
test distribution is $Q_\bv$. Otherwise,
let $h$ be its output and query $p=\Pr_{\x\sim Q}[h(\x)=1]$
with tolerance $\eps$.

If $Q=\cN(\vec 0,\vec I)$, completeness guarantees
that the learner accepts, and soundness gives
\[
p\le
\Pr_{\x\sim\cN(\vec 0,\vec I)}[h(\x)\ne f_\bv(\x)]
+\rho
\le 2\eps.
\]
If $Q=Q_\bv$, soundness implies that,
whenever the learner accepts,
\[
p\ge
\Pr_{\x\sim Q_\bv}[f_\bv(\x)=1]
-\Pr_{\x\sim Q_\bv}[h(\x)\ne f_\bv(\x)]
\ge 19\eps.
\]
Consequently, thresholding the SQ estimate
at $10\eps$ distinguishes the two cases.

For PQ learning, let
$h:\R^d\to\{\pm1,\bot\}$ be the output
and instead query $p=\Pr_{\x\sim Q}[h(\x)\ne-1]$
with tolerance $\eps$.
When $Q=\cN(\vec 0,\vec I)$, the coverage and
soundness guarantees give
\[
\begin{aligned}
p
\le
\Pr_{\x\sim\cN(\vec 0,\vec I)}[h(\x)=\bot]+
\Pr_{\x\sim\cN(\vec 0,\vec I)}
[h(\x)=1,f_\bv(\x)=-1]
+\rho\le 3\eps.
\end{aligned}
\]
When $Q=Q_\bv$, soundness gives
\[
\begin{aligned}
p
\ge
\Pr_{\x\sim Q_\bv}[f_\bv(\x)=1]-
\Pr_{\x\sim Q_\bv}
[h(\x)=-1,f_\bv(\x)=1]
\ge 19\eps.
\end{aligned}
\]
Again, one SQ query of tolerance $\eps$
distinguishes the two cases.

In either case, we obtain an SQ distinguisher
for NGCA using at most $T+1$ queries to the
unknown distribution, all of tolerance
at least $\min\{\tau,\eps\}$.
Applying the NGCA lower bound of
\Cref{fact:ngca-sq}, we conclude that either
\[
T+1\ge 2^{d^{\Omega(1)}}
\quad\text{or}\quad
\min\{\tau,\eps\}\le d^{-\Omega(r)}.
\]
Since $r=\Theta(1/\eps^2)$, this gives
the claimed SQ lower bound in the stated
parameter regime.
\end{proof}

\bibliographystyle{alphaabbr}
\bibliography{mydb}

\pagebreak

\appendix

\section*{Appendix} \label{sec:appendix}
The appendix is structured as follows.
\Cref{sec:omitted-preliminaries} contains additional preliminaries.
\Cref{app:omitted-facts} collects standard facts and supporting
technical proofs used in the main text.
\Cref{app:partitioning-details} contains the omitted technical
proofs for our partitioning algorithm (\cref{sec:relaxed-gaussian-partitioning}).
 \Cref{app:agnostic-details} contains the extensions
to Gaussian mixtures and unknown Gaussian reference means.
\Cref{sec:multiplicative-sq} presents an SQ lower bound for agnostic learning with
a multiplicative approximation guarantee.
{Finally, in \Cref{sec:testable}, we prove our testable
agnostic learning results.}

\section{Omitted Preliminaries} \label{sec:omitted-preliminaries}
\textbf{Gaussian space:} 
We denote by $L_2(\cN(\vec{0},\bI_d))$ the vector space of all functions
$f:\R^d\to\R$ such that
$\E_{\x\sim\cN(\vec{0},\bI_d)}[f(\x)^2]<\infty$.
We use Hermite polynomials normalized with respect to the standard Gaussian measure. Let \(\mathrm{He}_m\) denote the probabilists’ Hermite polynomial of degree \(m\), and define
\[
h_m(t):=\frac{\mathrm{He}_m(t)}{\sqrt{m!}}
=\frac{(-1)^m}{\sqrt{m!}}e^{t^2/2}
\frac{d^m}{dt^m}e^{-t^2/2}.
\]
In particular, \(h_0(t)=1\), \(h_1(t)=t\), and \(h_2(t)=(t^2-1)/\sqrt2\). For a multi-index \(\beta\in\mathbb Z_{\ge0}^d\), define
$
H_\beta(\x):=\prod_{r=1}^d h_{\beta_r}(\x_r),
|\beta|:=\sum_{r=1}^d\beta_r.
$
These polynomials form an orthonormal basis of \(L_2(\cN(\vec{0},\bI_d))\), i.e.,
$
\E_{\x\sim\cN(\vec{0},\bI_d)}
[H_\beta(\x)H_\gamma(\x)]
=\Ind\{\beta=\gamma\}.
$
We also use the order-\(m\) symmetric Hermite tensor \(\mathcal H_m(\x)\), whose entries are
\[
(\mathcal H_m(\x))_{j_1,\ldots,j_m}
:=\sqrt{\frac{\beta!}{m!}}\,H_\beta(\x),
\qquad
\beta_r:=|\{s\in[m]:j_s=r\}|,
\]
where \(\beta!:=\prod_{r=1}^d\beta_r!\). This normalization gives
\(\langle\mathcal H_m(\x),\bu^{\otimes m}\rangle=h_m(\bu\cdot\x)\) for every unit vector \(\bu\). Moreover, for symmetric order-\(m\) tensors \(\vec{S},\bT\),
$
\E_{\x\sim\cN(\vec{0},\bI_d)}
\!\left[
\langle\vec{S},\mathcal H_m(\x)\rangle
\langle\bT,\mathcal H_m(\x)\rangle
\right]
=\langle\vec{S},\bT\rangle_F.
$
\begin{definition}[Gaussian surface area]
\label{def:gaussian-surface-area}
For a Borel set $A \subseteq \mathbb{R}^d$ and $t>0$, let
$
A_t := \{\x \in \mathbb{R}^d : \operatorname{dist}(\x,A) \le t\}.
$
The Gaussian surface area of $A$ is defined as
\[
\Gamma(A) :=
\liminf_{t \downarrow 0}
\frac{ \Pr_{\x\sim \cN(\vec{0},\bI_d)}[\x\in A_t \setminus A]}{t}.
\]
For a Borel measurable Boolean function
$f:\mathbb{R}^d \to \{-1,+1\}$, let
$A_b := \{\x : f(\x)=b\}$ and define
$
\Gamma(f) := \Gamma(A_{-1})+\Gamma(A_{+1}).
$
For a class $\mathcal{F}$ of such functions, define
$
\Gamma(\mathcal{F}) := \sup_{f \in \mathcal{F}} \Gamma(f).
$
\end{definition}
\begin{definition}[Gaussian noise operator]
For $\rho\in(0,1)$ and a bounded measurable function
$f:\R^d\to\R$, define
\[
T_\rho f(\x)
=
\E_{\bz\sim\cN(\vec{0},\bI_d)}
\left[
f\left(\sqrt{1-\rho^2}\,\x+\rho\bz\right)
\right].
\]
\end{definition}

\begin{fact}[Properties of Gaussian smoothing]
Let $\rho\in(0,1)$. The following properties hold:
\begin{enumerate}
    \item For every bounded measurable
    $f:\R^d\to\R$, the function $T_\rho f$
    is differentiable and
    \[
    \|\nabla T_\rho f(\x)\|_2
    \leq \frac{\|f\|_\infty}{\rho}.
    \]

    \item For every
    $f:\R^d\to\{\pm1\}$ with finite
    Gaussian surface area $\Gamma(f)$,
    \[
    \E_{\x\sim\cN(\vec{0},\bI_d)}
    \left[|f(\x)-T_\rho f(\x)|\right]
    \leq 2\sqrt{\pi}\,\rho\,\Gamma(f).
    \]
\end{enumerate}
\end{fact}
\paragraph{Basics on SQ Model}
SQ algorithms are a broad class of algorithms that, instead of having direct
access to samples, are allowed to query expectations of bounded functions
of the distribution.

\begin{definition}[SQ algorithms]
Let $D$ be a distribution on $\mathcal X$. A statistical query is a bounded
function $q:\mathcal X\to[-1,1]$. 
For $\tau>0$, the oracle $\mathrm{STAT}(\tau)$ responds to a query
$q$ with a value $v$ satisfying
$|v-\E_{\x\sim D}[q(\x)]|\leq\tau$.
We call $\tau$ the \emph{tolerance} of the statistical query.
A \emph{Statistical Query} algorithm is an algorithm whose objective is
to learn some information about an unknown distribution $D$ by making
adaptive calls to the corresponding oracle.
\end{definition}

The SQ model was introduced in \cite{Kearns93} as a natural restriction of
the PAC model \cite{Val84}. Subsequently, the model has been extensively
studied in a range of contexts, see, e.g., \cite{Feldman16b}. The class of
SQ algorithms is broad and captures a range of known supervised learning
algorithms. More broadly, several known algorithmic techniques in machine
learning are known to be implementable using SQs
(see, e.g., \cite{FGR+13,FeldmanGV17}).

\paragraph{Testable Learning.}
Testable learning \cite{rubinfeld2023testing} is a popular model in which
a learner may test the distributional assumptions on the input
distribution. For every input distribution, with high probability,
whenever the test passes, the learner returns a hypothesis with
the prescribed accuracy. Moreover, the learner must accept with
high probability whenever the distributional assumptions hold.
\begin{definition}[Testable learning]
\label{def:testable-learning}
Let $\mathcal F\subseteq\{\R^d\to\{\pm1\}\}$ be a concept class,
and let $\mathcal D$ be a family of distributions over
$\R^d$.
We say that an algorithm $\mathcal A$ testably agnostically learns
$\mathcal F$ to error $\tau+\eps$ with probability $1-\delta$ with respect to $\mathcal D$
if the following holds.

{For every distribution $D$ over $\R^d\times\{\pm1\}$,} 
on input $\epsilon,\delta\in(0,1/2)$ and
$m(\epsilon,\delta)$ i.i.d.\ samples from $D$,
the algorithm outputs either $\textsc{Reject}$ or
$(\textsc{Accept},h)$, where $h:\R^d\to\{\pm1\}$,
satisfying:
\begin{enumerate}
    \item \emph{Soundness.}
    The probability that $\mathcal A$ accepts and outputs
    a hypothesis $h$ with $\err_D(h)>\tau +\eps$ is at most $\delta$.

    \item \emph{Completeness.}
    If $D_\x\in\mathcal D$, then $\mathcal A$ accepts with
    probability at least $1-\delta$.
\end{enumerate}
\end{definition}

\section{Omitted Facts and Technical Proofs}
\label{app:omitted-facts}
In this section, we collect standard facts used in our proofs.

The following fact shows that Hellinger distance cannot
decrease under refinement. We use it in the analysis of
our partitioning algorithm.
\begin{fact}[Squared Hellinger distance under refinement]
\label{fact:hellinger-refinement}
Let $P=(p_i)_{i=1}^j$ and $Q=(q_i)_{i=1}^j$ be probability
distributions, and let $I\subseteq[j]$.
If the two distributions obtained by grouping coordinates
into $I$ and $[j]\setminus I$ have squared Hellinger distance
at least $\kappa$, then
$
    \sum_{i=1}^j\sqrt{p_iq_i}\leq1-\kappa.
$
\end{fact}

\begin{proof}
By Cauchy--Schwarz applied separately to the two groups,
\begin{align*}
    \sum_{i=1}^j\sqrt{p_iq_i}
    &\leq
    \sqrt{\left(\sum_{i\in I}p_i\right)
          \left(\sum_{i\in I}q_i\right)}
    +
    \sqrt{\left(\sum_{i\notin I}p_i\right)
          \left(\sum_{i\notin I}q_i\right)}\\
    &\leq1-\kappa,
\end{align*}
where the last inequality follows from the assumed
squared Hellinger distance between the grouped distributions.
\end{proof}
The following standard fact bounds the difference between the
true and empirical cumulative distribution functions uniformly
over all thresholds.

\begin{fact}[Dvoretzky--Kiefer--Wolfowitz (DKW) inequality
{\cite{DKW56}}]
\label{fact:dkw}
Let $D$ be a distribution over $\R$, and let
$x_1,\ldots,x_n$ be independent samples from $D$.
Define the cumulative distribution function
$F(t)=\Pr_{x\sim D}[x\leq t]$ and its empirical counterpart
\[
\widehat F_n(t)
=\frac{1}{n}\sum_{i=1}^n \Ind\{x_i\leq t\}.
\]
For every $\eps>0$,
\[
\Pr\left[
\sup_{t\in\R}\left|\widehat F_n(t)-F(t)\right|>\eps
\right]
\leq 2\exp(-2n\eps^2).
\]
In particular, for $\eps,\delta\in(0,1)$, if
$n\geq \frac{1}{2\eps^2}\log\left(\frac{2}{\delta}\right)$,
then, with probability at least $1-\delta$,
$\left|\widehat F_n(t)-F(t)\right|\leq\eps$
simultaneously for all $t\in\R$.
\end{fact}

The following fact shows that $L_1$ polynomial regression
and thresholding efficiently produce a classifier whose
error is controlled by the polynomial approximation error.

\begin{fact}[$L_1$ polynomial regression {\cite{kalai2008agnostically}}]
\label{fact:polynomial-regression}
Let $D$ be a distribution over $\R^d\times\{\pm1\}$ and let $p$ be a degree-$\ell$ polynomial. For every $\eps,\zeta>0$, using $d^{O(\ell)}\poly(1/\eps,\log(1/\zeta))$ time and samples, one can output a degree-$\ell$ polynomial $q$ and a threshold $t\in[-1,1]$ such that, with probability at least $1-\zeta$, $\Pr_D[\sign(q(\x)-t)\neq y]\leq \E_D[|y-p(\x)|]/2+\eps$.
\end{fact}
The following fact shows that projecting both distributions
onto the same subspace preserves smoothness with the same
parameters. This fact is used in the analysis of our partitioning algorithm.
\begin{fact}[Smoothness under projection]
\label{fact:smoothness-projection}
Let $P,Q$ be distributions on $\R^d$, and suppose that $Q$
is $(q,\sigma)$-smooth with respect to $P$. For every subspace
$W\subseteq\R^d$, the projection of $Q$ onto $W$ is
$(q,\sigma)$-smooth with respect to the projection of $P$ onto $W$.
\end{fact}

\begin{proof}
Let $\boldsymbol{\Pi}=\proj_W$. For any measurable $A\subseteq W$, apply
smoothness to $\boldsymbol{\Pi}^{-1}(A)=\{\x:\boldsymbol{\Pi}\x\in A\}$ to obtain
\[
 \Pr_{\x\sim Q}[\boldsymbol{\Pi}\x\in A]
 \leq \frac{\Pr_{\x\sim P}[\boldsymbol{\Pi}\x\in A]^{1/q}}{\sigma}.
\]
This is the required smoothness inequality for the projections.
\end{proof}
We use the following notion of the $\delta$-surface of a concept.

\begin{definition}[$\delta$-surface]
For a Boolean function $f$ on $\mathbb{R}^d$ and $\delta>0$, define
\[
    S_\delta(f)
    \eqdef
    \left\{
        \bu\in\mathbb{R}^d:
        \exists\bv\in\mathbb{R}^d,\;
        \|\bu-\bv\|_2\leq\delta,\;
        f(\bu)\neq f(\bv)
    \right\}.
\]
\end{definition}
We next state facts that bound the Gaussian sandwiching
degree in terms of a noise sensitivity quantity.

\begin{fact}[Gaussian sandwiching from boundary smoothness
{\cite[Theorem~3.2]{klivans2026sandwiching}}] %
\label{fact:gaussian-sandwiching-smooth-boundary}
Let $\mathcal{F}$ be a class of $k$-MIMs over $\R^d$. 
Define
$
\sigma_{\mathcal{F}}
:=
\max\left\{
1,\,
\sup_{f\in\mathcal{F}}\sup_{\delta>0}(\Pr_{\x\sim \cN(\vec{0},\bI)}[\x\in S_\delta(f)]/\delta)
\right\}.
$
If $\sigma_{\mathcal{F}}<\infty$, then, for every
$\epsilon\in(0,1)$, every $f\in\mathcal{F}$ admits $\eps$-sandwiching polynomials 
 of degree
at most
$\ell(\eps)=
\widetilde{O}\left(
\sigma_{\mathcal{F}}^2k^3/\epsilon^4
\right)
$ with respect to the Gaussian distribution.
\end{fact}
The following fact states known bounds for the quantity $\sigma(\mathcal{F})$ for well studied function classes. This helps us apply our results to concrete function families. 
\begin{fact}[Gaussian boundary smoothness and surface area]
\label{fact:gaussian-boundary-smoothness}
For a Boolean concept class $\mathcal F$ over $\R^d$, let
$\sigma(\mathcal F)$ denote the quantity of \Cref{fact:gaussian-sandwiching-smooth-boundary}.
Note that, by definition,
$\Gamma(\mathcal F)\leq\sigma(\mathcal F)$.
Moreover, for $K,q\in\Z_+$ with $K\leq d$, the following bounds
hold (see e.g., \cite{klivans2026sandwiching} Lemmas 4.7, 4.5, 4.12 and 4.14):
\begin{enumerate}
    \item If $\mathcal F$ is the class of convex concepts with
    intrinsic dimension at most $K$, then
    $\sigma(\mathcal F)=O(K\log(2K))$.

    \item If $\mathcal F$ is the class of polynomial threshold
    functions of degree at most $r$ and intrinsic dimension
    at most $K$, then $\sigma(\mathcal F)=O(r^3K)$.
    \item If $\mathcal F$ is the class of intersections of
    at most $k$ halfspaces, then
    $\sigma(\mathcal F)=O(\sqrt{\log(2k)})$.

    \item If $\mathcal F$ is the class of arbitrary Boolean
    functions of at most $k$ halfspaces, then
    $\sigma(\mathcal F)=O(k)$.
\end{enumerate}
\end{fact}
The following fact translates a bound on directional moment
discrepancies into a bound on the difference in expectations
of a low-degree polynomial under the two distributions.
\begin{fact}[$L_1$ polynomials under moment matching]
\label{fact:moment-matching-polynomial-expectations}
Let $P=\cN(\vec{0},\bI_d)$, and let $Q$ be a distribution over $\R^d$ such that
$P$ and $Q$ match their first $k$ moments with slack $\Delta$ according to
\Cref{def:moment-matching}. Let $p:\R^d\to\R$ be a polynomial of degree at
most $k$ satisfying
$
    \|p\|_{L_1(P)}\le M.
$
Then
\[
    \left|\E_{\x\sim P}[p(\x)]-\E_{\x\sim Q}[p(\x)]\right|
    \le (dk)^{O(k)}M\Delta.
\]
\end{fact}

\begin{proof}
We first relate the Gaussian $L_1$ norm of $p$ to its coefficients. Expand
$p$ in the normalized multivariate Hermite basis as
\[
    p(\x)=\sum_{|\alpha|\le k}\widehat p_\alpha H_\alpha(\x).
\]
The number of terms in this expansion is at most
$\binom{d+k}{k}\le(d+k)^k$, and orthonormality gives
\[
    \sum_{|\alpha|\le k}\widehat p_\alpha^2
    =\|p\|_{L_2(P)}^2.
\]
Every normalized Hermite polynomial $H_\alpha$ of degree at most $k$ has
the sum of the absolute values of its monomial coefficients bounded by
$k^{O(k)}$. Consequently, if
\[
    p(\x)=\sum_{|\beta|\le k}c_\beta \x^\beta
\]
is the monomial expansion of $p$, Cauchy--Schwarz implies
\begin{equation}\label{eq:coefficient-l1-from-gaussian-l2}
    \sum_{|\beta|\le k}|c_\beta|
    \le (dk)^{O(k)}\|p\|_{L_2(P)}.
\end{equation}

We next bound the Gaussian $L_2$ norm by the Gaussian $L_1$ norm. Gaussian
hypercontractivity gives $\|p\|_{L_4(P)}\le 3^{k/2}\|p\|_{L_2(P)}$.
Applying the Paley--Zygmund inequality to $p(\x)^2$ therefore yields
\[
    \Pr_{\x\sim P}\left[
        |p(\x)|\ge \frac{\|p\|_{L_2(P)}}{\sqrt{2}}
    \right]
    \ge \frac{1}{4\cdot 9^k}.
\]
It follows that
\[
    \|p\|_{L_2(P)}
    \le 4\sqrt{2}\,9^k\|p\|_{L_1(P)}
    \le 4\sqrt{2}\,9^k M.
\]
Together with \eqref{eq:coefficient-l1-from-gaussian-l2}, this shows that
\begin{equation}\label{eq:coefficient-l1-from-gaussian-l1}
    \sum_{|\beta|\le k}|c_\beta|
    \le (dk)^{O(k)}M.
\end{equation}

It remains to use moment matching. Fix a monomial $\x^\beta$ of degree
$r=|\beta|\in[k]$, and choose indices $i_1,\ldots,i_r\in[d]$, with
repetitions, such that $\x^\beta=\prod_{j=1}^r \x_{i_j}$. By the polarization
identity,
\[
    \prod_{j=1}^r \x_{i_j}
    =\frac{1}{2^r r!}
    \sum_{\varepsilon\in\{\pm1\}^r}
    \left(\prod_{j=1}^r\varepsilon_j\right)
    \left(\sum_{j=1}^r\varepsilon_j\x_{i_j}\right)^r.
\]
For each $\varepsilon\in\{\pm1\}^r$, the vector
$\bw_\varepsilon=\sum_{j=1}^r\varepsilon_j \be_{i_j}$ satisfies
$\|\bw_\varepsilon\|_2\le r$. Applying moment matching to
$\bw_\varepsilon/\|\bw_\varepsilon\|_2$ when $\bw_\varepsilon\neq\vec{0}$ gives
\[
    \left|
    \E_{\x\sim P}[\x^\beta]-\E_{\x\sim Q}[\x^\beta]
    \right|
    \le \frac{r^r}{r!}\Delta
    \le e^k\Delta.
\]
The constant term has the same expectation under $P$ and $Q$. Hence, by
linearity, the triangle inequality, and
\eqref{eq:coefficient-l1-from-gaussian-l1},
\begin{align*}
    \left|\E_{\x\sim P}[p(\x)]-\E_{\x\sim Q}[p(\x)]\right|
    &\le e^k\Delta\sum_{1\le|\beta|\le k}|c_\beta|\\
    &\le (dk)^{O(k)}M\Delta,
\end{align*}
as claimed.
\end{proof}

\section{Omitted Details from \Cref{sec:relaxed-gaussian-partitioning}}
\label{app:partitioning-details}
We now prove \Cref{lem:relaxed-smoothness-truncation}, which shows
that the density ratio of a smooth distribution $Q$ with respect
to $P$ is bounded outside a set of small $Q$-mass.
\begin{proof}[Proof of \Cref{lem:relaxed-smoothness-truncation}]
By smoothness, $Q(E)\leq P(E)^{1/q}/\sigma$ for every
measurable event $E$.
In particular, $P(E)=0$ implies $Q(E)=0$, so the density ratio
$f=\mathrm dQ/\mathrm dP$ is well-defined $P$-almost everywhere.

If $q=1$, smoothness gives $f\leq1/\sigma$
$P$-almost everywhere.
Hence $G=\{f\leq1/\sigma\}$ has $Q$-mass one, and
$Q\mid G=Q$ is $(1,\sigma)$-smooth, hence also
$(1,(1-\varepsilon)\sigma)$-smooth.
This proves the lemma for $q=1$.

Suppose that $q>1$.
For $t>0$, define $E_t=\{\x:f(\x)>t\}$.
Since
\[
    Q(E_t)=\int_{E_t}f\,\mathrm dP\geq tP(E_t),
\]
smoothness gives
\[
    Q(E_t)\leq\frac{P(E_t)^{1/q}}{\sigma}
    \leq\frac{(Q(E_t)/t)^{1/q}}{\sigma}.
\]
If $Q(E_t)>0$, raising to the $q$th power and dividing by
$Q(E_t)$ yields $Q(E_t)^{q-1}\leq1/(\sigma^q t)$.
Taking $t=1/(\sigma^q\varepsilon^{q-1})$ therefore gives
$Q(E_t)\leq\varepsilon$.
This also holds if $Q(E_t)=0$.
Since $G=E_t^c$, we obtain $Q(G)\geq1-\varepsilon$.
Finally,
\[
    \frac{\mathrm d(Q\mid G)}{\mathrm dP}
    =\frac{f\Ind\{\x\in G\}}{Q(G)}
    \leq\frac{1}
    {(1-\varepsilon)\sigma^q\varepsilon^{q-1}},
\]
which proves the claimed smoothness of $Q\mid G$.
\end{proof}

\subsection{Confidence Amplification of Discrepancy Finder} In this subsection, we use standard amplification techniques
to increase the success probability of the discrepancy finder
in \Cref{thm:standalone-capped-finder}.
\begin{corollary}[Confidence amplification of the moment discrepancy finder]
\label{cor:relaxed-finder-amplification}
Under the hypotheses of
\Cref{thm:standalone-capped-finder}, for every
$\beta\in(0,1/2)$, \Cref{alg:amplified-relaxed-finder} draws
$
    d^k O(k^{O(k)}){V}/{\Delta^2}\log(2/\beta)
$
samples and runs in time
\[
    d^{O(k)}
    \exp\left(O(k^{O(k)})\frac{H^2}{\Delta^2}
                           \log\frac{H}{\Delta}\right)
    V^{O(1)}\log^{O(1)}\frac2\beta.
\]
It returns $\textsc{Reject}$, $\textsc{None}$, or a unit direction.
With probability at least $1-\beta$, all of the following hold:
\begin{enumerate}
 \item For every returned direction, there is a degree in $[k]$
    for which the corresponding directional moment differs from its
    Gaussian reference value by at least $\Delta/4$.
\item If the routine returns $\textsc{None}$, the distribution  matches directional
    moments with the standard Gaussian to slack $\Delta$.
    \item If every population Hermite moment tensor has Frobenius norm
    at most $H$, the routine does not return $\textsc{Reject}$.
\end{enumerate}
\end{corollary}

\begin{algorithm}[htbp]
    \caption{{\textsc{AmplifiedGaussianDiscrepancyFinder}}}
    \label{alg:amplified-relaxed-finder}
    {
    \begin{algorithmic}[1]
        \State \textbf{Input:} Sample access to $Q$ over $\R^d$;
        parameters $k,\Delta,V,H$ as in
        \Cref{thm:standalone-capped-finder}, and
        $\beta\in(0,1/2)$.
        \State \textbf{Output:} $\textsc{Reject}$,
        $\textsc{None}$, or a unit direction.

\medskip
        
        \State Set $J=\lceil C\log(2/\beta)\rceil$, where $C$ is
        a sufficiently large universal constant.
        \State Run \Cref{alg:standalone-moment-discrepancy-finder}
        independently $J$ times, with parameters $k,\Delta,V,H$.
        \If{a strict majority of the outputs are $\textsc{None}$}
            \State \Return $\textsc{None}$.
        \EndIf
        \State Let $\mathcal D$ be the list of returned directions.
        \If{$\mathcal D$ is empty}
            \State \Return $\textsc{Reject}$.
        \EndIf
        \State Set
        $n_0=\lceil C V(2k)^k/\Delta^2\rceil$ and
        $B=2\lceil C\log(2kJ/\beta)\rceil+1$.
        \State Draw one fresh batch of $Bn_0$ samples from $Q$,
        and divide it into $B$ blocks of size $n_0$.
        \For{each $\bv\in\mathcal D$ and $\ell\in[k]$}
            \State Let $\widehat m_{\bv,\ell}$ be the median of the
            $B$ block averages of $(\bv\cdot \x)^\ell$.
        \EndFor
        \If{some $\bv\in\mathcal D$ and $\ell\in[k]$ satisfy
        $\left|\widehat m_{\bv,\ell}
            -\E_{Z\sim\cN(0,1)}[Z^\ell]\right|
            \geq5\Delta/16$}
            \State \Return any such $\bv$.
        \Else
            \State \Return $\textsc{Reject}$.
        \EndIf
    \end{algorithmic}
    }
\end{algorithm}

\begin{proof}
{For each of the $J$ independent complete finder runs, call
the run successful if all three conclusions of
\Cref{thm:standalone-capped-finder} hold. Its success probability
is at least $2/3$, and these events are independent.
Hoeffding's inequality gives
\[
    \Pr\left[
       \text{at most $J/2$ complete runs are successful}
    \right]
    \leq e^{-J/18}\leq\beta/2
\]
for a sufficiently large universal $C$.
We will work on the complementary event, on which strictly more
than half of the complete runs are successful.

We next justify the fresh scalar validation stage. Condition on
all samples and outputs of the complete finder runs, so the list
$\mathcal D$ is fixed. It contains at most $J$ unit directions.
For every $\bv\in\mathcal D$ and $\ell\in[k]$, the polynomial
$(\bv\cdot \x)^\ell$ has degree at most $k$. The $V$-assumption gives
\[
    \E_Q[(\bv\cdot \x)^{2\ell}]
    \leq V\E_{Z\sim\cN(0,1)}[Z^{2\ell}]
    \leq V(2k)^k.
\]
In particular, its variance is at most $V(2k)^k$. If
$\overline m_{\bv,\ell}$ denotes its mean in one fresh block of
$n_0$ samples, Chebyshev's inequality gives
\[
    \Pr\left[
       |\overline m_{\bv,\ell}
           -\E_Q[(\bv\cdot \x)^\ell]|>\Delta/32
    \right]
    \leq\frac{32^2 V(2k)^k}{n_0\Delta^2}
    \leq\frac1{10},
\]
after increasing $C$.

For each fixed pair $(\bv,\ell)$ the blocks are independent.
Hoeffding's inequality therefore bounds the probability that at
least half of its block means have error greater than
$\Delta/32$ by $e^{-cB}$, for a universal $c>0$.
If fewer than half have such an error, their median has error at
most $\Delta/32$. There are at most $Jk$ pairs, so the choice of
$B$ and a union bound show that, with conditional probability at
least $1-\beta/2$,
\[
    \left|\widehat m_{\bv,\ell}
             -\E_Q[(\bv\cdot \x)^\ell]\right|
    \leq\Delta/32
    \qquad
    \text{for all $\bv\in\mathcal D$ and $\ell\in[k]$}.
\]
The same batch is used for every pair; independence between
different pairs is not needed. Integrating over the finder outputs
gives this validation guarantee unconditionally as well. If the
algorithm has already returned or $\mathcal D$ is empty, we take
the validation event to hold automatically.

By a union bound, the majority-success event and the validation
event hold together with probability at least $1-\beta$.
We prove all three conclusions on this joint event.

If a strict majority of the complete runs returns $\textsc{None}$,
at least one of those runs is successful. Its certificate proves
that every raw directional moment through degree $k$ differs from
its Gaussian reference value by less than $\Delta$. Thus the
amplified $\textsc{None}$ certificate is valid.

If a direction $\bv$ passes validation in degree $\ell$, its true
raw discrepancy is at least
\[
    \left|\E_Q[(\bv\cdot \x)^\ell]
              -\E_{Z\sim\cN(0,1)}[Z^\ell]\right|
    \geq\frac{5\Delta}{16}-\frac{\Delta}{32}
    =\frac{9\Delta}{32}\geq\frac{\Delta}{4}.
\]
This proves the returned-direction guarantee, even if the
direction came from an unsuccessful complete run.

Finally, suppose all the true Hermite moment tensors have
Frobenius norm at most $H$. No successful complete run can reject.
If a strict majority returns $\textsc{None}$, the amplified
algorithm returns $\textsc{None}$ as required. Otherwise, since
strictly more than half of the complete runs are successful,
at least one successful run must return a direction $\bv$.
For some degree $\ell\in[k]$, this direction has true discrepancy
at least $3\Delta/8$. Its estimated discrepancy is therefore at
least $3\Delta/8-\Delta/32=11\Delta/32>5\Delta/16$.
It passes validation, so the amplified routine cannot reject.
This proves the three asserted conclusions.

It remains to bound resources. The $J$ full finder runs use
\[
    d^k O(k^{O(k)})\frac{V}{\Delta^2}\log\frac2\beta
\]
samples. The shared validation batch has size
\[
    Bn_0
    =O\left(
       \frac{V(2k)^k}{\Delta^2}
       \log\frac{2kJ}{\beta}\right).
\]
Since $J=O(\log(2/\beta))$ and $\beta<1/2$,
\[
    \log\frac{2kJ}{\beta}
    =O\left(\log(2k)+\log\frac2\beta\right).
\]
Factors depending only on $k$ are absorbed into
$O(k^{O(k)})$, so the validation batch is bounded by the same
displayed sample bound. Crucially, this is one common batch,
rather than a separate batch for every proposal.

The $J$ full runs multiply the base running time by
$O(\log(2/\beta))$. Computing all block means and medians requires
at most $\poly(d,k,J,B,n_0)$ additional arithmetic operations.
Using $k\geq1$, $d\geq1$, $V,H\geq1$, and
$\Delta<1/2$, these operations are absorbed into
\[
    d^{O(k)}
    \exp\left(O(k^{O(k)})\frac{H^2}{\Delta^2}
                           \log\frac{H}{\Delta}\right)
    V^{O(1)}\log^{O(1)}\frac2\beta.
\]
This proves the sample and time bounds.}
\end{proof}

\section{Omitted Details from \Cref{sec:agnostic}}
\label{app:agnostic-details}
\subsection{Applications to Well-Studied Distribution Families}
\label{sec agnostic mixture}

\Cref{thm:main-learning} gives an agnostic learning algorithm
for marginal distributions that are $(q,\sigma)$-smooth with
respect to a known Gaussian reference.
In this subsection, we show that this assumption holds for
natural distribution families.

First, we show that mixtures of Gaussians with identity covariance
are smooth with respect to the standard Gaussian, with a smoothness
parameter that depends on the distances of their means from the origin.
\begin{lemma}[Smoothness of shifted Gaussians and their mixtures]
\label{lem:gaussian-mixture-smoothness}
Let $G=\cN(\vec{0},\bI_d)$ and let
$Q=\sum_{i=1}^s\lambda_i\cN(\boldsymbol{\mu}_i,\bI_d)$,
where $\lambda_i\geq0$ and $\sum_{i=1}^s\lambda_i=1$.
Then $Q$ is $(2,\sigma)$-smooth with respect to $G$, where
\[
    \sigma:=
    \left(\sum_{i=1}^s
        \lambda_i e^{\|\boldsymbol{\mu}_i\|_2^2/2}
    \right)^{-1}.
\]
In particular, $\cN(\boldsymbol{\mu},\bI_d)$ is
$(2,e^{-\|\boldsymbol{\mu}\|_2^2/2})$-smooth with respect to $G$.
If $\|\boldsymbol{\mu}_i\|_2\leq\mu$ for every $i$, then the mixture
$Q$ is $(2,e^{-\mu^2/2})$-smooth with respect to $G$.
\end{lemma}

\begin{proof}[Proof of \Cref{lem:gaussian-mixture-smoothness}]
The density ratio of $\cN(\boldsymbol{\mu}_i,\bI_d)$ with respect
to $G$ is
\[
    r_i(\x)=
    \exp\left(\boldsymbol{\mu}_i\cdot\x
                  -\frac{\|\boldsymbol{\mu}_i\|_2^2}{2}\right).
\]
Completing the square in the Gaussian exponent gives 
$\E_{\x\sim G}[r_i(\x)^2]=e^{\|\boldsymbol{\mu}_i\|_2^2}$.
Consequently, for every measurable event $E\subseteq\R^d$,
Cauchy--Schwarz gives
\begin{align*}
    Q(E)
    &=\sum_{i=1}^s\lambda_i
        \E_{\x\sim G}[r_i(\x)\Ind\{\x\in E\}]\\
    &\leq\sum_{i=1}^s\lambda_i
        e^{\|\boldsymbol{\mu}_i\|_2^2/2}G(E)^{1/2}
    =\frac{G(E)^{1/2}}{\sigma}.
\end{align*}
This proves the smoothness claim. The single-Gaussian case follows
by taking $s=1$, and the uniform norm bound follows from
$\sum_i\lambda_i e^{\|\boldsymbol{\mu}_i\|_2^2/2}\leq e^{\mu^2/2}$.
\end{proof}

By \Cref{lem:gaussian-mixture-smoothness}, our agnostic learner applies to Gaussian mixtures with parameters controlled by the norms of their means. Moreover, for classes closed under translations, the choice of origin for the Gaussian reference is essentially without loss of generality: an unknown reference mean can be handled by first estimating the mean of the observed feature distribution.

\begin{corollary}[Gaussian mixtures and unknown Gaussian reference means]
\label{cor:mixtures-unknown-gaussian-mean}
Let $\mathcal F$ be an $(\ell,K)$-well-behaved MIM class, and fix $c,\eps\in(0,1/2)$ and
$\sigma\in(0,1]$. Write $\ell_c=\ell(c'c)$ for a sufficiently small
universal constant $c'>0$. The following hold.
\begin{enumerate}
    \item Suppose that
    $D_\x\leq Q/\sigma$, where
    $Q=\sum_{i=1}^s\lambda_i\cN(\boldsymbol{\mu}_i,\bI_d)$,
    $\lambda_i\geq0$, $\sum_i\lambda_i=1$, and
    $\|\boldsymbol{\mu}_i\|_2\leq M$ for every $i$.
    There is an algorithm using at most
    \[
        N=
        d^{O(\ell_c)}
        \exp\left(
            (K\ell_c)^{O(\ell_c)}
            \left(M^2+\log\frac{2}{c\sigma\eps}\right)^{O(\ell_c)}
            \log(2+\Gamma(\mathcal F))
        \right)
    \]
    samples from $D$. 

    \item Suppose additionally that $\mathcal F$ is closed under
    translations. Let $q\geq1$, and suppose that $D_\x$ is
    $(q,\sigma)$-smooth with respect to
    $\cN(\boldsymbol{\mu},\bI_d)$ for an unknown
    $\boldsymbol{\mu}\in\R^d$.
    There is an algorithm using at most
    \[
        N=
        d^{O(\ell_c)}
        \exp\left(
            (qK\ell_c)^{O(\ell_c)}
            \log^{O(\ell_c)}\frac{2}{c\sigma\eps}
            \log(2+\Gamma(\mathcal F))
        \right)
    \]
    samples from $D$, without knowing $\boldsymbol{\mu}$ or having
    sample access to $\cN(\boldsymbol{\mu},\bI_d)$.
\end{enumerate}
In both cases, the algorithm runs in time polynomial in $N$ and,
with probability at least $2/3$, outputs a hypothesis $H$ satisfying
$\err_D(H)\leq(1+c)\opt_D(\mathcal F)+\eps$.
\end{corollary}

\begin{proof}[Proof of \Cref{cor:mixtures-unknown-gaussian-mean}]
For the first assertion, \Cref{lem:gaussian-mixture-smoothness}
gives, for every measurable event $E$,
\[
    D_\x(E)
    \leq\frac{Q(E)}{\sigma}
    \leq\frac{e^{M^2/2}}{\sigma}
        \cN(\vec{0},\bI_d)(E)^{1/2}.
\]
Thus $D_\x$ is $(2,\sigma e^{-M^2/2})$-smooth with respect to the
standard Gaussian. Applying \Cref{thm:main-learning} with these
parameters proves the first assertion.

For the second assertion, smoothness and the Gaussian tail bound
give, for every unit vector $\bv$ and $t>0$,
\[
    \Pr_{\x\sim D_\x}\left[
        |\bv\cdot(\x-\boldsymbol{\mu})|>t
    \right]
    \leq
    \min\left\{1,\frac{2}{\sigma}e^{-t^2/(2q)}\right\}.
\]
Integrating this bound yields
$\|\E_{\x\sim D_\x}[\x]-\boldsymbol{\mu}\|_2
=O(\sqrt{q\log(2/\sigma)})$ and
$\E_{\x\sim D_\x}[\|\x-\boldsymbol{\mu}\|_2^2]
=O(dq\log(2/\sigma))$.
Consequently, using $O(dq\log(2/\sigma))$ independent examples,
the empirical mean $\vec{m}$ satisfies
$\|\vec{m}-\E_{\x\sim D_\x}[\x]\|_2\leq1$ with probability at least
$99/100$. On this event,
\[
    \|\boldsymbol{\mu}-\vec{m}\|_2^2
    \leq Cq\log(2/\sigma)
\]
for a universal constant $C>0$.

Condition on such an estimate, and let $D'$ be the distribution
of $(\x-\vec{m},y)$ for $(\x,y)\sim D$. Write
$\boldsymbol{\mu}'=\boldsymbol{\mu}-\vec{m}$.
The centered feature distribution is $(q,\sigma)$-smooth with
respect to $\cN(\boldsymbol{\mu}',\bI_d)$. Hence, by
\Cref{lem:gaussian-mixture-smoothness},
\begin{align*}
    D'_\x(E)
    &\leq
    \frac{\cN(\boldsymbol{\mu}',\bI_d)(E)^{1/q}}{\sigma}\\
    &\leq
    \frac{e^{\|\boldsymbol{\mu}'\|_2^2/(2q)}}{\sigma}
        \cN(\vec{0},\bI_d)(E)^{1/(2q)}.
\end{align*}
Therefore $D'_\x$ is $(2q,\tau)$-smooth for a known choice
$\tau=c_0\sigma^{C_0}$, where $c_0>0$ and $C_0>1$ are universal
constants.

Apply \Cref{thm:main-learning} to fresh centered examples, using
smoothness parameters $(2q,\tau)$. Since
$\log(2/\tau)=O(\log(2/\sigma))$, its sample complexity has the
stated form. Translation closure gives
$\opt_{D'}(\mathcal F)=\opt_D(\mathcal F)$, so translating the
output back by setting $H(\x)=h(\x-\vec{m})$ preserves the error
guarantee. Standard confidence amplification, with a smaller
internal accuracy parameter, accounts for the mean-estimation
failure probability and gives overall success probability at
least $2/3$. The mean-estimation and validation costs are absorbed
into the displayed bound.
\end{proof}

\section{Multiplicative Approximation SQ Lower Bound}\label{sec:multiplicative-sq}
We next prove near-optimality of our agnostic learning algorithm as a simple consequence of
\cite{DKPZ21}. First, we define the $L_1$ approximation degree that
quantifies the complexity of the lower bound. For a Boolean function
$f:\R^K\to\{\pm1\}$ and $\eps\in(0,1)$, define
\[
\deg_{1,\eps}(f)
:=
\min\left\{
 r:\ \exists\text{ a polynomial }p:\R^K\to\R,\ \deg(p)\le r,
 \ \E_{\bz\sim\cN(\vec{0},\bI_K)}[|p(\bz)-f(\bz)|]<\eps
\right\}.
\]

\begin{fact}[{\cite[Theorem~4]{DKPZ21}}]
\label{fact:DKPZ21-generic-sq-lower-bound}
Fix a constant $a\in(0,1/2)$. There is a sufficiently small constant
$b>0$ for which the following holds. Let $K\le d^a$ and
$\eps\ge d^{-b}$, fix $f:\R^K\to\{\pm1\}$, and set
$r=\deg_{1,2\eps}(f)$. Suppose that
$\mathcal F\subseteq\{\R^d\to\{\pm1\}\}$ contains every orthogonal
embedding of $f$, namely $\{f(\vec{P}\x):\vec{P}\in\R^{K\times d}, \vec{P} \vec{P}^\top=\bI_K \}\subseteq  \mathcal F$.
Then any SQ algorithm that agnostically learns $\mathcal F$ under
$\cN(\vec{0},\bI_d)$ to error $\opt+\eps$ either makes a query of tolerance at
most $d^{-\Omega(r)}$ or makes at least $2^{d^{\Omega(1)}}$ queries.
\end{fact}

The preceding fact immediately gives the desired tradeoff for a
multiplicative approximation guarantee.

\begin{theorem}[Multiplicative Approximation SQ Lower Bound]
\label{thm:multiplicative-sq-lower-bound}
Fix a constant $a\in(0,1/2)$. There is a sufficiently small constant
$b>0$ for which the following holds.
Fix $c\in(0,1/4)$ and $\eps\in(0,c]$, and assume $2c\ge d^{-b}$. Let
$K\le d^a$, fix
$f:\R^K\to\{\pm1\}$, and let 
$\mathcal F\subseteq\{\R^d\to\{\pm1\}\}$ such that $\{f(\vec{P}\x):\vec{P}\in\R^{K\times d}, \vec{P} \vec{P}^\top=\bI_K \}\subseteq  \mathcal F$. 
Set $r:=\deg_{1,4c}(f)$. Then any SQ algorithm that agnostically learns $\mathcal{F}$ under $\cN(\vec{0},\bI_d)$ to error $(1+c)\opt +\eps$
either makes a query of tolerance at most $d^{-\Omega(r)}$ or makes at
least $2^{d^{\Omega(1)}}$ queries.
\end{theorem}

\begin{proof}
Assume that there exists an algorithm that agnostically learns $\mathcal{F}$ to error $(1+c)\opt +\eps$. 
Since $\opt\le 1$ and $\eps\le c$,
\[
(1+c)\opt+\eps
\le \opt+c+\eps
\le \opt+2c.
\]
Thus such an algorithm also agnostically learns $\mathcal F$ to additive
error $\opt+2c$. Applying
\Cref{fact:DKPZ21-generic-sq-lower-bound} completes the proof of the claim.
\end{proof}

\section{Testable Learning of MIMs from Gaussian Partitioning}
\label{sec:testable}
In this section, we use our Gaussian partitioning primitive to
design a testable agnostic learner for well-behaved MIM classes
under smooth distributions with respect to the Gaussian.
Recall the definition of testable learning in \Cref{def:testable-learning}.
The proofs share several technical similarities with those of our agnostic learning results, hence we present it here in the appendix.

The intuition is as follows. Under our sandwiching assumption,
a partition satisfying the guarantees of
\Cref{thm:relaxed-gaussian-partitioning} suffices to obtain a weak learner. 
This holds without distributional assumptions (subject to a caveat
discussed below), since sandwiching polynomials certify the approximation
error on regions that match the required moments.

We can check the cell diameters directly from the partition
description and test the cell masses and conditional moments
empirically using fresh samples. Thus, testing the output
partition certifies the properties needed by the weak learner.
Boosting then converts this testable weak learner into a
testable learner with a relative approximation guarantee.

A caveat is that our original weak learner also uses smoothness
to transfer a Gaussian noise sensitivity bound to $Q$.
This allows us to freeze the constrained coordinates within
each cell and reduce to functions depending only on the
orthogonal directions. 

We achieve this by assuming that, on sufficiently thin cells,
the target admits accurate sandwiching polynomials depending
only on the orthogonal directions. This eliminates the need for the perturbation argument (\Cref{lem:pertubation}). This property seems natural, and it
holds for functions of halfspaces (\Cref{lem:functions-halfspaces-cell-sandwiching}).

One way to picture this is the following. For each cell $R$, define the bounding functions on $V_R^\perp$ by
$
    f_R^+(\x):=
    \sup_{\mathbf y\in\proj_{V_R}(R)}f(\x+\mathbf y),
 $ $   f_R^-(\x):=
    \inf_{\mathbf y\in\proj_{V_R}(R)}f(\x+\mathbf y).
$
An upper sandwiching polynomial for $f_R^+$ and a lower
sandwiching polynomial for $f_R^-$ automatically sandwich
every restriction $f(\x+\mathbf y)$ with
$\mathbf y\in\proj_{V_R}(R)$.
Thus, it suffices that these bounding functions admit accurate
Gaussian sandwiching polynomials and that their Gaussian
disagreement probability, averaged over cells with weights
$P(R)$, becomes small as the cell diameters decrease.

We first present our new well-behaved definition with this additional assumption:

\begin{definition}[Testable well-behaved MIMs]
\label{def:testable-wellbehaved}
Let $\ell:(0,1]\to\Z_+$ be non-increasing, let $K\in\Z_+$,
and fix $A>0$ and $\Lambda\geq1$.
We say that a class $\mathcal F$ of Boolean $K$-MIMs is
$(\ell,K,A,\Lambda)$-testable well-behaved if the following hold.
\begin{enumerate}
    \item \emph{Closure under affine restrictions.}
    For every $f\in\mathcal F$, every subspace $W\subseteq\R^d$,
    and every $\bb\in W^\perp$, the function
    $\x\mapsto f(\bb+\proj_W\x)$ belongs to $\mathcal F$.

    \item \emph{Sandwiching on thin cells.}
    Let $f\in\mathcal F$ and fix $\delta,\tau\in(0,1)$.
    For every finite polyhedral partition $\mathcal P$ of $\R^d$
    and every $\mathcal G\subseteq\mathcal P$ satisfying
    $\diam(R)\leq\delta$ for all $R\in\mathcal G$,
    there exist polynomials $p_R^-,p_R^+:V_R^\perp\to\R$,
    one pair for each $R\in\mathcal G$, of degree at most
    $\ell(\tau)$, both depending on a common subspace of dimension
    at most $K$, such that
    \[
        p_R^-(\x)\leq f(\x+\mathbf y)\leq p_R^+(\x)
        \qquad
        \text{for every $\x\in V_R^\perp$ and
        $\mathbf y\in\proj_{V_R}(R)$}.
    \]
    Furthermore, these polynomials satisfy
    \[
        \sum_{R\in\mathcal G}Q(R)\,
        \E_{\x\sim\cN(\vec{0},\bI_{V_R^\perp})}
        [p_R^+(\x)-p_R^-(\x)]
        \leq A\Lambda\delta+\tau
    \]
    for every probability distribution $Q$ satisfying
    $Q(R)\leq\Lambda
    \Pr_{\x\sim\cN(\vec{0},\bI_d)}[\x\in R]$
    for all $R\in\mathcal G$.
\end{enumerate}
\end{definition}

Using this definition, we modify our previous agnostic learning
algorithm to obtain the following result.
{For simplicity of notation, let $\mathcal D_{q,\sigma}$ denote
the family of distributions over $\R^d$ that are
$(q,\sigma)$-smooth with respect to $\cN(\vec{0},\bI_d)$,
where $q\geq1$ and $\sigma\in(0,1]$.}

\begin{theorem}[Testable agnostic learning of MIMs]
\label{thm:testable-main}
Let $\mathcal F$ be an $(\ell,K,A,\Lambda)$-testable well-behaved
class for every $\Lambda\geq1$.
Fix $q\geq1$, $\sigma\in(0,1]$, and
$c,\epsilon\in(0,1/2)$.
For a sufficiently large universal constant $C$, let
$\ell_c=\ell(c/C)$.

There is an algorithm that, given samples from any distribution
$D$ on $\R^d\times\{\pm1\}$, testably learns $\mathcal F$
{under $\mathcal D_{q,\sigma}$} to error
$(1+c)\opt_D(\mathcal F)+\epsilon$ with probability at least
$99\%$, using
\[
    N=
    d^{O(\ell_c)}
    \exp\left(
        (qK\ell_c)^{O(\ell_c)}
        \log^{O(\ell_c)}\left(\frac{2}{c\sigma\epsilon}\right)
        \log(2+A)
    \right)
\]
samples from $D$ and running in time $\poly(N)$.
\end{theorem}
Our results are summarized in \Cref{tab:testable-learning-results}.
\begin{table}[htbp]
\centering
\renewcommand{\arraystretch}{1.3}
\begin{tabular}{@{}p{0.62\linewidth}p{0.30\linewidth}@{}}
\hline
\textbf{Concept class}
& \textbf{Dimension dependence} \\
\hline
$(\ell,K,A,\Lambda)$-testable well-behaved MIMs
& $d^{O(\ell_c)}$ \\

Intersections of $k$ halfspaces
& $d^{\widetilde O(k^3/c^4)}$ \\

 Boolean functions of $k$ halfspaces
& $d^{\widetilde O(k^5/c^4)}$ \\
\hline
\end{tabular}
\caption{Testable learning under $\mathcal D_{q,\sigma}$
to error $(1+c)\opt_D(\mathcal F)+\epsilon$ with confidence
$99\%$. The table displays the dependence on $d$ in the
sample complexity and running time; the remaining factors
are given in \Cref{thm:testable-main}.
Here $\ell_c=\ell(c/C)$, and the general class is assumed
to be testable well-behaved.}
\label{tab:testable-learning-results}
\end{table}

The remainder of this section is organized as follows.
In \Cref{sec:testable-weak-learner}, we present our testable weak
learner and prove its guarantees.
In \Cref{sec:gaussian-testable-learner}, we prove our main testable
learning result and apply it to intersections of halfspaces.

\subsection{Testable Weak Learner}
\label{sec:testable-weak-learner}

We obtain a testable weak learner by checking the partition on
fresh samples before running polynomial regression.
The sandwiching condition in \Cref{def:testable-wellbehaved}
replaces the freezing argument in the proof of \Cref{lem:weaklearner}, so it suffices to check cell
masses, diameters, and conditional moments. The full algorithm is given in \Cref{alg:testable-weak-learner}.

\begin{theorem}[Testable weak learner]
\label{thm:testable-weak-learner}
Let $\mathcal F$ be an $(\ell,K,A,\Lambda)$-testable well-behaved
class for every $\Lambda\geq1$.
Fix $q\geq1$, $\sigma\in(0,1]$, and $c,\beta\in(0,1/2)$.
For a sufficiently large universal constant $C$, let
$\ell_c=\ell(c/C)$.

There is an algorithm that, given samples from any distribution
$D$ on $\R^d\times\{\pm1\}$, testably learns $\mathcal F$
{under $\mathcal D_{q,\sigma}$} with probability at least $1-\beta$,
to error $1/2-c/12$ whenever
$\opt_D(\mathcal F)\leq1/(2(1+c))$.
The algorithm uses
\[
    n=d^{O(\ell_c)}
    \exp\left(
        (qK\ell_c)^{O(\ell_c)}
        \log^{O(\ell_c)}(2/(c\sigma))
        \log(2+A)
    \right)\log^{O(1)}(2/\beta)
\]
samples from $D$ and runs in time $\poly(n)$. {Its completeness guarantee holds for every $D$ with
$D_\x\in\mathcal D_{q,\sigma}$, regardless of
$\opt_D(\mathcal F)$.}
\end{theorem}

\begin{algorithm}[htbp]
    \caption{\textsc{TestableWeakLearner}}
    \label{alg:testable-weak-learner}
    \begin{algorithmic}[1]
        \State \textbf{Input:}
        Sample access to $D$, parameters $q,\sigma,c$, and
        class parameters $\ell,K,A$.

        \State \textbf{Output:}
        $\textsc{Reject}$ or a hypothesis
        $h:\R^d\to\{\pm1\}$.

        \medskip

        \State Set $\eta,\tau,\ell_c,\Delta,\Lambda,\delta,n_0$
        as in the proof of \Cref{thm:testable-weak-learner}.

        \State Run \Cref{alg:relaxed-gaussian-partitioning}
        with parameters $(q,\sigma,\eta/10,\delta,\ell_c,\Delta/10)$
        and its prescribed sample and time budgets.
        Denote its output by $(\mathcal P,\mathcal G_0)$,
        and let $B,H$ be its cell bound and Hermite norm threshold.

       \State \textbf{if} \Cref{alg:relaxed-gaussian-partitioning} does not finish within its budgets or $|\mathcal P|>B$ \textbf{then} \Return $\textsc{Reject}$.

        \State Draw a fresh labeled sample $S$ of size $n_0$
        from $D$, and let $\widehat Q$ be its empirical
        $\x$-marginal.

        \State Set $\mathcal G=\emptyset$ and
        $P=\cN(\vec{0},\bI_d)$.

        \For{$R\in\mathcal G_0$}
            \If{$\diam(R)\leq\delta$,
                $\widehat Q(R)\geq1/(2B)$, and
                $\widehat Q(R)\leq\Lambda P(R)$} \label{line:testable-cell-check}

                \State {If $V_R^\perp\neq\{\vec{0}\}$, run
\Cref{alg:standalone-moment-discrepancy-finder}
using the $V_R^\perp$ projections of $S\cap R$ as its
supplied sample, with degree $\ell_c$, slack $\Delta$,
and Hermite cap $H$.}

                \State If $V_R^\perp=\{\vec{0}\}$,
                take the output to be $\textsc{None}$.

                \State \textbf{if} the output is $\textsc{None}$ \textbf{then} add $R$ to $\mathcal G$.\label{line:testable-moment-check}
            \EndIf
        \EndFor

       \State \textbf{if} $\sum_{R\in\mathcal G}\widehat Q(R)<1-\eta$ \textbf{then} \Return $\textsc{Reject}$.

        \For{$R\in\mathcal G$}
            \State Using the same sample $S\cap R$,
            find a polynomial $q_R$ on $V_R^\perp$
            of degree at most $\ell_c$ whose empirical
            $L_1$ loss is within $\tau$ of the minimum.

            \State Choose $t_R\in[-1,1]$ minimizing the
            empirical classification error of
            $\sign(q_R(\proj_{V_R^\perp}\x)-t_R)$
            on $S\cap R$.

            \State Set
            $h_R(\x)=\sign(q_R(\proj_{V_R^\perp}\x)-t_R)$.
        \EndFor

        \State Define $h(\x)=h_R(\x)$ when
        $\x\in R\in\mathcal G$, and set $h(\x)=1$
        on the remaining cells.

        \State \Return $h$.
    \end{algorithmic}
\end{algorithm}

\begin{proof}[Proof of \Cref{thm:testable-weak-learner}]
We first show that \Cref{alg:testable-weak-learner} satisfies the guarantees of
the theorem with constant success probability, and then
amplify this probability to $1-\beta$.
Let $Q=D_\x$ and $P=\cN(\vec{0},\bI_d)$.
Let $\eta=\tau=c/C$ and $\ell_c=\ell(\tau)$.
Set $\Delta=(2K\ell_c)^{-C\ell_c}$,
$\Lambda=2/(\sigma^q(\eta/100)^{q-1})$, and
$\delta=\tau/((1+A)\Lambda)$.
Run \Cref{alg:relaxed-gaussian-partitioning} with parameters
$(q,\sigma,\eta/10,\delta,\ell_c,\Delta/10)$, obtaining a partition
$\mathcal P$ and candidate certified cells $\mathcal G_0$.
Let $B$ be the bound supplied by
\Cref{thm:relaxed-gaussian-partitioning}.
We reject if the partitioning procedure exceeds its stated
sampling or running-time budget, or returns more than $B$ cells.

Set $n_0=d^{O(\ell_c)}\poly(B,\Lambda,1/\sigma,1/\tau)$,
with sufficiently large implicit constants.
Draw a fresh labeled sample $S$ of size $n_0$, independently
of the partition, and denote its empirical distribution by
$\widehat D$, with  $\x$-marginal $\widehat Q$.
Use the available Gaussian cell probabilities $P(R)$.
For each $R\in\mathcal G_0$, retain $R$ only if:
\begin{enumerate}
    \item $\diam(R)\leq\delta$,
    $\widehat Q(R)\geq1/(2B)$, and
    $\widehat Q(R)\leq\Lambda P(R)$
    (Line~\ref{line:testable-cell-check}).
    \item The conditional empirical distribution, projected
    onto $V_R^\perp$, passes the moment discrepancy test
    with slack $\Delta$
    (Line~\ref{line:testable-moment-check}).
\end{enumerate}
The diameter is checked from the recorded orthogonal cuts.
{For the moment test, when $V_R^\perp\neq\{\vec{0}\}$, run
\Cref{alg:standalone-moment-discrepancy-finder}
using the $V_R^\perp$ projections of $S\cap R$ as its
supplied sample, with degree $\ell_c$, slack $\Delta$,
and Hermite cap $H$.}
Retain $R$ only when the output is $\textsc{None}$.
If $V_R^\perp=\{\vec{0}\}$, the moment test passes automatically.

Let $\mathcal G$ be the retained collection.
Reject if $\widehat Q(\bigcup_{R\in\mathcal G}R)<1-\eta$.
Otherwise, run degree-$\ell_c$ empirical $L_1$ regression on each
$S\cap R$, followed by threshold selection as in
\Cref{alg weak learner}, with accuracy $\tau$.
Label the remaining cells arbitrarily and return the resulting
hypothesis $h$.

We first show Completeness, i.e., if $Q$ is smooth the
algorithm returns a hypothesis with constant probability.
\begin{claim}[Completeness]
\label{claim:testable-weak-completeness}
If {$Q\in\mathcal D_{q,\sigma}$}, then \Cref{alg:testable-weak-learner} accepts with probability at least
$1/2$.
\end{claim}

\begin{proof}[Proof of \Cref{claim:testable-weak-completeness}]
By \Cref{thm:relaxed-gaussian-partitioning}, with probability
at least $2/3$, the cells in $\mathcal G_0$ have total
$Q$-mass at least $1-\eta/10$, and every $R\in\mathcal G_0$
has $Q$-mass at least $1/B$, diameter at most $\delta$,
and conditional moments on $V_R^\perp$ matching the Gaussian
moments with slack $\Delta/10$.

Applying smoothness to unions of cells shows that the discrete
distribution $(Q(R))_{R\in\mathcal P}$ is $(q,\sigma)$-smooth
with respect to $(P(R))_{R\in\mathcal P}$.
Now applying \Cref{lem:relaxed-smoothness-truncation} to these
distributions shows that cells satisfying
$Q(R)>(\Lambda/2)P(R)$ have total $Q$-mass at most $\eta/100$.

Conditional on the partition, Hoeffding's inequality and a
union bound over at most $B$ cells ensure, with probability
at least $7/8$, that
$|\widehat Q(R)-Q(R)|\leq\eta/(10B)$ for every cell $R$.
The candidate cells satisfying
$Q(R)\leq(\Lambda/2)P(R)$ have total $Q$-mass at least
$1-11\eta/100$, and each satisfies $Q(R)\geq1/B$.
For every such cell,
$\widehat Q(R)\geq Q(R)-\eta/(10B)\geq1/(2B)$ and
$\widehat Q(R)\leq Q(R)+\eta/(10B)\leq2Q(R)\leq\Lambda P(R)$.
Thus these cells pass the mass checks on
Line~\ref{line:testable-cell-check}.
Since there are at most $B$ such cells, their total mass
estimation error is at most $\eta/10$, so their total
empirical mass is at least $1-21\eta/100$.

By
\Cref{lem:conditional-density-averaging,lem:relaxed-polynomial-control},
all but $\eta/100$ of the $Q$-mass lies in cells $R$ satisfying
$\|\E_{\x\sim Q\mid R}
[\mathcal H_i(\proj_{V_R^\perp}\x)]\|_F\leq H$
for every $i\in[\ell_c]$, and
$\E_{Q\mid R}[p^2]\leq V\|p\|_{L_2(P)}^2$
for every polynomial $p$ of degree at most $\ell_c$ depending
only on $V_R^\perp$, where
$V=Cq^2(8q-1)^{\ell_c}/(\sigma\eta)$ for a sufficiently
large universal constant $C$.

{For each candidate cell satisfying the bounds above,
conditional on its sample count, the projected observations
in $S\cap R$ are independent samples from the projected
distribution $Q\mid R$. By the tensor-estimation bound in the
proof of \Cref{claim:standalone-capped-estimation}, taking $n_0$
sufficiently large makes the probability that a cell contains
at least $n_0/(2B)$ observations but fails the finder's
estimation event at most $1/(8B)$. A union bound gives
simultaneous success for all cells with this many observations
with probability at least $7/8$.
On the cell-mass event above, all relevant cells have this many
observations. The proof of \Cref{thm:standalone-capped-finder}
then shows that the finder does not reject. It cannot return
a direction either, since any returned direction would have
population moment discrepancy at least $3\Delta/8$,
contradicting the conditional moment-matching bound
$\Delta/10$. Hence every such cell returns $\textsc{None}$.}
Together with the cell-mass event above, both events hold
with conditional probability at least $3/4$.
On this event, cells of total $Q$-mass at least
$1-12\eta/100$ pass both tests, and their empirical mass is
at least $1-22\eta/100\geq1-\eta$.
Thus the algorithm accepts.
This completes the proof of \Cref{claim:testable-weak-completeness}.
\end{proof}

\begin{claim}[Soundness]
\label{claim:testable-weak-soundness}
For every distribution $D$, with probability at least $11/12$
over the fresh sample $S$, whenever the algorithm accepts,
its output satisfies
\[
    \err_D(h)\leq\opt_D(\mathcal F)+7\tau.
\]
\end{claim}

\begin{proof}[Proof of \Cref{claim:testable-weak-soundness}]
Fix $f\in\mathcal F$ with
$\err_D(f)\leq\opt_D(\mathcal F)+\tau$.
By the cell-mass test, acceptance gives
$\widehat Q(R)\leq\Lambda P(R)$ for every $R\in\mathcal G$.
Apply \Cref{def:testable-wellbehaved} to this partition and
$f$, and write $s_R:=p_R^+-p_R^-\geq0$.
The resulting polynomials satisfy
\[
    \sum_{R\in\mathcal G}\widehat Q(R)\,
    \E_{G_R}[s_R]
    \leq A\Lambda\delta+\tau
    \leq2\tau\;,
\]
where $G_R=\cN(\vec{0},\bI_{V_R^\perp})$.
Let $\widehat Q_R$ denote the projection of
$\widehat Q\mid R$ onto $V_R^\perp$.
{Since the finder uses the exact empirical moments,
its $\textsc{None}$ certificate ensures that $\widehat Q_R$
and $G_R$ match directional moments through degree $\ell_c$
with slack $\Delta$.}
Since $s_R$ depends on at most $K$ directions, empirical moment
matching and \Cref{fact:moment-matching-polynomial-expectations}
give
\begin{align*}
    \sum_{R\in\mathcal G}\widehat Q(R)\,
        \E_{\widehat Q_R}[s_R]
    \leq
      \bigl(1+(K\ell_c)^{O(\ell_c)}\Delta\bigr)
      \sum_{R\in\mathcal G}\widehat Q(R)\,
        \E_{G_R}[s_R]\leq4\tau.
\end{align*}
{Conditional on $S$, apply \Cref{fact:polynomial-regression}
to the projected labeled empirical distribution in each
$R\in\mathcal G$. A union bound gives probability at least
$23/24$ that all regression calls satisfy
$\err_{\widehat D\mid R}(h_R)
\leq \E_{\widehat D\mid R}
[|y-p_R^+(\x_{V_R^\perp})|]/2+\tau$.
The pointwise sandwich and the preceding gap bound therefore
give $\err_{\widehat D}(h)
\leq\err_{\widehat D}(f)+\eta+3\tau$,
where $\eta$ accounts for the empirical mass outside
$\mathcal G$. }

Conditional on the independently constructed partition,
piecewise degree-$\ell_c$ polynomial threshold hypotheses have
VC dimension at most
$(d+\ell_c)^{O(\ell_c)}B$.
Taking $S$ sufficiently large ensures, with probability at
least $11/12$, that
$\err_D(h)\leq\err_{\widehat D}(h)+\tau$ uniformly over these
hypotheses, and
$\err_{\widehat D}(f)\leq\err_D(f)+\tau$ for the fixed $f$.
On this event, whenever the algorithm accepts,
\begin{align*}
    \err_D(h)
    \leq \err_D(f)+\eta+5\tau
    \leq \opt_D(\mathcal F)+\eta+6\tau
    =\opt_D(\mathcal F)+7\tau.
\end{align*}
This completes the proof of \Cref{claim:testable-weak-soundness}.
\end{proof}

Under the assumption
$\opt_D(\mathcal F)\leq1/(2(1+c))$,
\Cref{claim:testable-weak-soundness} gives
\[
    \err_D(h)
    \leq \frac{1}{2(1+c)}+\frac{7c}{C}
    \leq \frac12-\frac c{12}
\]
for a sufficiently large universal constant $C$.

The partition construction, empirical moment tests, regression,
and generalization require
$d^{O(\ell_c)}\poly(B,\Lambda,1/\sigma,1/\tau)$ samples and time
polynomial in this quantity.

To obtain failure probability at most $\beta$, use
$O(\log(1/\beta))$ independent runs and return the first accepted
hypothesis, rejecting if every run rejects.
Choose the fresh sample sizes so that the sum of the soundness
failure probabilities across these runs is at most $\beta/2$.
By \Cref{claim:testable-weak-completeness}, {if $Q\in\mathcal D_{q,\sigma}$},
the probability that all runs reject is at most $\beta/2$.
Thus both guarantees hold with probability at least $1-\beta$,
with overhead polynomial in $\log(1/\beta)$.
\end{proof}
\subsection{Testable Learner for MIMs under Smooth Distributions}
\label{sec:gaussian-testable-learner}
In this section, we prove our main testable learning result,
\Cref{thm:testable-main}, and apply it to intersections and
arbitrary Boolean functions of $k$ halfspaces.

The proof applies the agnostic boosting fact to the testable
weak learner from \Cref{thm:testable-weak-learner}. The only
modification to the boosting algorithm is that it rejects
whenever a weak-learning call rejects.

\begin{algorithm}[htbp]
    \caption{\textsc{TestableBoosting}}
    \label{alg:testable-boosting}
    \begin{algorithmic}[1]
        \State \textbf{Input:}
        Sample access to $D$, parameters $q,\sigma,c,\epsilon$,
        and class parameters $\ell,K,A$.

        \State \textbf{Output:}
        $\textsc{Reject}$ or a hypothesis
        $h:\R^d\to\{\pm1\}$.

        \medskip

        \State Set $\alpha=c/(2(1+c))$ and $\gamma=c/12$.

        \State Let $T=\poly(1/c,1/\epsilon)$ and
        $\Gamma=O(1/\epsilon)$ be the bounds from the
        agnostic boosting fact on the number of weak-learning
        calls and on $dD_t/dD$, respectively.

        \State Set
        $\sigma_*=\sigma/\Gamma$ and
        $\beta=1/(1000T)$.

        \State Run the boosting algorithm from
        \Cref{fact:boosting} with parameters
        $(\alpha,\gamma,\epsilon)$ and confidence $1-1/1000$,
        answering its weak-learning queries as follows.

        \For{each weak-learning query on a distribution $D_t$}
            \State Apply \Cref{thm:testable-weak-learner}
            to fresh samples from $D_t$ with parameters
            $(q,\sigma_*,c)$ and failure probability $\beta$.

            \If{the weak learner returns $\textsc{Reject}$}
                \State \Return $\textsc{Reject}$.
            \EndIf

            \State Supply the returned hypothesis $h_t$
            to the boosting algorithm.
        \EndFor

        \State \Return the hypothesis produced by the
        boosting algorithm.
    \end{algorithmic}
\end{algorithm}

\begin{proof}[Proof of \Cref{thm:testable-main}]
Let $Q=D_\x$ and $P=\cN(\vec{0},\bI_d)$.
We use \Cref{fact:boosting} with
$\alpha=c/(2(1+c))$, $\gamma=c/12$, and confidence $1-1/1000$.
These choices give $1/2-\alpha=1/(2(1+c))$ and
$1/(1-2\alpha)=1+c$.
Thus \Cref{thm:testable-weak-learner} supplies exactly the
weak-learning guarantee required by the boosting fact
whenever a call accepts.

Apply \Cref{thm:testable-weak-learner} with failure probability
$\beta=1/(1000T)$ for each weak-learning query.
By the union bound, all weak-learning calls are successful
with probability at least $99\%$.

For soundness, assume that all weak-learning calls and the
boosting procedure satisfy their guarantees.
If any weak-learning call rejects, the algorithm rejects.
Thus, if the algorithm outputs a hypothesis, every call has
returned a hypothesis satisfying the required weak-learning
guarantee. By \Cref{fact:boosting},
\[
    \err_D(h)
    \leq
    \frac{\opt_D(\mathcal F)}{1-2\alpha}+\epsilon
    =
    (1+c)\opt_D(\mathcal F)+\epsilon.
\]
For completeness, suppose that {$Q\in\mathcal D_{q,\sigma}$}.
The boosting fact guarantees that every query distribution
satisfies $dD_t/dD\leq\Gamma$ for $\Gamma=O(1/\eps)$.
Consequently, its  $\x$-marginal $Q_t$ satisfies
$
    Q_t(E)\leq\Gamma Q(E),
    \text{for every measurable $E\subseteq\R^d$}.
$
Therefore, {$Q_t\in\mathcal D_{q,\sigma/\Gamma}$}.
Hence all weak-learning calls {accept} with probability $99\%$,
so the boosting algorithm finishes and returns a hypothesis.

Finally, the weak learner is run with
$\sigma_*=\sigma/\Gamma$ and degree $\ell_c=\ell(c/C)$.
Since $\Gamma=O(1/\epsilon)$, substituting $\sigma_*$
into the bound of \Cref{thm:testable-weak-learner} gives
\[
    d^{O(\ell_c)}
    \exp\left(
        (qK\ell_c)^{O(\ell_c)}
        \log^{O(\ell_c)}\left(\frac{2}{c\sigma\epsilon}\right)
        \log(2+A)
    \right)
\]
samples per call.
The boosting fact uses $\poly(1/c,1/\epsilon)$ calls and
polynomial sampling and computational overhead.
This proves the stated sample complexity and running time, which completes the proof of \Cref{thm:testable-main}.
\end{proof}

We next show that functions of halfspaces satisfy Condition~2
of \Cref{def:testable-wellbehaved}, yielding a testable learning
algorithm for this class.

For each thin cell $R$, we construct two functions of halfspaces
$h_R^\pm$ on $V_R^\perp$ such that
$h_R^-(\x_{V_R^\perp})\leq h(\x)\leq h_R^+(\x_{V_R^\perp})$
for every $\x\in R$, where $h$ is the target {function}.
We obtain these bounding functions by adjusting each halfspace's threshold
according to its variation along $V_R$ within $R$.
The two thresholds differ by at most the cell diameter.
Gaussian anti-concentration then bounds the Gaussian disagreement
between $h_R^-$ and $h_R^+$, averaged over the partition.
Finally, applying Gaussian sandwiching to these two functions of halfspaces
gives the required polynomials.

\begin{lemma}[Sandwiching functions of halfspaces on thin cells]
\label{lem:functions-halfspaces-cell-sandwiching}
For every $k\in\N_+$ and $\Lambda\geq1$, the class of arbitrary
Boolean functions of at most $k$ affine halfspaces satisfies
Condition~2 of \Cref{def:testable-wellbehaved} with parameters
$(\ell,k,2k,\Lambda)$, where
$\ell(\tau)=\widetilde O(k^5/\tau^4)$.
{Moreover, intersections of at most $k$ halfspaces satisfy
the same condition with $\ell(\tau)=\widetilde O(k^3/\tau^4)$.}
\end{lemma}

\begin{proof}[Proof of \Cref{lem:functions-halfspaces-cell-sandwiching}]
Fix $\delta,\tau\in(0,1)$ and let $P=\cN(\vec{0},\bI_d)$.
Let $\mathcal P$ be a finite polyhedral partition and let
$\mathcal G\subseteq\mathcal P$ consist of cells of diameter
at most $\delta$ in their constraint subspaces.
For each cell, write $A_R:=\proj_{V_R}(R)$ and
$G_R:=\cN(\vec{0},\bI_{V_R^\perp})$, so that
$R=A_R+V_R^\perp$ and $\diam(A_R)\leq\delta$ for
$R\in\mathcal G$.
We may omit empty cells, assigning zero polynomials to them.

Let $f$ be a Boolean function of at most $k$ halfspaces.
If $f$ is constant, take $p_R^-=p_R^+=f$.
Otherwise, absorbing constant halfspaces into the Boolean link
and repeating the remaining halfspaces if necessary, write
$f(\bu)=F(h_1(\bu),\ldots,h_k(\bu))$, where
$F:\{\pm1\}^k\to\{\pm1\}$ and
$h_j(\bu)=2\Ind\{\bw_j\cdot\bu\leq t_j\}-1$
with $\|\bw_j\|_2=1$.
For each nonempty $R\in\mathcal G$, define
$\alpha_{R,j}:=\inf_{\mathbf y\in A_R}\bw_j\cdot\mathbf y$
and
$\beta_{R,j}:=\sup_{\mathbf y\in A_R}\bw_j\cdot\mathbf y$.
Then $0\leq\beta_{R,j}-\alpha_{R,j}\leq\delta$.
Define shifted halfspaces on $V_R^\perp$ by
\begin{align*}
    h_{R,j}^-(\x)
    &:=2\Ind\{\bw_j\cdot\x\leq t_j-\beta_{R,j}\}-1,\\
    h_{R,j}^+(\x)
    &:=2\Ind\{\bw_j\cdot\x\leq t_j-\alpha_{R,j}\}-1.
\end{align*}
For every $\x\in V_R^\perp$ and $\mathbf y\in A_R$,
$h_{R,j}^-(\x)\leq h_j(\x+\mathbf y)\leq h_{R,j}^+(\x)$.

Let $E_R$ be the set of $\x\in V_R^\perp$ for which some
pair $h_{R,j}^-(\x),h_{R,j}^+(\x)$ disagrees, and define
\[
    h_R^\pm(\x):=
    \begin{cases}
        F(h_{R,1}^-(\x),\ldots,h_{R,k}^-(\x)),
            & \x\notin E_R,\\
        \pm1, & \x\in E_R.
    \end{cases}
\]
Outside $E_R$, every underlying halfspace has the same value
for all $\mathbf y\in A_R$.
On $E_R$, the two bounds are $-1$ and $1$.
Hence $h_R^-(\x)\leq f(\x+\mathbf y)\leq h_R^+(\x)$
for every $\x\in V_R^\perp$ and $\mathbf y\in A_R$.

Both functions are Boolean functions of the same $2k$ shifted
halfspaces and depend only on
$W_R:=\operatorname{span}\{\proj_{V_R^\perp}\bw_j:j\in[k]\}$,
which has dimension at most $k$.
{By
\Cref{fact:gaussian-sandwiching-smooth-boundary,fact:gaussian-boundary-smoothness},
applied on $W_R$ with error $\tau/2$, we obtain polynomials
of degree at most $\ell(\tau)=\widetilde O(k^5/\tau^4)$.
Extending them by projection gives $p_R^\pm$, both depending
on $W_R$, with $p_R^-\leq h_R^-$, $h_R^+\leq p_R^+$, and}
\[
    \E_{G_R}[h_R^--p_R^-]\leq\tau/2,
    \qquad
    \E_{G_R}[p_R^+-h_R^+]\leq\tau/2.
\]
If $W_R=\{\vec{0}\}$, take the corresponding constant
polynomials instead.
Thus the required pointwise sandwich holds.
Writing $b_R:=G_R(E_R)$, we have
$\E_{G_R}[p_R^+-p_R^-]\leq2b_R+\tau$.

We now bound the weighted disagreement.
If $\bu=\x+\mathbf y\in R$ with $\x\in E_R$,
then for some $j\in[k]$,
$t_j-\beta_{R,j}<\bw_j\cdot\x\leq t_j-\alpha_{R,j}$.
Together with
$\alpha_{R,j}\leq\bw_j\cdot\mathbf y\leq\beta_{R,j}$,
this implies $|\bw_j\cdot\bu-t_j|\leq\delta$.
Independence of the orthogonal Gaussian projections gives
\[
    P(R)b_R
    =\Pr_{\bu\sim P}
      [\bu\in R,\ \proj_{V_R^\perp}\bu\in E_R].
\]
By disjointness of the cells and Gaussian anti-concentration,
\[
    \sum_{R\in\mathcal G}P(R)b_R
    \leq\sum_{j=1}^k
       \Pr_{\bu\sim P}[|\bw_j\cdot\bu-t_j|\leq\delta]
    \leq k\delta.
\]
The polynomials have been constructed independently of $Q$.
For any distribution $Q$ satisfying $Q(R)\leq\Lambda P(R)$
on $\mathcal G$, we therefore obtain
\begin{align*}
    \sum_{R\in\mathcal G}Q(R)\,
        \E_{G_R}[p_R^+-p_R^-]
    &\leq2\sum_{R\in\mathcal G}Q(R)b_R
          +\tau\sum_{R\in\mathcal G}Q(R)\\
    &\leq2\Lambda\sum_{R\in\mathcal G}P(R)b_R+\tau\\
    &\leq2\Lambda k\delta+\tau.
\end{align*}
This proves Condition~2 with $A=2k$.

{If $f$ is an intersection, choose the $h_j$ as its defining
halfspaces and use instead
\begin{align*}
    h_R^-(\x)
    &:=2\Ind\{\bw_j\cdot\x\leq t_j-\beta_{R,j}
               \text{ for every }j\in[k]\}-1,\\
    h_R^+(\x)
    &:=2\Ind\{\bw_j\cdot\x\leq t_j-\alpha_{R,j}
               \text{ for every }j\in[k]\}-1.
\end{align*}
These functions satisfy the same pointwise sandwich, and
their disagreement is contained in $E_R$, so the preceding
weighted gap bound applies.
Since both functions are intersections of at most $k$
halfspaces, the same two facts give Gaussian sandwiching
polynomials of degree $\widetilde O(k^3/\tau^4)$ with error
$\tau/2$ on their common subspace $W_R$.
The same argument therefore proves Condition~2 with this
smaller degree.}
\end{proof}

{
For intersections, the  sandwiching degree along with our main theorem gives the
following testable learning guarantee.
\begin{corollary}[Testable learning of intersections of halfspaces]
\label{cor:testable-intersections}
Let $\mathcal F_k$ be the class of intersections of at most
$k$ halfspaces in $\R^d$, where $k\in\N_+$.
Fix $q\geq1$, $\sigma\in(0,1]$, and
$c,\eps\in(0,1/2)$.
There is an algorithm that, given samples from any distribution
$D$ on $\R^d\times\{\pm1\}$, testably learns $\mathcal F_k$
under $\mathcal D_{q,\sigma}$ to error
$(1+c)\opt_D(\mathcal F_k)+\eps$ with probability at least
$99\%$, using
\[
    N=d^{\widetilde O(k^3/c^4)}
    \exp\left(
        \left(\frac{qk}{c}\right)^{\widetilde O(k^3/c^4)}
        \log^{\widetilde O(k^3/c^4)}
        \left(\frac{2}{c\sigma\eps}\right)
    \right)
\]
samples from $D$ and running in time $\poly(N)$.
\end{corollary}

\begin{proof}[Proof of \Cref{cor:testable-intersections}]
The class $\mathcal F_k$ consists of $k$-MIMs and is closed
under affine restrictions.
By \Cref{lem:functions-halfspaces-cell-sandwiching}, it is
$(\ell,k,2k,\Lambda)$-testable well-behaved for every
$\Lambda\geq1$, with $\ell(\tau)=\widetilde O(k^3/\tau^4)$.
Applying \Cref{thm:testable-main} with $K=k$, $A=2k$, and
$\ell_c=\ell(c/C)=\widetilde O(k^3/c^4)$ gives the result.
\end{proof}
}
Using the sandwiching degree for functions of halfspaces we get a testable learning result for that class. 
\begin{corollary}[Testable learning of functions of halfspaces]
\label{cor:testable-functions-halfspaces}
Let $\mathcal H_k$ be the class of arbitrary Boolean functions
of at most $k$ affine halfspaces in $\R^d$, where $k\in\N_+$.
Fix $q\geq1$, $\sigma\in(0,1]$, and
$c,\epsilon\in(0,1/2)$.
There is an algorithm that, given samples from any distribution
$D$ on $\R^d\times\{\pm1\}$, testably learns $\mathcal H_k$
under $\mathcal D_{q,\sigma}$ to error
$(1+c)\opt_D(\mathcal H_k)+\epsilon$ with probability at least
$99\%$, using
\[
    N=d^{\widetilde O(k^5/c^4)}
    \exp\left(
        \left(\frac{qk}{c}\right)^{\widetilde O(k^5/c^4)}
        \log^{\widetilde O(k^5/c^4)}
        \left(\frac{2}{c\sigma\epsilon}\right)
    \right)
\]
samples from $D$ and running in time $\poly(N)$.
\end{corollary}

\begin{proof}[Proof of \Cref{cor:testable-functions-halfspaces}]
Every member of $\mathcal H_k$ depends only on the span of its
halfspace normals and is therefore a $k$-MIM.
Affine restrictions preserve the class: each halfspace remains
a halfspace or becomes constant, and constant inputs can be
absorbed into the Boolean link.
By \Cref{lem:functions-halfspaces-cell-sandwiching},
$\mathcal H_k$ is consequently
$(\ell,k,2k,\Lambda)$-testable well-behaved for every
$\Lambda\geq1$, where
$\ell(\tau)=\widetilde O(k^5/\tau^4)$.
Applying \Cref{thm:testable-main} with $K=k$, $A=2k$, and
$\ell_c=\ell(c/C)=\widetilde O(k^5/c^4)$ gives the result.
\end{proof}

\end{document}